\documentclass[12pt]{article}
\usepackage{bbm,latexsym}
\usepackage{amssymb,amsmath}
\usepackage{epsfig}
\usepackage{color}
\newcommand{\figurescale}{0.42}
\newcommand{\figurescaletwo}{0.46}
\begin{document}
\renewcommand{\thefootnote}{\fnsymbol{footnote}}

\title{
\normalsize \hfill UWThPh-2012-30 \\[10mm]
\LARGE Correlations of the elements of\\
 the neutrino mass matrix}

\author{
W.~Grimus\thanks{E-mail: walter.grimus@univie.ac.at} \
and P.O.~Ludl\thanks{E-mail: patrick.ludl@univie.ac.at}
\\[5mm]
\small University of Vienna, Faculty of Physics \\
\small Boltzmanngasse 5, A--1090 Vienna, Austria
}

\date{20 December 2012}

\maketitle

\begin{abstract}
Assuming Majorana nature of neutrinos, 
we re-investigate, in the light of the recent measurement of the
reactor mixing angle, the allowed ranges for the absolute values of the
elements of the neutrino mass matrix in the basis where the
charged-lepton mass matrix is diagonal. Apart from the derivation of
upper and lower bounds on the values of the matrix elements, we also
study their correlations. Moreover, we analyse the sensitivity of
bounds and correlations to the global fit results of the neutrino
oscillation parameters which are available in the
literature.
\end{abstract}

\newpage

\renewcommand{\thefootnote}{\arabic{footnote}}

\section{Introduction}

Our knowledge of the neutrino oscillation parameters has enormously
improved in the recent years. 
The experimental results of the Double Chooz, Daya Bay and RENO
Collaborations~\cite{daya-reno} 
have
impressively confirmed the earlier hints~\cite{T2K-MINOS} for a
non-zero reactor mixing angle. Taking 
these novel results into account, the recent global
fits~\cite{forero,fogli} establish $\theta_{13}>0$ 
at a confidence level of $\sim 10\sigma$.
Also the values of the solar mixing angle and the 
mass-squared
differences are 
now
known with good accuracy.
While for a long time $\mathrm{sin}^2\theta_{23}\approx 1/2$
has been a very good approximation for the best fit value, the recent
global fits~\cite{forero, fogli} hint towards 
a deviation from maximal atmospheric neutrino mixing. However, even at
$1\sigma$ it is not clear from the recent fits 
in which octant $\theta_{23}$ lies. The global 
fit of~\cite{forero} 
allows $\theta_{23}=45^\circ$ at $2\sigma$ 
and that of~\cite{fogli}
allows maximal atmospheric mixing within the $3\sigma$ range. 
The least known mixing parameter is the 
CP phase $\delta$: it is unconstrained at $2\sigma$, but
the best fit values of~\cite{fogli} hint towards $\delta\approx\pi$.

With all these improved results, it is 
worthwhile to perform an investigation 
of the allowed ranges for the elements of the 
neutrino mass matrix $\mathcal{M}_\nu$, 
as was done 
by Merle and Rodejohann in their seminal paper~\cite{merle}.
In the context of textures of $\mathcal{M}_\nu$, the
correlations of the elements 
$\left( \mathcal{M}_\nu \right)_{\alpha\beta}$ 
$(\alpha, \beta = e, \mu, \tau)$
of the neutrino 
mass matrix are of particular interest. Therefore, the main goal of our
paper is the construction of plots correlating the absolute values of
the elements of the 
neutrino mass matrix with each other. 
We will assume Majorana nature of neutrinos in this paper.

The structure of our paper is as follows. In section~\ref{section-mnu}
we investigate 
analytically 
upper and lower bounds on the absolute values of the elements of the
neutrino mass matrix. In 
section~\ref{section-numerical} we will outline the numerical methods
we will apply in our analysis, and the results 
will be presented in section~\ref{section-results}.
Finally we will draw conclusions in section~\ref{section-conclusions}.

\section{The elements of the neutrino mass matrix}\label{section-mnu}

Assuming neutrinos to be of Majorana nature, the neutrino mass matrix
is a complex 
symmetric $3\times 3$ matrix, which can be diagonalized as
\begin{equation}\label{mnudiag}
U^T \mathcal{M}_\nu U = \mathrm{diag} \left(m_1,\,m_2,\,m_3 \right),
\end{equation}
where the $m_i$ are the neutrino masses and $U$ is a unitary
matrix. In the following 
we will always assume to work in the basis in which the charged lepton
mass matrix 
is given by
\begin{equation}
\mathcal{M}_\ell=\mathrm{diag}(m_e,\,m_\mu,\,m_\tau),
\end{equation}
which implies $U=U_{\text{PMNS}}$. Consequently, $U$ can be decomposed as
\begin{equation}
U = D_1 V D_2,
\end{equation}
where 
$D_1=\mathrm{diag} \left(
e^{i\varphi_e},\,e^{i\varphi_\mu},\,e^{i\varphi_\tau}
\right)$
is a diagonal phase matrix, $V$ is the mixing matrix in the
parameterization 
suggested
in~\cite{rpp}, 
and $D_2=\mathrm{diag}(e^{i\sigma_1},\,e^{i\sigma_2},\,e^{i\sigma_3})$ is the matrix
of Majorana phases. 
Without loss of generality, 
we assume $\sigma_3=0$.
Inserting this into equation~(\ref{mnudiag}) leads to
\begin{equation}
(\mathcal{M}_\nu)_{\alpha\beta} = 
\left( U^\ast \mathrm{diag}(m_1,\,m_2,\,m_3) U^\dagger \right)_{\alpha\beta}=
e^{-i(\varphi_\alpha +\varphi_\beta)} \sum_{k=1}^3 m_k \,
e^{-2i\sigma_k} V_{\alpha k}^\ast V_{\beta k}^\ast.
\end{equation}
The absolute values of the elements of the neutrino mass matrix
\begin{equation}\label{absmnu}
|(\mathcal{M}_\nu)_{\alpha\beta}| =
|(\mathcal{M}_\nu^\ast)_{\alpha\beta}| = 
\left| \sum_k m_k\, e^{2i\sigma_k} V_{\alpha k} V_{\beta k}\right|
\end{equation}
depend on nine real parameters, namely the three neutrino masses
$m_i$, the three mixing angles 
$\theta_{12},\,\theta_{23}$ and $\theta_{13}$, the Dirac CP phase
$\delta$ and the two Majorana phases $\sigma_1$ and $\sigma_2$.

From equation~(\ref{absmnu}) we can deduce an upper bound on
$|(\mathcal{M}_\nu)_{\alpha\beta}|$ as follows. 
Rewriting the absolute value as a scalar product
\begin{equation}
|(\mathcal{M}_\nu)_{\alpha\beta}| =
\left| \sum_k m_k\, e^{2i\sigma_k} V_{\alpha k} V_{\beta k}\right| =
\left| \sum_k \underbrace{\sqrt{m_k} e^{i\sigma_k} V_{\alpha k}}_{A_k^\ast}
\underbrace{\sqrt{m_k} e^{i\sigma_k} V_{\beta k}}_{B_k} \right| 
\equiv 
|\langle A|B\rangle|,
\end{equation}
we can use Cauchy-Schwarz's inequality to find
\begin{equation}
|(\mathcal{M}_\nu)_{\alpha\beta}|\leq |A|\,|B|.
\end{equation}
Due to the unitarity of $V$, we have 
$\sum_k |V_{\alpha k}|^2=1$, and thus
\begin{equation}
|A|\leq \sqrt{ \max_k m_k},\quad |B|\leq \sqrt{ \max_k m_k},
\end{equation}
which leads to the final result
\begin{equation}\label{mnu-upperbound}
|(\mathcal{M}_\nu)_{\alpha\beta}|\leq \max_k m_k,
\end{equation}
i.e.\ the absolute value of an element of the neutrino mass matrix is
smaller than the largest neutrino mass. 

We can also construct a lower bound on
$|(\mathcal{M}_\nu)_{\alpha\beta}|$. Defining
\begin{equation}
a_k \equiv m_k |V_{\alpha k}| |V_{\beta k}|
\end{equation}
and taking into account that the Majorana phases are not constrained
by experiment up to now, we find 
\begin{equation}\label{3}
|(\mathcal{M}_\nu)_{\alpha\beta}| = 
\left| \sum_{k=1}^3 e^{i\rho_k} a_k \right|,
\end{equation}
where 
the $\rho_k$ 
are unconstrained phases. 
Now we have to distinguish two cases. If the three numbers $a_k$ are
such that they can be conceived as the lengths of the sides of a
triangle, then the right-hand side of equation~(\ref{3}) can become
zero and $|(\mathcal{M}_\nu)_{\alpha\beta}|$ as well. It is easy to
show that it is possible to construct a triangle with side lengths
$a_k$, if and only if\footnote{A different 
  but equivalent inequality
  has been derived in~\cite{CP}.}  
\begin{equation}
2\max_k a_k - \sum_k a_k \leq 0.
\end{equation}
Therefore, we end up with the 
inequality\footnote{If the right-hand side of inequality~(\ref{ineq1}) is negative,
then the resulting bound is $|(\mathcal{M}_\nu)_{\alpha\beta}| \geq 0$.}
\begin{equation}\label{ineq1}
|(\mathcal{M}_\nu)_{\alpha\beta}| \geq 
2\max_k a_k - \sum_k a_k.
\end{equation}
In this way, one gets rid of the Majorana phases and 
this inequality may be used to rule out single texture zeros in the
neutrino mass matrix. 

\section{Numerical analysis}\label{section-numerical}

As already discussed in the previous section, the absolute values of the 
elements of $\mathcal{M}_\nu$ depend on nine variables, two of which---namely
the two Majorana phases $\sigma_1$ and $\sigma_2$---are totally
unconstrained.
In~\cite{merle} the then available experimental data were used
to produce plots of the $|(\mathcal{M}_\nu)_{\alpha\beta}|$ versus the smallest neutrino
mass $m_0$. The goal of the present paper is to produce---using the
results of the latest global fits of neutrino oscillation experiments---plots
of $|(\mathcal{M}_\nu)_{\alpha\beta}|$ versus
$|(\mathcal{M}_\nu)_{\alpha'\beta'}|$. Since 
the knowledge of the neutrino oscillation parameters has improved considerably
in the recent years,
we also redo the numerical analysis of~\cite{merle} and
show the plots of
$|(\mathcal{M}_\nu)_{\alpha\beta}|$ versus $m_0$.

Let us now turn to our numerical strategy. Concerning the desired
plots, a first attempt 
 of creating scatterplots was, unfortunately, doomed to failure
because even random point numbers as high as
$10^9$ were not sufficient to fathom enough of the allowed parameter space
to achieve appealing plots. Therefore we follow a different strategy.
From the scatterplots we can guess the shapes of the areas which would be filled
in the limit of infinitely many points. In particular, we find that the allowed
areas have no ``holes,'' from where it
becomes clear that 
it is sufficient to construct their boundaries. We do this in the following way.
Consider a plot showing $|(\mathcal{M}_\nu)_{\alpha\beta}|$ or $m_0$
on the $x$-axis 
and $|(\mathcal{M}_\nu)_{\alpha'\beta'}|$ on the $y$-axis. Then we
start by pinning the 
quantity on the $x$-axis
to some given value $x_0$. Then we minimize and maximize $y$ for fixed
$x=x_0$ and obtain two points $(x_0,y_{min})$ and $(x_0,y_{max})$ of the boundary.
Afterwards we do the same for fixed $y=y_0$ which leads to the two points
$(x_{min},y_0)$ and $(x_{max},y_0)$. Repeating this procedure with a suitable number
of different values for $x_0$ and $y_0$ finally yields the desired allowed area.

As described above, we need an algorithm which allows to minimize and maximize
real functions of nine variables.\footnote{Analysing the predictions
  of the ``best fit'' results 
  of the global fits would reduce the number of parameters. However, 
  instead of adapting our program 
  to a smaller number of parameters, we let the parameters which shall
  assume their best fit values 
  $b$ vary in the interval $(b-\epsilon |b|,b+\epsilon |b|)$, with
  $\epsilon=10^{-6}$.} 
For this purpose we choose the Nelder--Mead algorithm
(downhill simplex method)~\cite{nelder-mead}, which is a direct search method for finding
a local minimum of a given function. However, since all functions $f$ we will consider
are non-negative, by minimizing $-f$ we can use the Nelder--Mead algorithm to find
a local maximum of $f$. Since we are interested in the global minima (maxima) of $f$,
single runs of the algorithm are not sufficient. Thus for
every minimization (or maximization) we start with 200000 different
random start simplices and also perform perturbations
of candidates for a good minimum (maximum). 

Since there are nine parameters in $\mathcal{M}_\nu$, the
domain of the Nelder--Mead algorithm is, by construction, the full
parameter space $\mathbbm{R}^9$.
In order to restrict the search
to some domain $D\subset \mathbbm{R}^9$, we 
decided for the following procedure.
In the physical region, i.e. for an absolute neutrino mass
scale of the order of at most $\sim \text{eV}$ (see the discussion
at the end of this section), also the values of the function $f$ to be minimized
will be of the order of at most $\sim\text{eV}$.
Therefore, instead of minimizing $f(p)$, we minimize $f(p)+\Pi_D(p)$, where
$\Pi_D(p)$ is the characteristic function
\begin{equation}
\Pi_D(p):=
\begin{cases}
0\enspace\text{eV} & \text{for }p\in D,\\
10^6\,\text{eV} & \text{for }p\not\in D.
\end{cases}
\end{equation}
Maximization of $f$ is then equivalent
to minimization of $\Pi_D-f$. The vector $p$ collects the nine parameters of
the $|(\mathcal{M}_\nu)_{\alpha\beta}|$.
The pinning of $x$ to $x_0$ is also achieved by
adding a characteristic function, namely
\begin{equation}
\Pi_I(x(p)):=
\begin{cases}
0\enspace\text{eV} & \text{for }x(p)\in I,\\
10^6\,\text{eV} & \text{for }x(p)\not\in I,
\end{cases}
\end{equation}
where $I$ is an interval containing $x_0$. The pinning of $y$ to $y_0$ is done
in the same way.

We also want to evaluate the lower bound~(\ref{ineq1}) derived in section~\ref{section-mnu}.
For this purpose, we will use the Nelder--Mead algorithm on the function
\begin{equation}
2\max_k a_k -\sum_k a_k + \Pi_D(p')
\end{equation}
of the seven real variables $p'=(m_0, \Delta m_{21}^2, \Delta m_{31}^2,
\theta_{12}, \theta_{23}, \theta_{13}, \delta)$.

Next let us discuss the domain $D$ of the parameters $p$.
For the mass-squared differences and the sines squared of the mixing angles
we use the best fit as well as the $n\sigma$-ranges ($n=1,2,3$) provided by~\cite{forero,fogli}. 
The Majorana phases $\sigma_1$ and $\sigma_2$ are unconstrained and can thus vary between zero and $2\pi$.
The situation for $\delta$ is a bit ambiguous because the best fit values are very different in 
the two global fits~\cite{forero} (version 3) and~\cite{fogli}; while for the normal ordering of the
neutrino mass spectrum both fits favour a value of $\delta$ which is roughly $\pi$,
the best fit values in the case of an inverted spectrum are $\sim 0$ and $\sim \pi$ for~\cite{forero} (version 3)
and~\cite{fogli}, respectively. For this reason, in the ``best fit''-plot we do not fix $\delta$ to its
best fit value, but allow it to vary in its $1\sigma$-range.
Also for the $1\sigma$-plots $\delta$ is allowed to vary in the respective $1\sigma$-ranges; however,
$\delta$ is unconstrained
at the two and three sigma level.

The last parameter which remains to be discussed here, is the smallest neutrino mass $m_0$.
The strongest constraints on the absolute values of the neutrino masses comes from cosmology,
where the sum of the light neutrino masses---in the form of the relic neutrino energy
density $\Omega_\nu$---is one of the parameters of the standard model of cosmology~\cite{rpp}.
There is no unique consensus on the resulting upper bound on $\sum_\nu m_\nu$, however
most constraints are of the order of $\sum_{\nu} m_\nu<\mathcal{O}(1~\text{eV})$~\cite{rpp}.
Therefore, we allow $m_0$ to vary between zero and $0.3~\text{eV}$. In the limit
of $m_0\rightarrow 0.3~\text{eV}$, this implies an
upper bound on the largest neutrino mass of
\begin{equation}
\max_k m_k \lesssim m_0+\frac{|\Delta m_{31}^2|}{2 m_0} \approx 0.304~\text{eV}.
\end{equation}
According to equation~(\ref{mnu-upperbound}), this directly translates to an upper bound of
$\approx 0.3~\text{eV}$ on the absolute values of the elements of the neutrino mass matrix.

Assuming that the ``standard mechanism'' (induced by the
Majorana mass term) dominates
neutrinoless double beta decay,
its non-observation
gives upper bounds on $|(\mathcal{M}_\nu)_{ee}|$, the current bounds
being of the order of~\cite{rodejohann-betabeta,EXO-200}
\begin{equation}\label{betabetabound}
|(\mathcal{M}_\nu)_{ee}|\lesssim0.4~\text{eV},
\end{equation}
which is comparable to the cosmological bound. Thus,
unfortunately, the bound~(\ref{betabetabound}) will be fulfilled
automatically in our numerical analysis 
and will, therefore, give no
additional constraint on $|(\mathcal{M}_\nu)_{ee}|$.
However, the bound~(\ref{betabetabound})
implies an upper bound on the absolute neutrino mass scale~\cite{rodejohann-betabeta}.
Using the $3\sigma$-ranges for the oscillation parameters 
one can estimate the bound to be \cite{rodejohann-betabeta} 
\begin{equation}
m_0\lesssim 1.9\,\text{eV}.
\end{equation}
Here the constraints from cosmology are much stronger. 

The bound on
$m_0$ coming from 
tritium decay is of almost the same size as the one from neutrinoless
double beta decay, namely 
$m_0 < 2\,\text{eV}$~\cite{rpp}, and will therefore
not constrain the results of our analysis.

\section{Results}\label{section-results}

\paragraph{Lower bounds on $|(\mathcal{M}_\nu)_{\alpha\beta}|$:}
Using inequality~(\ref{ineq1}) and the global fits of~\cite{forero,fogli}
shows that the only elements of $|(\mathcal{M}_\nu)_{\alpha\beta}|$ which 
have a non-trivial lower bound
are $|(\mathcal{M}_\nu)_{ee}|$ in the case of an
inverted neutrino mass spectrum  and $|(\mathcal{M}_\nu)_{\tau\tau}|$
in the case of normal ordering of the neutrino masses. The resulting
bounds can be found in table~\ref{lowerbounds}.

Let us compare these bounds with the results of the analysis of Merle and Rodejohann~\cite{merle}.
From the plots presented there one
can read off that the only non-trivial lower bound on an element of $\mathcal{M}_\nu$ is the
bound
\begin{equation}
|(\mathcal{M}_\nu)_{ee}|\geq 7\times 10^{-3}\,\text{eV}
\end{equation}
in the case of an inverted neutrino mass spectrum.
Note that with the present data this bound has improved by a factor of two (see table~\ref{lowerbounds}),
corresponding to a factor four in the lifetime of neutrinoless double beta decay.

\begin{table}
\begin{center}
\begin{tabular}{llccc}
 & & $1\sigma$ & $2\sigma$ & $3\sigma$\\
$|(\mathcal{M}_\nu)_{ee}|$ (inv. spectrum) & Forero \textit{et al.} & $1.52\times 10^{-2}$ & $1.36\times 10^{-2}$ & $1.14\times 10^{-2}$\\
                                           &  Fogli \textit{et al.} & $1.62\times 10^{-2}$ & $1.44\times 10^{-2}$ & $1.24\times 10^{-2}$\\
$|(\mathcal{M}_\nu)_{\tau\tau}|$ (norm. spectrum) & Forero \textit{et al.} & $0$ & $0$ & $0$\\
                                                  &  Fogli \textit{et al.} & $1.86\times 10^{-2}$ & $1.27\times 10^{-2}$ & $0$\\

\end{tabular}
\end{center}
\caption{Numerical results for the lower bounds on $|(\mathcal{M}_\nu)_{ee}|$ (inverted spectrum) and 
$|(\mathcal{M}_\nu)_{\tau\tau}|$ (normal spectrum)
in units of eV  using the
global fits of Forero \textit{et al.}\ (version 3)~\cite{forero} and Fogli \textit{et al.}~\cite{fogli}.
The lower bounds for all other $|(\mathcal{M}_\nu)_{\alpha\beta}|$ are zero.
\label{lowerbounds}}
\end{table}

\paragraph{How to read the plots:}
Before we discuss the resulting plots, we have to
explain how they are to be read. This may best be done by means of an example.
The properties we will outline in the following, hold for all plots in this paper.

\begin{figure}
\begin{center}
\includegraphics[angle=-90,keepaspectratio=true,width=0.6\textwidth]
{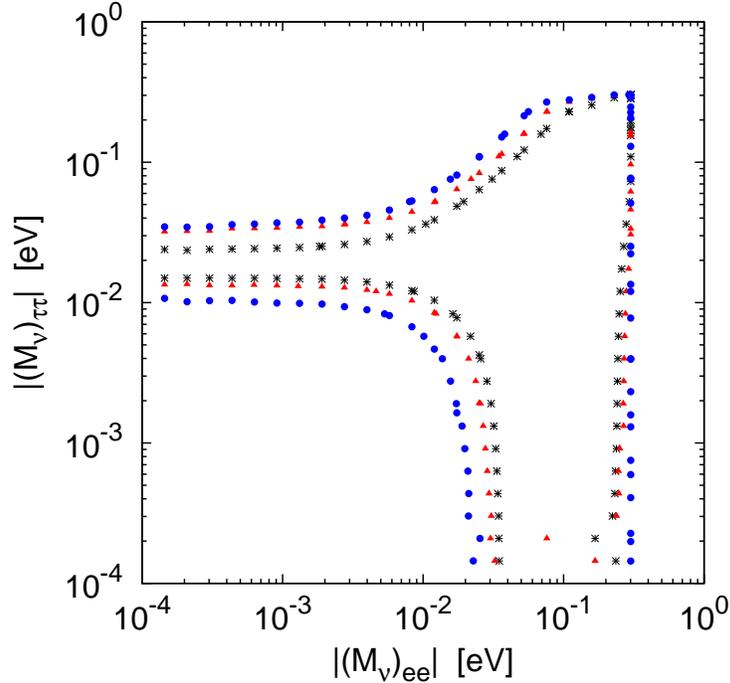}
\caption{Allowed ranges for $|(\mathcal{M}_\nu)_{ee}|$ vs.\ 
$|(\mathcal{M}_\nu)_{\tau\tau}|$
in the case of a normal mass spectrum, using the
global fit of Forero~\textit{et al.}\ (version 3).
The best fit area is bounded by black
stars, the $1\sigma$ area by red triangles, and the $3\sigma$
area by blue points.}\label{m11-m33}
\end{center}
\end{figure}

Figure~\ref{m11-m33} shows the allowed areas for
$|(\mathcal{M}_\nu)_{ee}|$ 
vs.\ $|(\mathcal{M}_\nu)_{\tau\tau}|$ in the case of a normal mass spectrum.
The plot consists of three types of points, each
one describing the boundary of an allowed area. The best fit area is bounded by black
stars~\begin{large}$\ast$\end{large},
the $1\sigma$ area is bounded by red triangles~\textcolor{red}{$\blacktriangle$}, and finally the $3\sigma$
area is bounded by blue points~\textcolor{blue}{$\bullet$}.
Note that also the plot axes comprise parts of the boundaries of the allowed areas, as can be
seen in figure~\ref{m11-m33}. This feature is very pronounced for instance in figure~\ref{m13-m33}.

In appendix~\ref{fogli} we present the plots based on the global fit of Fogli~\textit{et al.}
The plots based on the global fits of Forero~\textit{et al.} (versions 2 and 3) are shown
in appendices~\ref{forero2} and~\ref{forero3}, respectively.

\begin{figure}
\begin{center}
\includegraphics[angle=-90,keepaspectratio=true,width=0.6\textwidth]
{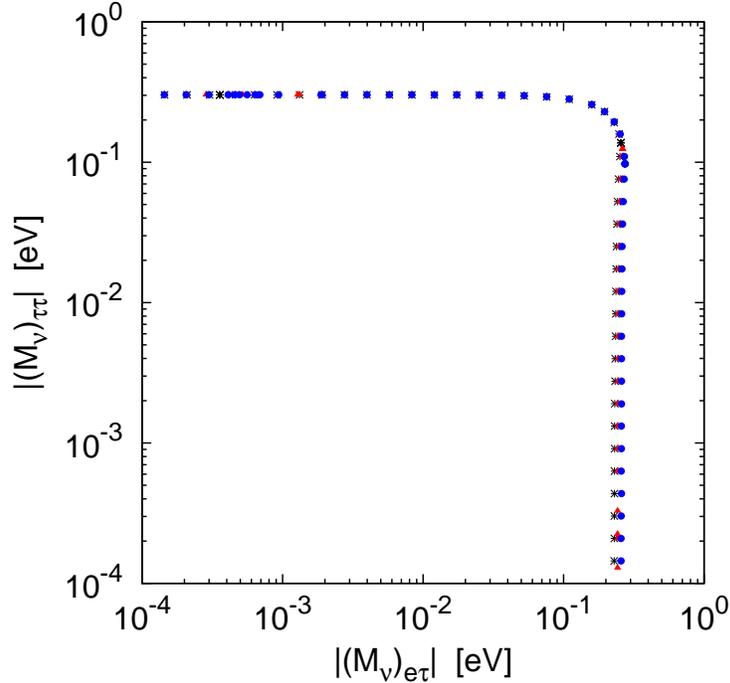}
\caption{Allowed ranges for $|(\mathcal{M}_\nu)_{e\tau}|$ vs.\ 
$|(\mathcal{M}_\nu)_{\tau\tau}|$
in the case of a normal mass spectrum, using the
global fit of Forero~\textit{et al.}\ (version 3).
For more information \textit{cf.}\ figure~\ref{m11-m33} and section~\ref{section-results}.}\label{m13-m33}
\end{center}
\end{figure}

\paragraph{Plots of the smallest neutrino mass versus $|(\mathcal{M}_\nu)_{\alpha\beta}|$:}

Concerning the plots of $|(\mathcal{M}_\nu)_{\alpha\beta}|$ as a function of the smallest neutrino mass,
an interesting point is whether they have changed since the original analysis of
Merle and Rodejohann~\cite{merle} in 2006. The main difference is that at the time when
reference~\cite{merle}
was written, only an upper bound on the size of $\theta_{13}$ was known, which is the
reason that~\cite{merle} contains plots for different sizes of $\mathrm{sin}^2\,2\theta_{13}$.
Comparing our plots to the ones of~\cite{merle} with $\mathrm{sin}^2\,2\theta_{13}=0.1$,
we find that at the $3\sigma$ level the plots in~\cite{merle} are still in good agreement
also with the new results.
However, for the best fit only the plots of the smallest neutrino mass versus
$|(\mathcal{M}_\nu)_{ee}|$ and $|(\mathcal{M}_\nu)_{\mu\tau}|$
are in good agreement with the plots obtained from the new data. The $1\sigma$ regions
are not indicated in~\cite{merle}.

Finally, from the plots we can readily read off the lower bounds on the $|(\mathcal{M}_\nu)_{\alpha\beta}|$
and compare them with the bounds found from evaluating inequality~(\ref{ineq1})---see table~\ref{lowerbounds}.
We find full agreement, which provides a successful consistency check of our results.

\paragraph{Plots of $|(\mathcal{M}_\nu)_{\alpha\beta}|$ versus
  $|(\mathcal{M}_\nu)_{\alpha'\beta'}|$:} 

In this section we will discuss some of the conclusions one can draw from the
correlation plots.
These plots can be used to test the viability of two texture zeros in $\mathcal{M}_{\nu}$.
The results obtained from the plots are in perfect agreement with the results of a recent numerical
analysis provided in~\cite{grimus-ludl}, and may---at $3\sigma$---be condensed to the fact that
the seven two texture zeros originally presented in~\cite{FGM} are still viable with all
of the three global fits~\cite{forero,fogli}. For further details we refer the reader
to~\cite{grimus-ludl}.

Let us continue by discussing those correlations which appear manifest at the $3\sigma$ level.
Going through all the plots, we find the correlations
\begin{center}
\begin{tabular}{llll}
$|(\mathcal{M}_\nu)_{ee}|$ & vs. & $|(\mathcal{M}_\nu)_{\mu\mu}|$ & (normal spectrum)\\
$|(\mathcal{M}_\nu)_{ee}|$ & vs. & $|(\mathcal{M}_\nu)_{\mu\tau}|$ & (normal spectrum)\\
$|(\mathcal{M}_\nu)_{ee}|$ & vs. & $|(\mathcal{M}_\nu)_{\tau\tau}|$ & (normal spectrum)\\
$|(\mathcal{M}_\nu)_{\mu\mu}|$ & vs. & $|(\mathcal{M}_\nu)_{\mu\tau}|$ & (normal spectrum)\\
$|(\mathcal{M}_\nu)_{\mu\tau}|$ & vs. & $|(\mathcal{M}_\nu)_{\tau\tau}|$ & (normal spectrum)\\
\end{tabular}
\end{center}
most stringent. All these five correlations may be subsumed as follows:
``If one matrix element is small, the other one must be large,'' as can, for example, be seen from
figure~\ref{m11-m33}. However, there are also matrix elements which appear to be
totally uncorrelated. A good example for this case is $|(\mathcal{M}_\nu)_{e\tau}|$ vs.\
$|(\mathcal{M}_\nu)_{\tau\tau}|$ in the case of a normal spectrum---see figure~\ref{m13-m33}.

While at the $3\sigma$ level all plots produced using the three different
global fits of Forero~\textit{et al.}\ and Fogli~\textit{et al.}\ agree very well, at the $1\sigma$
level this situation changes completely. Instead of presenting a confusing list showing all differences,
let us just mention the most important point which is the fact that the
fit results of Fogli~\textit{et al.}\
no longer allow a vanishing matrix element $|(\mathcal{M}_\nu)_{\tau\tau}|$ at $1\sigma$ in the case
of a normal neutrino mass spectrum,
while the fit results of Forero~\textit{et al.}\ still do.
This evidently also has strong consequences on all correlation plots
including $|(\mathcal{M}_\nu)_{\tau\tau}|$, a fine example being the plot of
$|(\mathcal{M}_\nu)_{\mu\tau}|$ vs.\ $|(\mathcal{M}_\nu)_{\tau\tau}|$,
which is shown in figure~\ref{m23-m33}. See also the corresponding plot in appendix~\ref{differences}.

Let us finally discuss the sensitivity of the correlation plots to the data.
As already pointed out, correlations which contain $|(\mathcal{M}_\nu)_{\tau\tau}|$
are particularly sensitive, due to the strong constraint on
$|(\mathcal{M}_\nu)_{\tau\tau}|$ in the case
of a normal neutrino mass spectrum---see table~\ref{lowerbounds}.
However, also for an inverted mass spectrum the plots involving $|(\mathcal{M}_\nu)_{\tau\tau}|$
differ visibly for the fits by Forero~\textit{et al.} (version 3) and Fogli~\textit{et al.}
Other correlation plots which
are quite sensitive to the data at $1\sigma$ are:
\begin{center}
\begin{tabular}{llll}
$|(\mathcal{M}_\nu)_{ee}|$ & vs. & $|(\mathcal{M}_\nu)_{e\tau}|$ & (normal spectrum)\\
$|(\mathcal{M}_\nu)_{ee}|$ & vs. & $|(\mathcal{M}_\nu)_{\mu\mu}|$ & (inverted spectrum)\\
$|(\mathcal{M}_\nu)_{e\mu}|$ & vs. & $|(\mathcal{M}_\nu)_{\mu\mu}|$ & (normal and inv. spectrum)\\
$|(\mathcal{M}_\nu)_{e\tau}|$ & vs. & $|(\mathcal{M}_\nu)_{\mu\mu}|$ & (normal and inv. spectrum)\\
$|(\mathcal{M}_\nu)_{\mu\mu}|$ & vs. & $|(\mathcal{M}_\nu)_{\mu\tau}|$ & (inverted spectrum)\\
\end{tabular}
\end{center}
Thus in total 17 out of the 30 possible correlations are particularly sensitive to the
data at the $1\sigma$ level. In appendix~\ref{differences} we provide the 17 corresponding plots showing the allowed 
$1\sigma$ regions for both Forero~\textit{et al.}\ (version 3) and Fogli~\textit{et al.}

\begin{figure}
\begin{center}
\includegraphics[angle=-90,keepaspectratio=true,width=0.6\textwidth]
{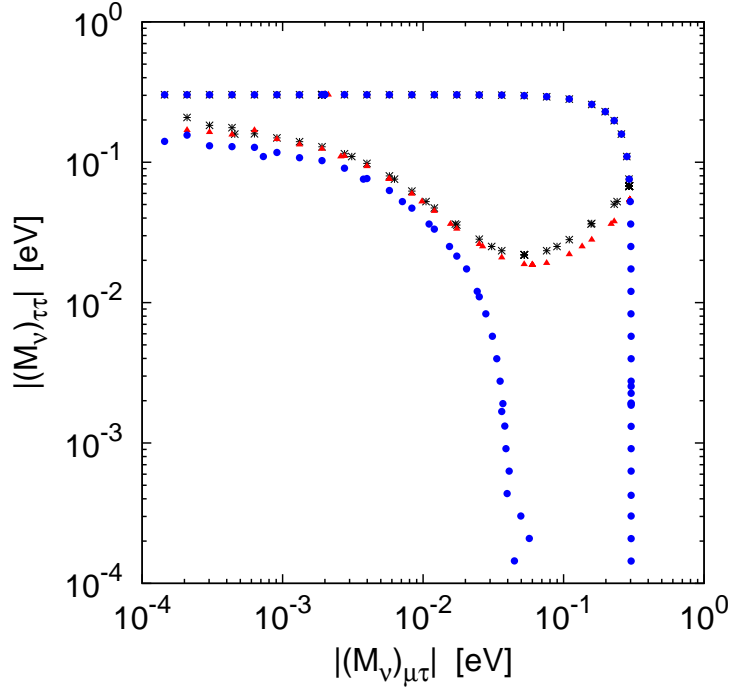}
\caption{Allowed ranges for $|(\mathcal{M}_\nu)_{\mu\tau}|$ vs.\ $|(\mathcal{M}_\nu)_{\tau\tau}|$
in the case of a normal mass spectrum, using the 
fit results of Fogli~\textit{et al.}
For more information \textit{cf.}\ figure~\ref{m11-m33} and section~\ref{section-results}.}\label{m23-m33}
\end{center}
\end{figure}

\section{Conclusions}\label{section-conclusions}

In this paper, assuming Majorana neutrinos, we re-investigated the
allowed ranges for the elements
of the neutrino mass matrix in the light of the results of the recent global fits.
In particular our analysis could profit from the fact that also the reactor mixing angle
is by now well determined. In contrast, at the time of the original analysis by Merle and
Rodejohann~\cite{merle} only an upper bound on $\mathrm{sin}^2\theta_{13}$ was known.

By means of Cauchy-Schwarz's inequality we could show that, in the basis where
the charged lepton mass matrix is diagonal, the absolute value of an element of $\mathcal{M}_\nu$
cannot exceed the largest neutrino mass. The most stringent bound on the absolute neutrino mass
scale coming from cosmology thus implies the bound
\begin{equation}
|(\mathcal{M}_\nu)_{\alpha\beta}|\lesssim 0.3\,\text{eV}.
\end{equation}
We could also derive lower bounds on the elements of the neutrino mass matrix.
Numerically evaluating these bounds on the basis of the global fits of oscillation data
revealed that only for $|(\mathcal{M}_\nu)_{ee}|$ (inverted spectrum) and
$|(\mathcal{M}_\nu)_{\tau\tau}|$ (normal spectrum) non-trivial lower bounds 
exist---see table~\ref{lowerbounds}. While the fact that $|(\mathcal{M}_\nu)_{ee}|$
is non-zero at $3\sigma$ was already clear in~\cite{merle}, the lower bound on
$|(\mathcal{M}_\nu)_{\tau\tau}|$ is a feature of the new fit of Fogli~\textit{et al.}~\cite{fogli}.
However the non-trivial bound is only valid at $2\sigma$ and the fits
of Forero~\textit{et al.}~\cite{forero} still allow zero $|(\mathcal{M}_\nu)_{\tau\tau}|$
even at $1\sigma$.

The second main point of our analysis was the creation of correlation
plots of the absolute values of the elements of $\mathcal{M}_\nu$.
We created plots based on three different global fits~\cite{forero,fogli}
and found that at the $3\sigma$ level there is no discrepancy between the different
global fits.
For every global fit we obtained 30 correlation plots (15 correlations, two spectra). 
Among the 30 possibilities we found only five stringent correlations at the $3\sigma$ level, namely:
\begin{center}
\begin{tabular}{llll}
$|(\mathcal{M}_\nu)_{ee}|$ & vs. & $|(\mathcal{M}_\nu)_{\mu\mu}|$ & (normal spectrum)\\
$|(\mathcal{M}_\nu)_{ee}|$ & vs. & $|(\mathcal{M}_\nu)_{\mu\tau}|$ & (normal spectrum)\\
$|(\mathcal{M}_\nu)_{ee}|$ & vs. & $|(\mathcal{M}_\nu)_{\tau\tau}|$ & (normal spectrum)\\
$|(\mathcal{M}_\nu)_{\mu\mu}|$ & vs. & $|(\mathcal{M}_\nu)_{\mu\tau}|$ & (normal spectrum)\\
$|(\mathcal{M}_\nu)_{\mu\tau}|$ & vs. & $|(\mathcal{M}_\nu)_{\tau\tau}|$ & (normal spectrum)\\
\end{tabular}
\end{center}
All these correlations may be subsumed as: ``If one matrix element is small, the
other one must be large.''

While at the $3\sigma$ level the different global fits all agree, this is not so when one
considers the $1\sigma$ level. There the most striking fact is that the fit of Fogli~\textit{et al.}~\cite{fogli}
does no longer allow a vanishing matrix element $(\mathcal{M}_\nu)_{\tau\tau}$ at $1\sigma$
in the case of a normal mass spectrum.

In summary, we conclude that correlations evident at the $3\sigma$ level exist only for the
normal mass spectrum. However, there are interesting features also at the $1\sigma$ level
which may be corroborated (or refuted) by future experimental results increasing the
accuracy of global fit data.

\paragraph{Acknowledgements:} This work is supported by the Austrian
Science Fund (FWF), Project No.\ P~24161-N16. P.O.L. thanks Andreas Singraber for
fruitful discussions and helpful comments on the numerical part of this work and
Helmut Moser for his tireless servicing of the institute's computer cluster
on which the
computations for this work were performed.

\begin{appendix}

\section{Plots based on Fogli \textit{et al.}}\label{fogli}

This appendix contains all 42 plots produced on the basis of
the global fit by Fogli \textit{et al.}~\cite{fogli}.

The figures in the left and right columns correspond to normal and inverted
mass ordering, respectively.
The plots consist of three types of points, each
one describing the boundary of an allowed area. The best fit area is bounded by black
stars~\begin{large}$\ast$\end{large},
the $1\sigma$ area is bounded by red triangles~\textcolor{red}{$\blacktriangle$}, and finally the $3\sigma$
area is bounded by blue points~\textcolor{blue}{$\bullet$}.
The same applies to the plots in appendices~\ref{forero2}
and~\ref{forero3}.

\begin{tabular}[t]{ll}
\includegraphics[angle=-90,keepaspectratio=true,scale=\figurescale]
{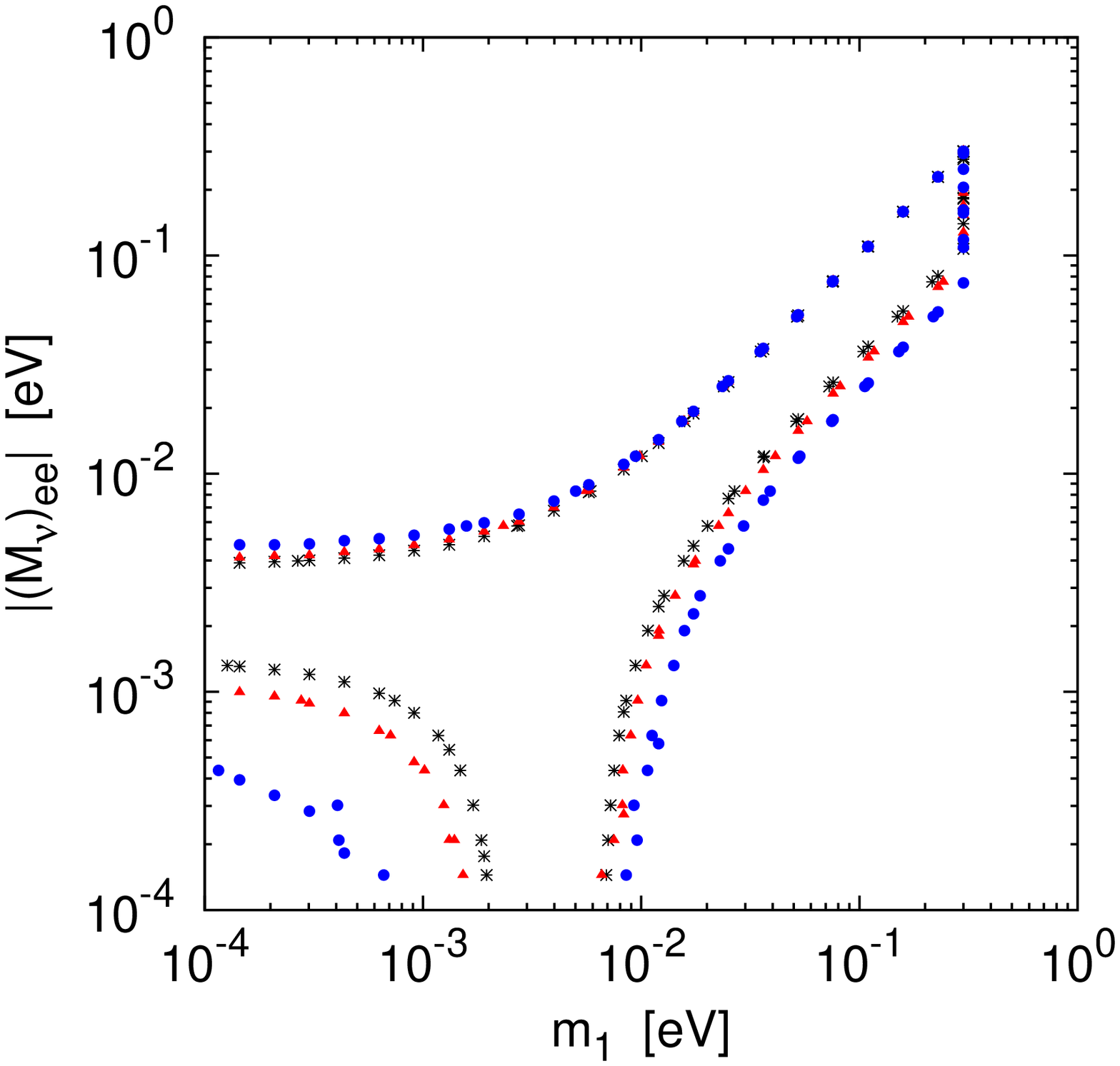} &
\includegraphics[angle=-90,keepaspectratio=true,scale=\figurescale]
{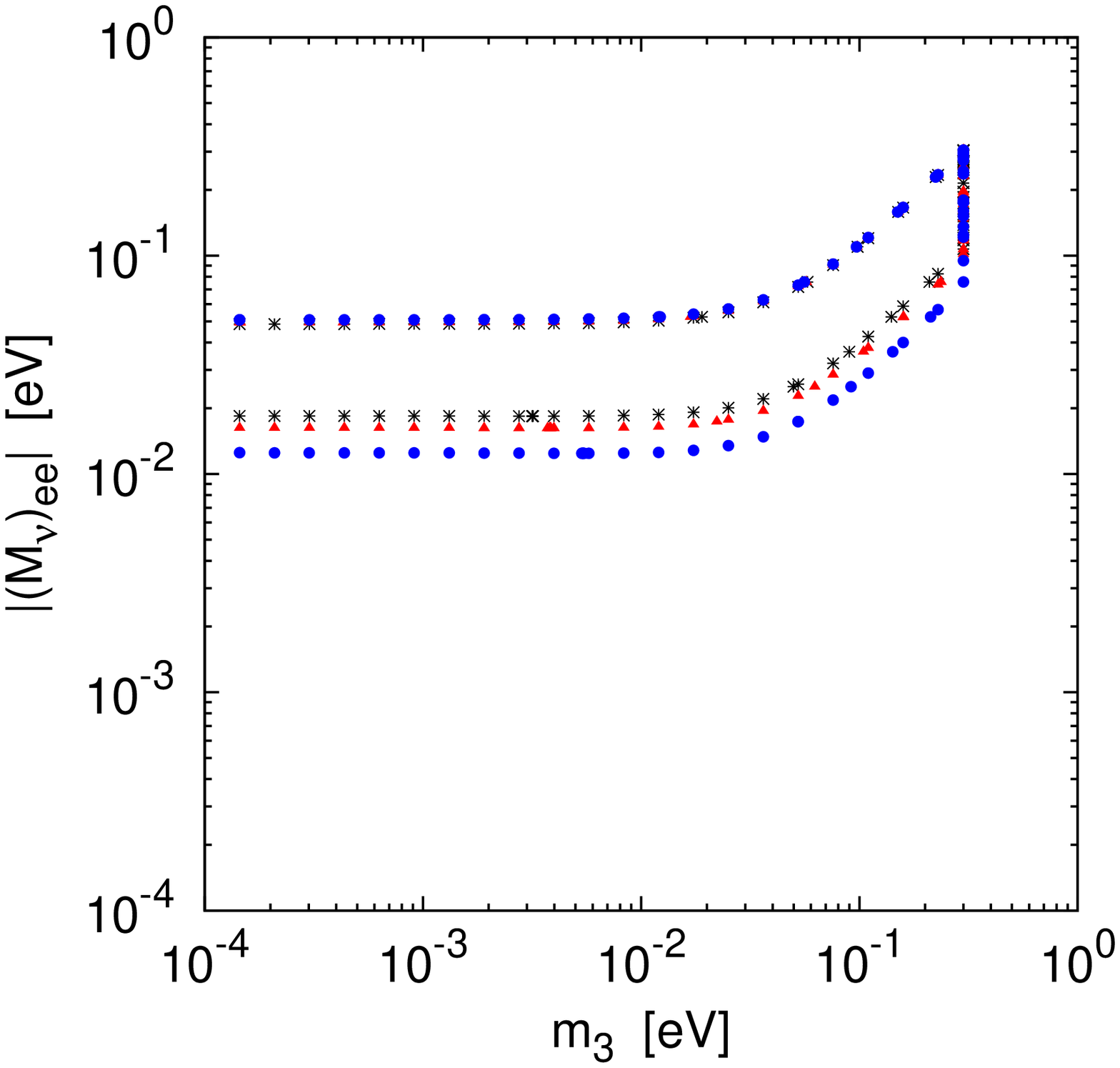}\\
\includegraphics[angle=-90,keepaspectratio=true,scale=\figurescale]
{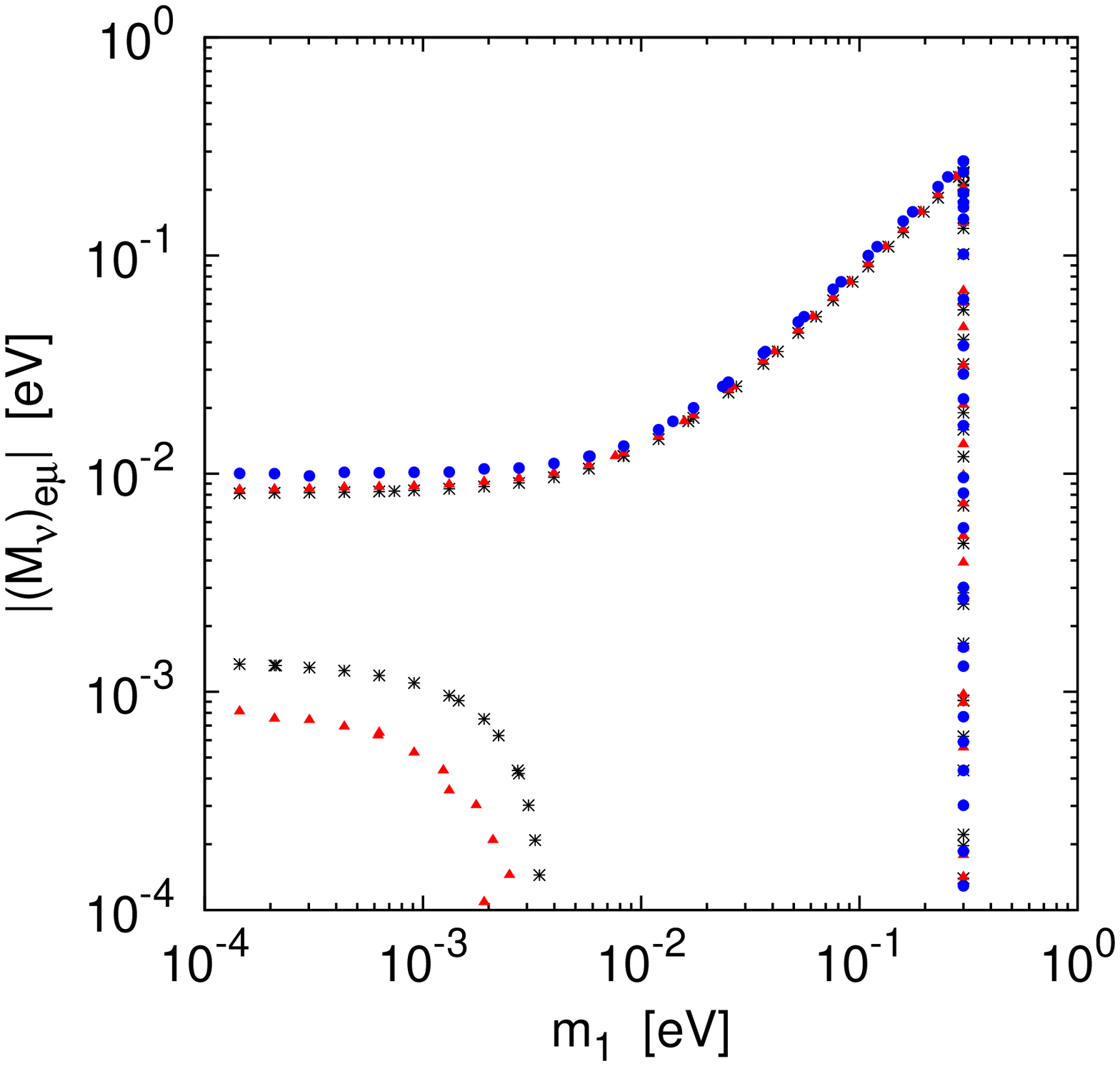} &
\includegraphics[angle=-90,keepaspectratio=true,scale=\figurescale]
{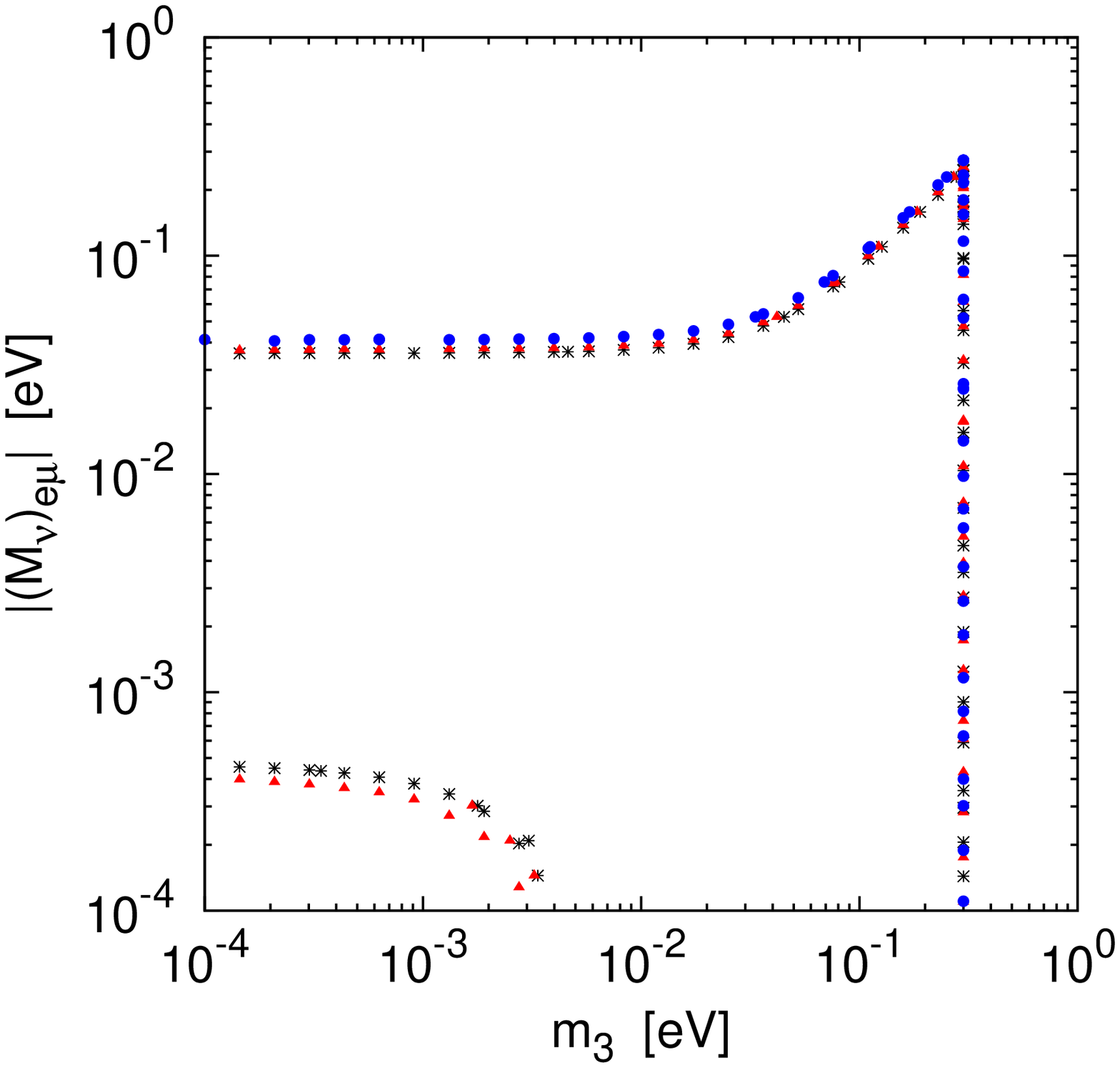}\\
\includegraphics[angle=-90,keepaspectratio=true,scale=\figurescale]
{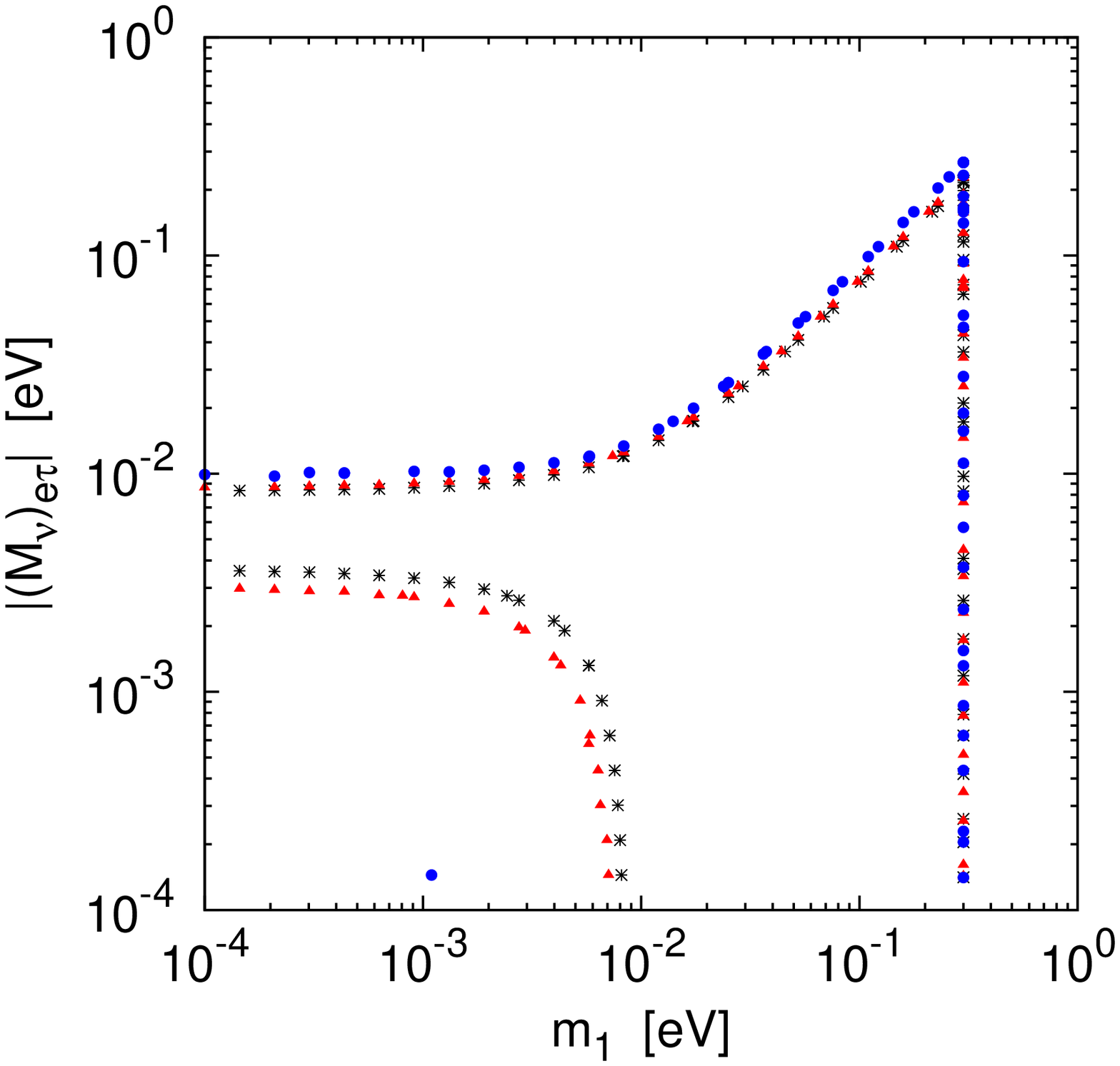} &
\includegraphics[angle=-90,keepaspectratio=true,scale=\figurescale]
{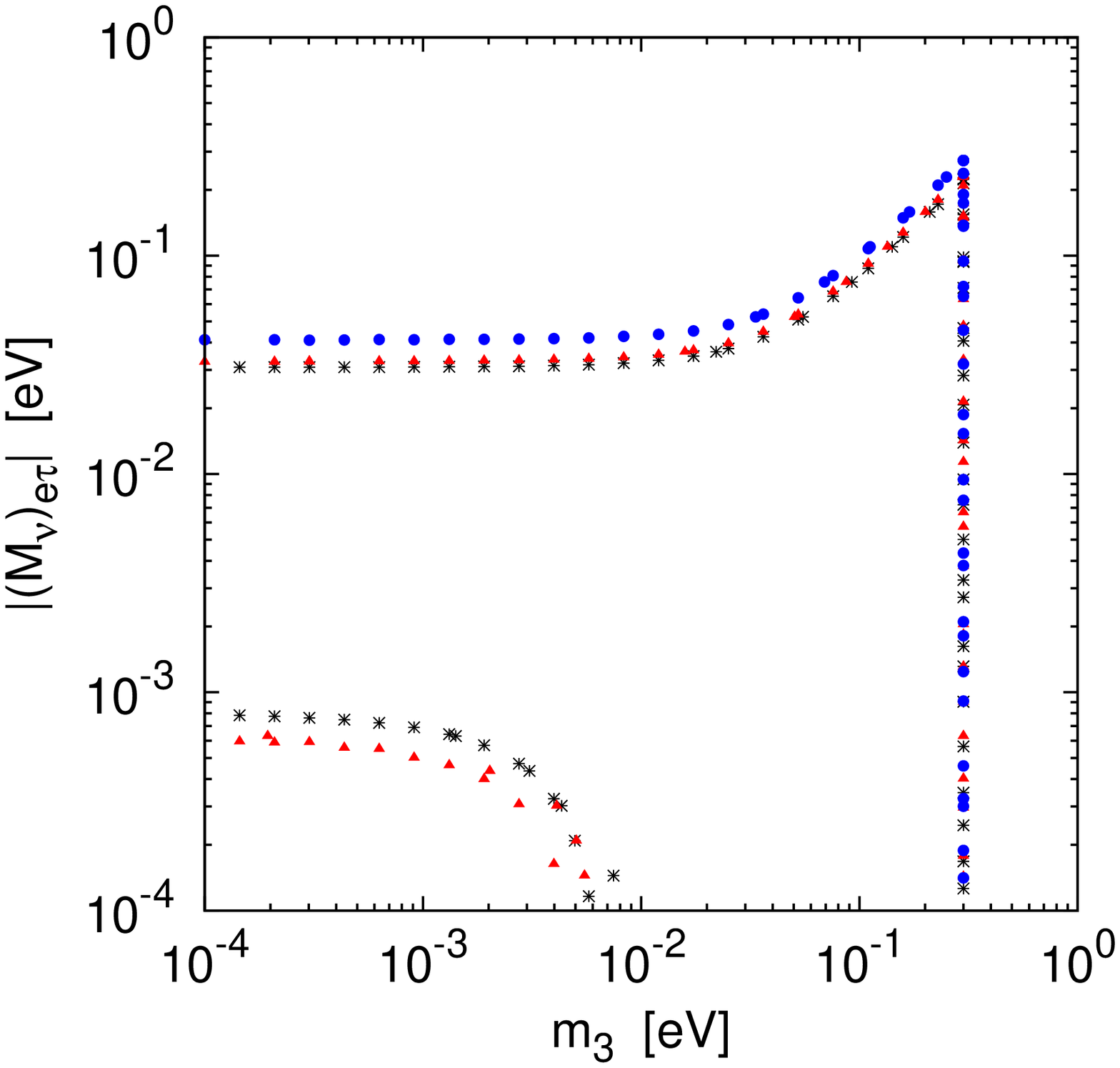}\\
\end{tabular}

\begin{tabular}[t]{ll}
\includegraphics[angle=-90,keepaspectratio=true,scale=\figurescale]
{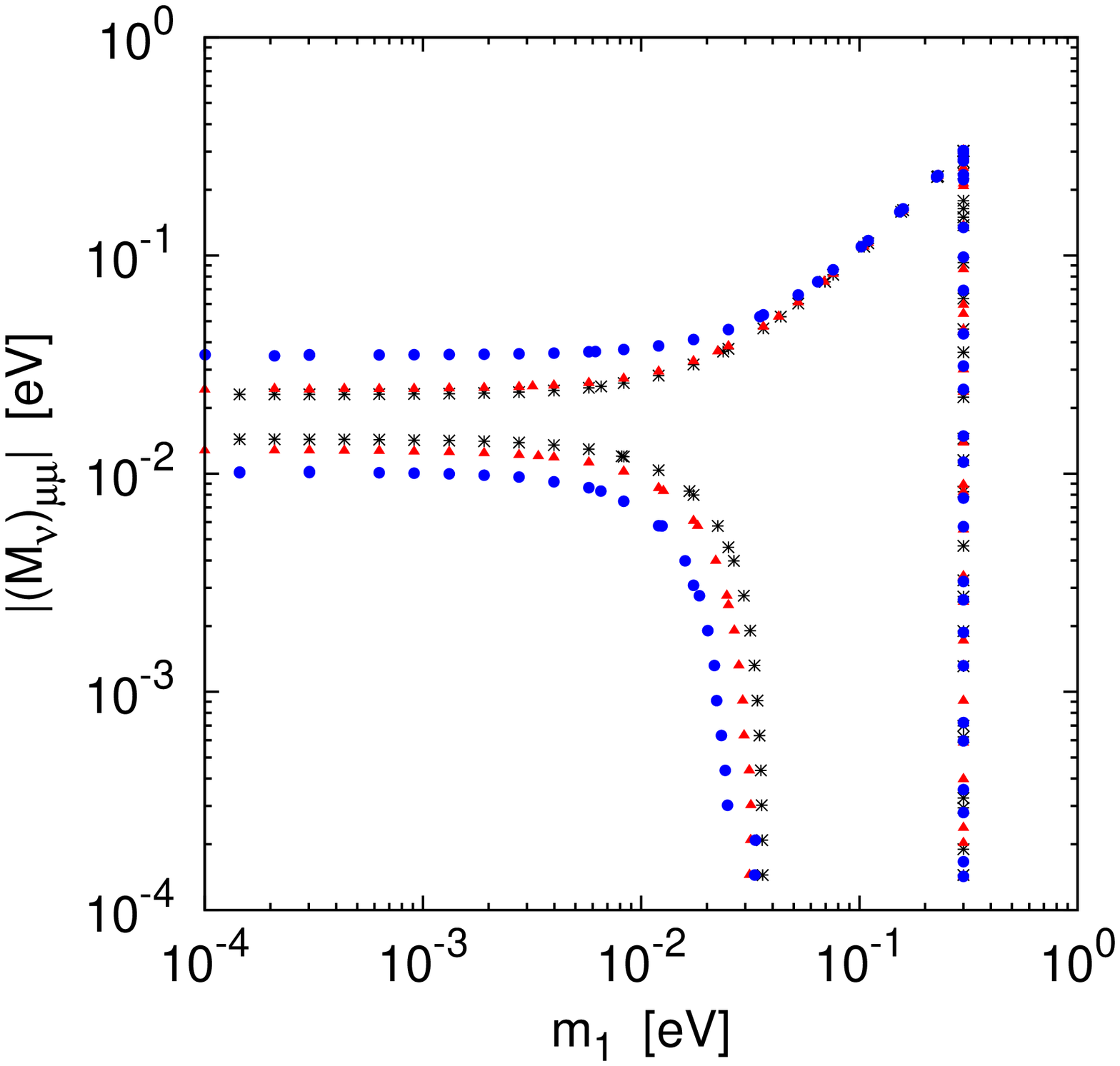} &
\includegraphics[angle=-90,keepaspectratio=true,scale=\figurescale]
{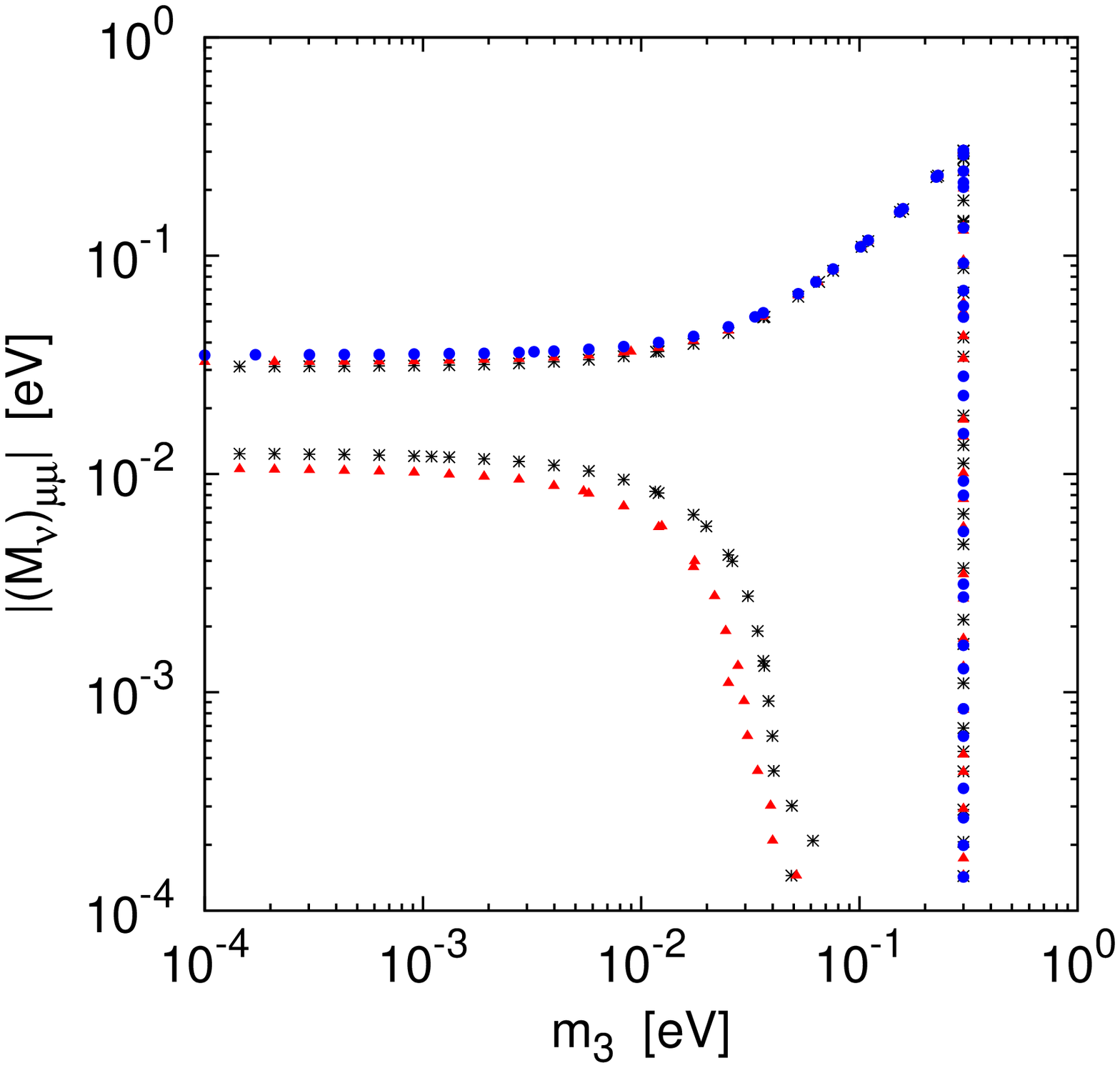}\\
\includegraphics[angle=-90,keepaspectratio=true,scale=\figurescale]
{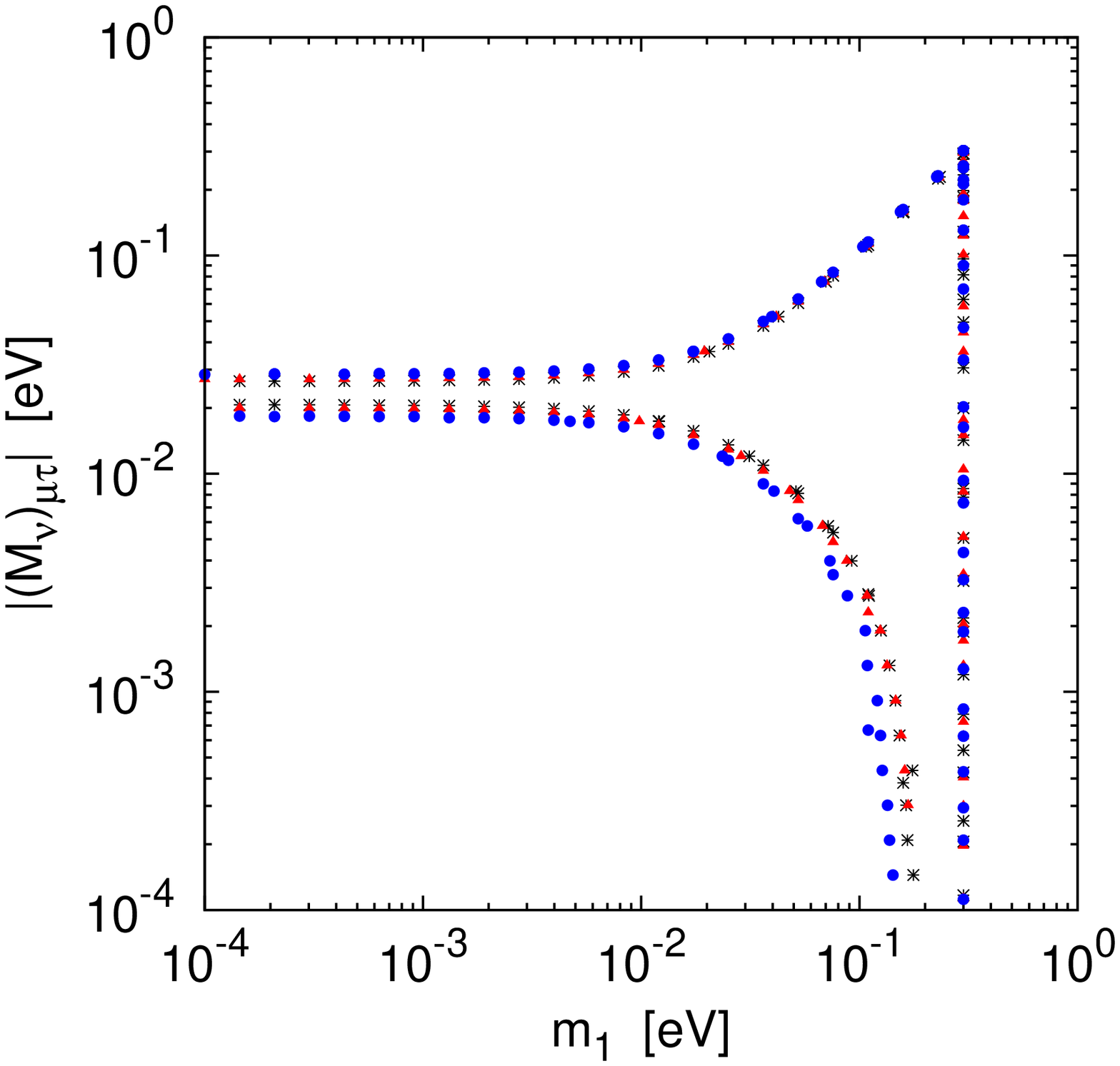} &
\includegraphics[angle=-90,keepaspectratio=true,scale=\figurescale]
{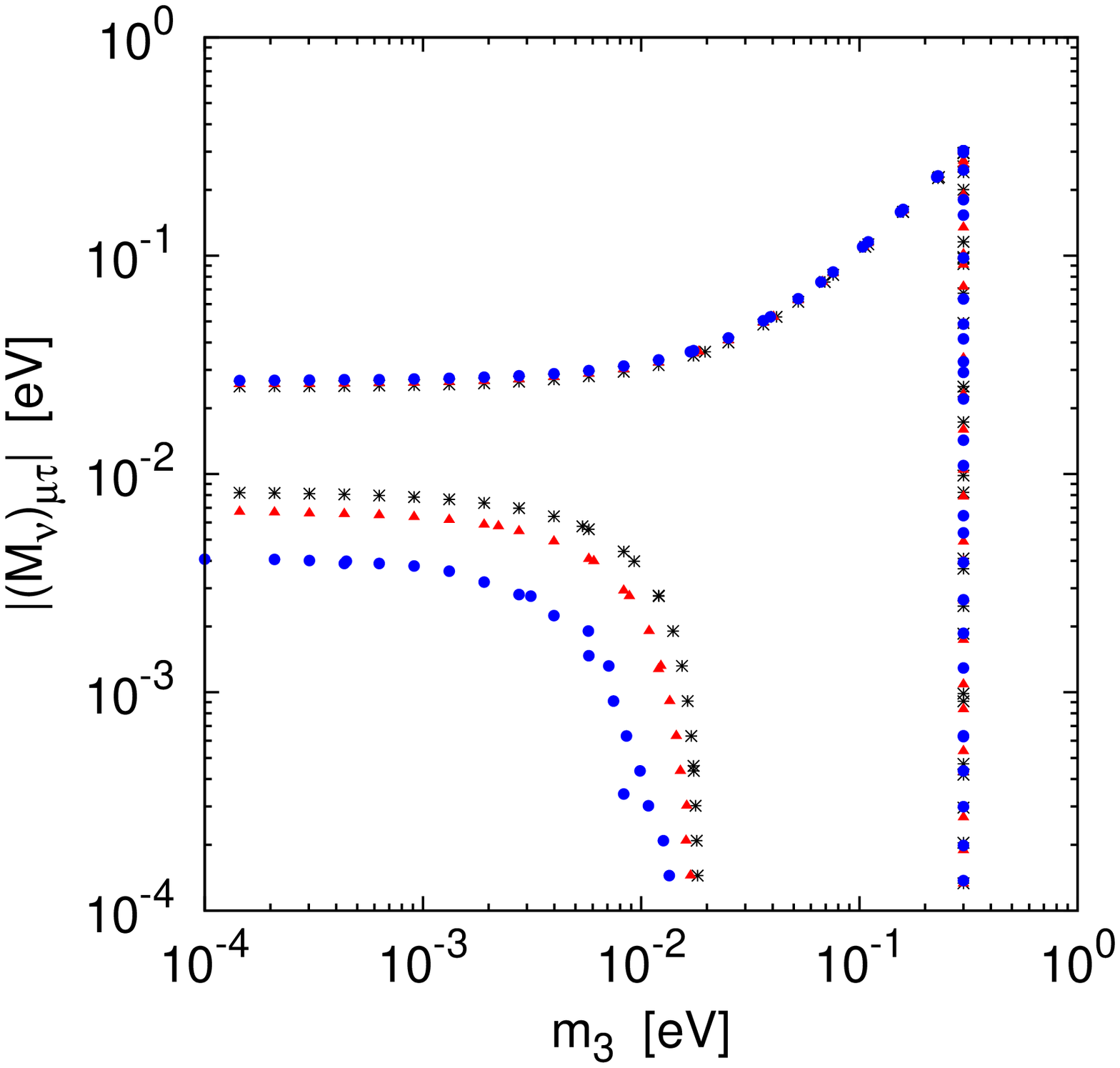}\\
\includegraphics[angle=-90,keepaspectratio=true,scale=\figurescale]
{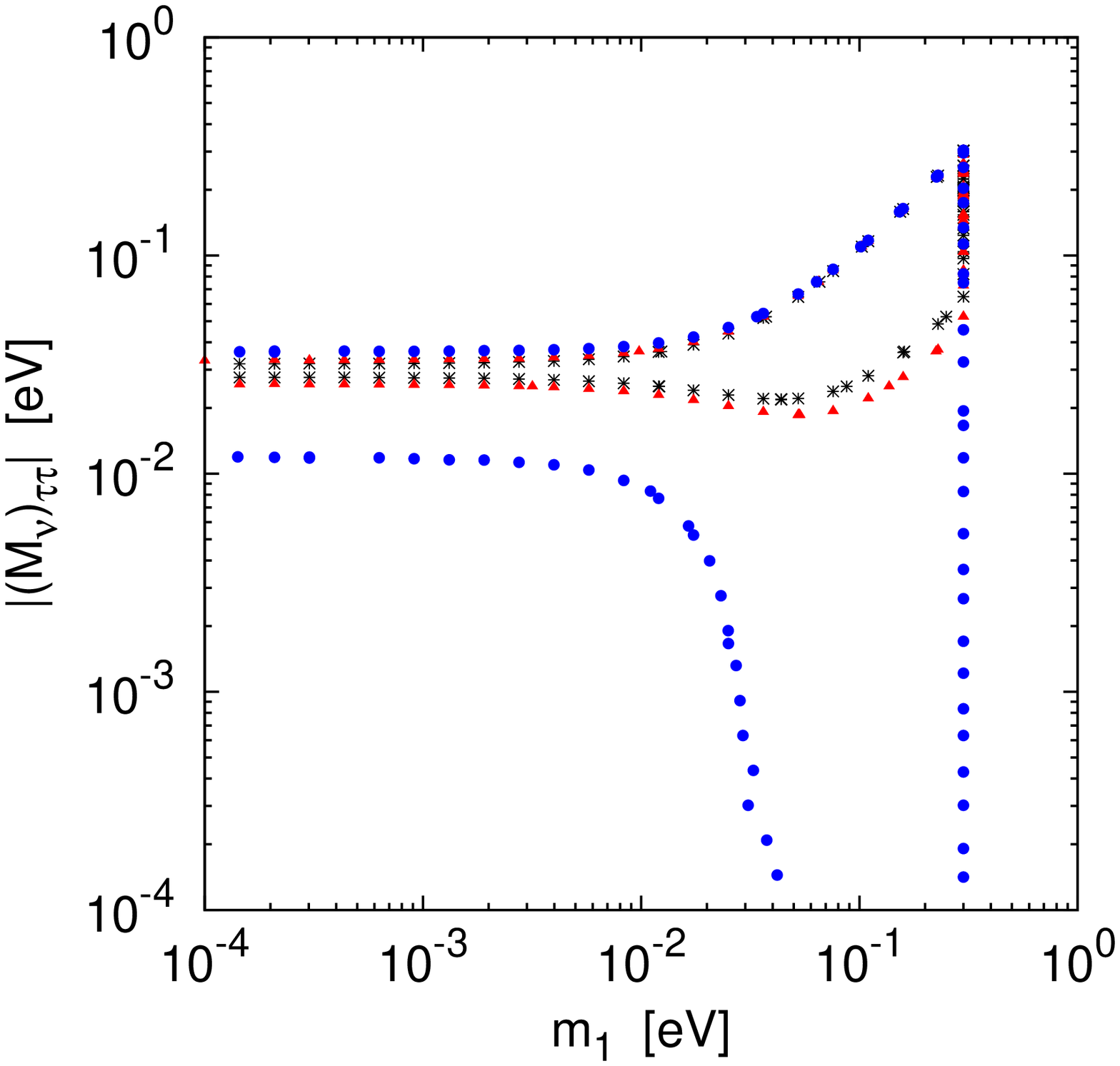} &
\includegraphics[angle=-90,keepaspectratio=true,scale=\figurescale]
{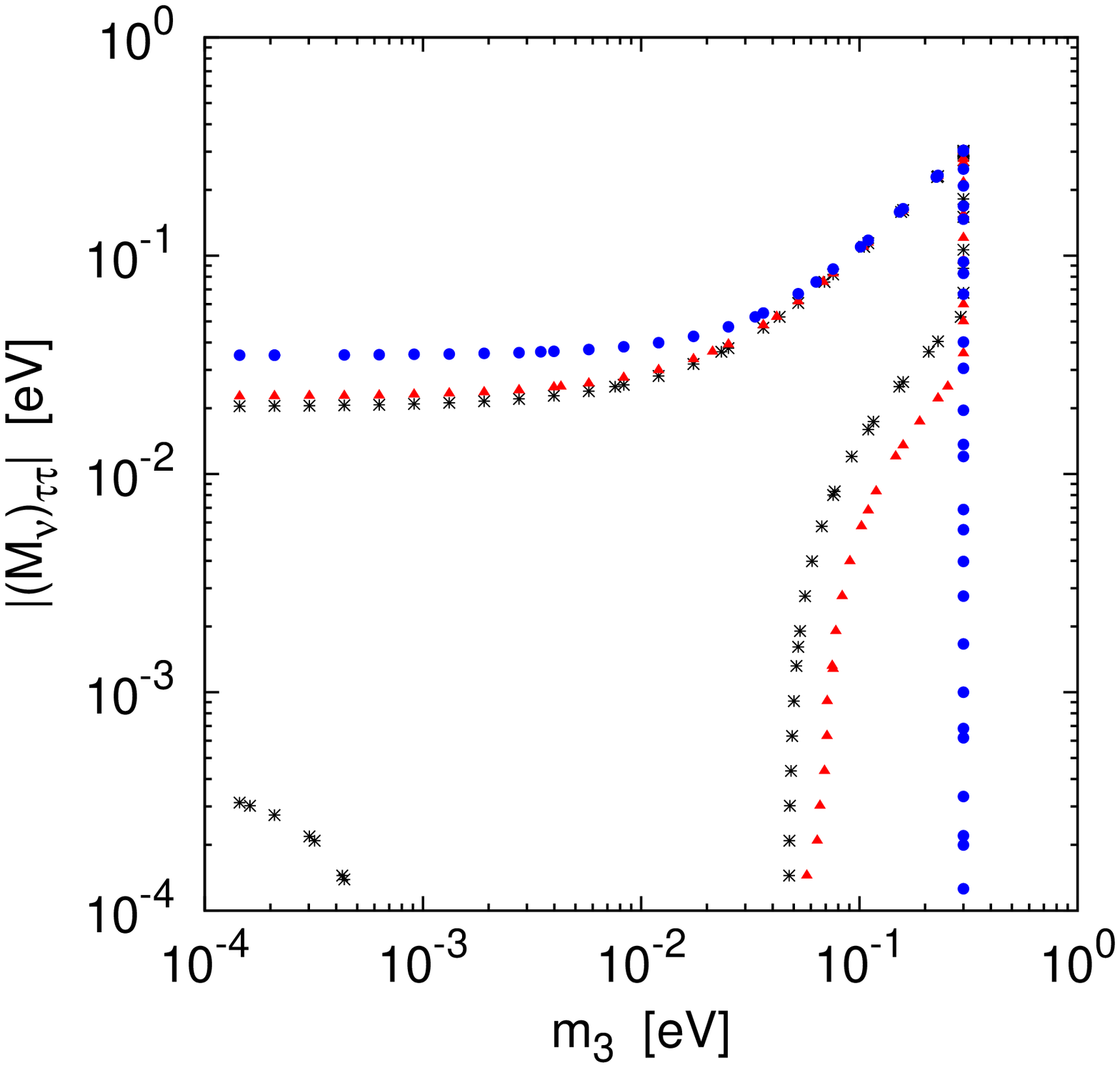}\\
\end{tabular}

\begin{tabular}[t]{ll}
\includegraphics[angle=-90,keepaspectratio=true,scale=\figurescale]
{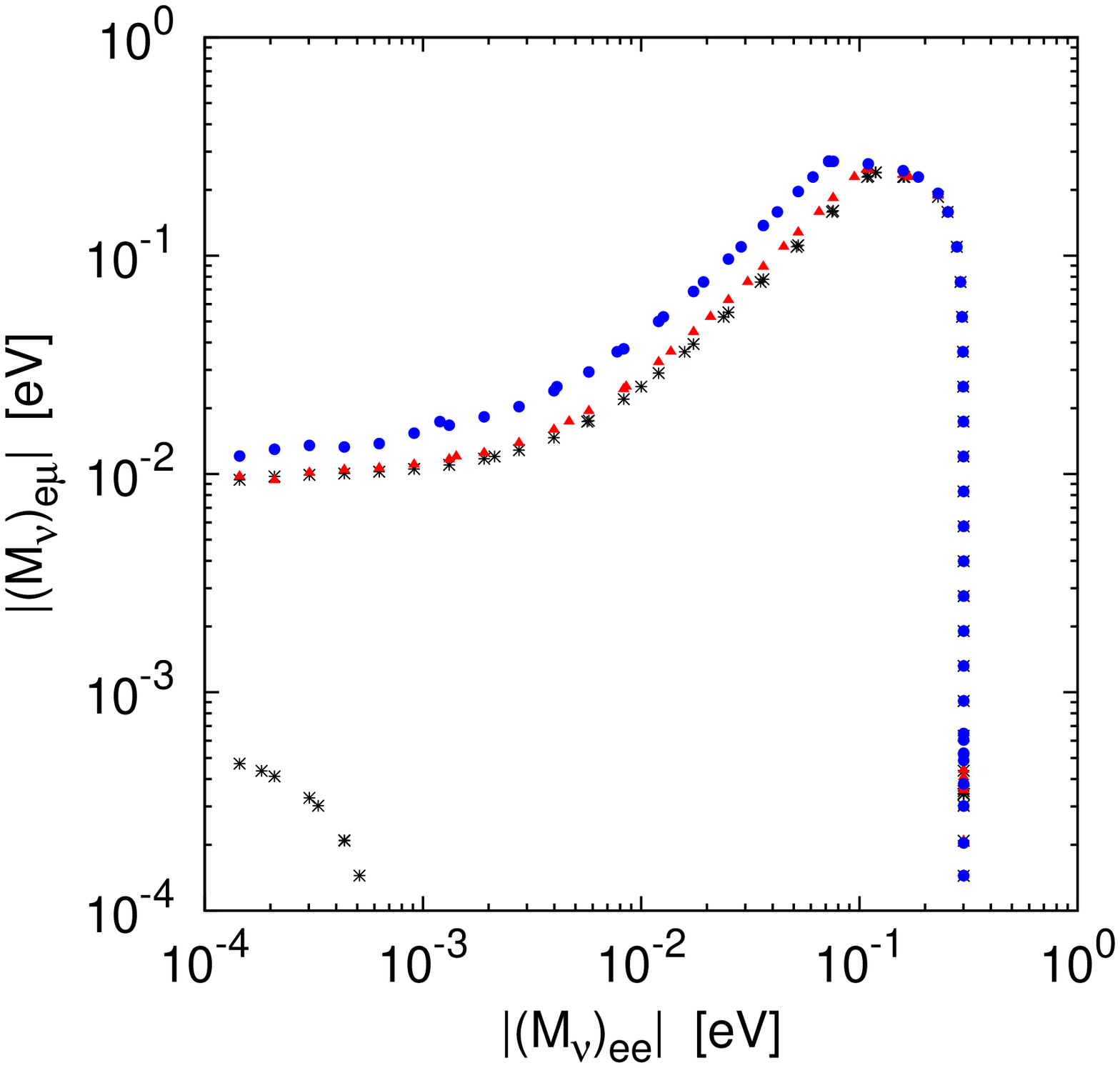} &
\includegraphics[angle=-90,keepaspectratio=true,scale=\figurescale]
{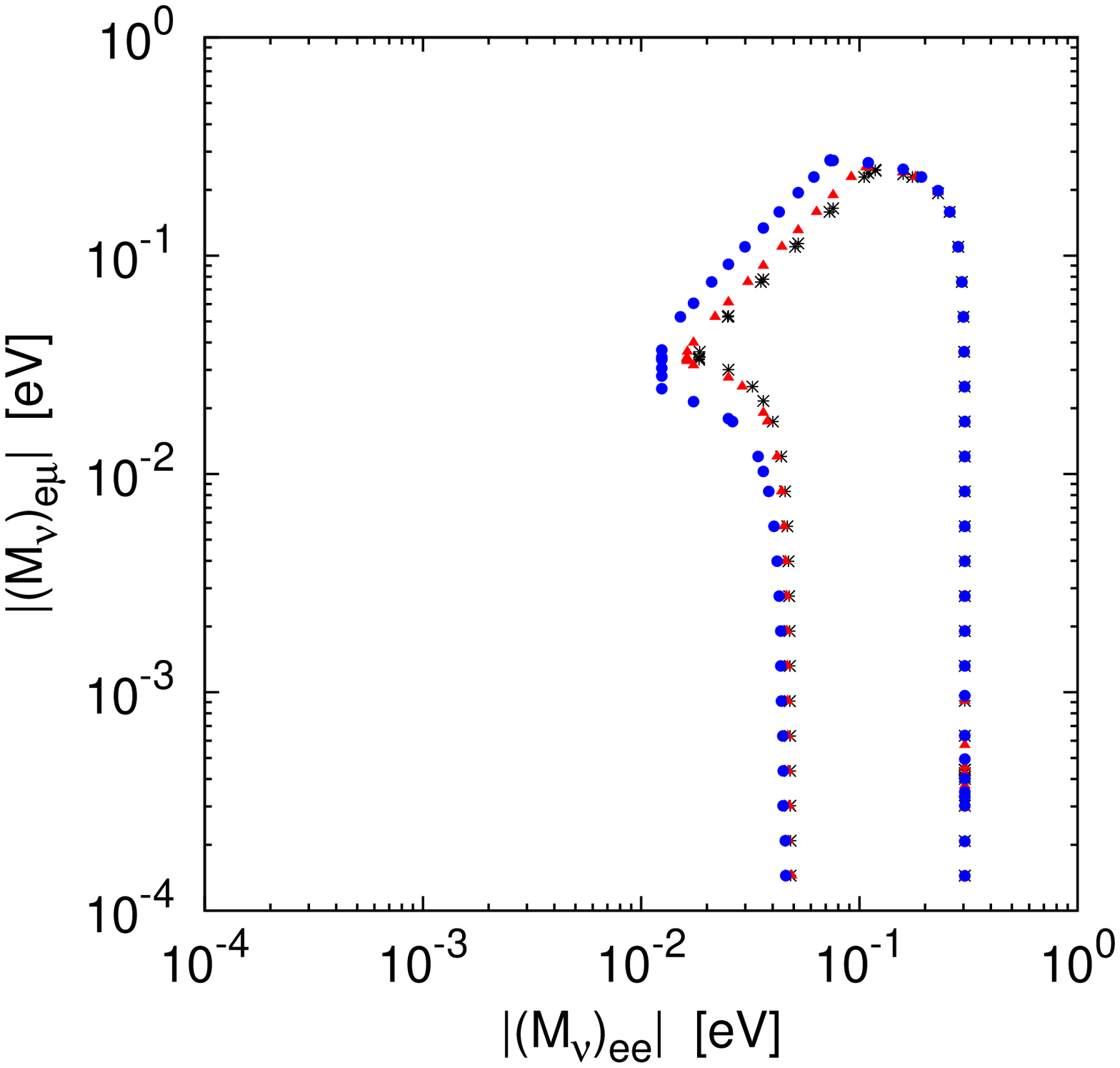}\\
\includegraphics[angle=-90,keepaspectratio=true,scale=\figurescale]
{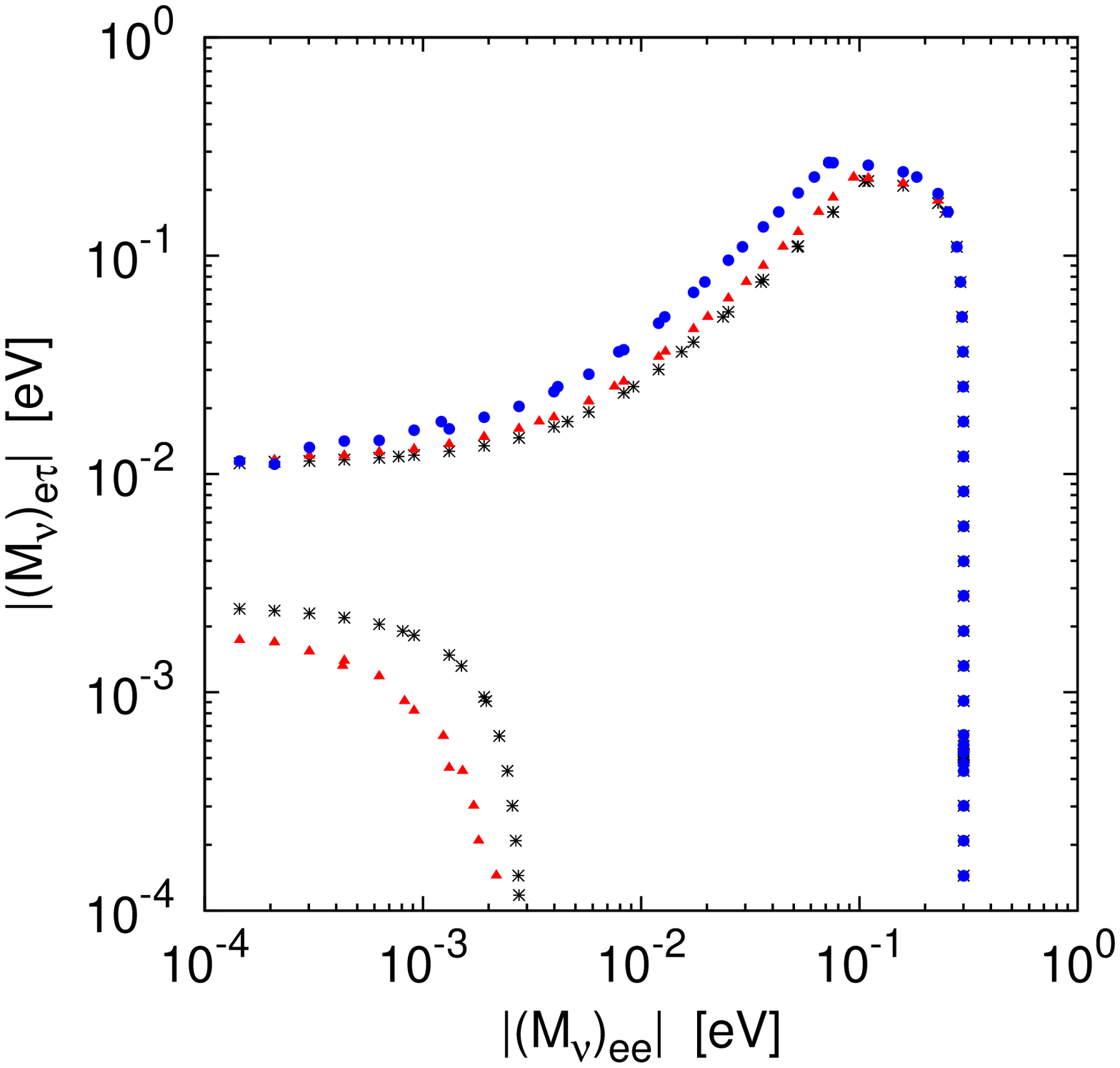} &
\includegraphics[angle=-90,keepaspectratio=true,scale=\figurescale]
{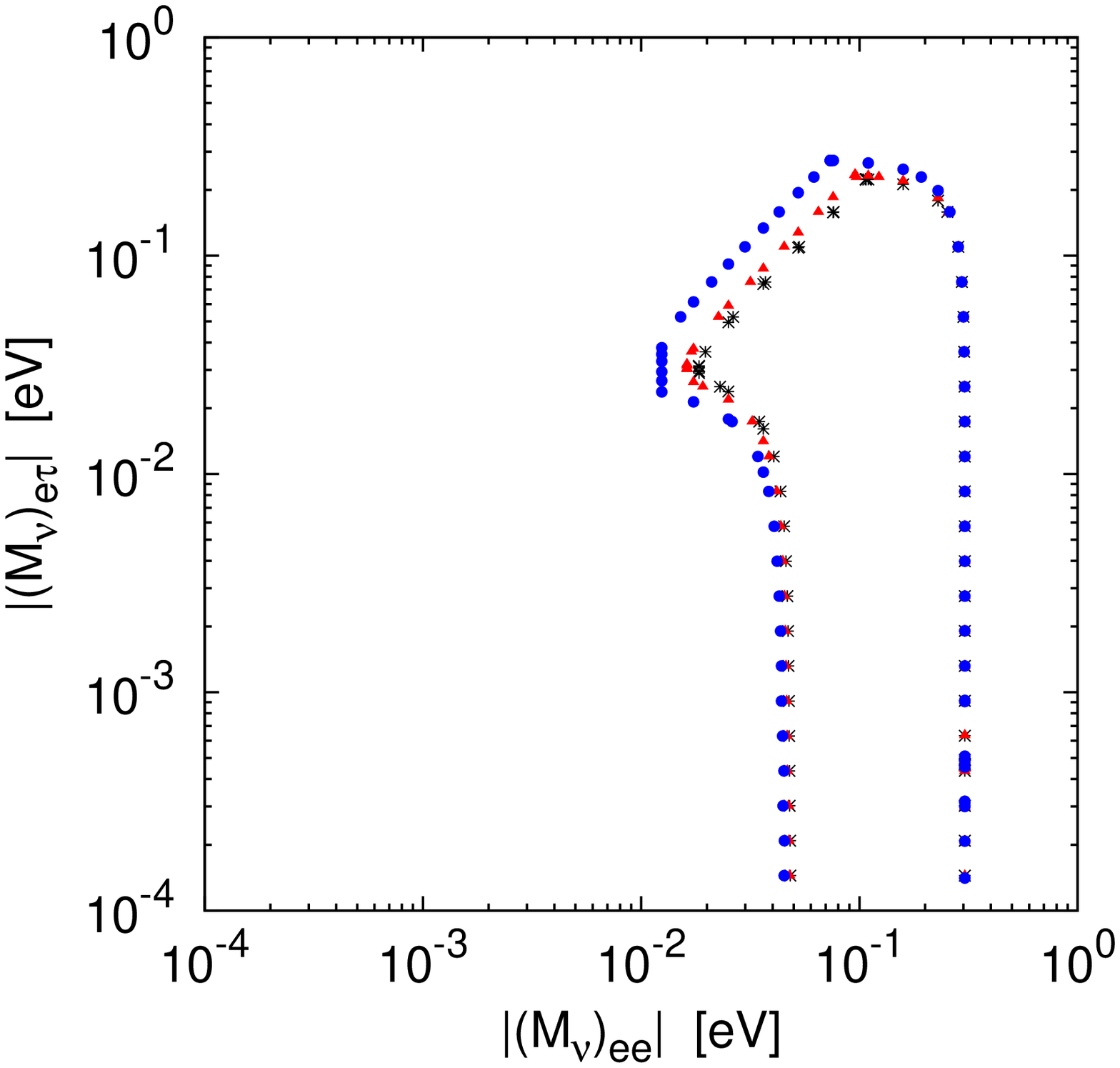}\\
\includegraphics[angle=-90,keepaspectratio=true,scale=\figurescale]
{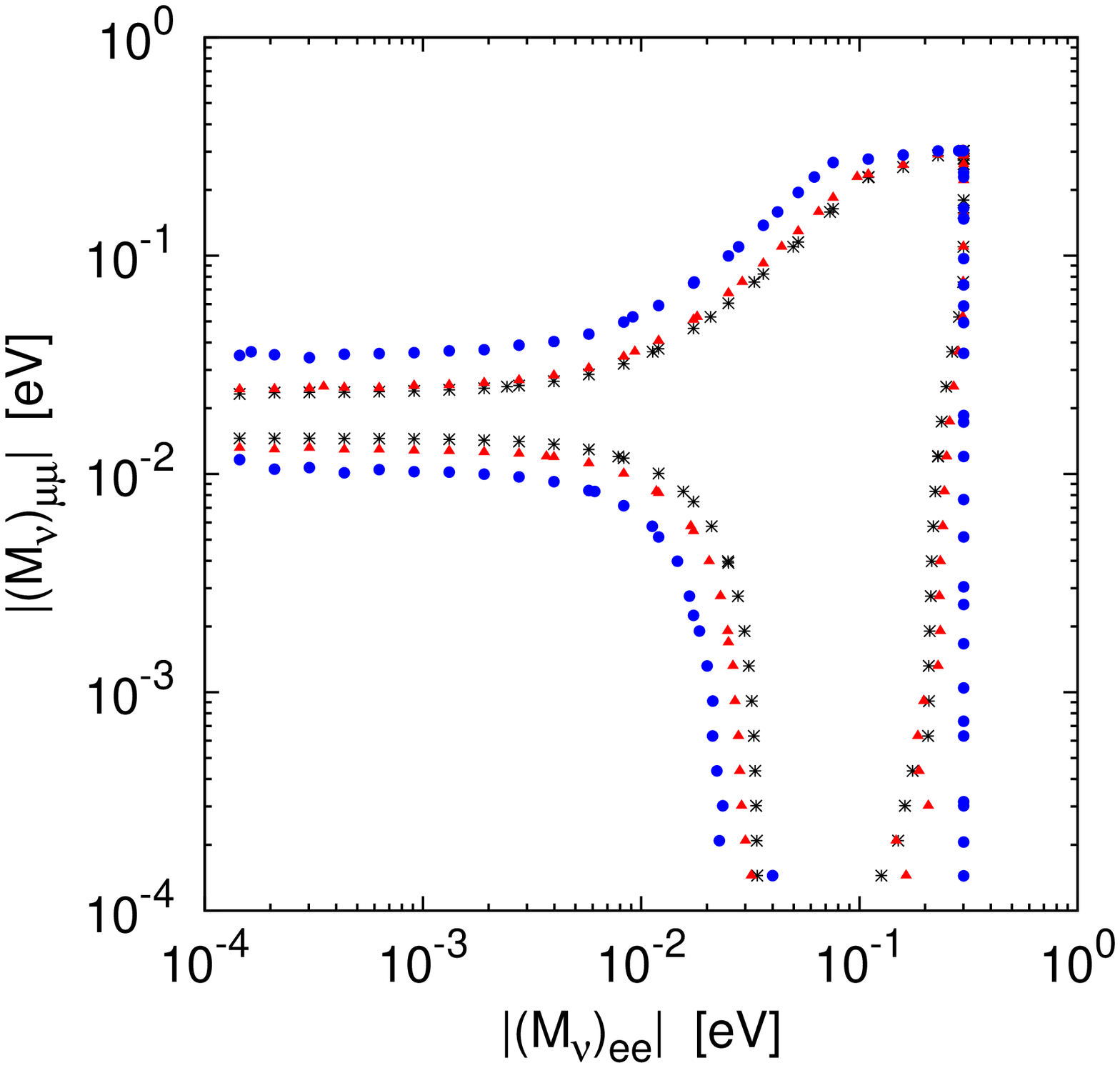} &
\includegraphics[angle=-90,keepaspectratio=true,scale=\figurescale]
{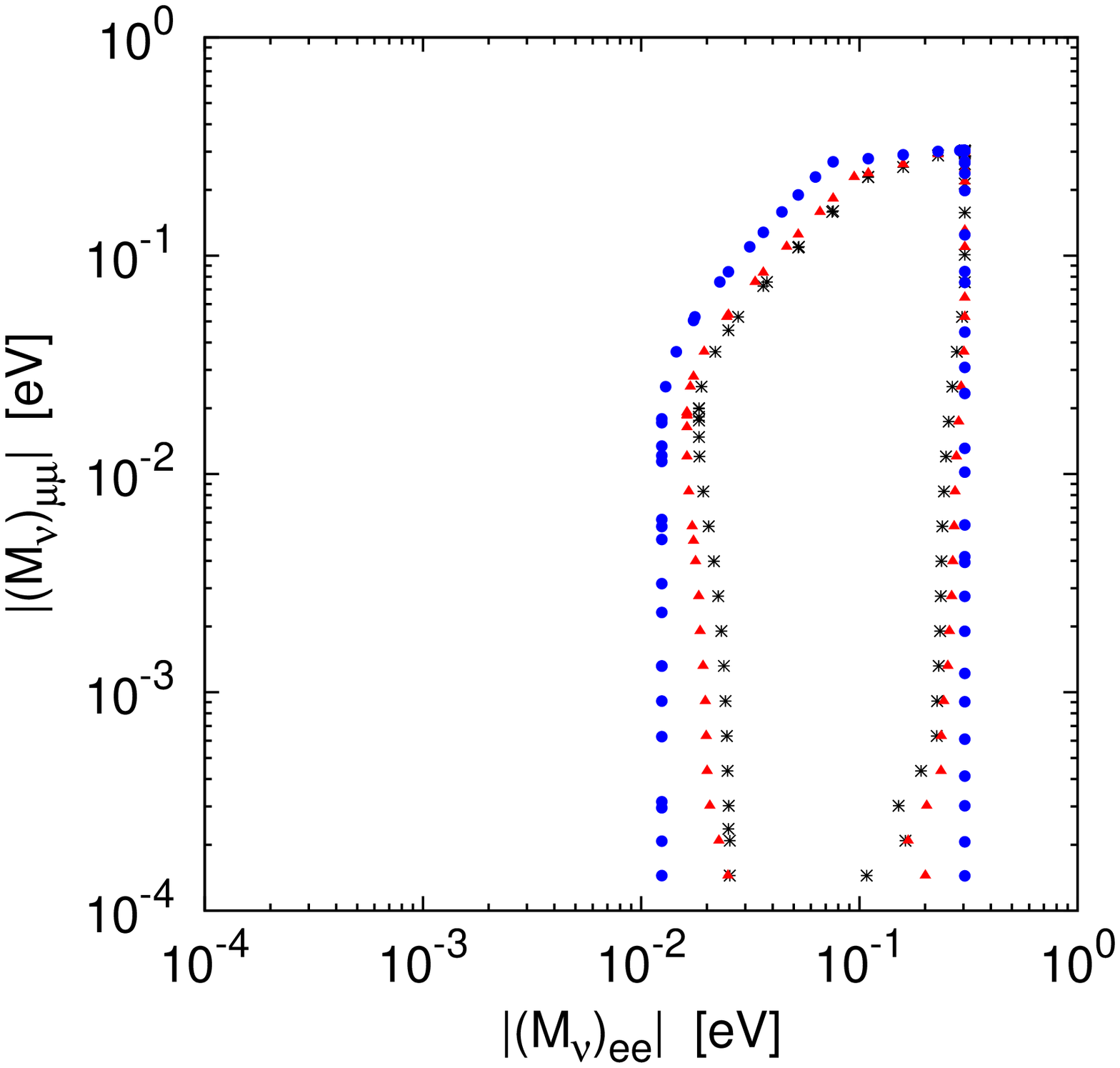}\\
\end{tabular}

\begin{tabular}[t]{ll}
\includegraphics[angle=-90,keepaspectratio=true,scale=\figurescale]
{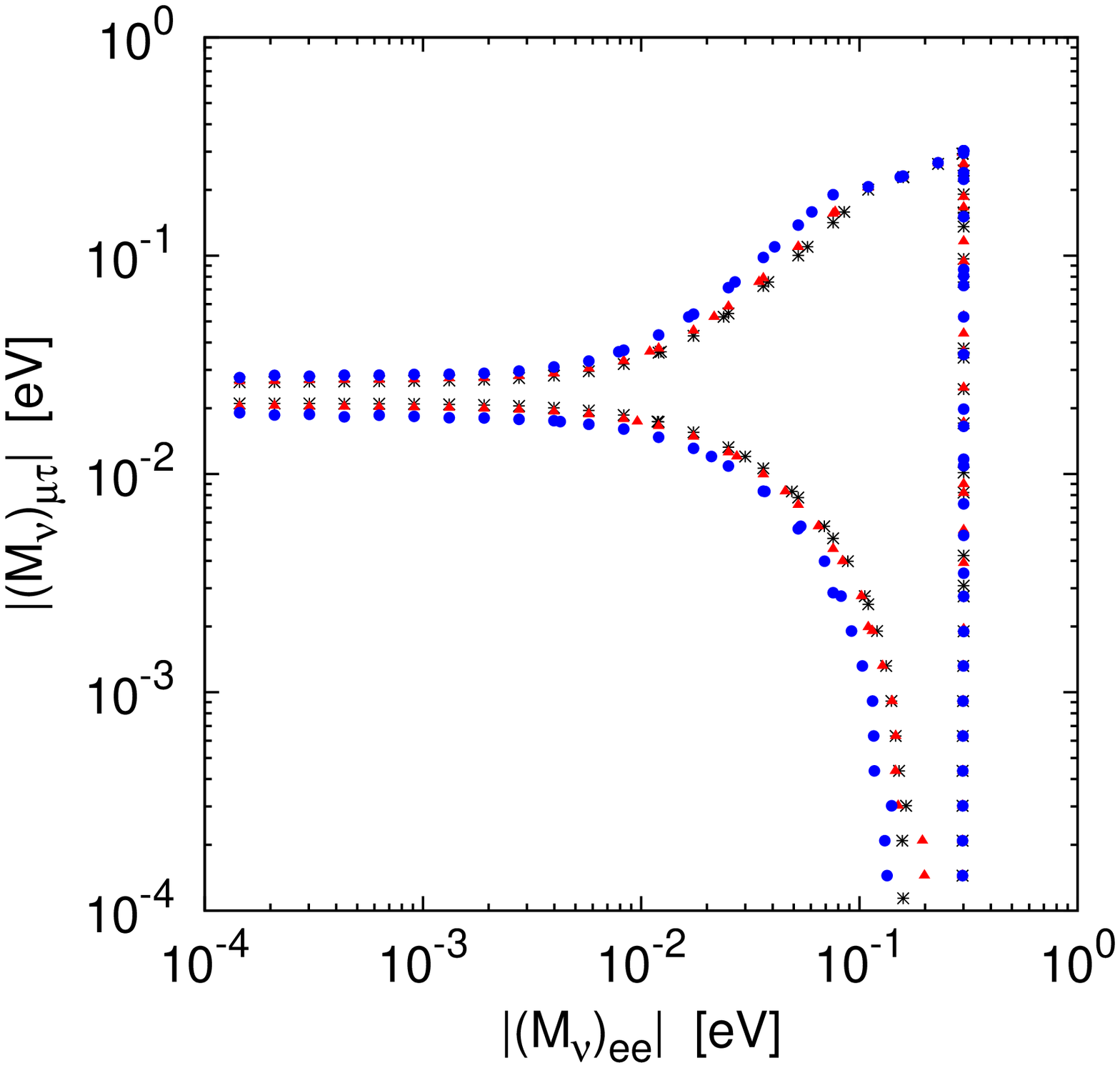} &
\includegraphics[angle=-90,keepaspectratio=true,scale=\figurescale]
{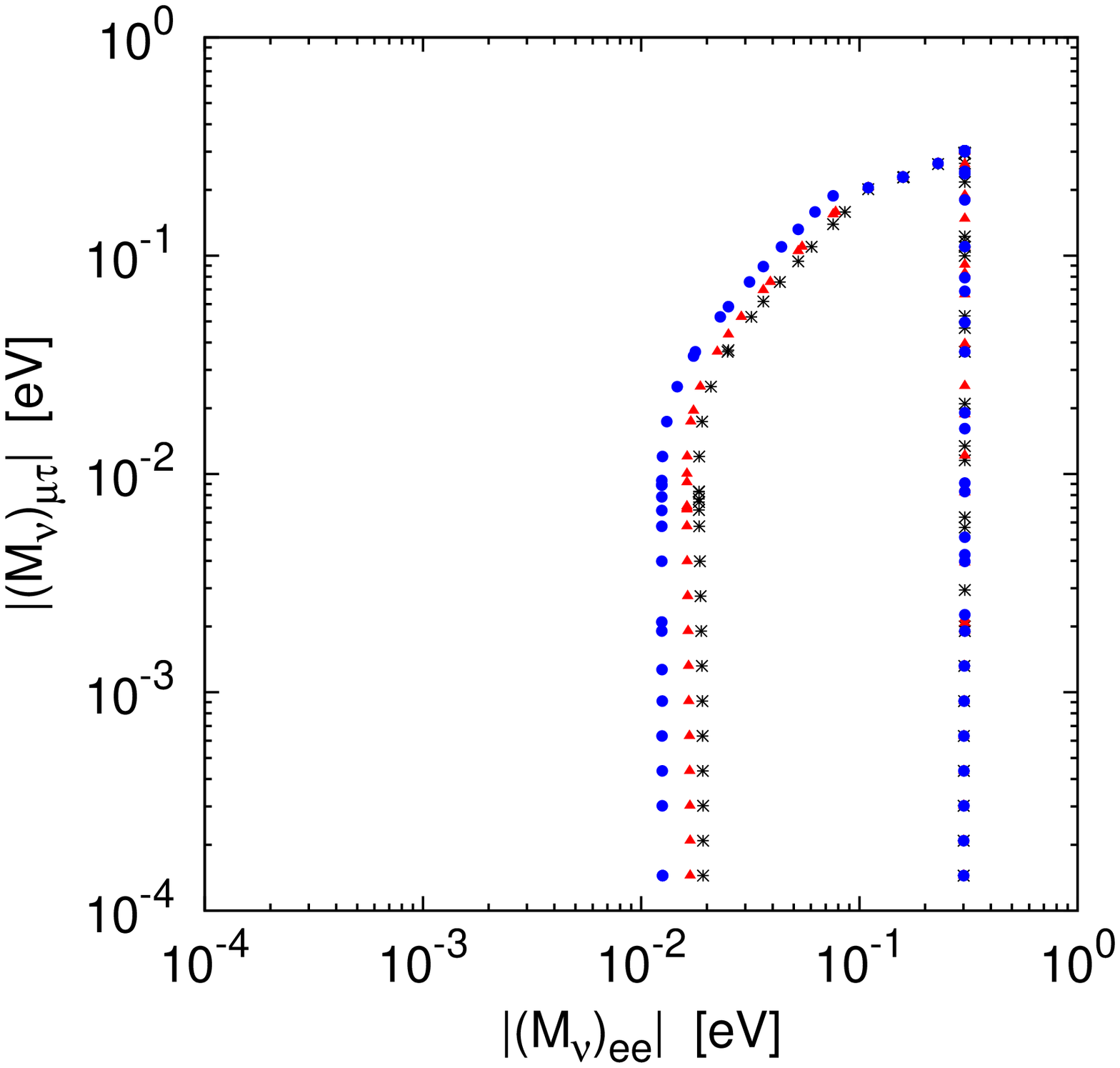}\\
\includegraphics[angle=-90,keepaspectratio=true,scale=\figurescale]
{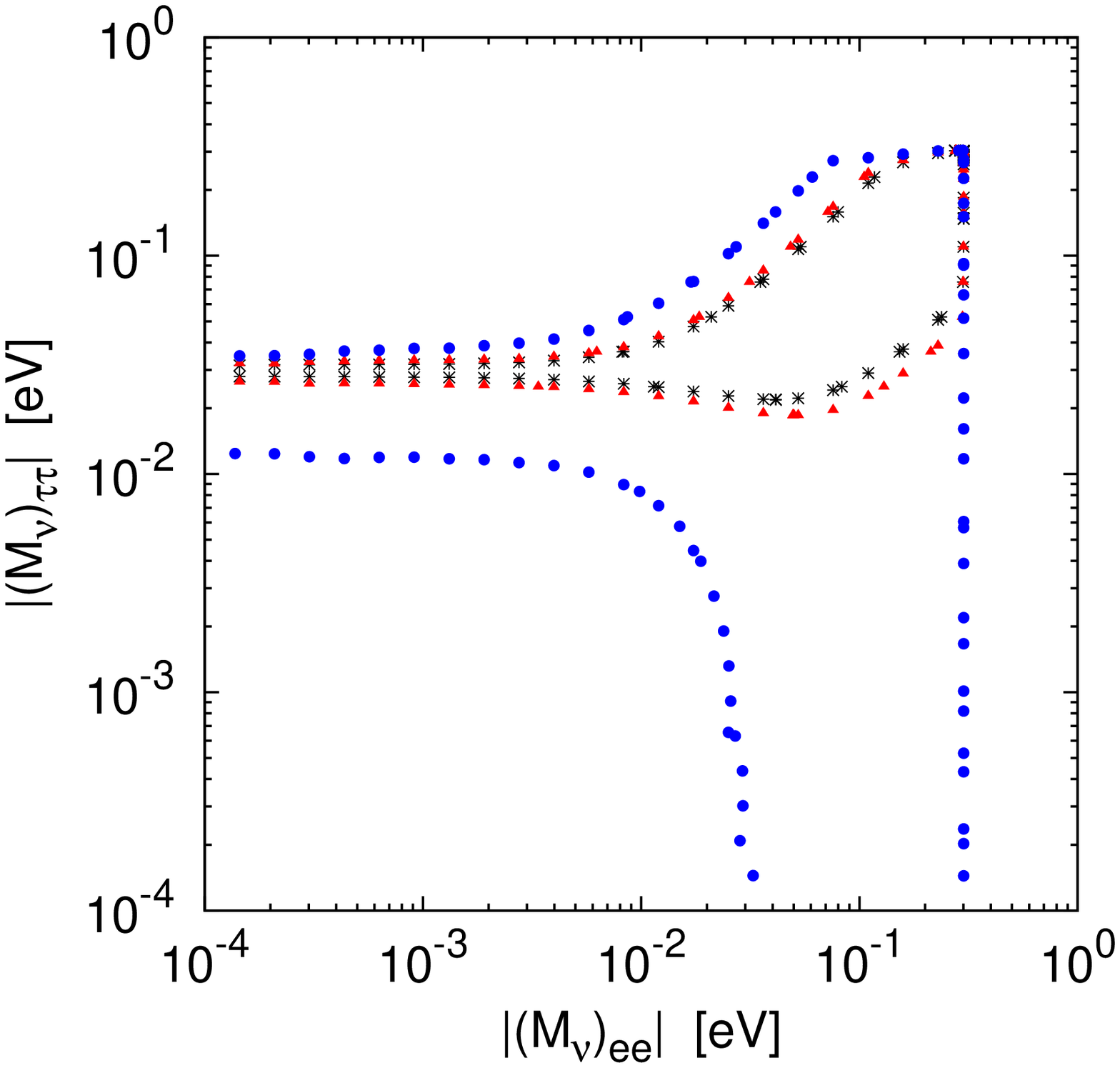} &
\includegraphics[angle=-90,keepaspectratio=true,scale=\figurescale]
{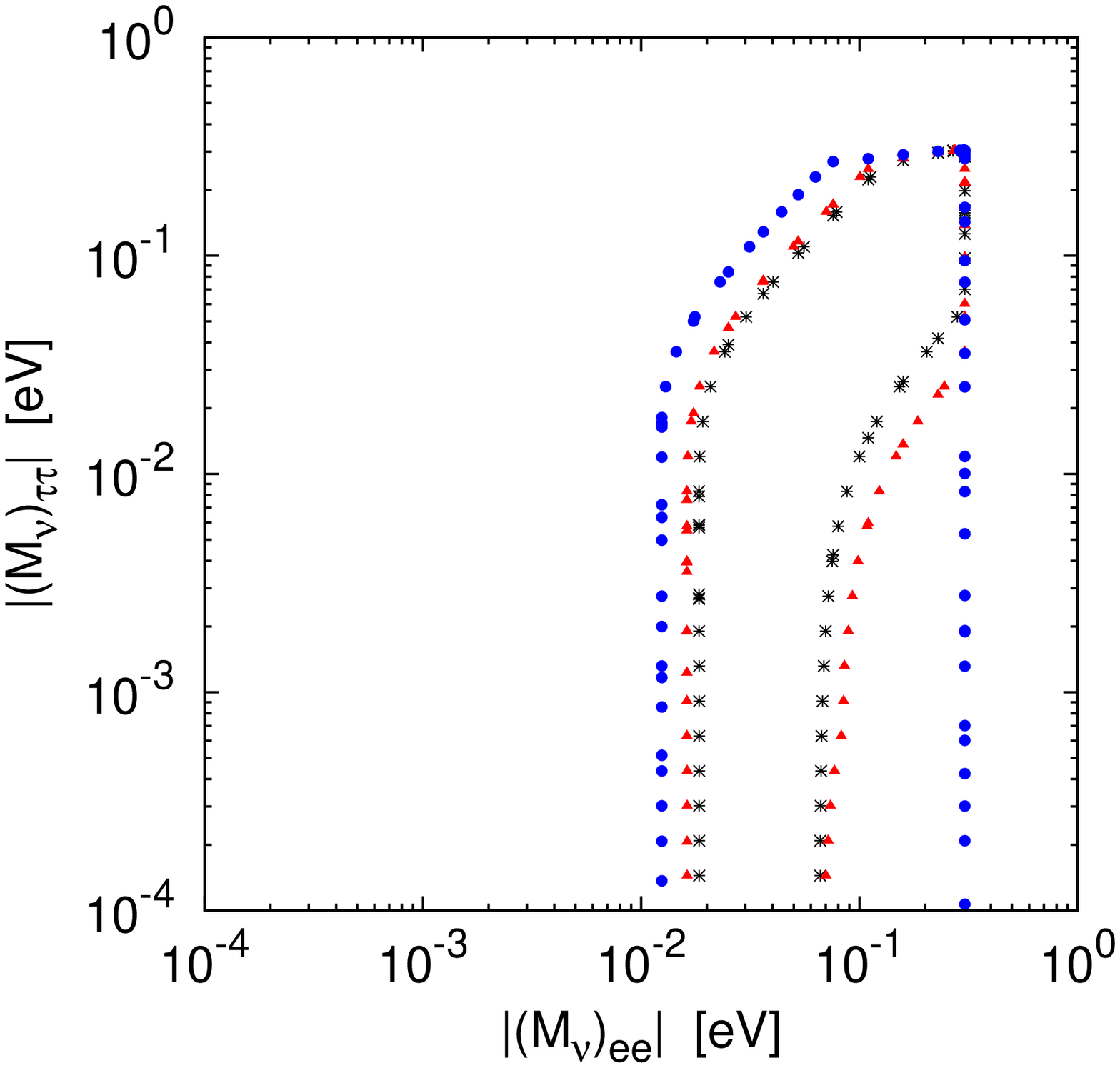}\\
\includegraphics[angle=-90,keepaspectratio=true,scale=\figurescale]
{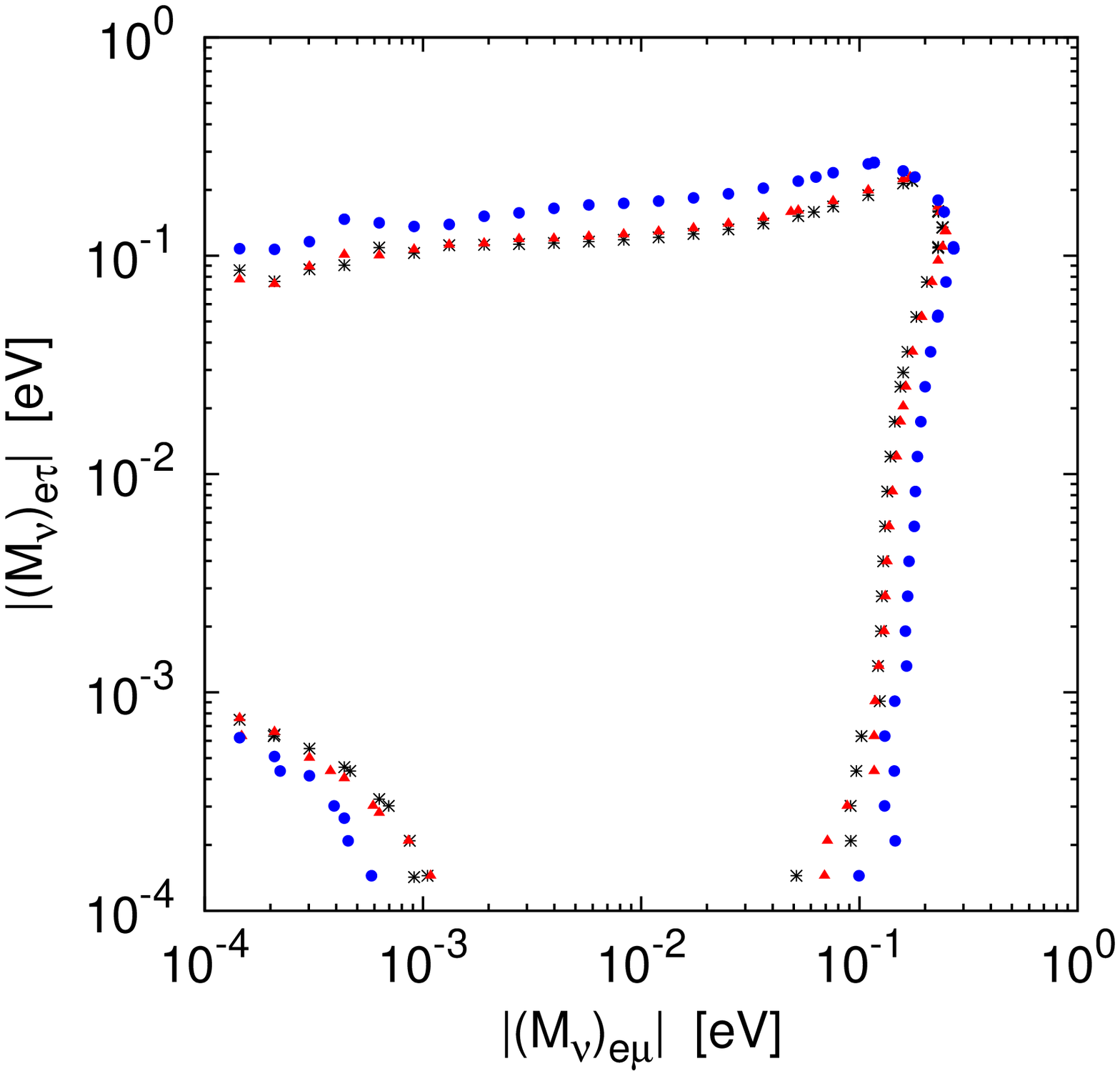} &
\includegraphics[angle=-90,keepaspectratio=true,scale=\figurescale]
{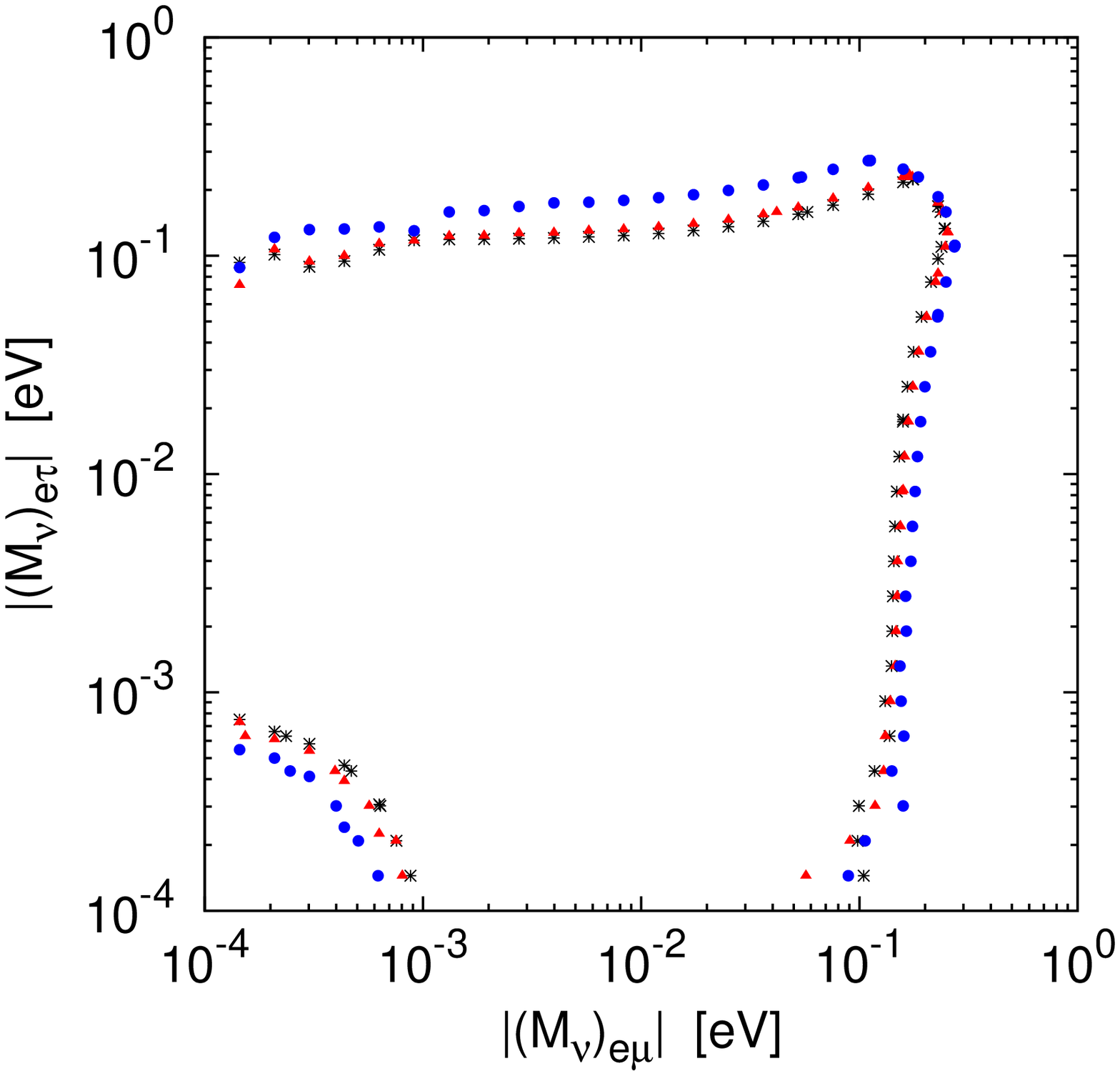}\\
\end{tabular}

\begin{tabular}[t]{ll}
\includegraphics[angle=-90,keepaspectratio=true,scale=\figurescale]
{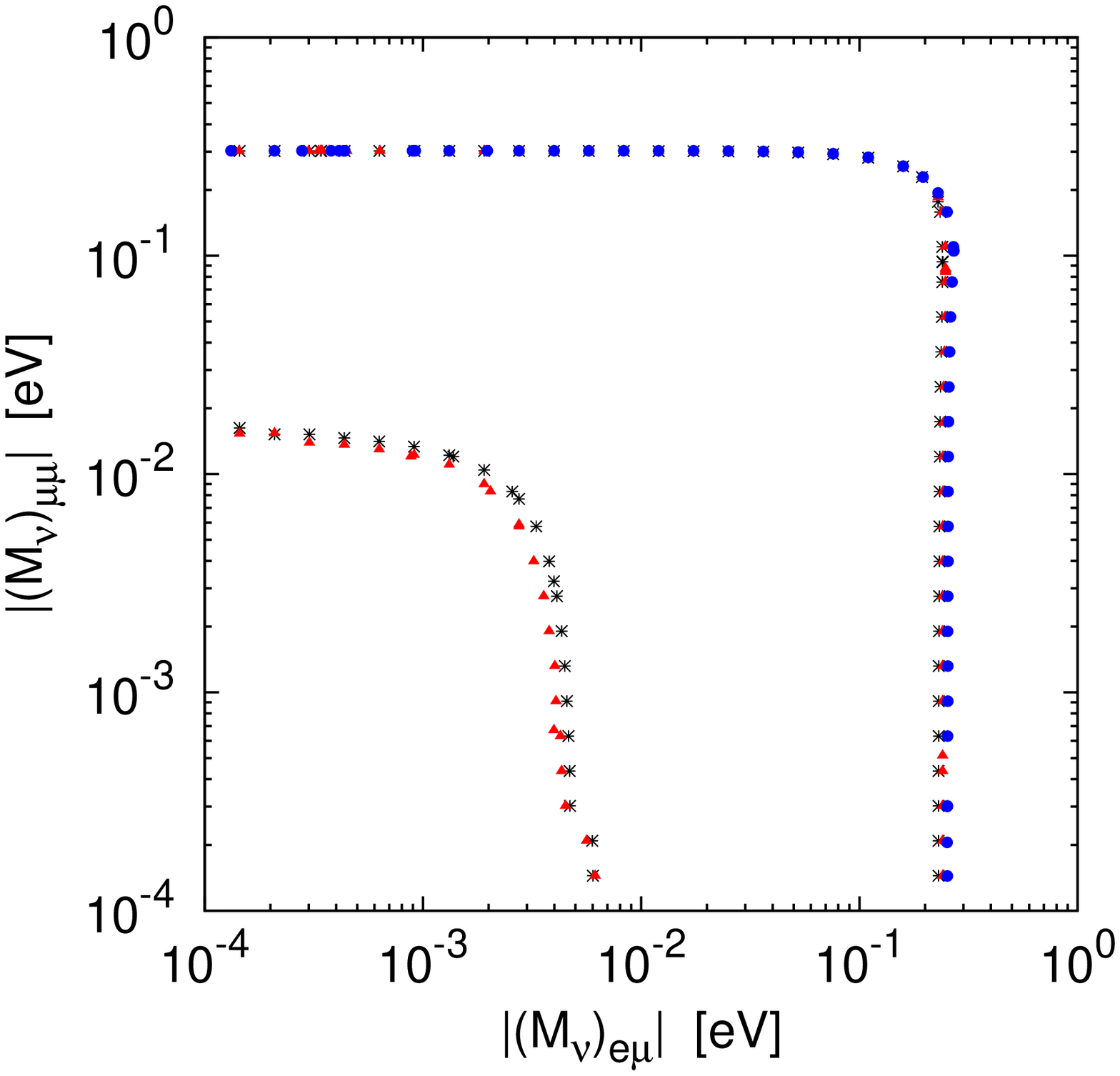} &
\includegraphics[angle=-90,keepaspectratio=true,scale=\figurescale]
{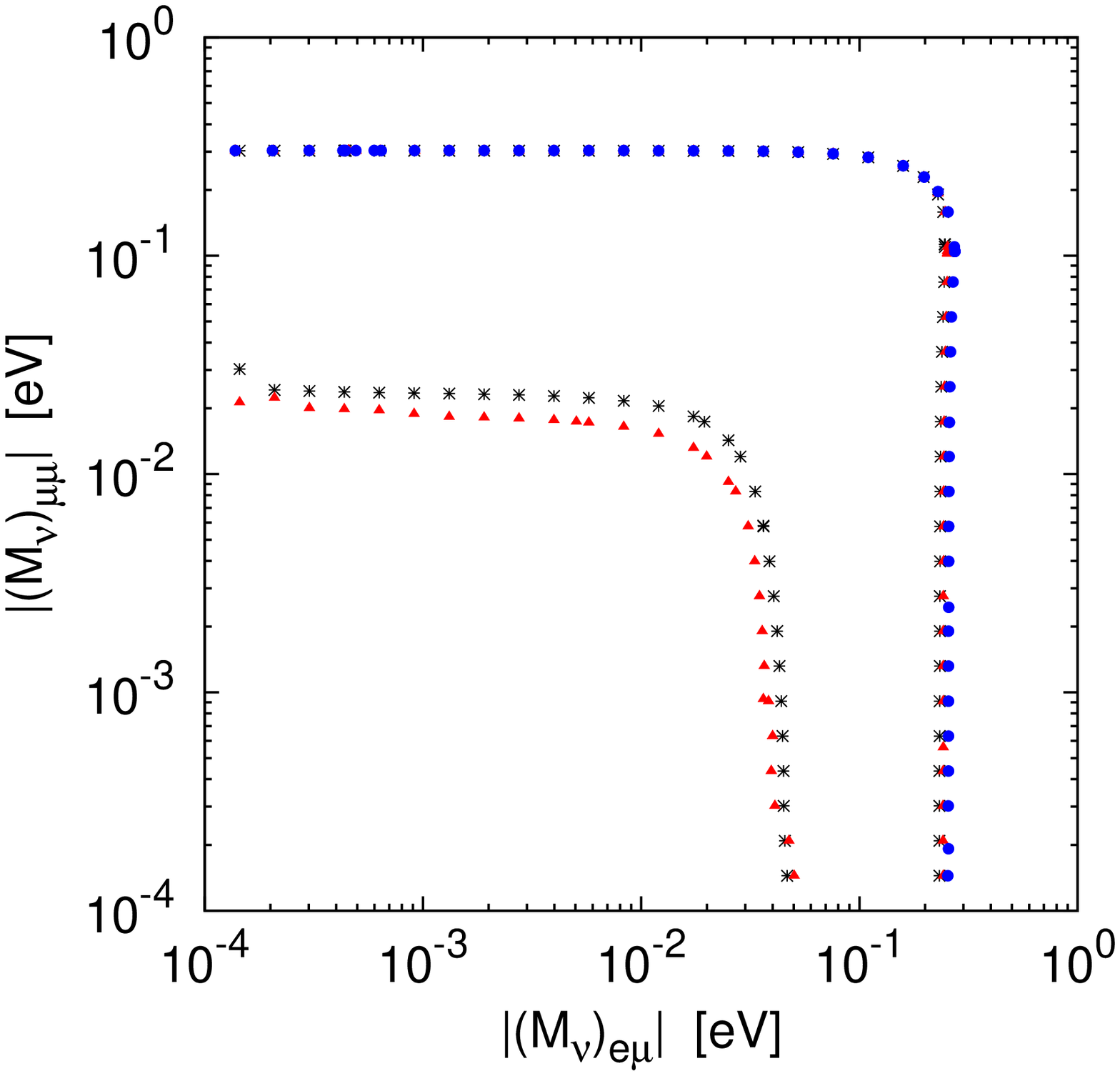}\\
\includegraphics[angle=-90,keepaspectratio=true,scale=\figurescale]
{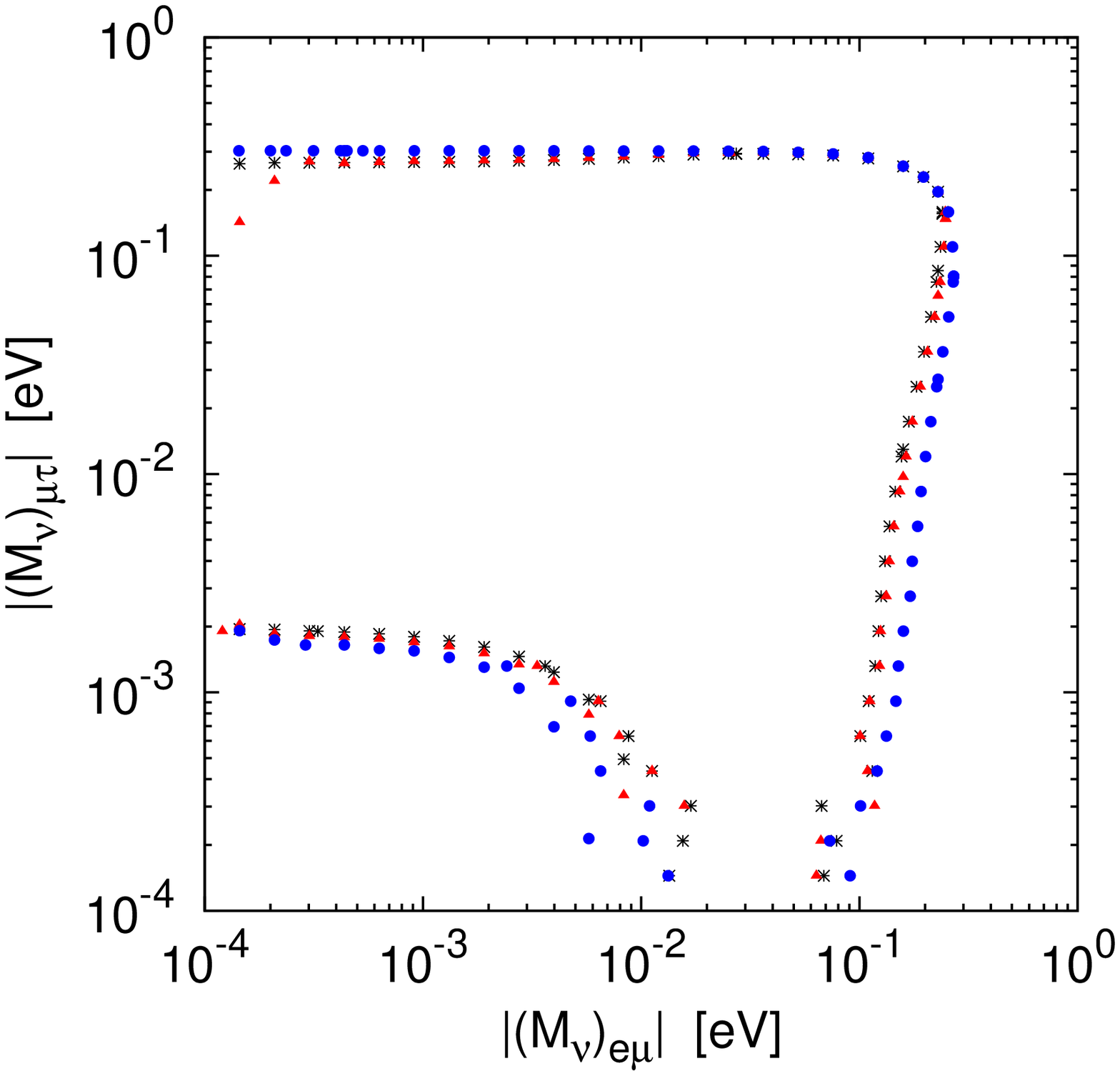} &
\includegraphics[angle=-90,keepaspectratio=true,scale=\figurescale]
{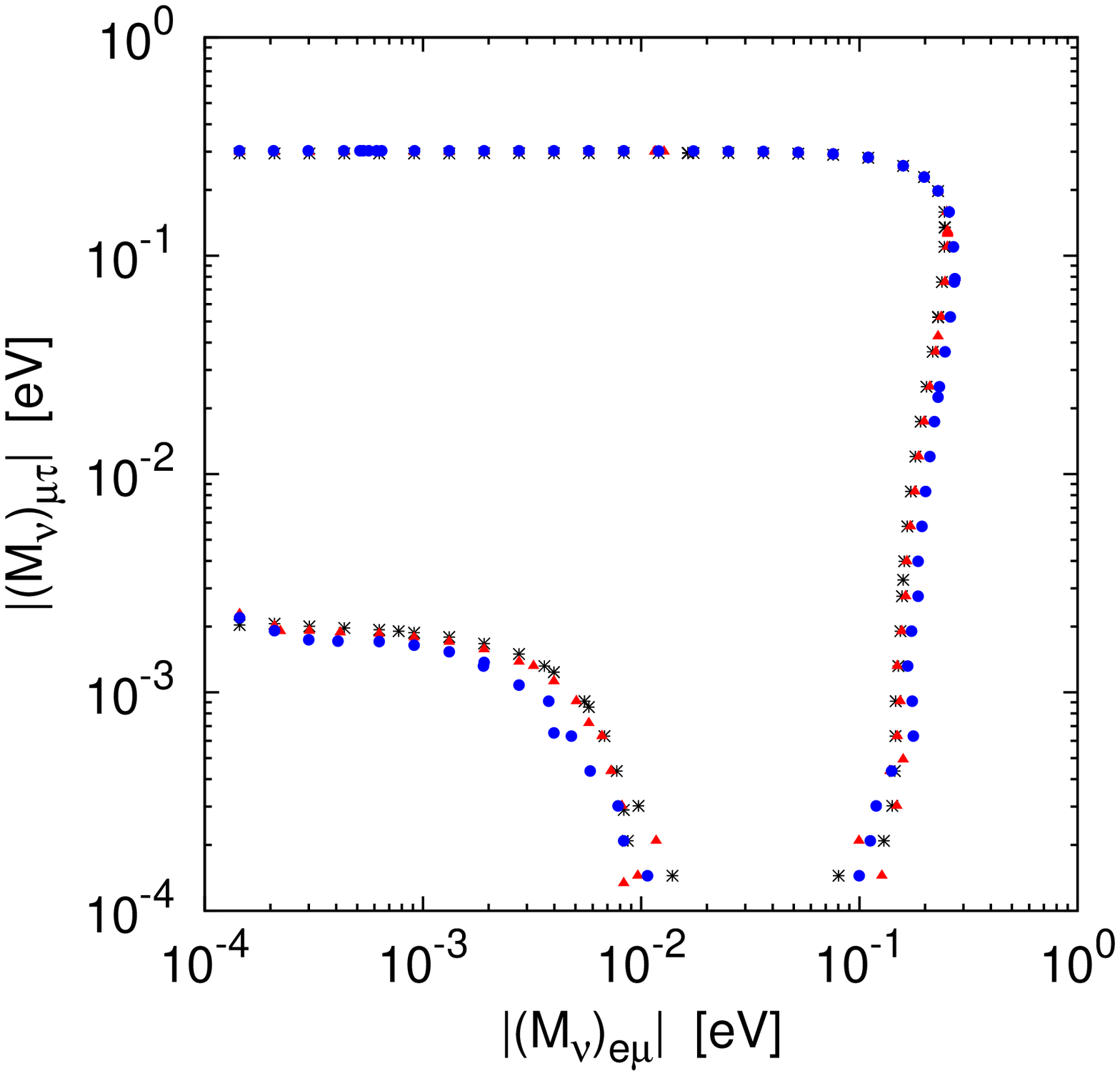}\\
\includegraphics[angle=-90,keepaspectratio=true,scale=\figurescale]
{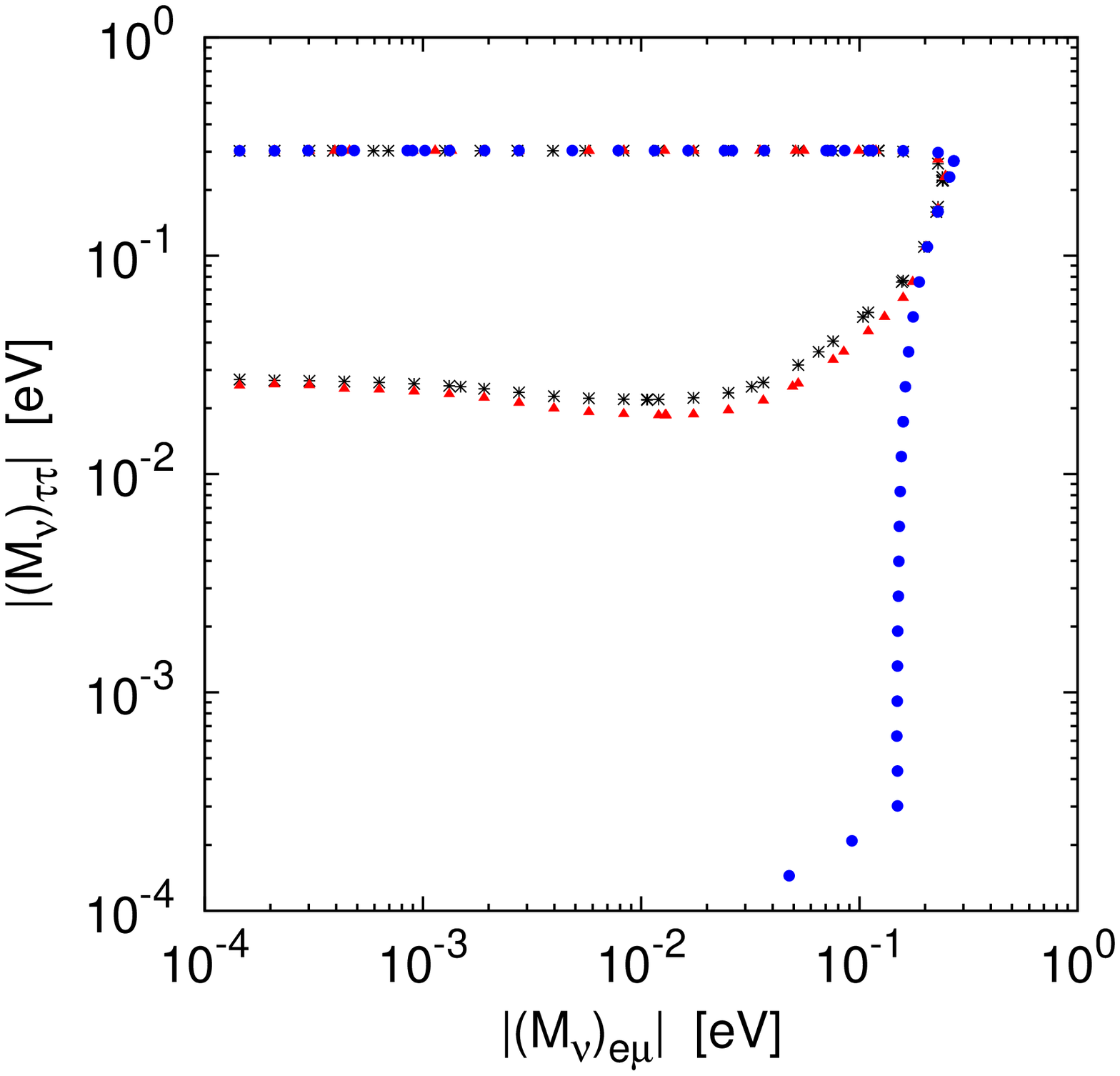} &
\includegraphics[angle=-90,keepaspectratio=true,scale=\figurescale]
{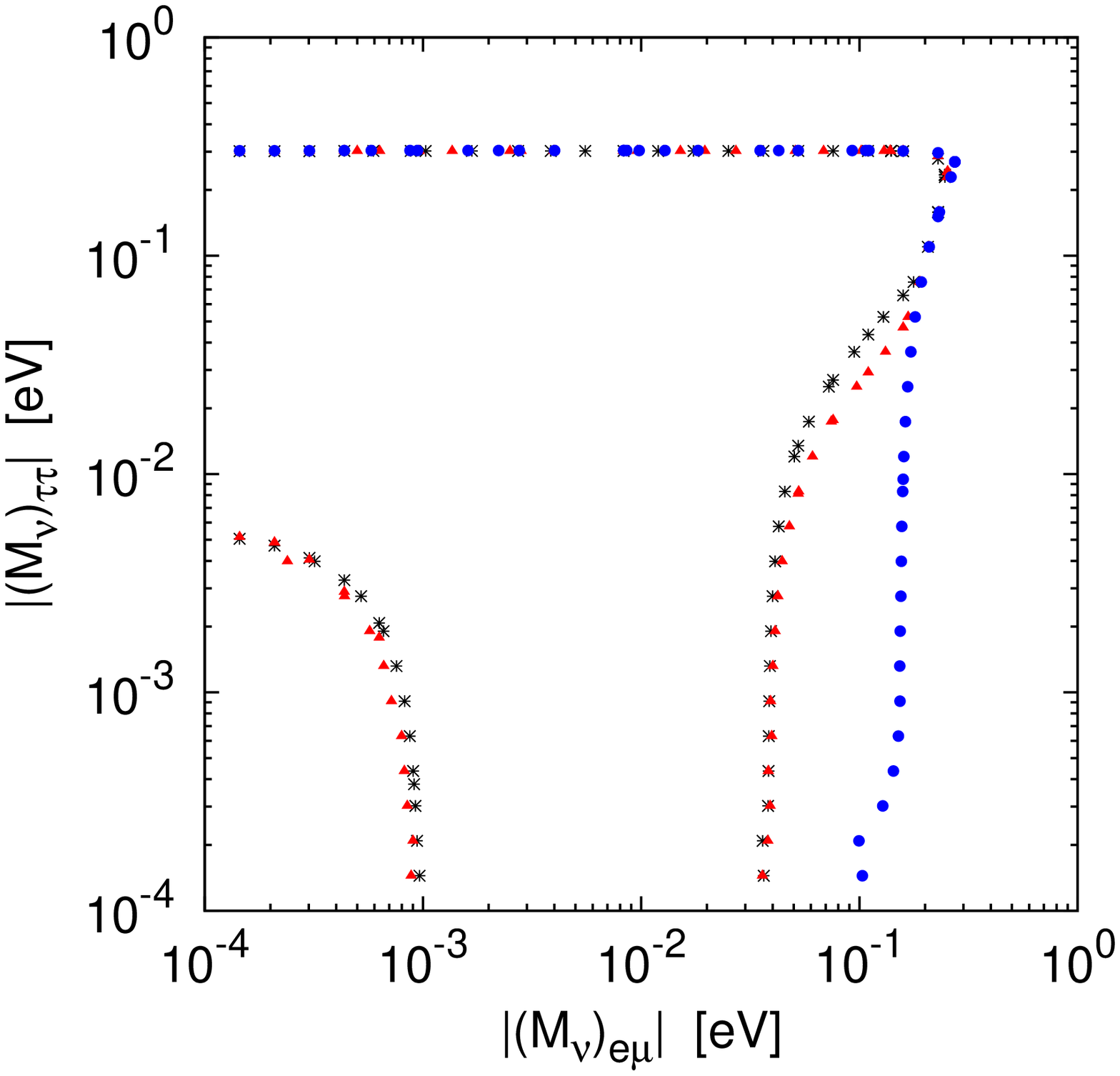}\\
\end{tabular}

\begin{tabular}[t]{ll}
\includegraphics[angle=-90,keepaspectratio=true,scale=\figurescale]
{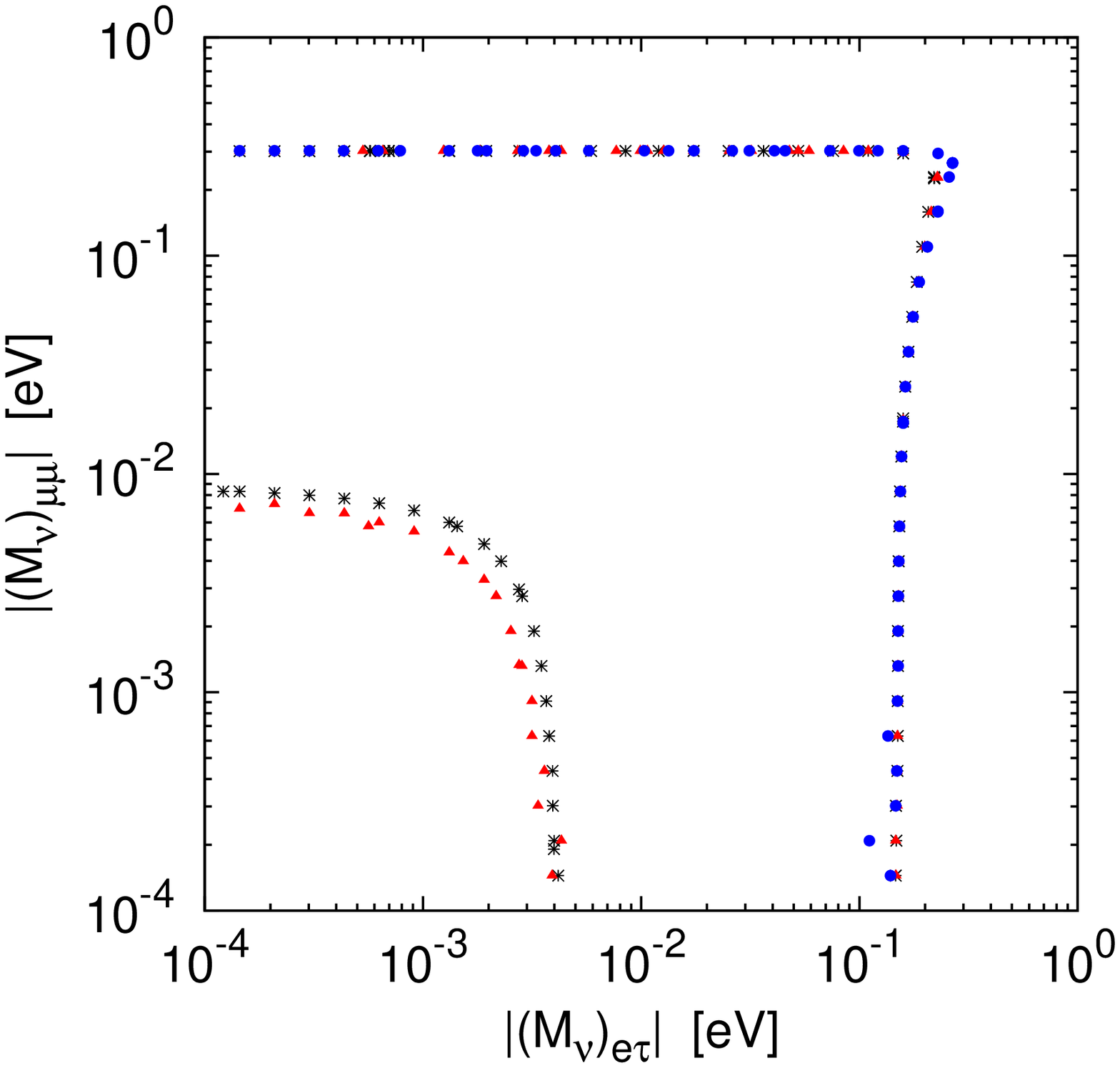} &
\includegraphics[angle=-90,keepaspectratio=true,scale=\figurescale]
{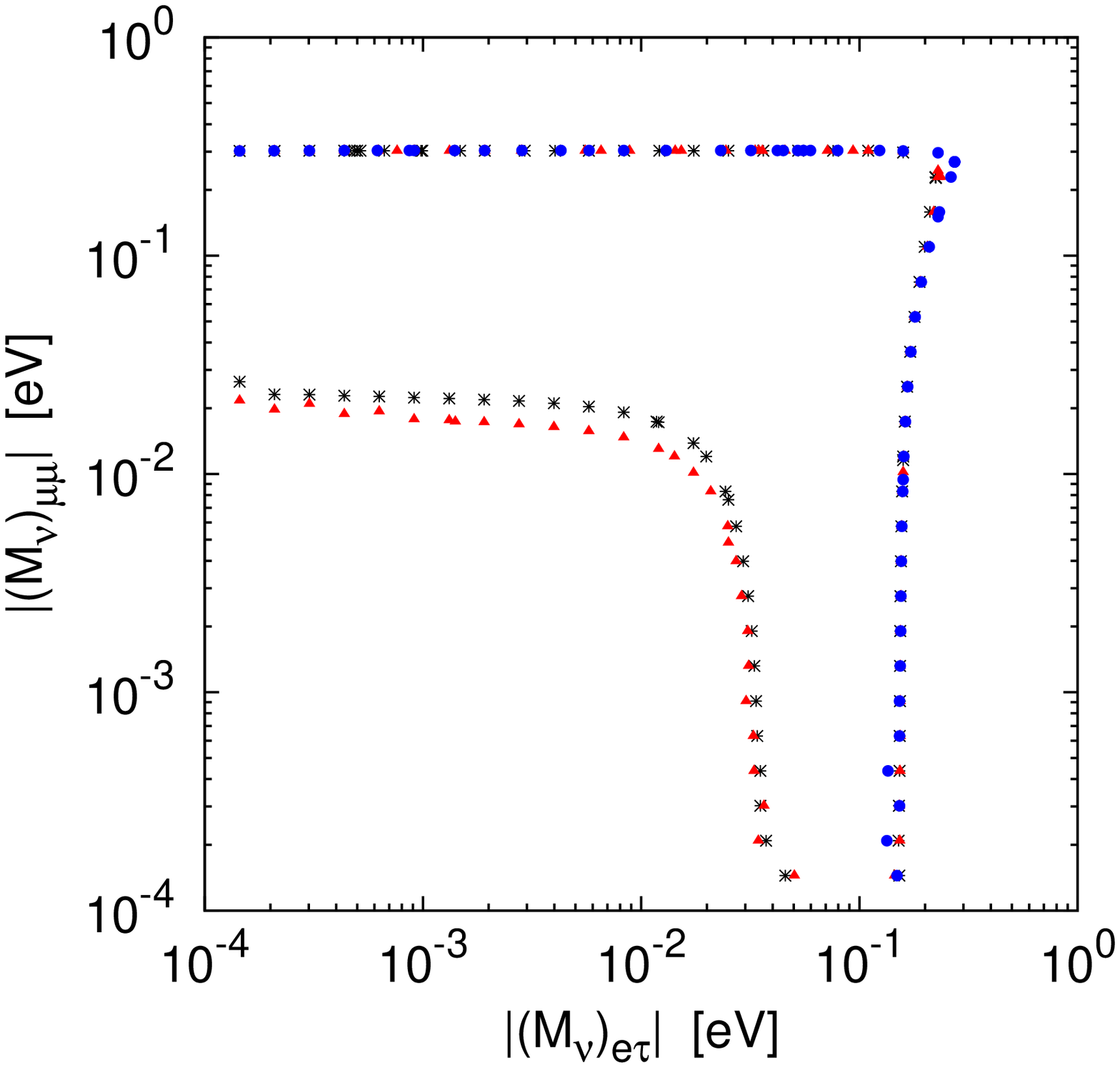}\\
\includegraphics[angle=-90,keepaspectratio=true,scale=\figurescale]
{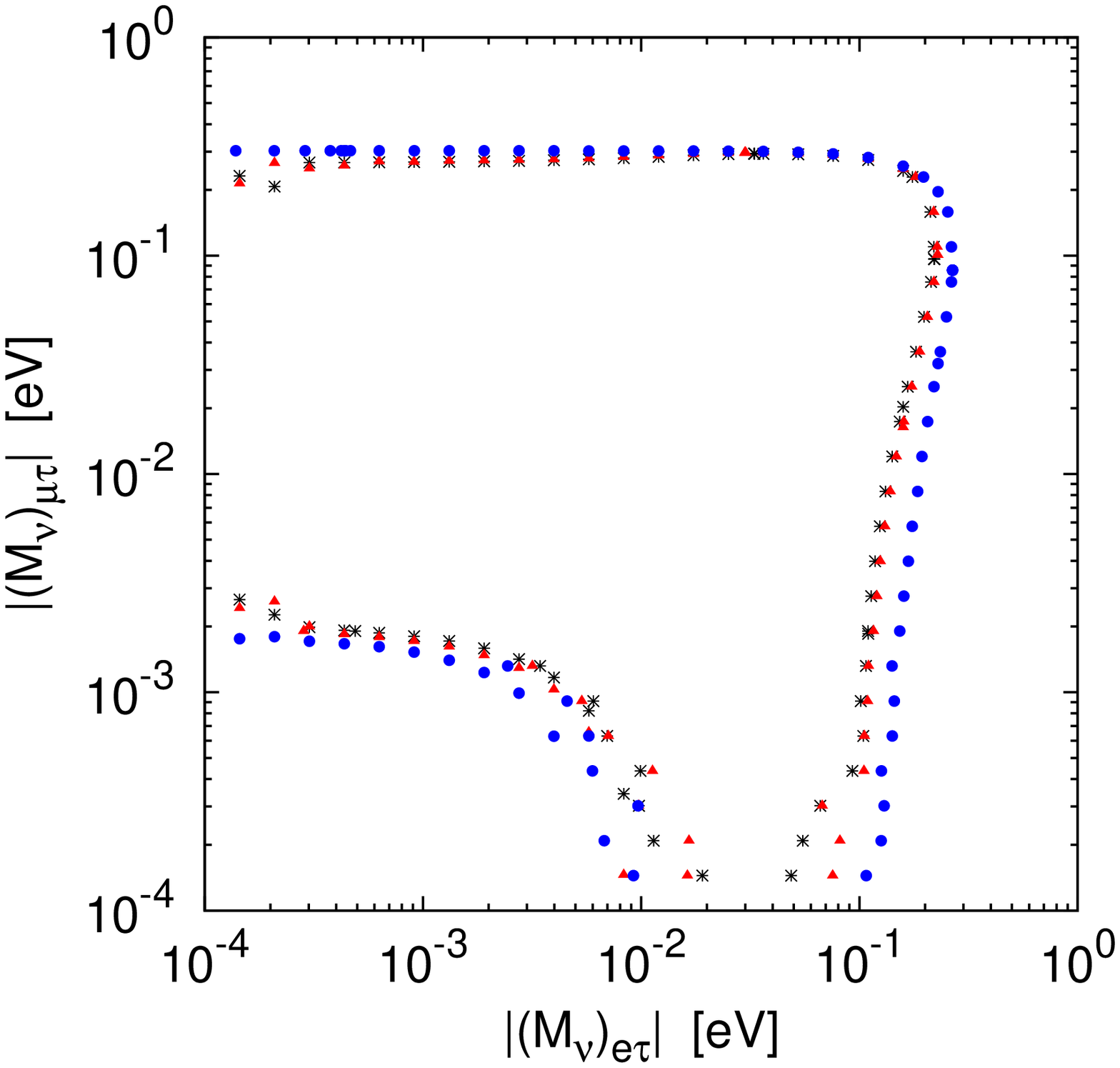} &
\includegraphics[angle=-90,keepaspectratio=true,scale=\figurescale]
{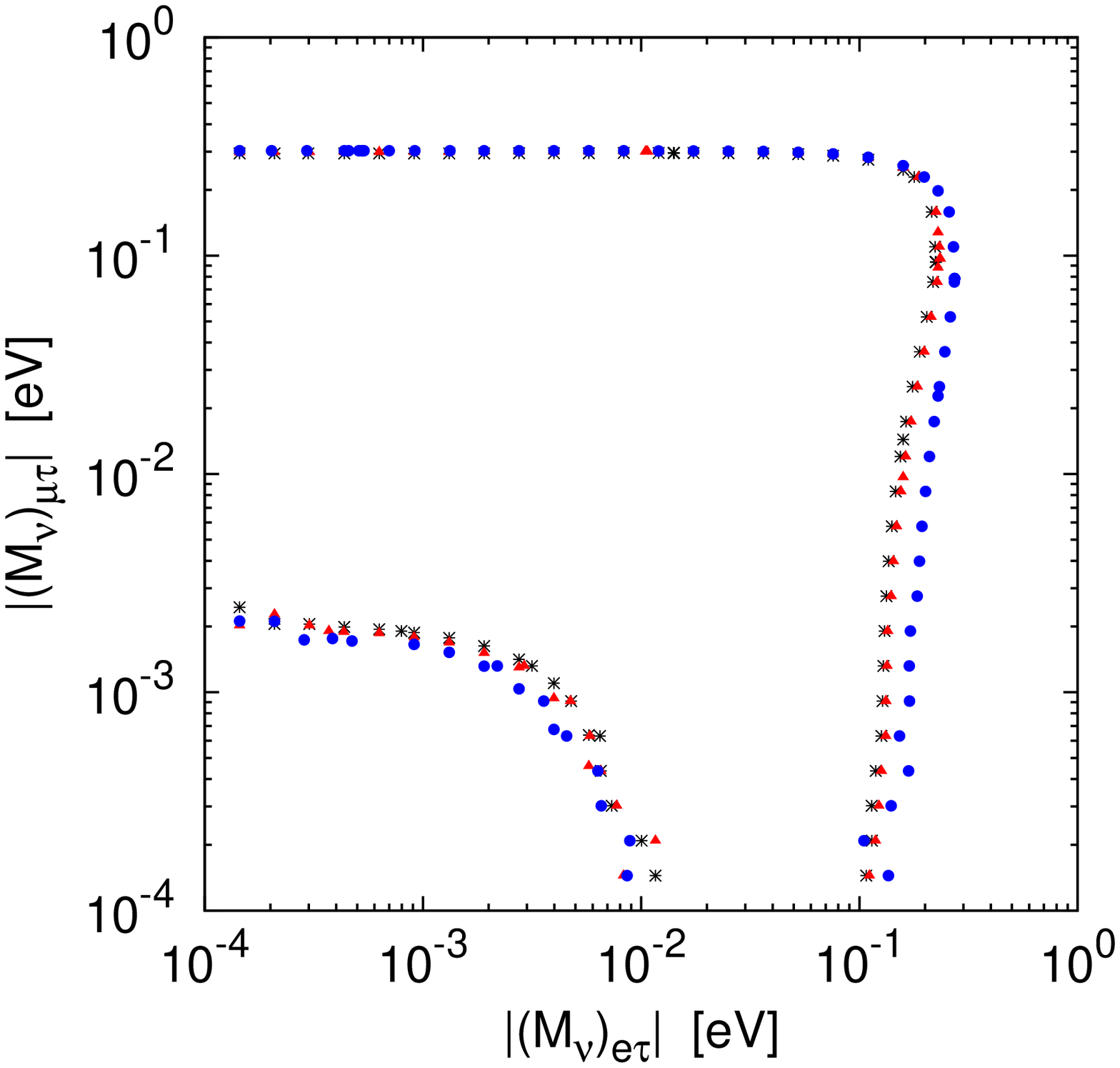}\\
\includegraphics[angle=-90,keepaspectratio=true,scale=\figurescale]
{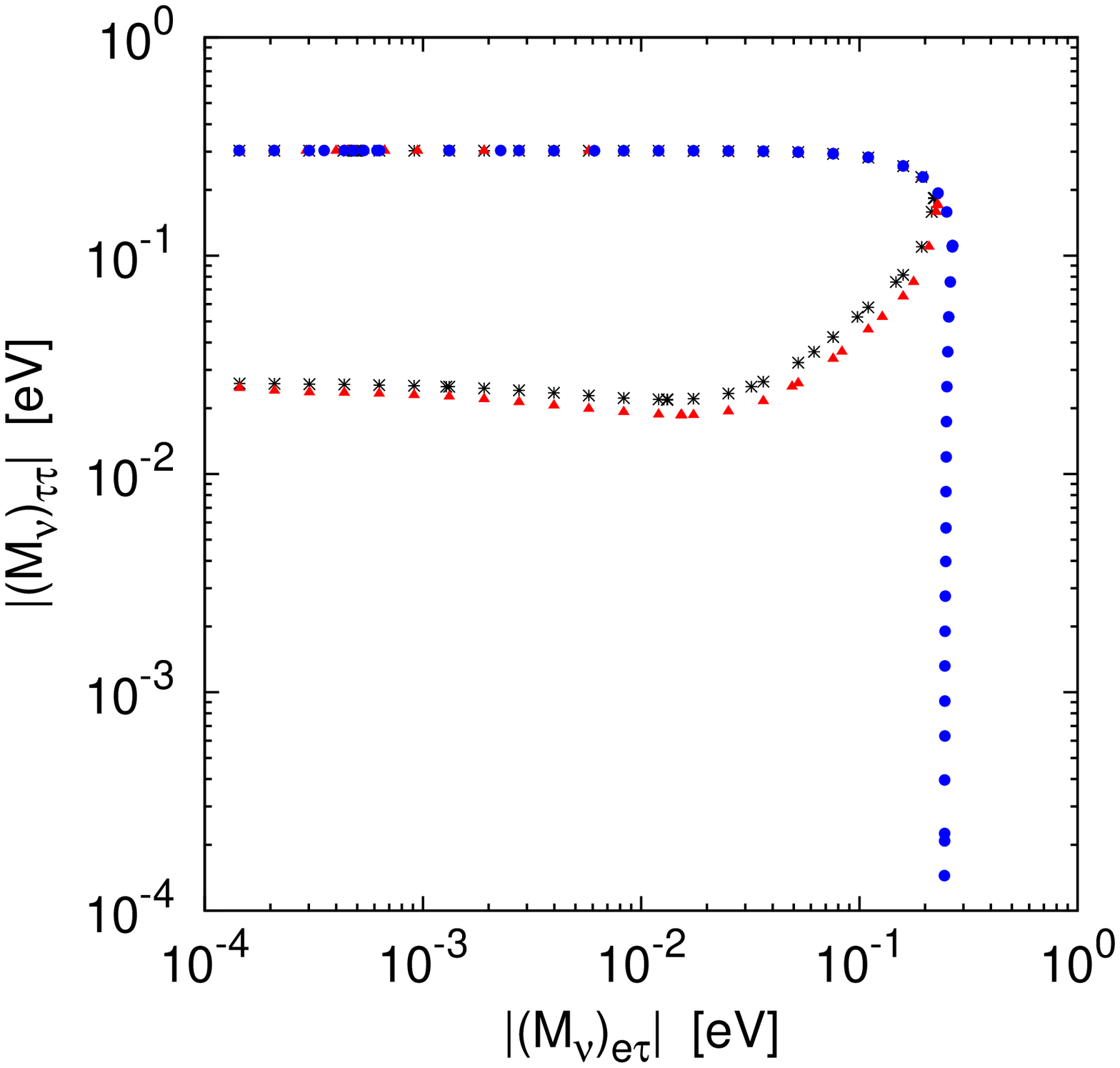} &
\includegraphics[angle=-90,keepaspectratio=true,scale=\figurescale]
{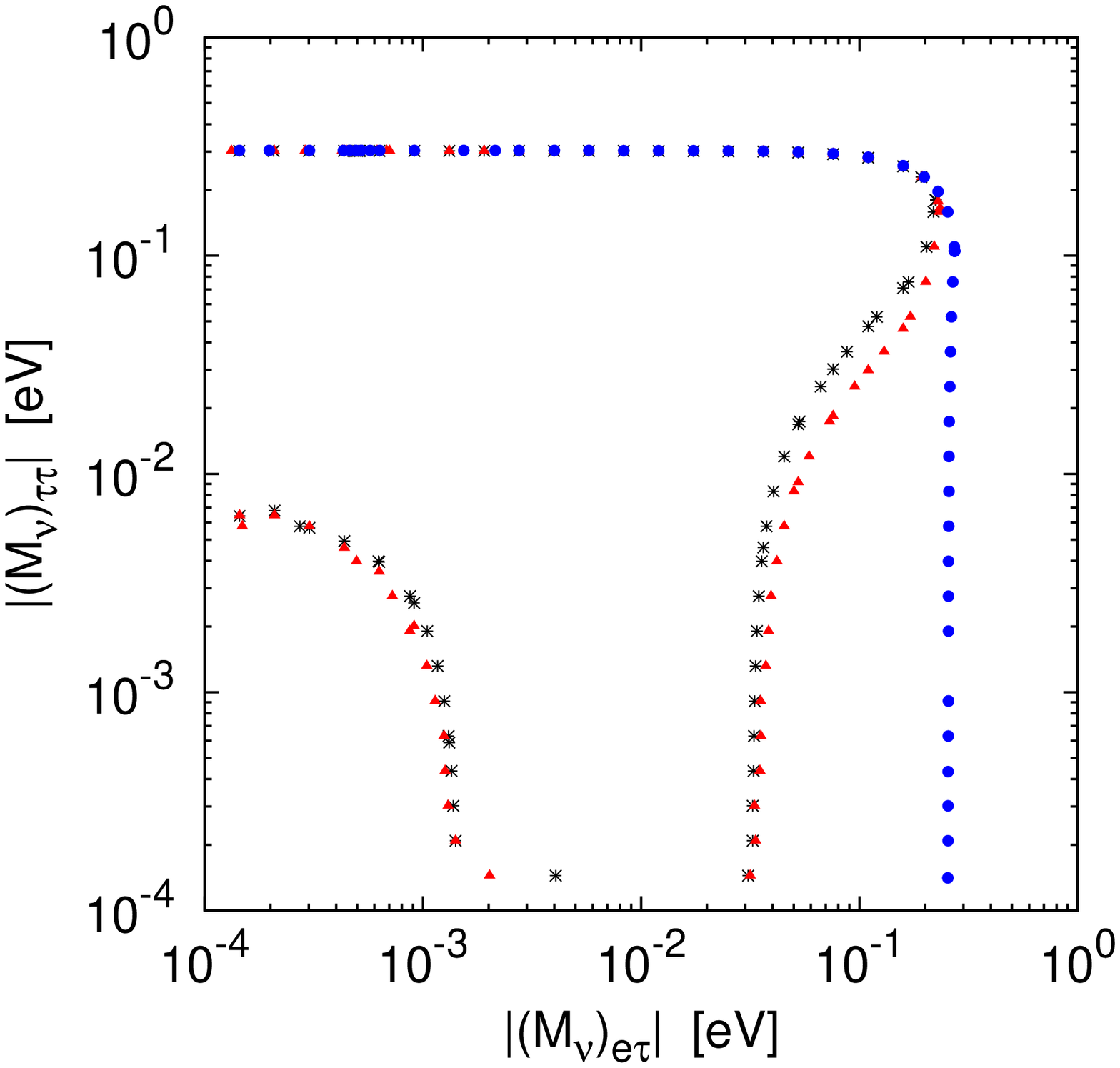}\\
\end{tabular}

\begin{tabular}[t]{ll}
\includegraphics[angle=-90,keepaspectratio=true,scale=\figurescale]
{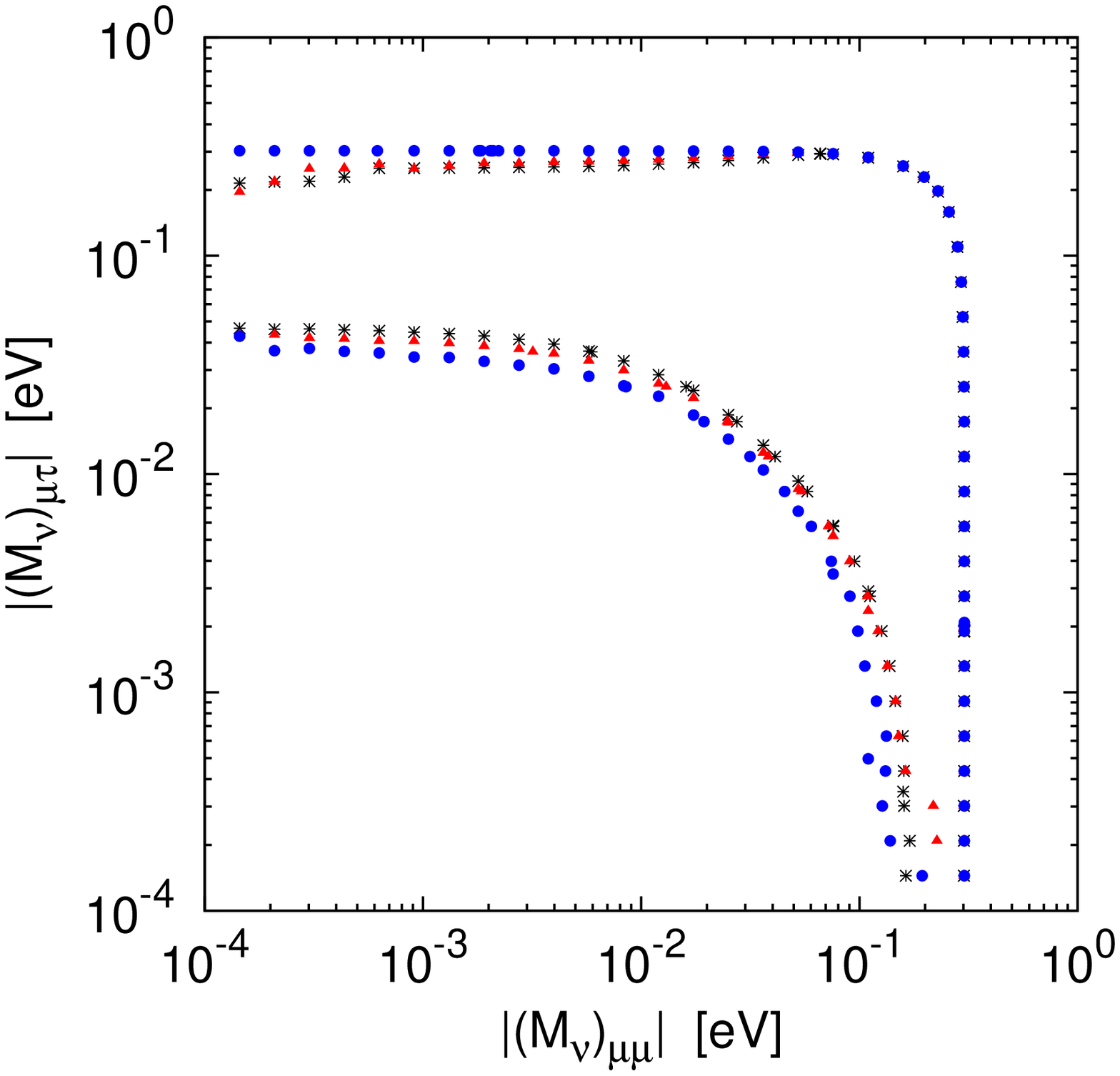} &
\includegraphics[angle=-90,keepaspectratio=true,scale=\figurescale]
{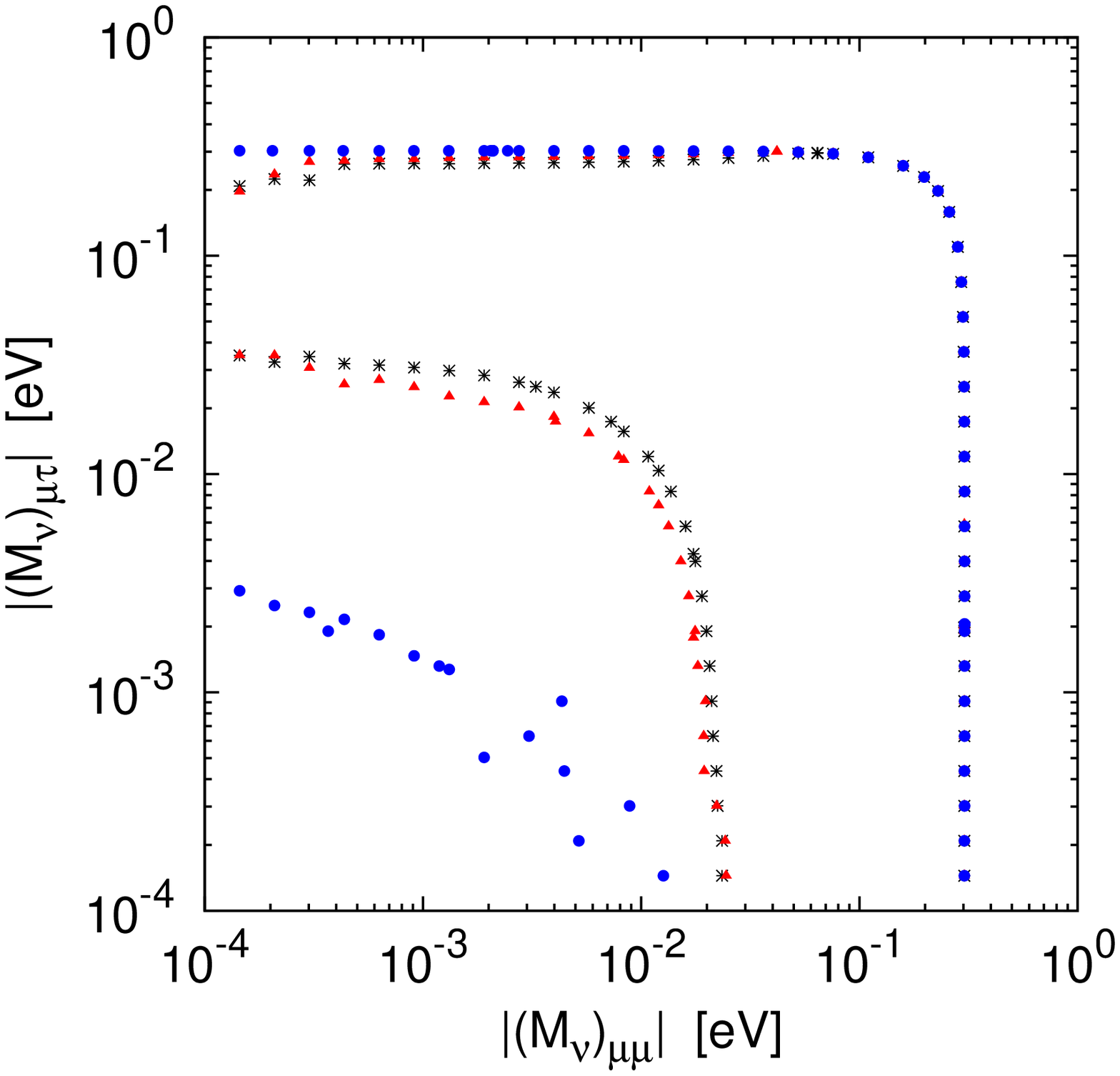}\\
\includegraphics[angle=-90,keepaspectratio=true,scale=\figurescale]
{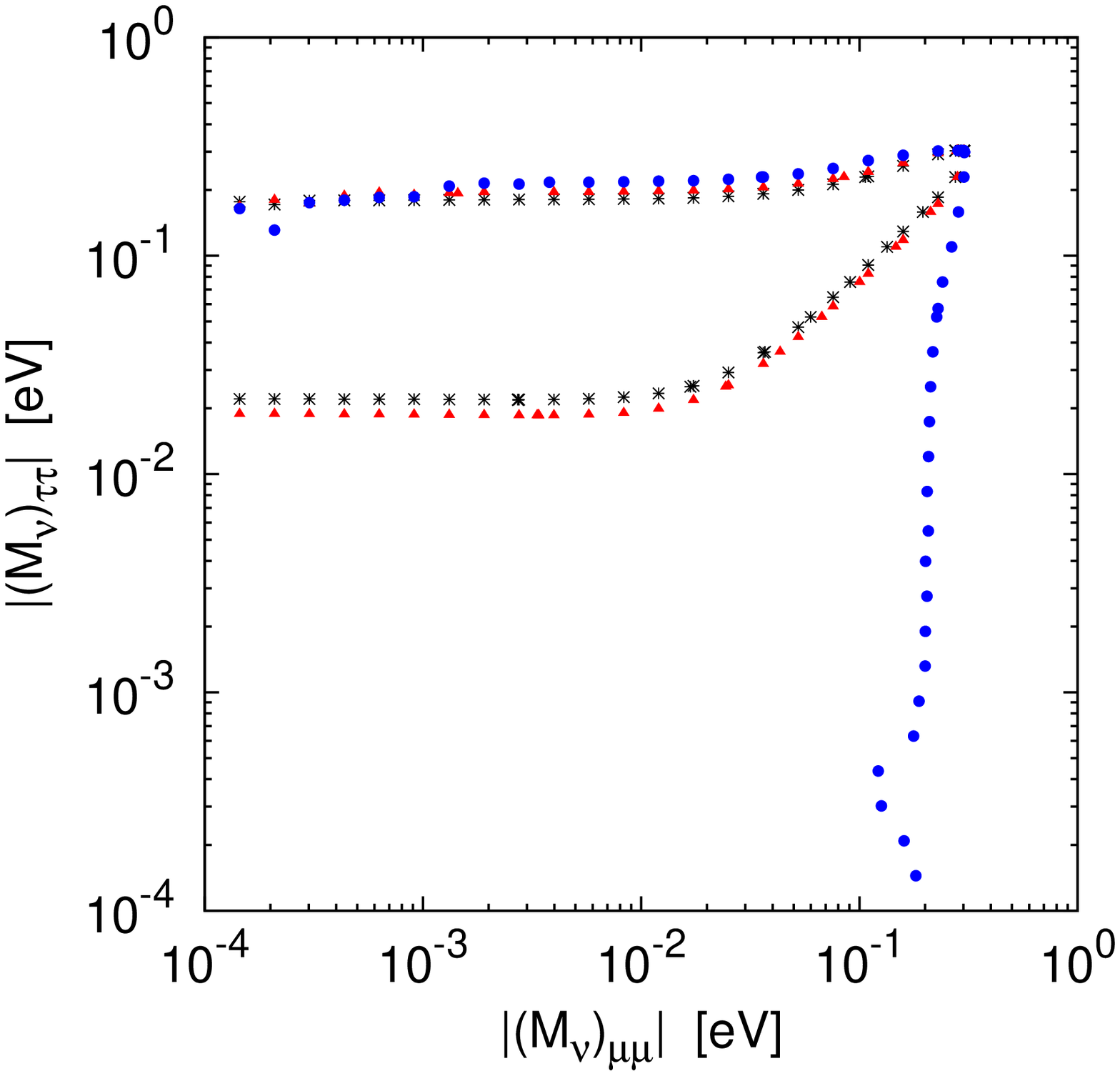} &
\includegraphics[angle=-90,keepaspectratio=true,scale=\figurescale]
{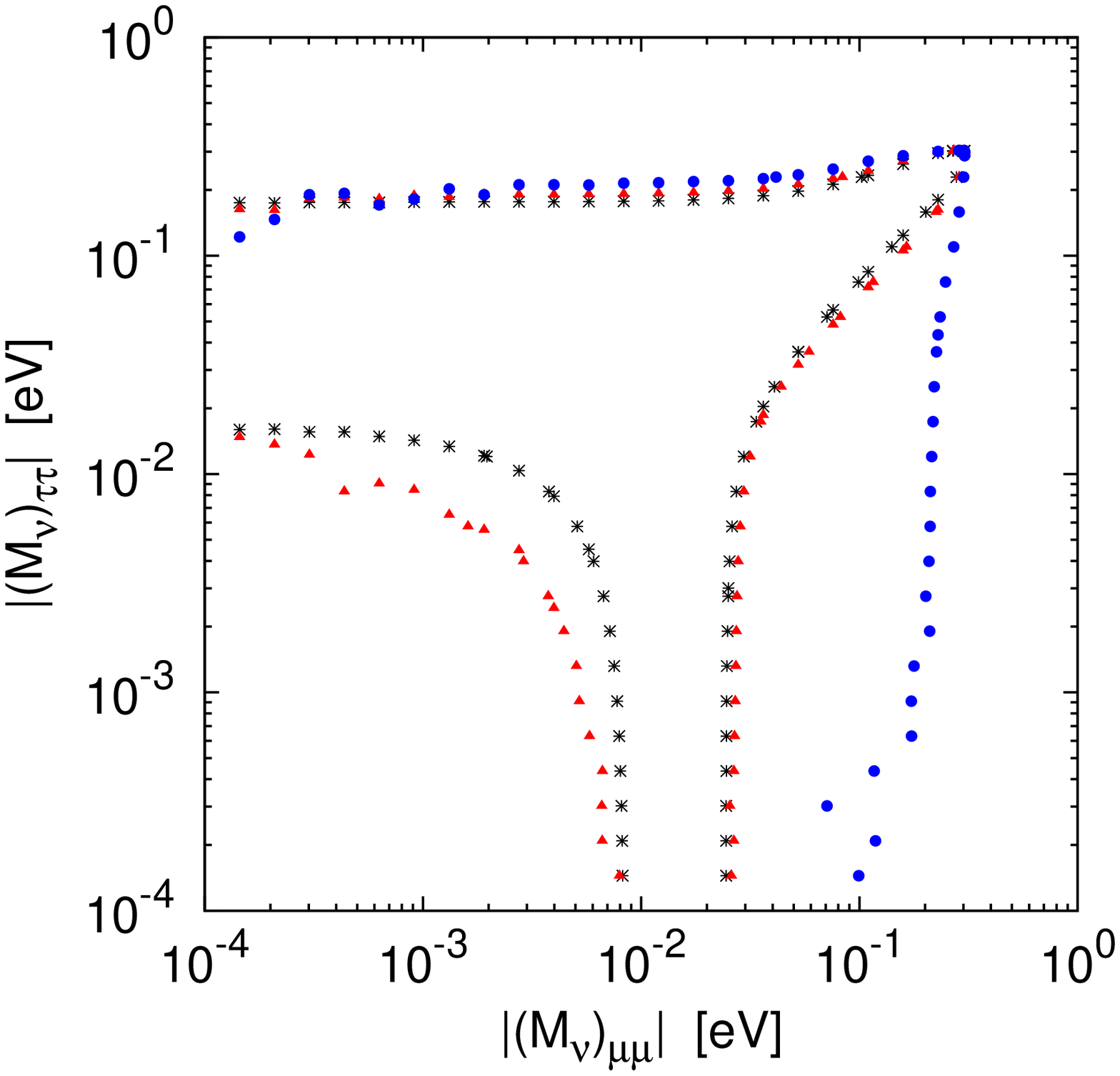}\\
\includegraphics[angle=-90,keepaspectratio=true,scale=\figurescale]
{figures/Fogli-normal/M23-M33normal.eps} &
\includegraphics[angle=-90,keepaspectratio=true,scale=\figurescale]
{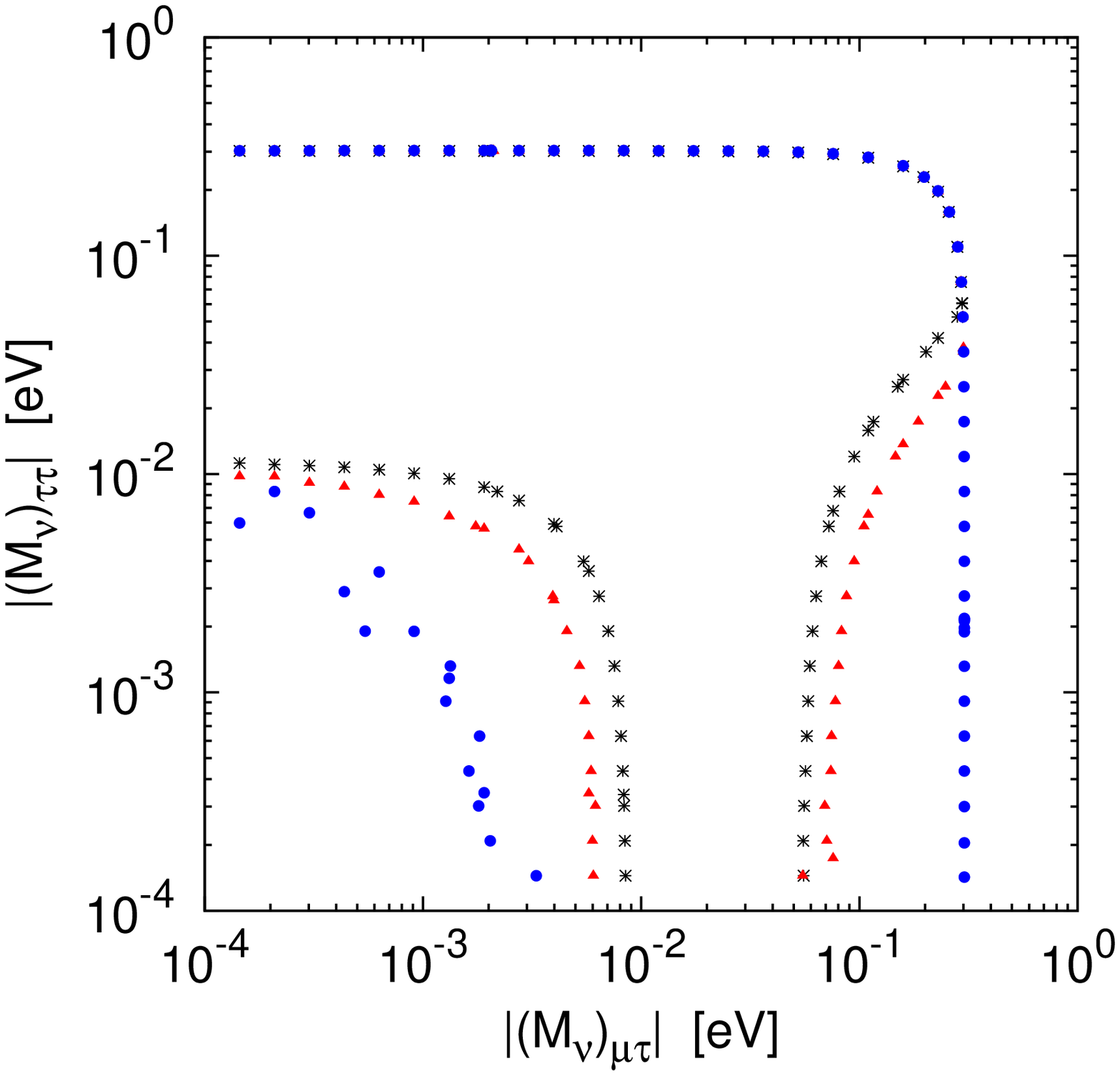}\\
\end{tabular}

\section{Differences between the plots based on Fogli~\textit{et al.}
and Forero~\textit{et al.} (version 3) at one sigma}\label{differences}

Here we present those correlations where there are notable differences between
the plots based on Fogli~\textit{et al.}~\cite{fogli} and Forero~\textit{et al.}
(version 3)~\cite{forero} at the $1\sigma$ level.
The following plots always contain the allowed $1\sigma$ areas for both
Fogli~\textit{et al.} (bounded by red crosses \textcolor{red}{$\times$})
as well as Forero~\textit{et al.} (bounded by black boxes $\boxdot$).
The plot title shows the neutrino mass spectrum.

We begin with the ten correlation plots involving $|(\mathcal{M}_\nu)_{\tau\tau}|$.
Afterwards we show the seven remaining plots.

\begin{center}
\begin{tabular}[t]{ll}
\includegraphics[angle=-90,keepaspectratio=true,scale=\figurescaletwo]
{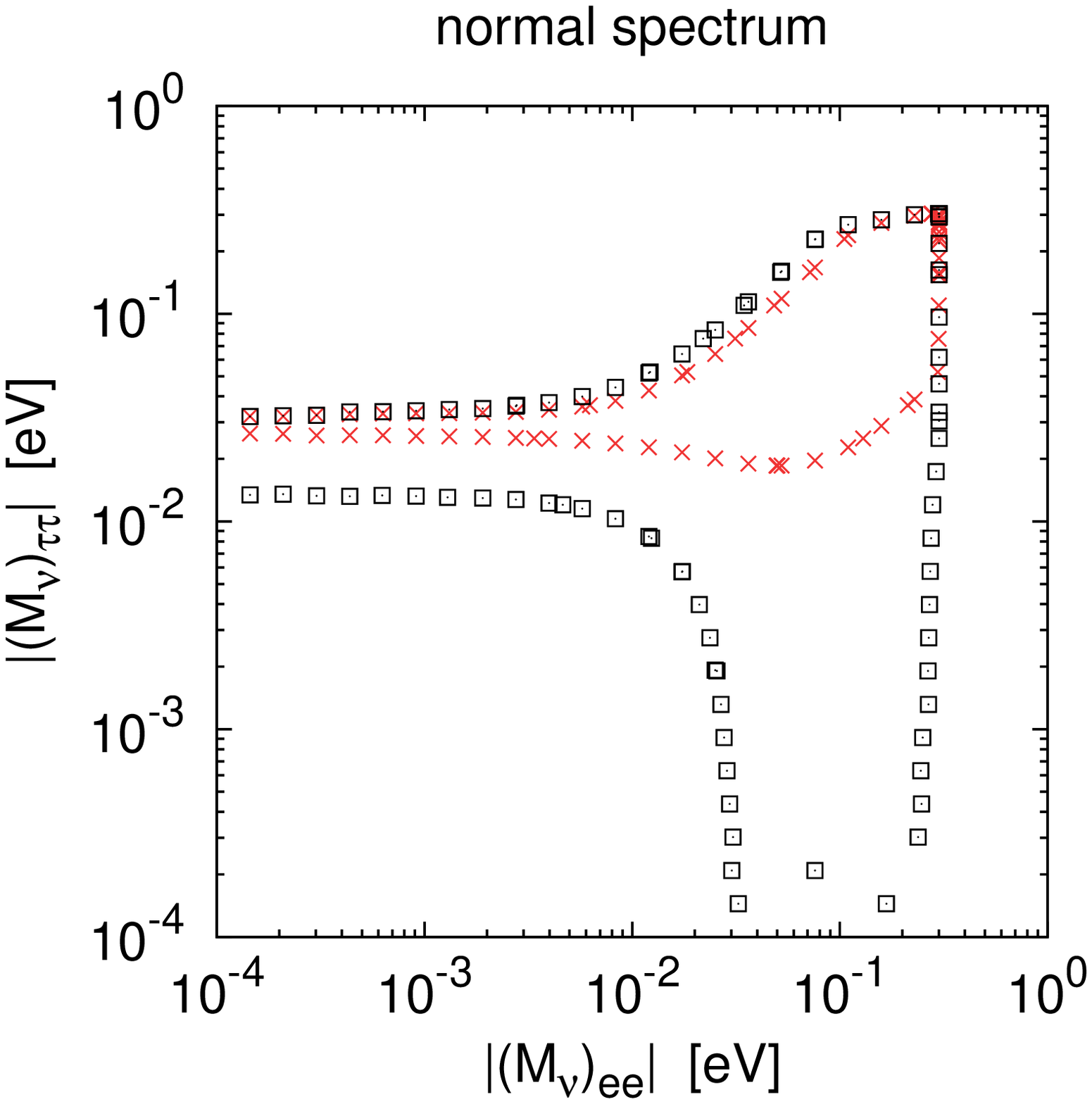} &
\includegraphics[angle=-90,keepaspectratio=true,scale=\figurescaletwo]
{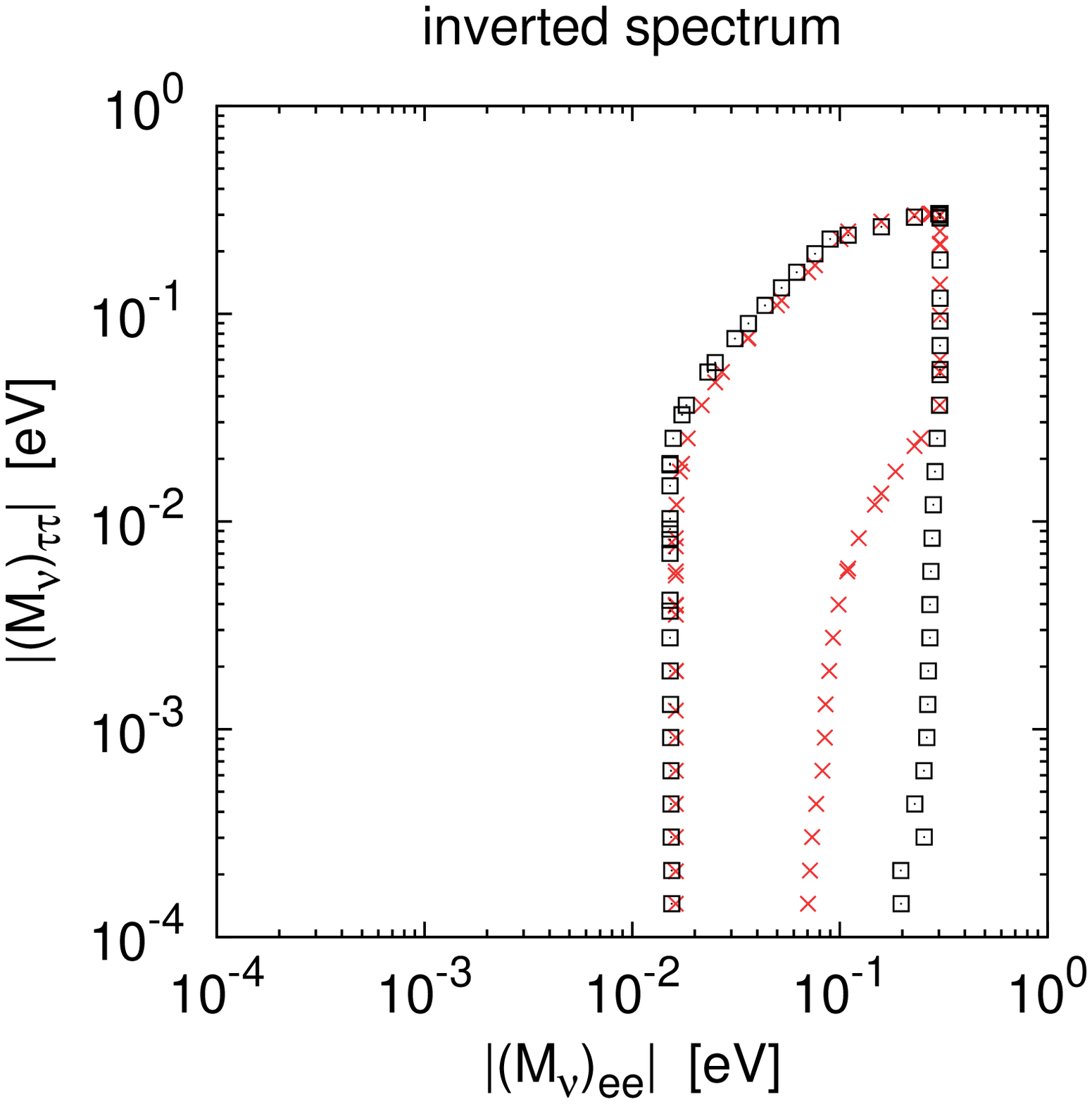}\\
\includegraphics[angle=-90,keepaspectratio=true,scale=\figurescaletwo]
{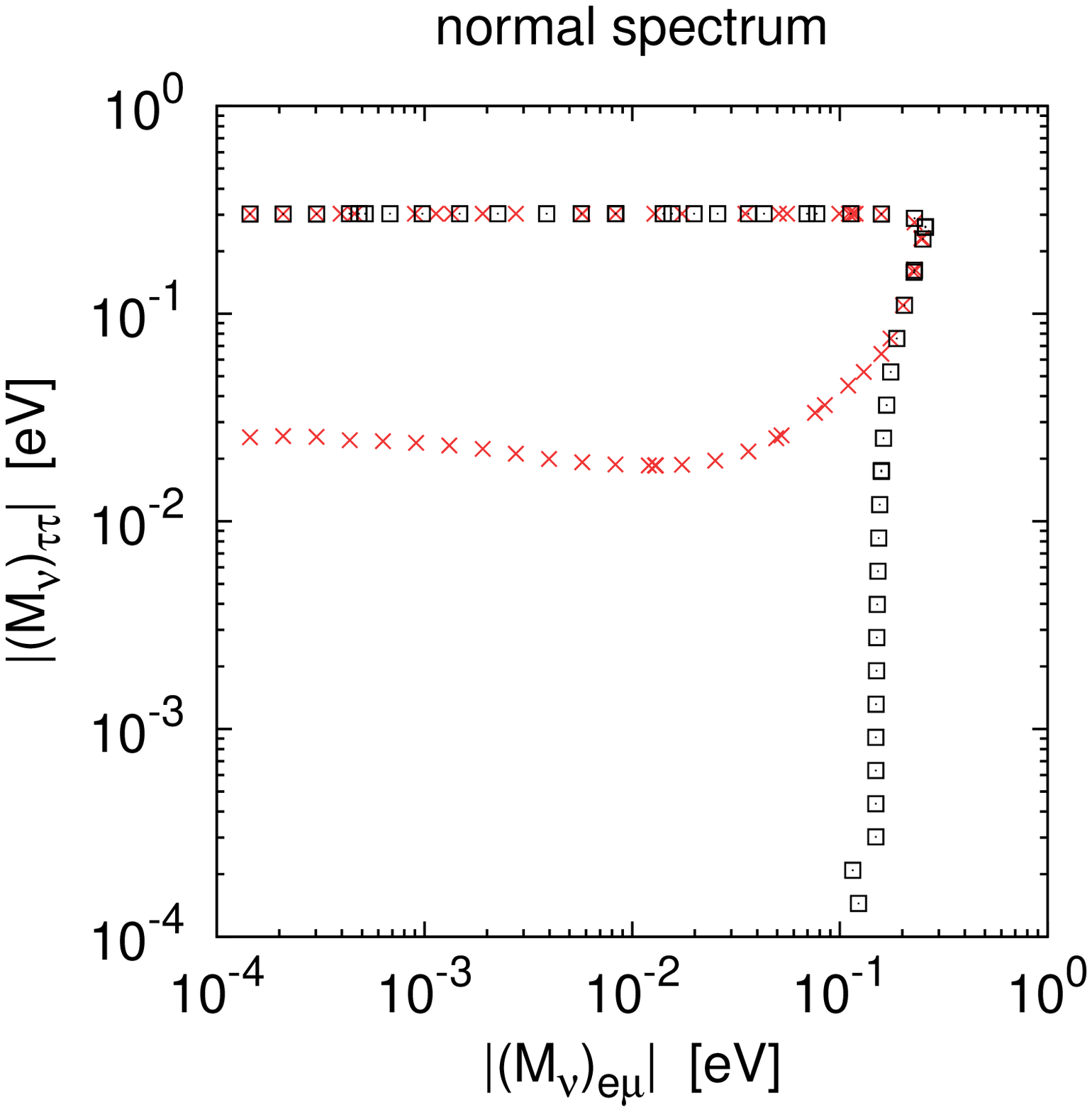} &
\includegraphics[angle=-90,keepaspectratio=true,scale=\figurescaletwo]
{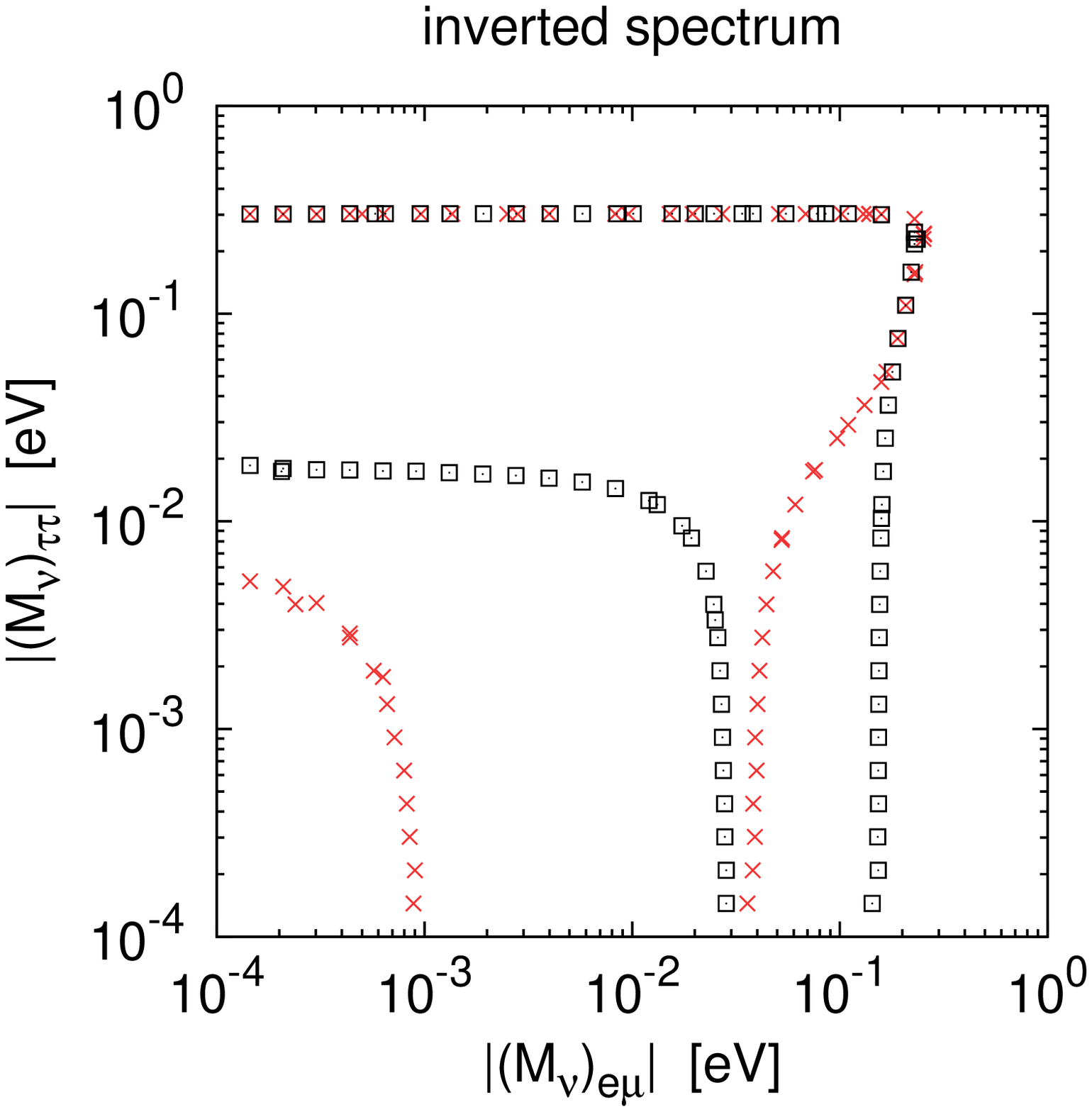}\\
\end{tabular}
\end{center}

\begin{center}
\begin{tabular}[t]{ll}
\includegraphics[angle=-90,keepaspectratio=true,scale=\figurescaletwo]
{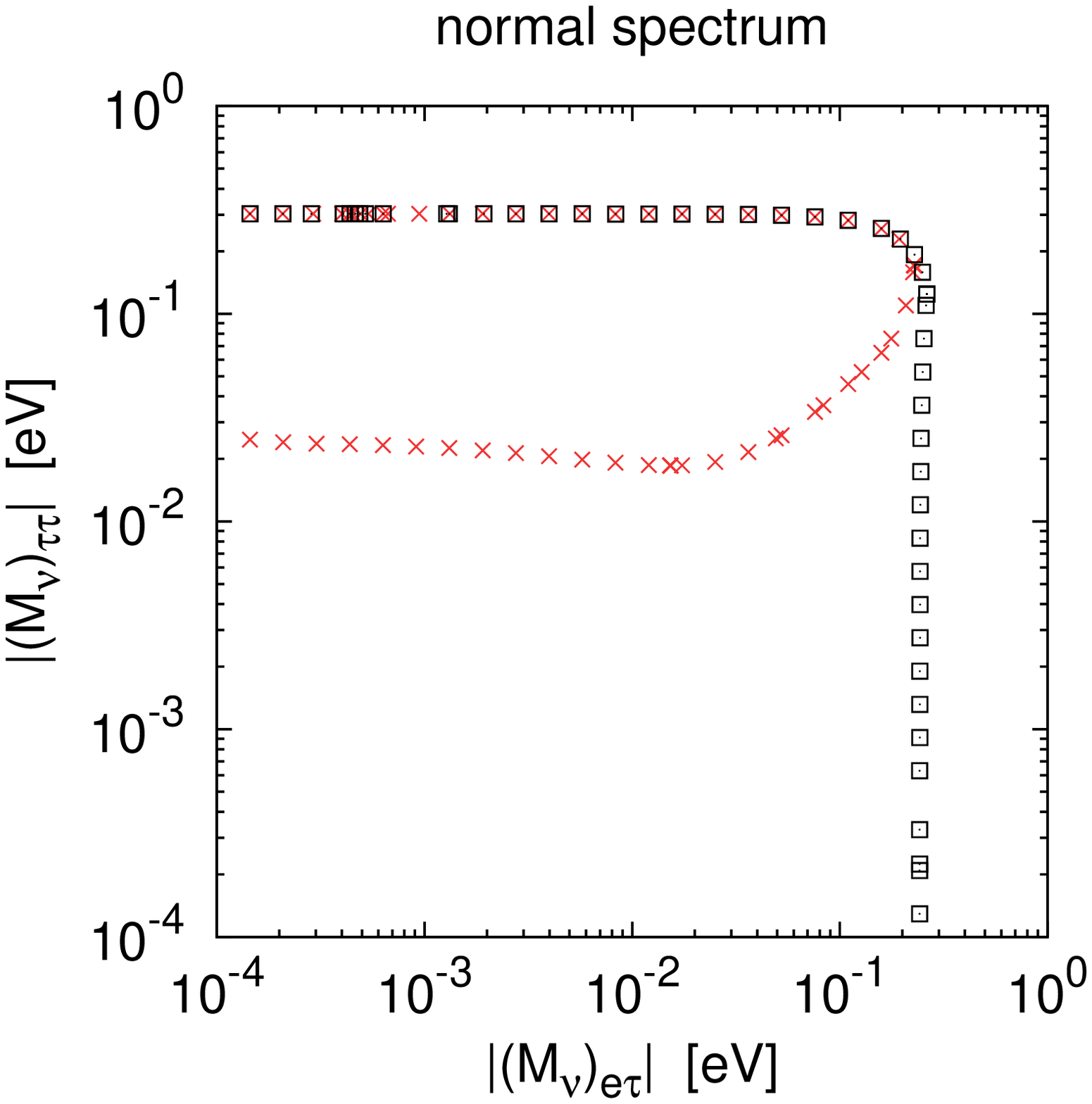} &
\includegraphics[angle=-90,keepaspectratio=true,scale=\figurescaletwo]
{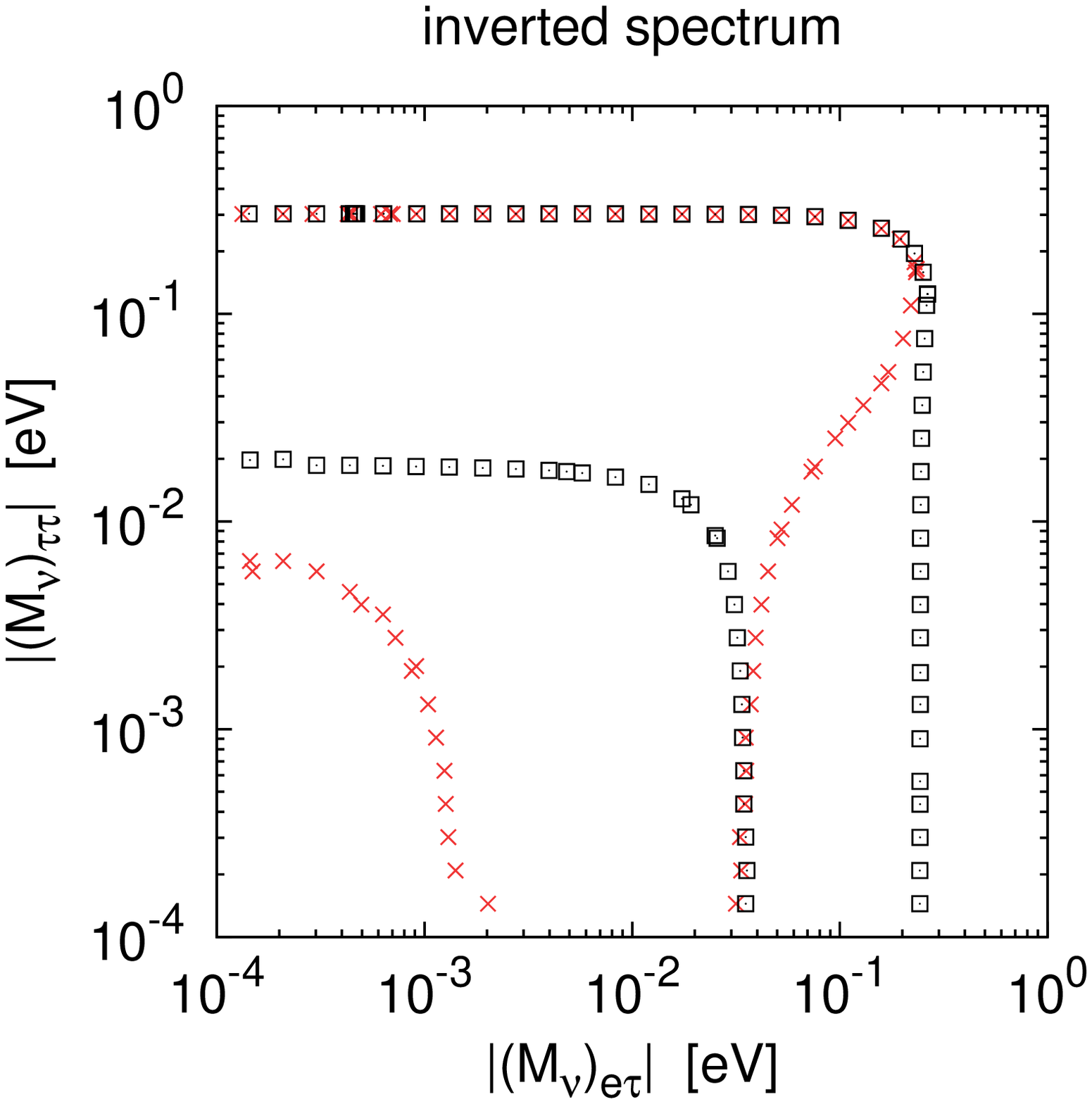}\\
\includegraphics[angle=-90,keepaspectratio=true,scale=\figurescaletwo]
{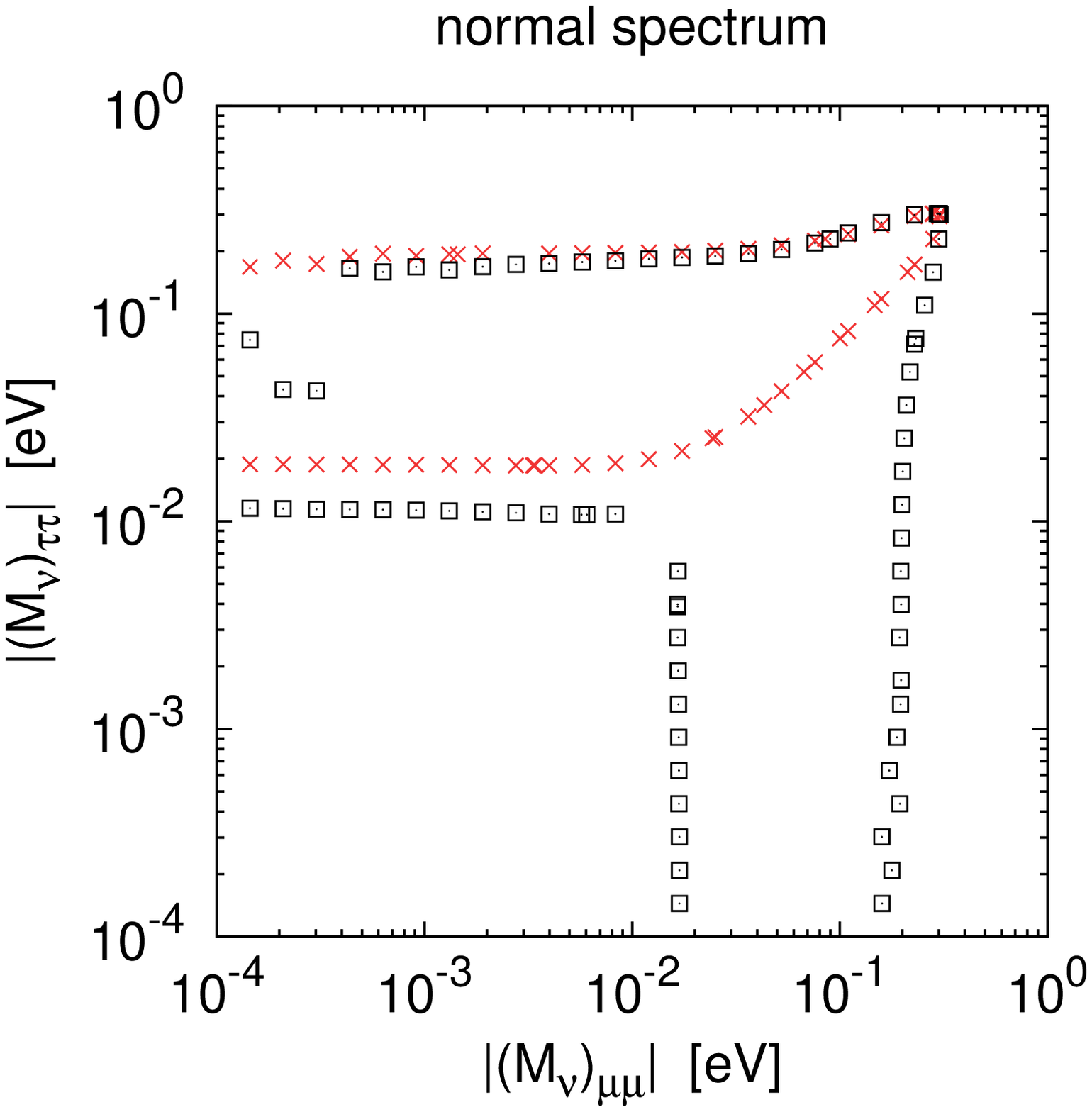} &
\includegraphics[angle=-90,keepaspectratio=true,scale=\figurescaletwo]
{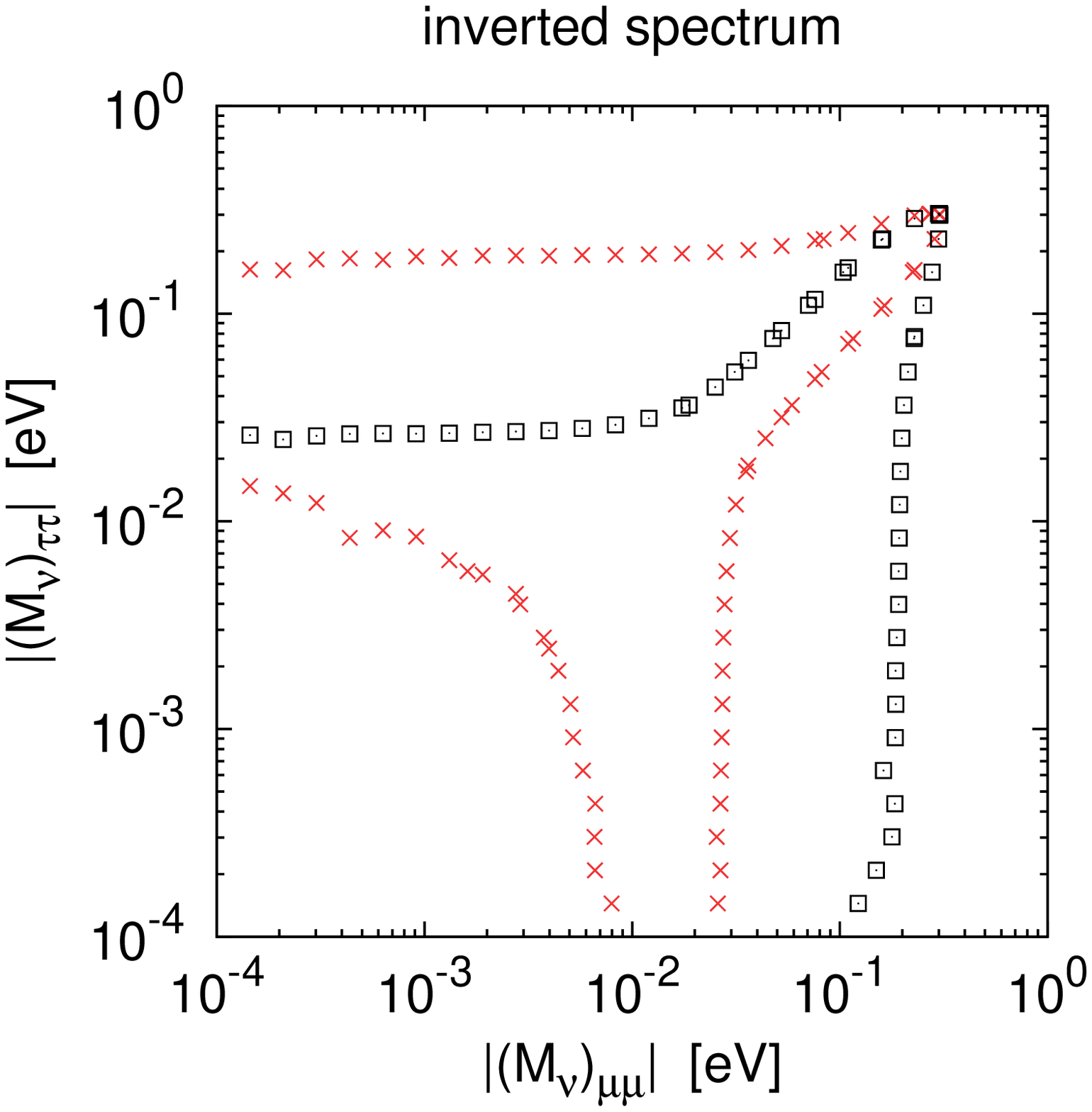}\\
\includegraphics[angle=-90,keepaspectratio=true,scale=\figurescaletwo]
{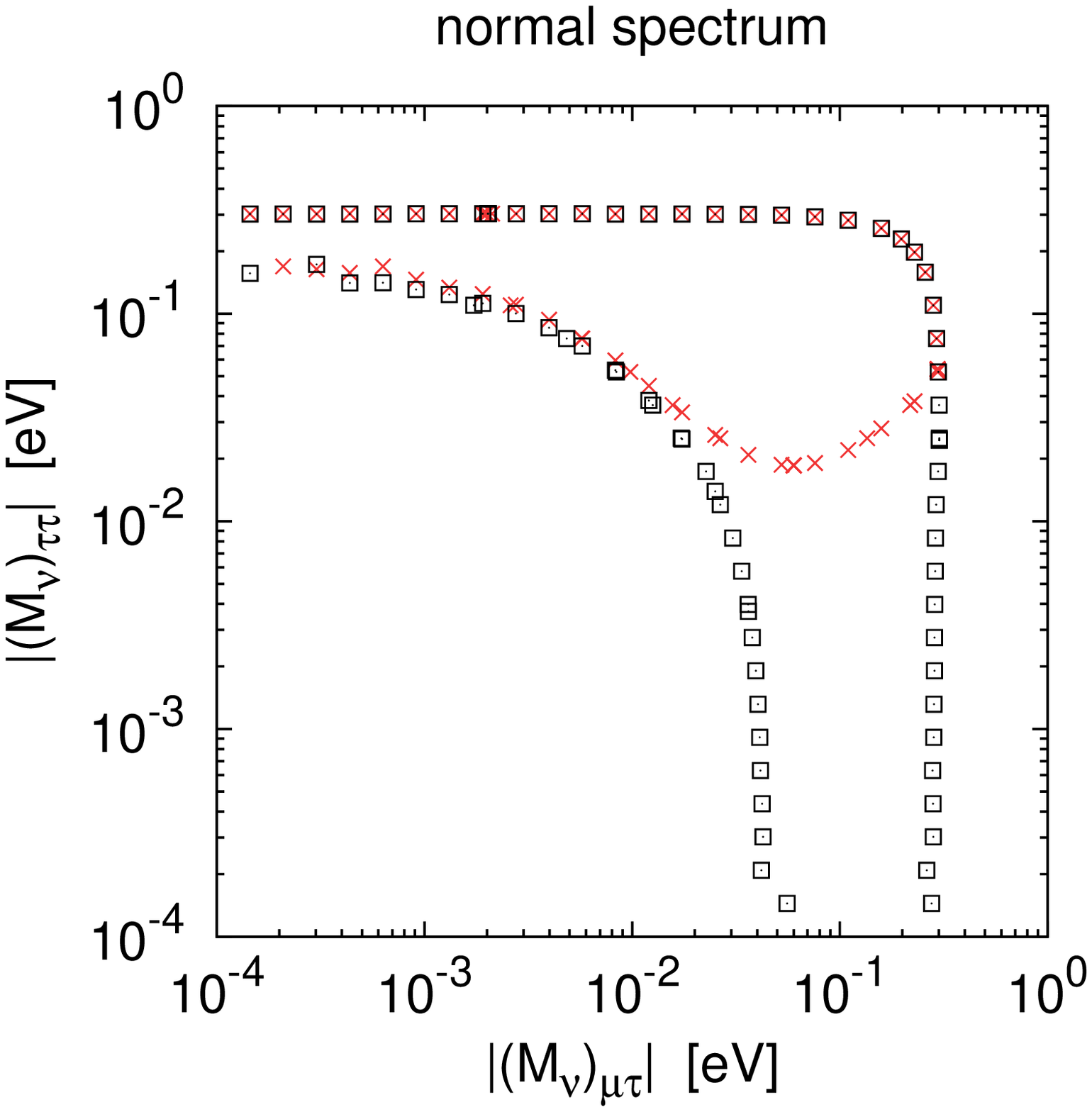} &
\includegraphics[angle=-90,keepaspectratio=true,scale=\figurescaletwo]
{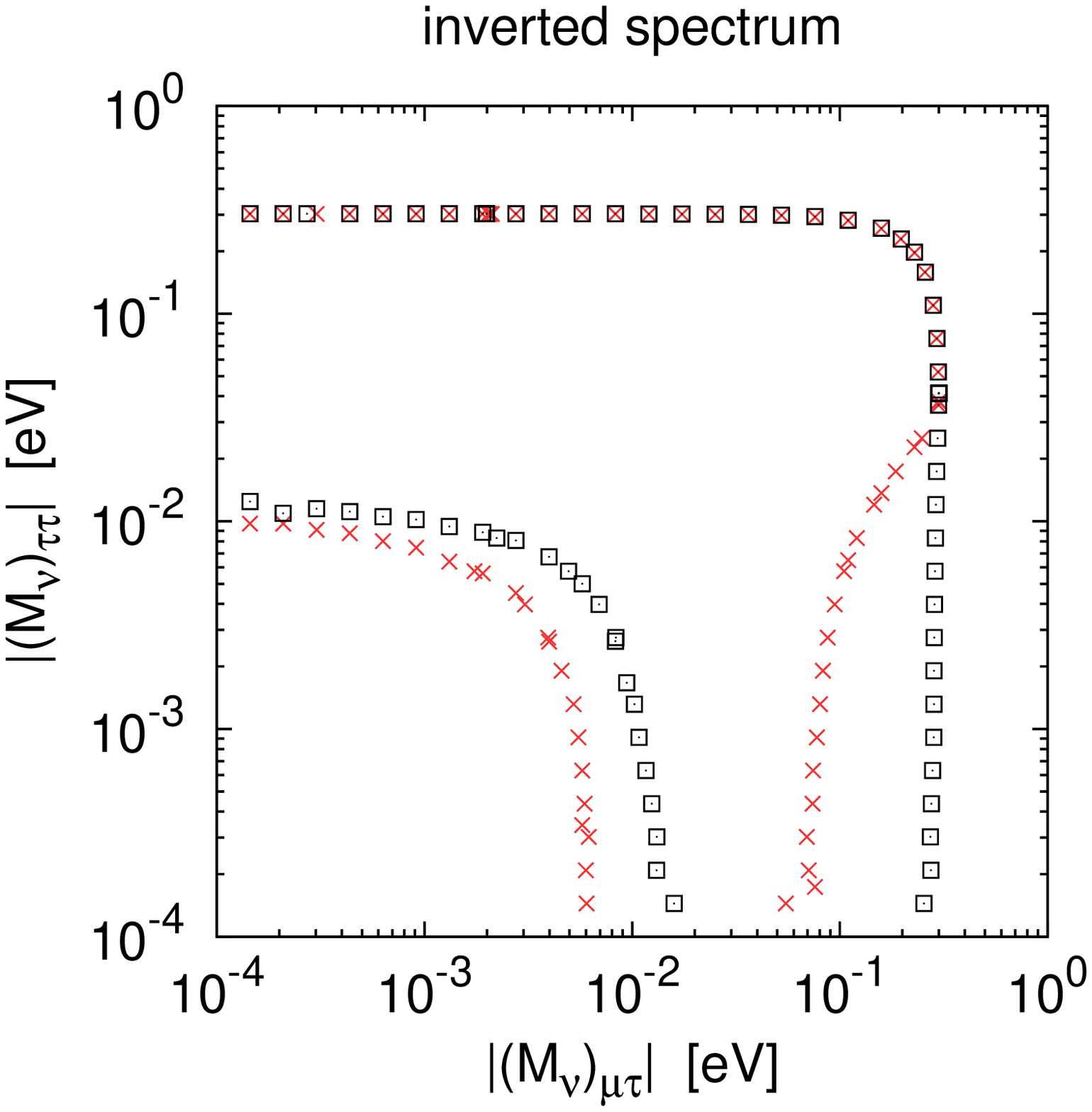}\\
\end{tabular}
\end{center}

\begin{center}
\begin{tabular}[t]{ll}
\includegraphics[angle=-90,keepaspectratio=true,scale=\figurescaletwo]
{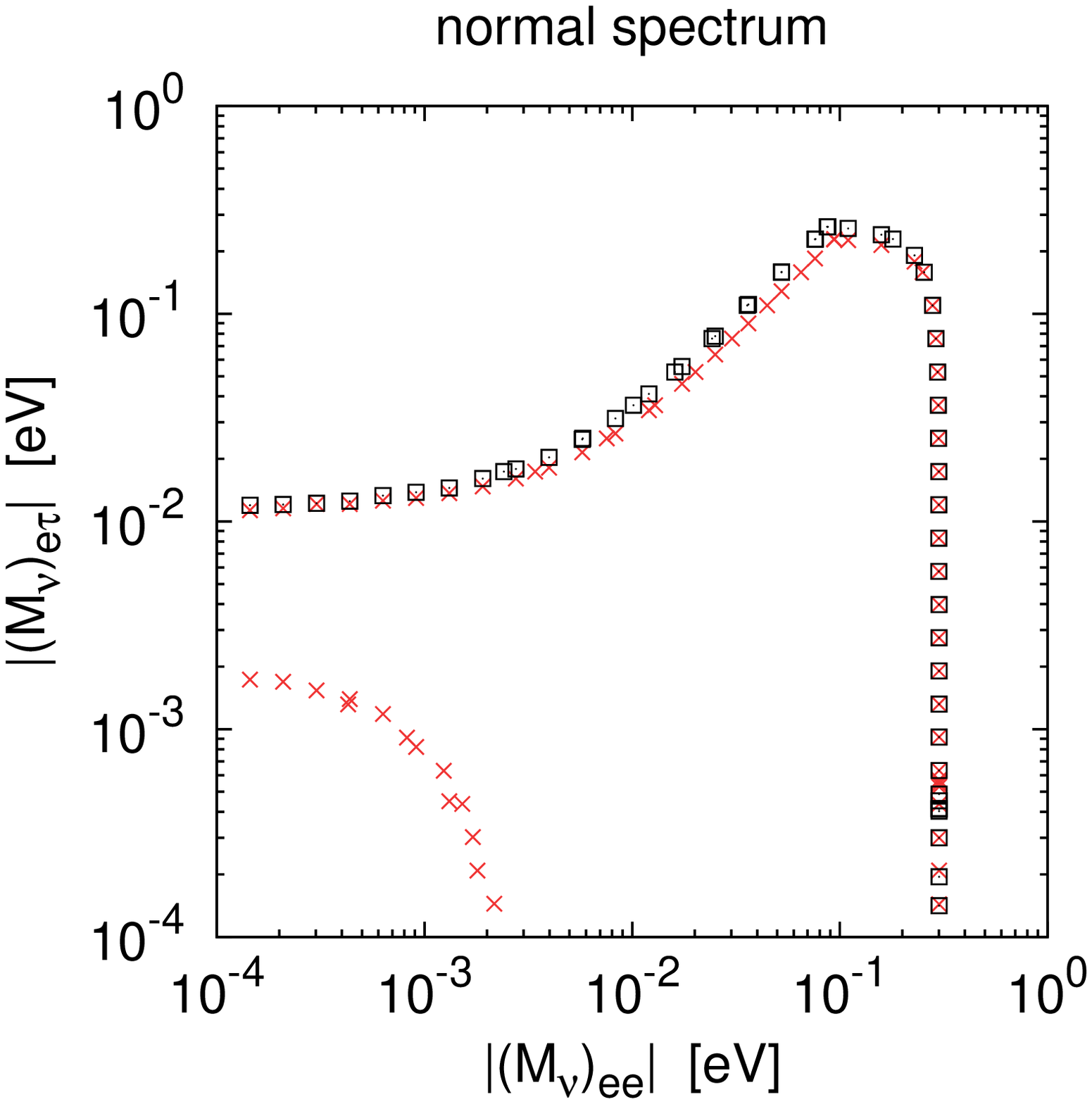} &
\includegraphics[angle=-90,keepaspectratio=true,scale=\figurescaletwo]
{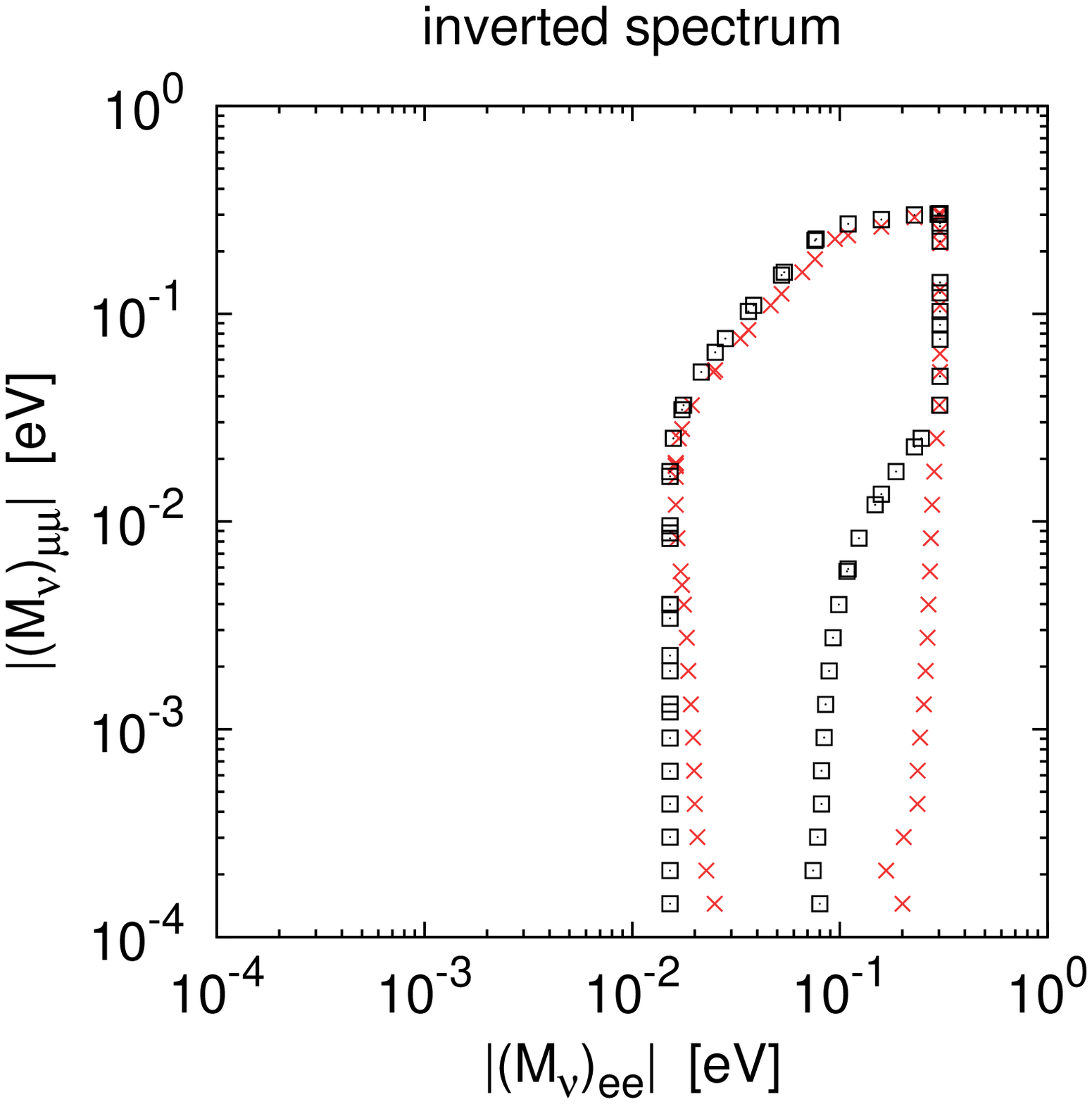}\\
\includegraphics[angle=-90,keepaspectratio=true,scale=\figurescaletwo]
{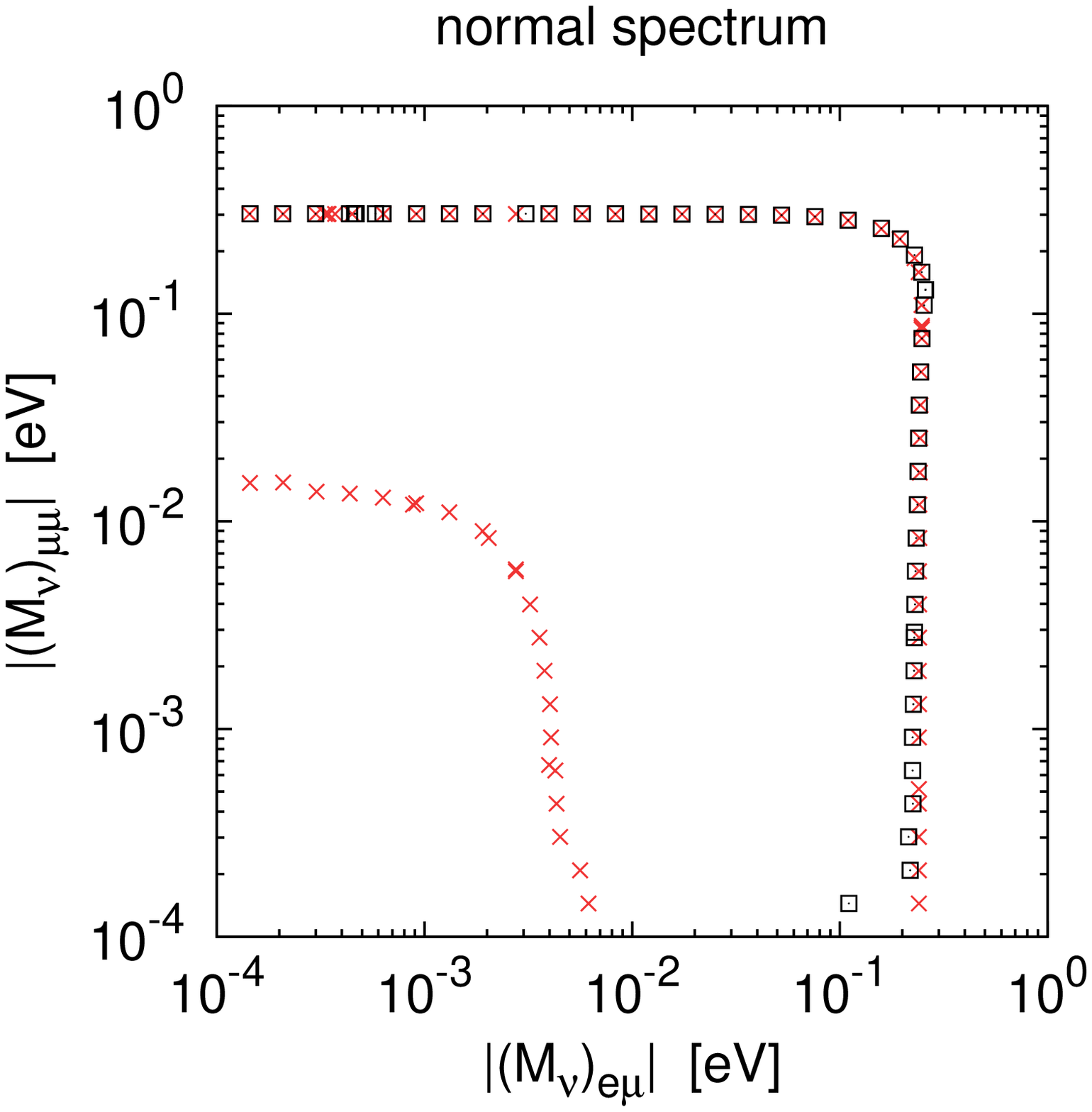} &
\includegraphics[angle=-90,keepaspectratio=true,scale=\figurescaletwo]
{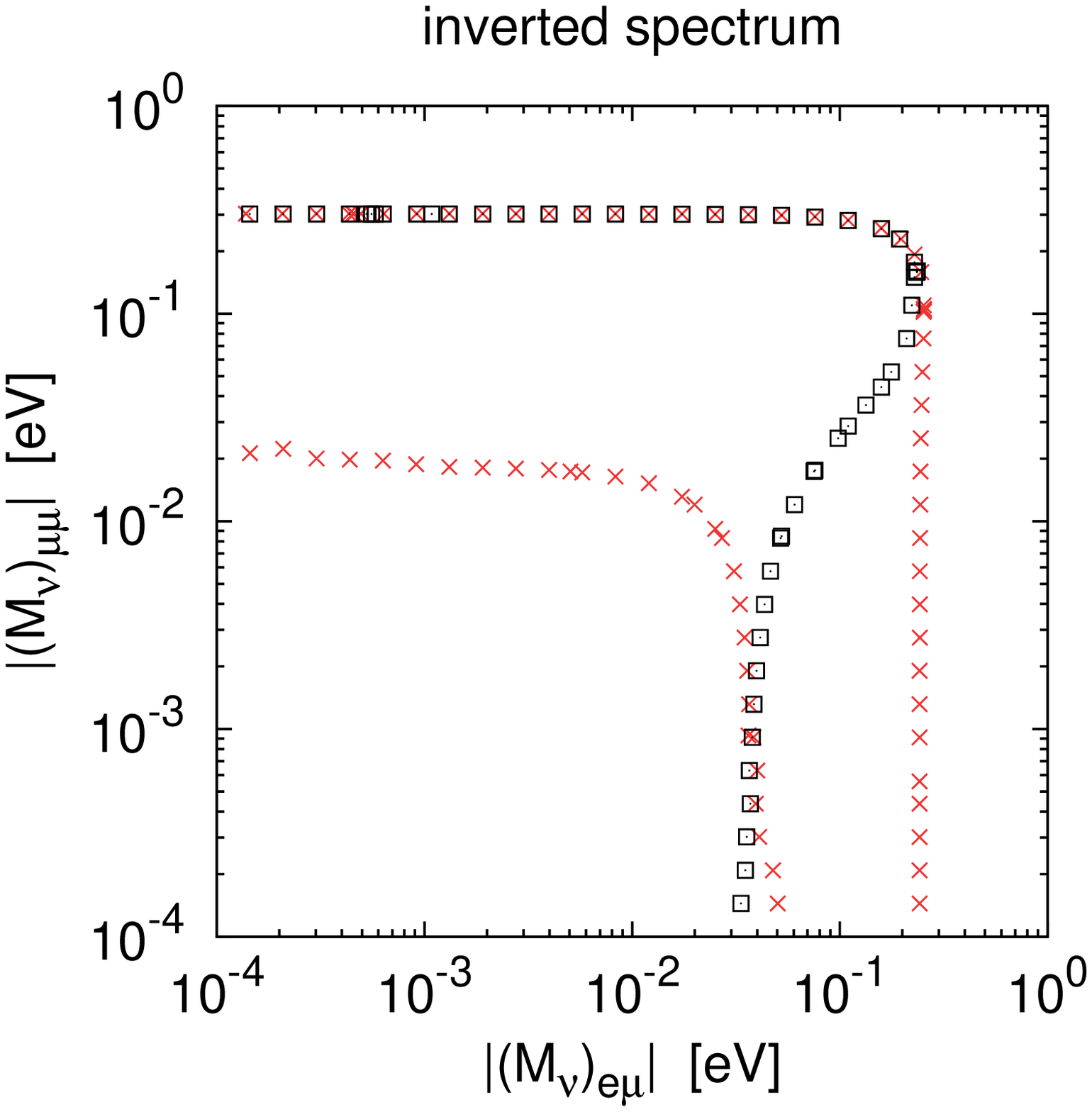}\\
\includegraphics[angle=-90,keepaspectratio=true,scale=\figurescaletwo]
{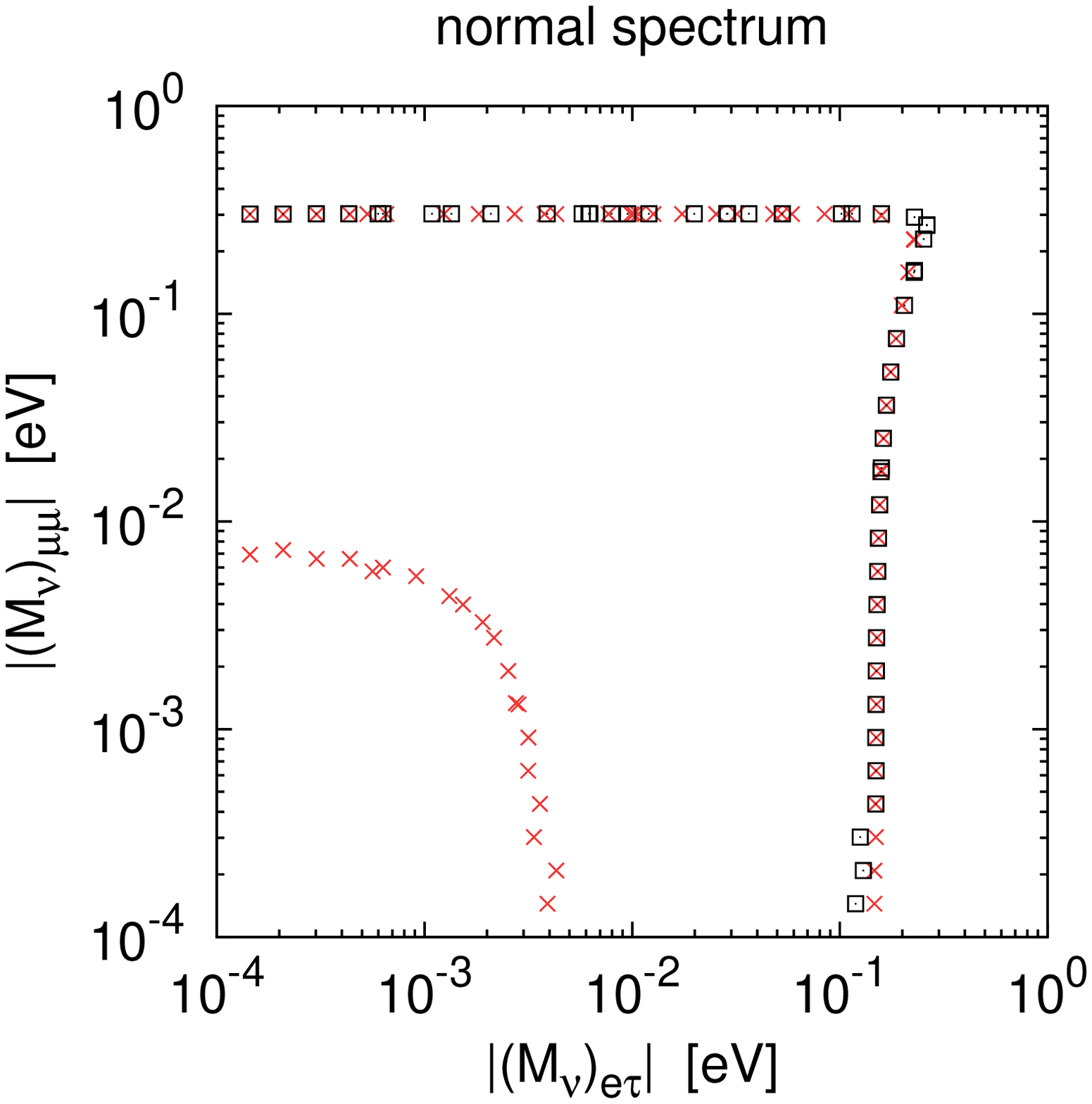} &
\includegraphics[angle=-90,keepaspectratio=true,scale=\figurescaletwo]
{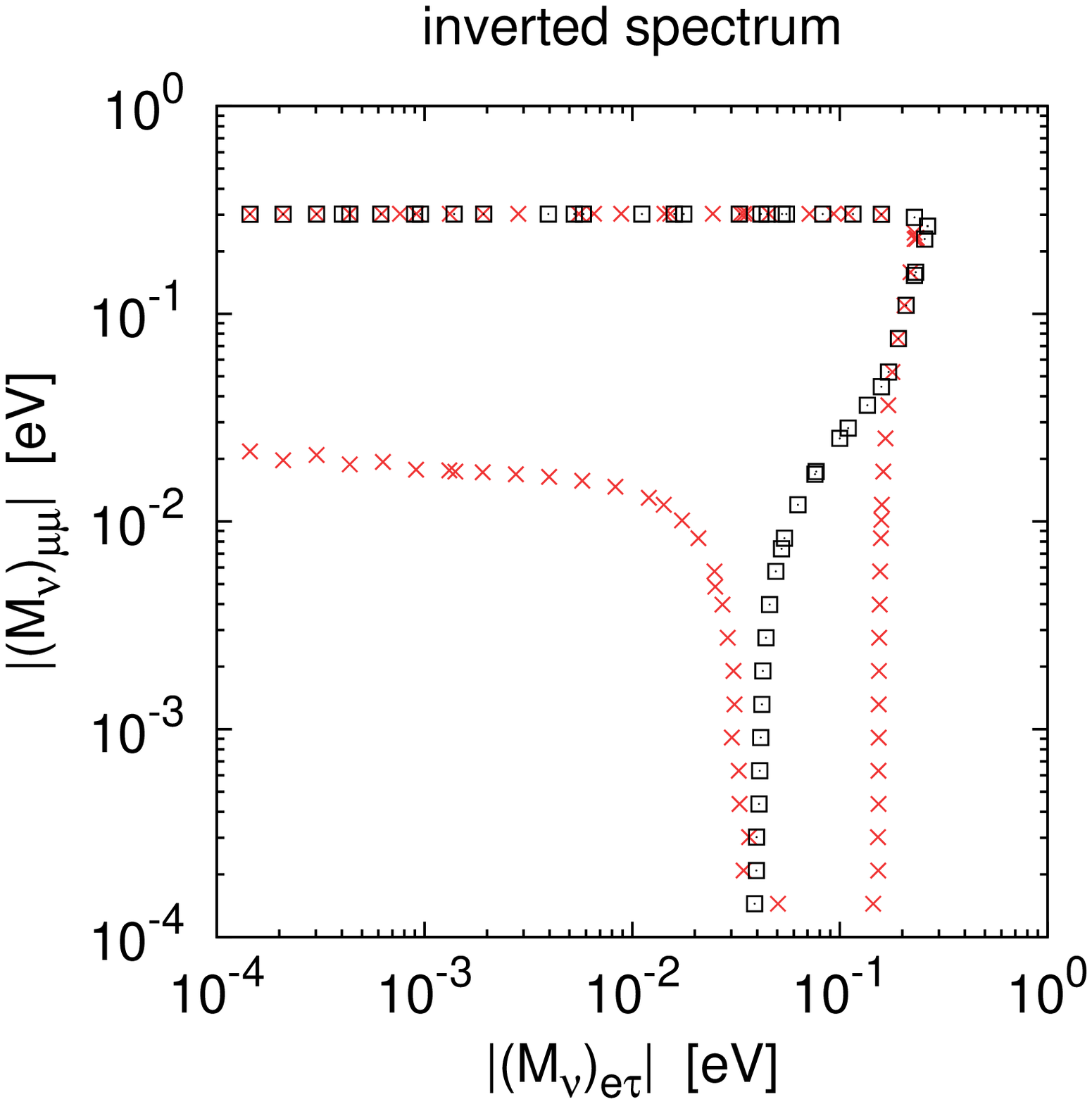}\\
\end{tabular}
\end{center}

\begin{center}
\begin{tabular}[t]{l}
\includegraphics[angle=-90,keepaspectratio=true,scale=\figurescaletwo]
{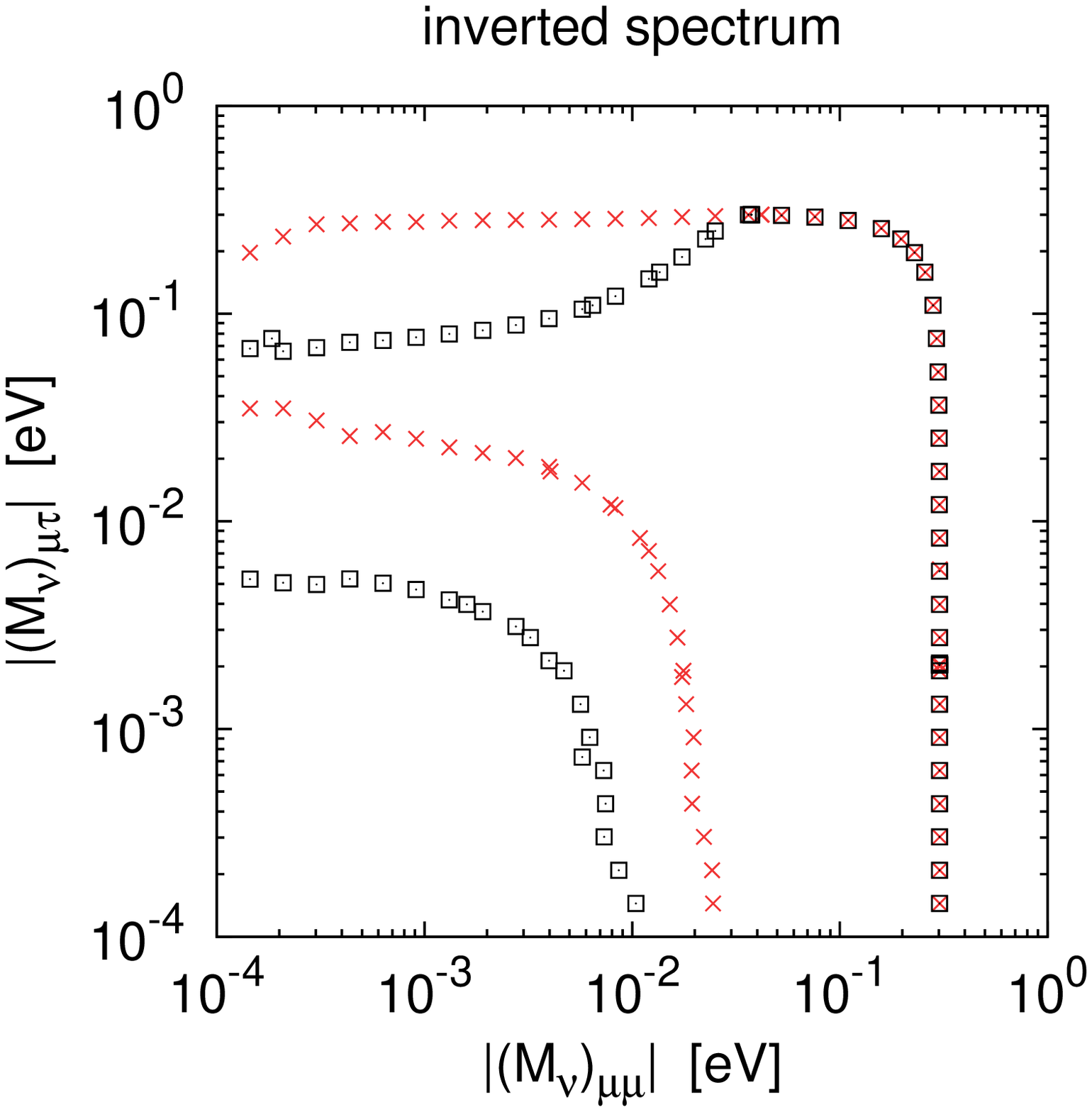}\\
\end{tabular}
\end{center}

\section{Plots based on Forero \textit{et al.} (version 2); \textit{cf.} app.~\ref{fogli}}\label{forero2}

\begin{tabular}[t]{ll}
\includegraphics[angle=-90,keepaspectratio=true,scale=\figurescale]
{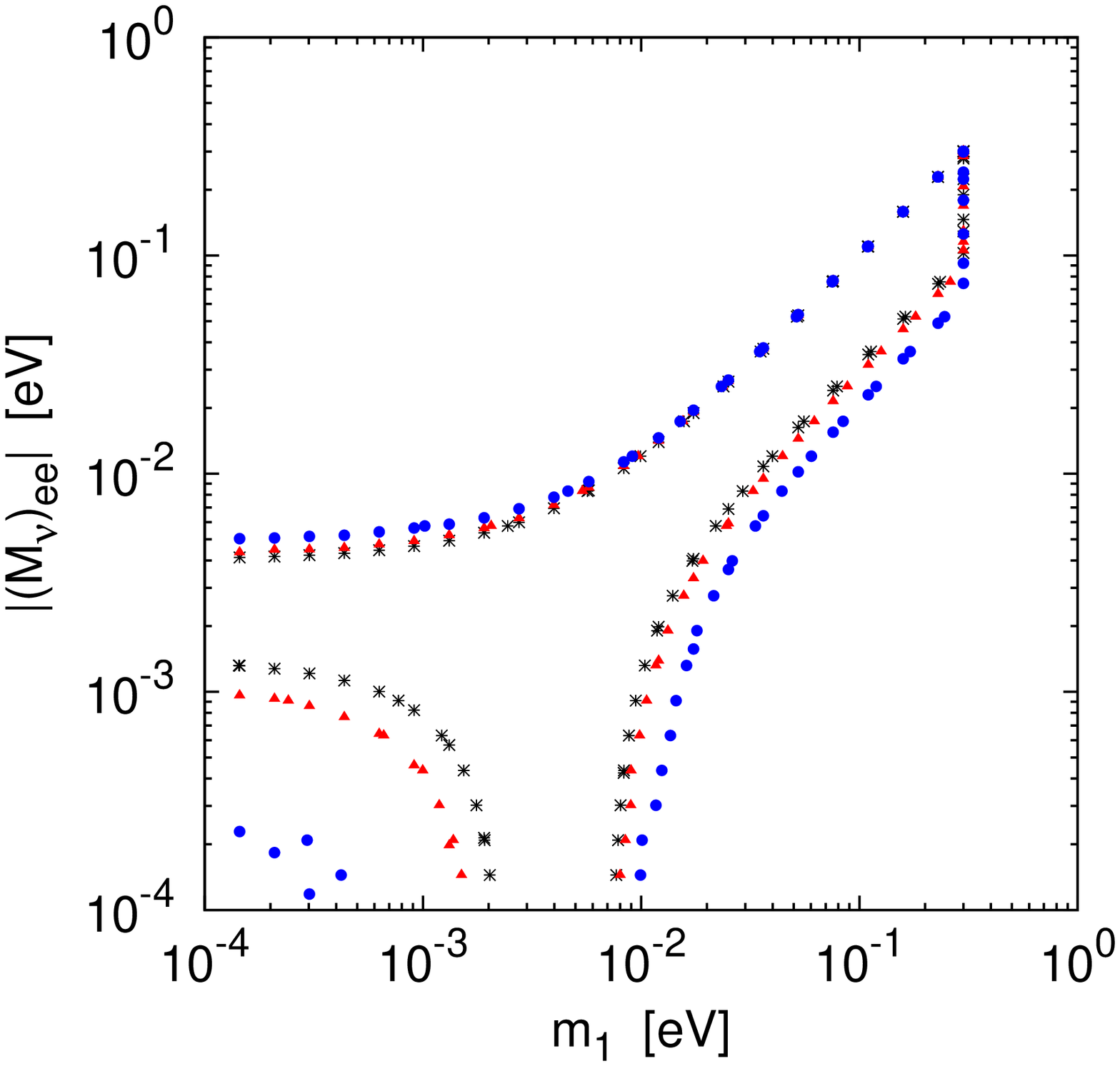} &
\includegraphics[angle=-90,keepaspectratio=true,scale=\figurescale]
{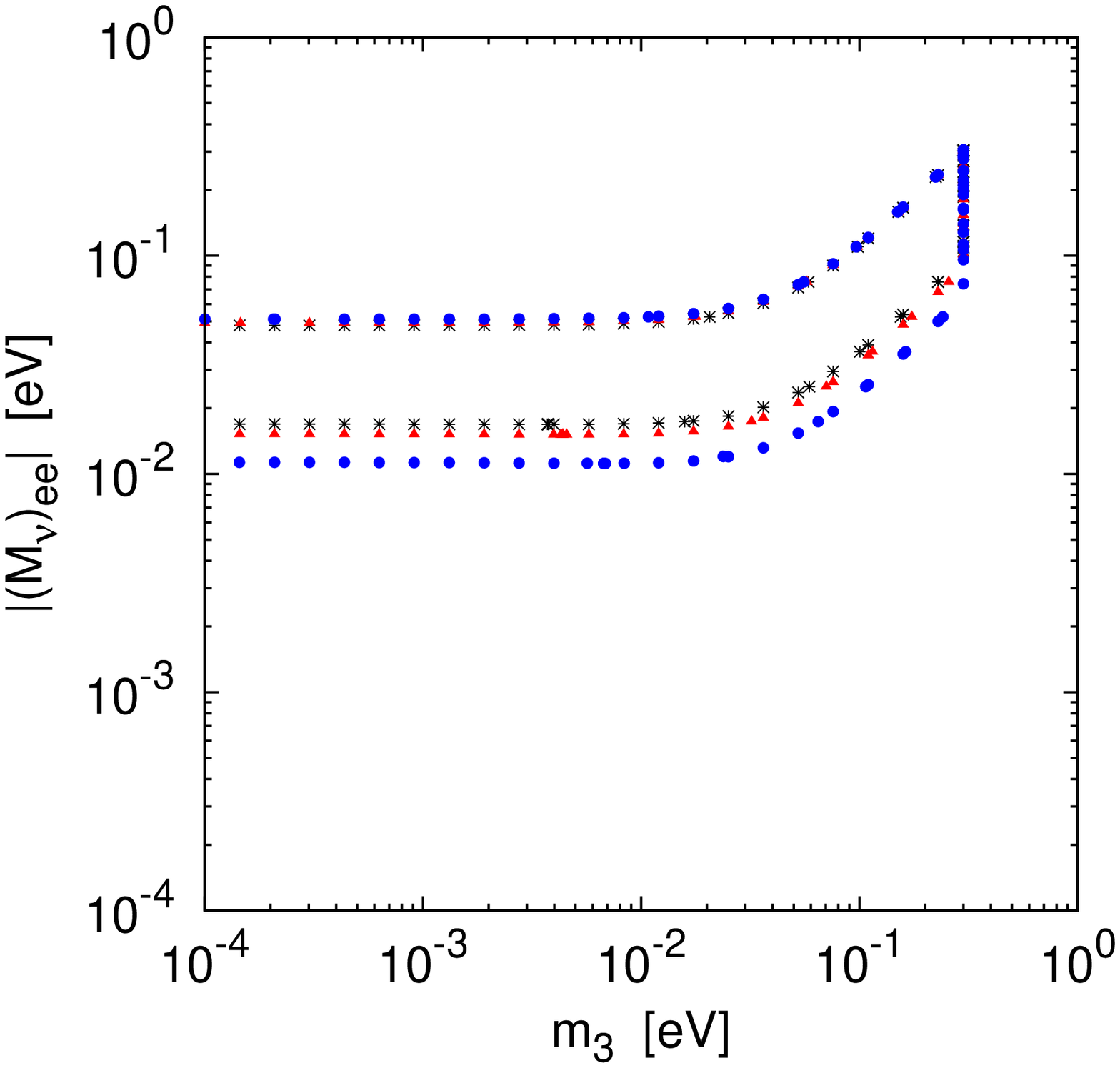}\\
\includegraphics[angle=-90,keepaspectratio=true,scale=\figurescale]
{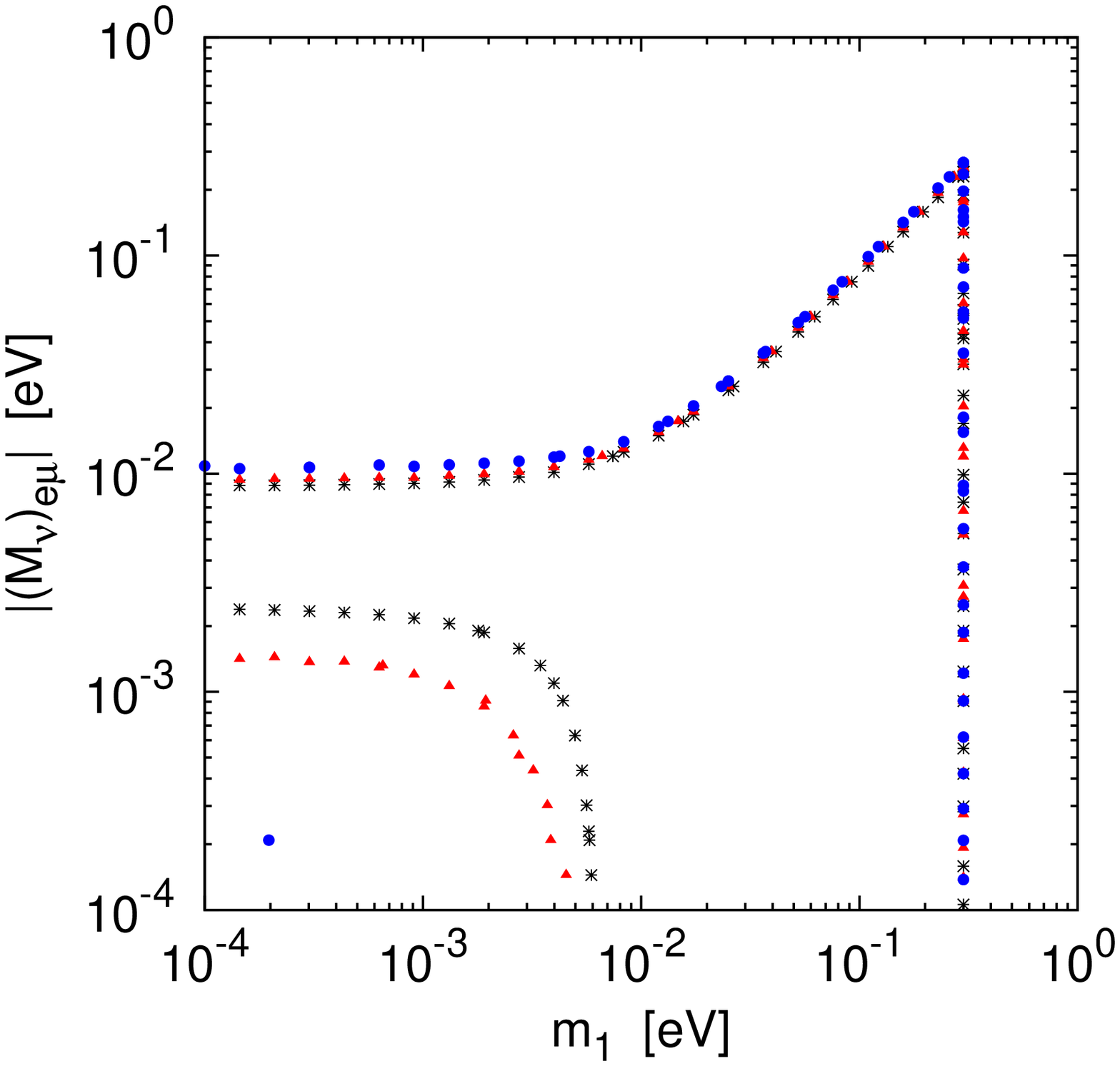} &
\includegraphics[angle=-90,keepaspectratio=true,scale=\figurescale]
{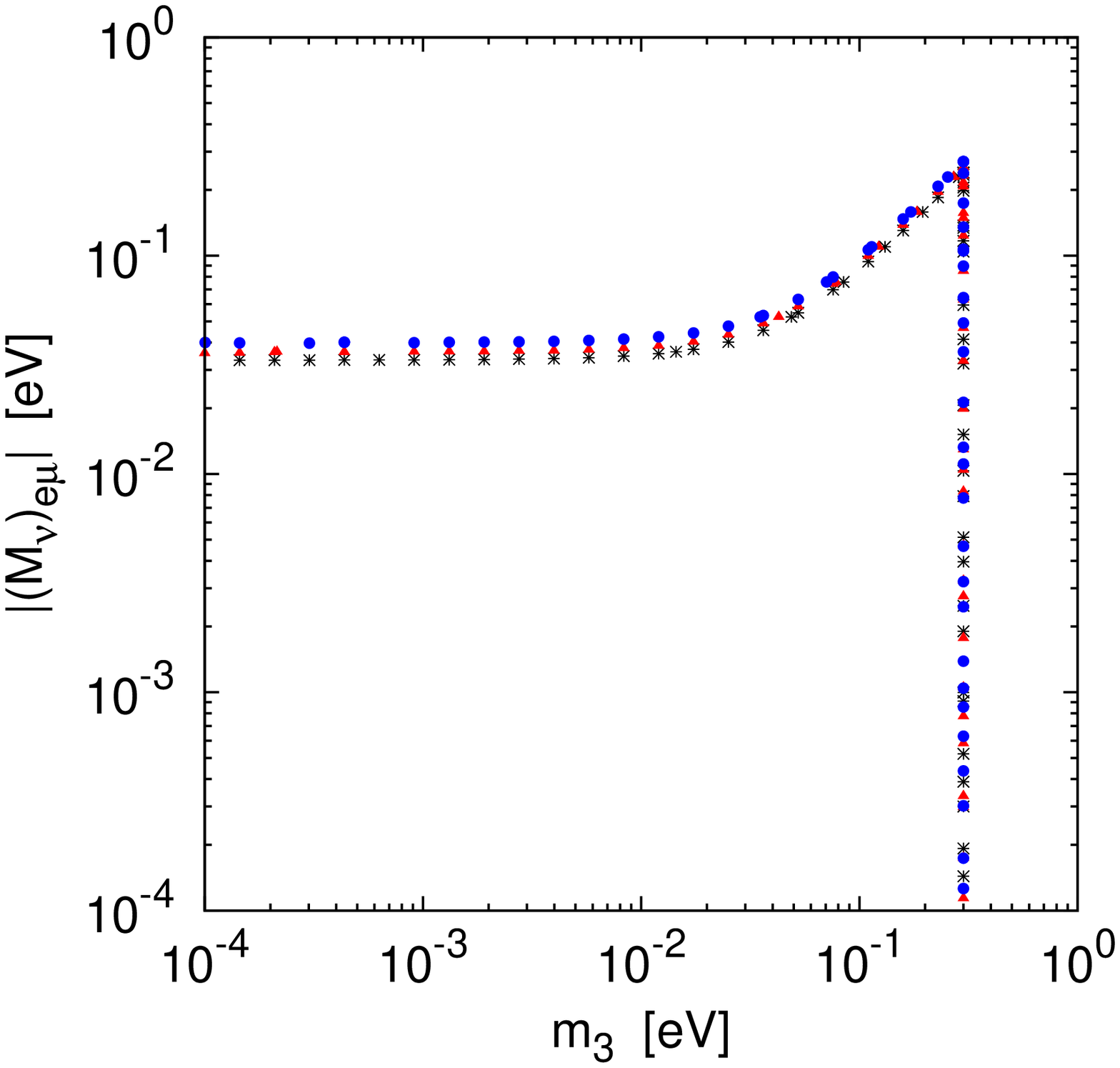}\\
\includegraphics[angle=-90,keepaspectratio=true,scale=\figurescale]
{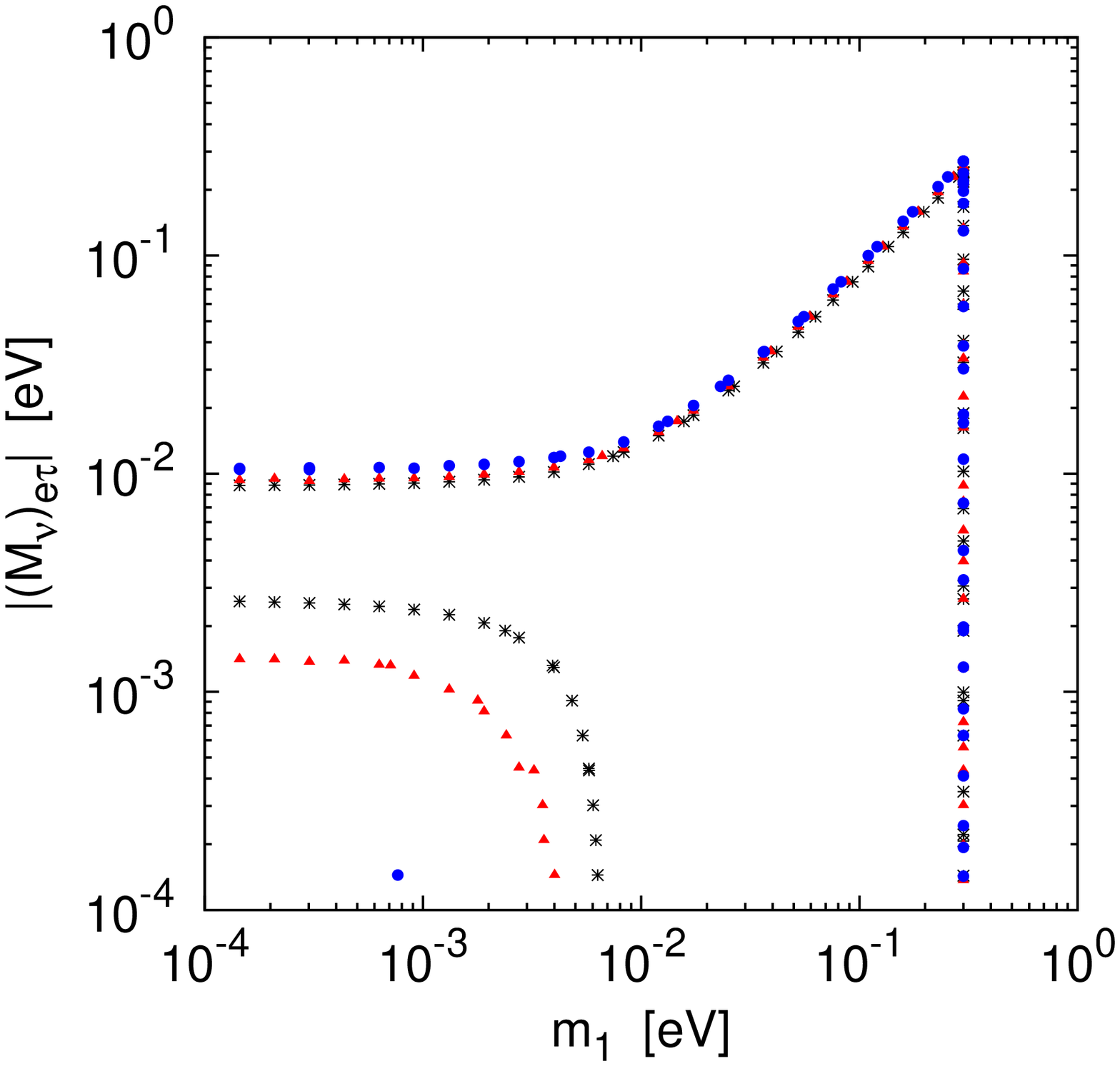} &
\includegraphics[angle=-90,keepaspectratio=true,scale=\figurescale]
{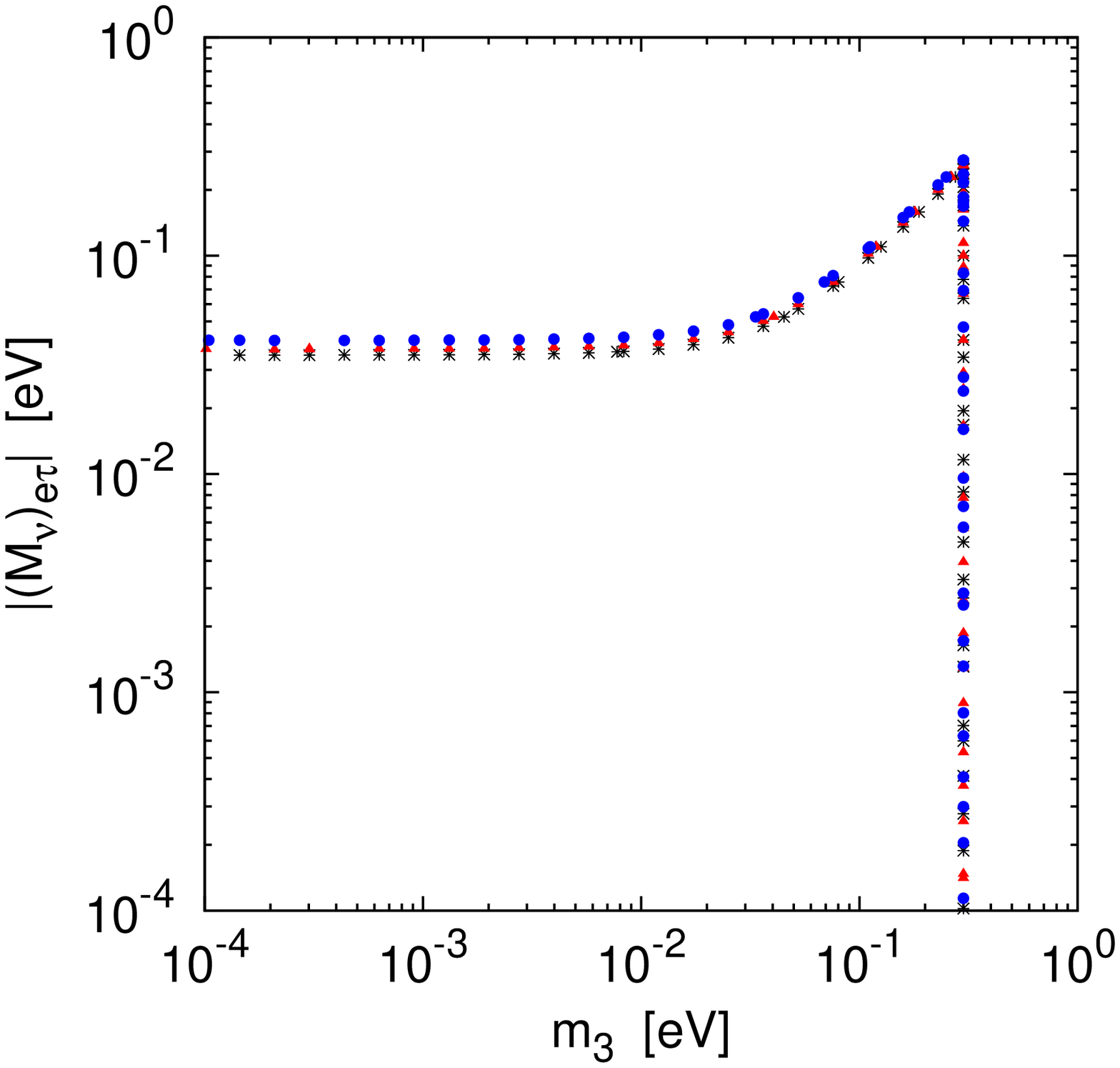}\\
\end{tabular}

\begin{tabular}[t]{ll}
\includegraphics[angle=-90,keepaspectratio=true,scale=\figurescale]
{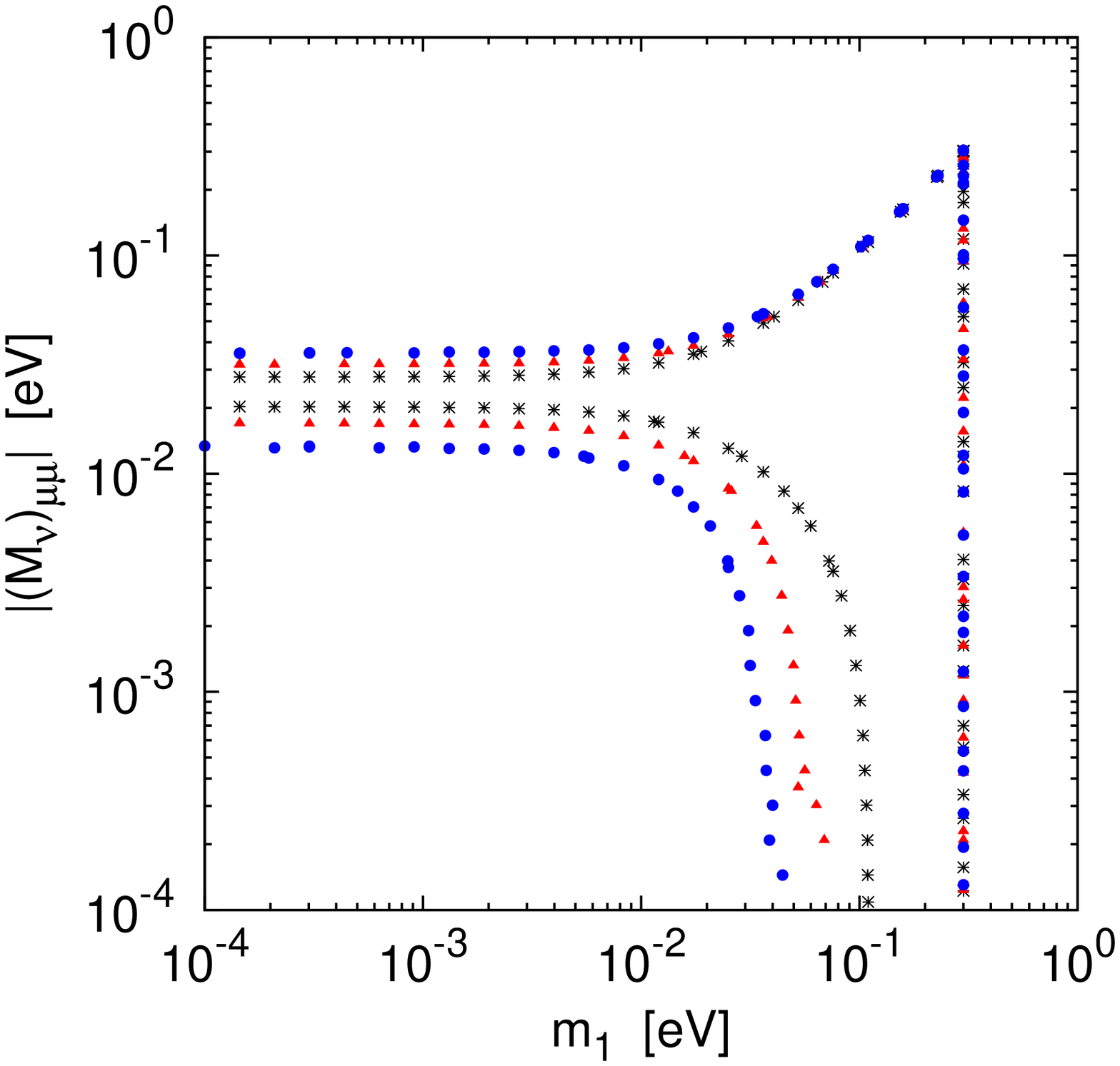} &
\includegraphics[angle=-90,keepaspectratio=true,scale=\figurescale]
{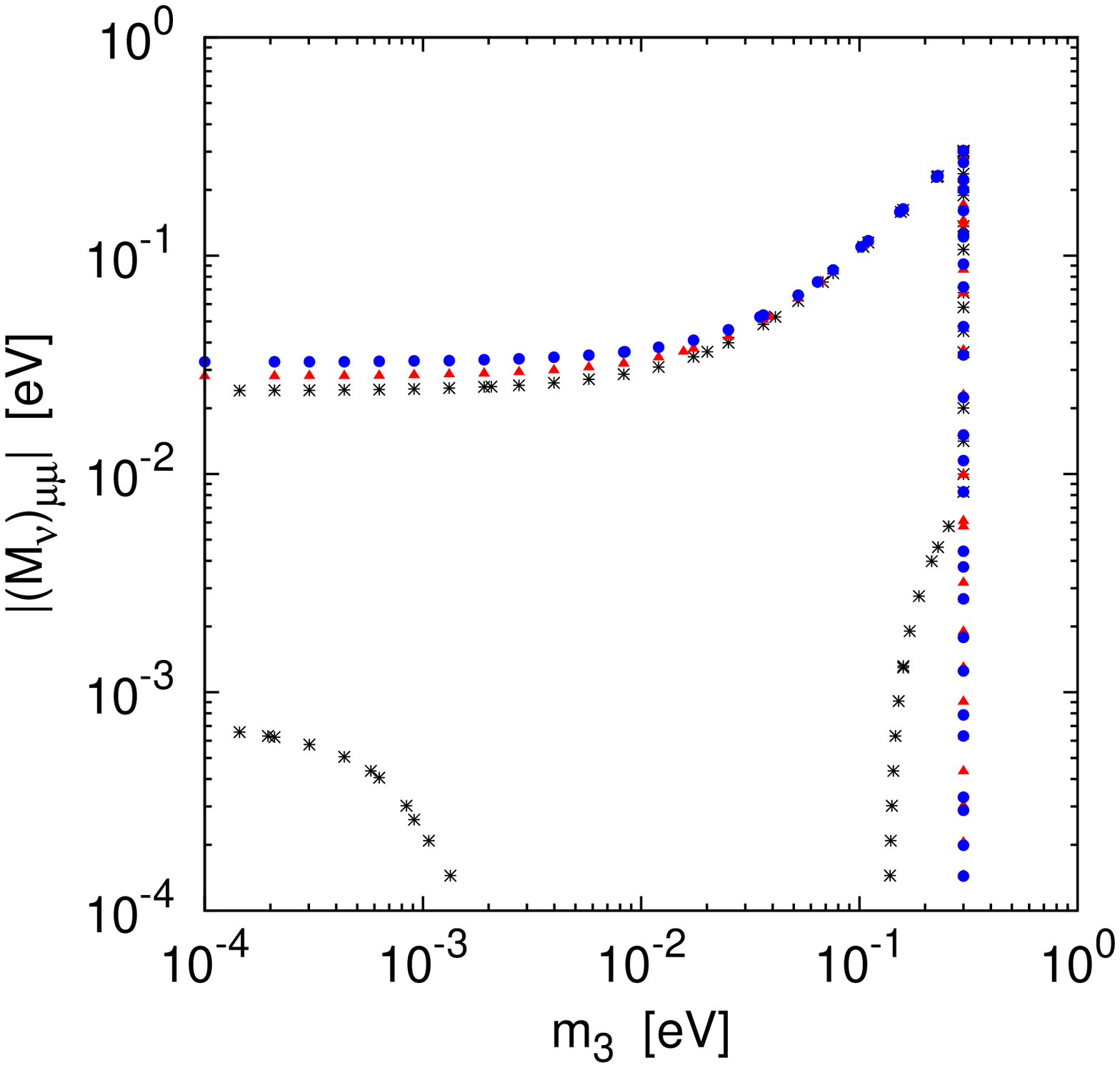}\\
\includegraphics[angle=-90,keepaspectratio=true,scale=\figurescale]
{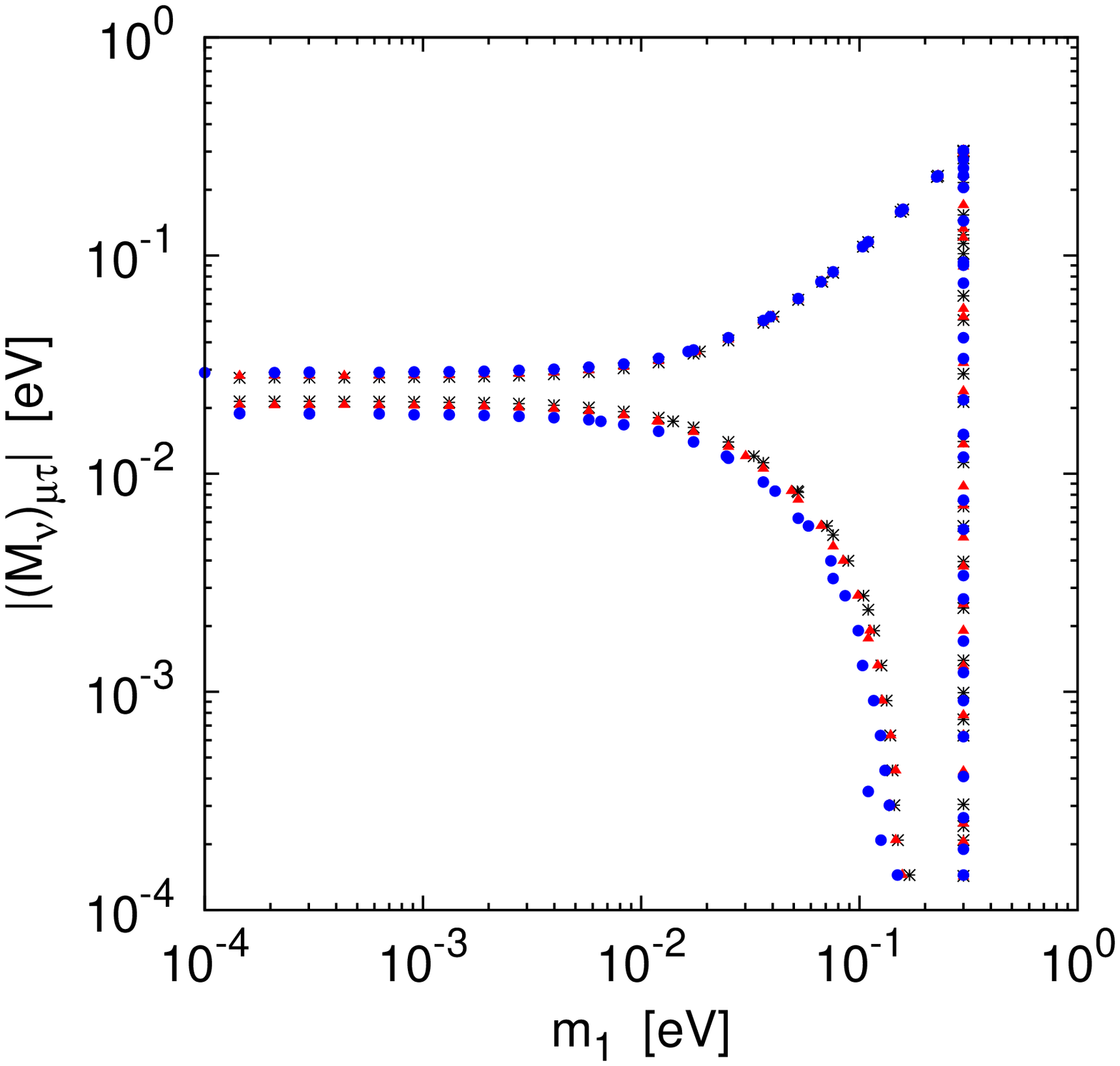} &
\includegraphics[angle=-90,keepaspectratio=true,scale=\figurescale]
{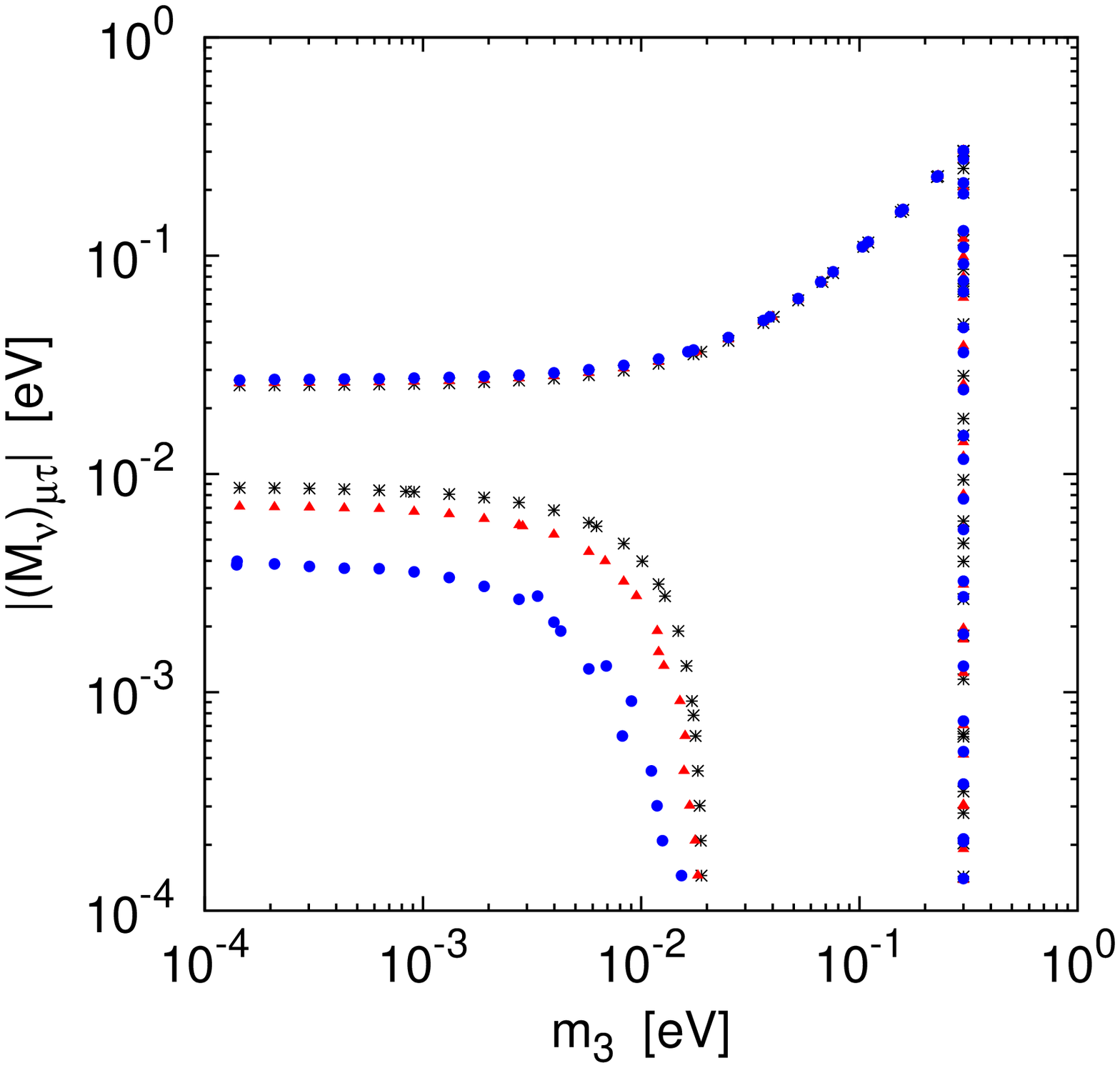}\\
\includegraphics[angle=-90,keepaspectratio=true,scale=\figurescale]
{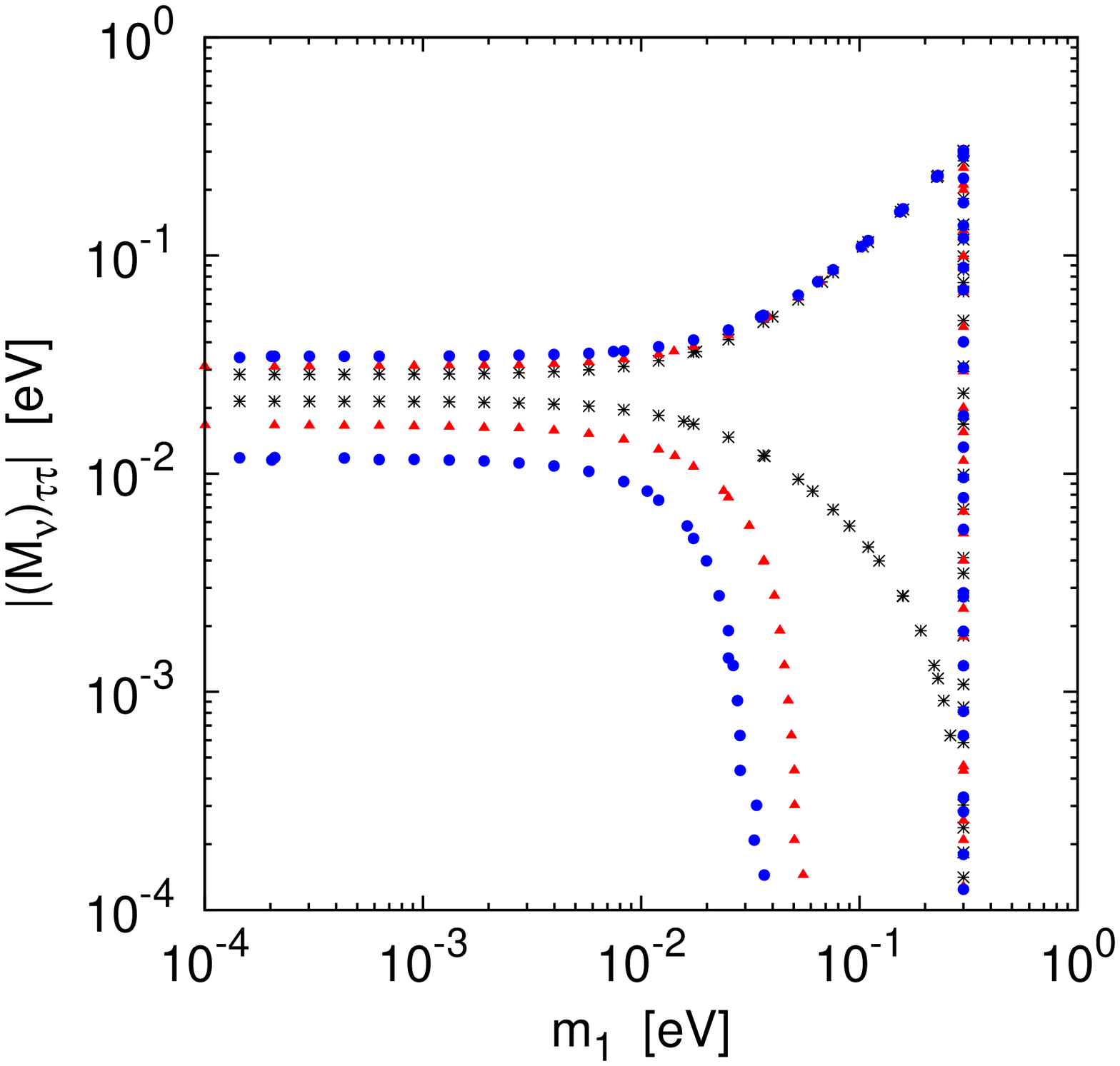} &
\includegraphics[angle=-90,keepaspectratio=true,scale=\figurescale]
{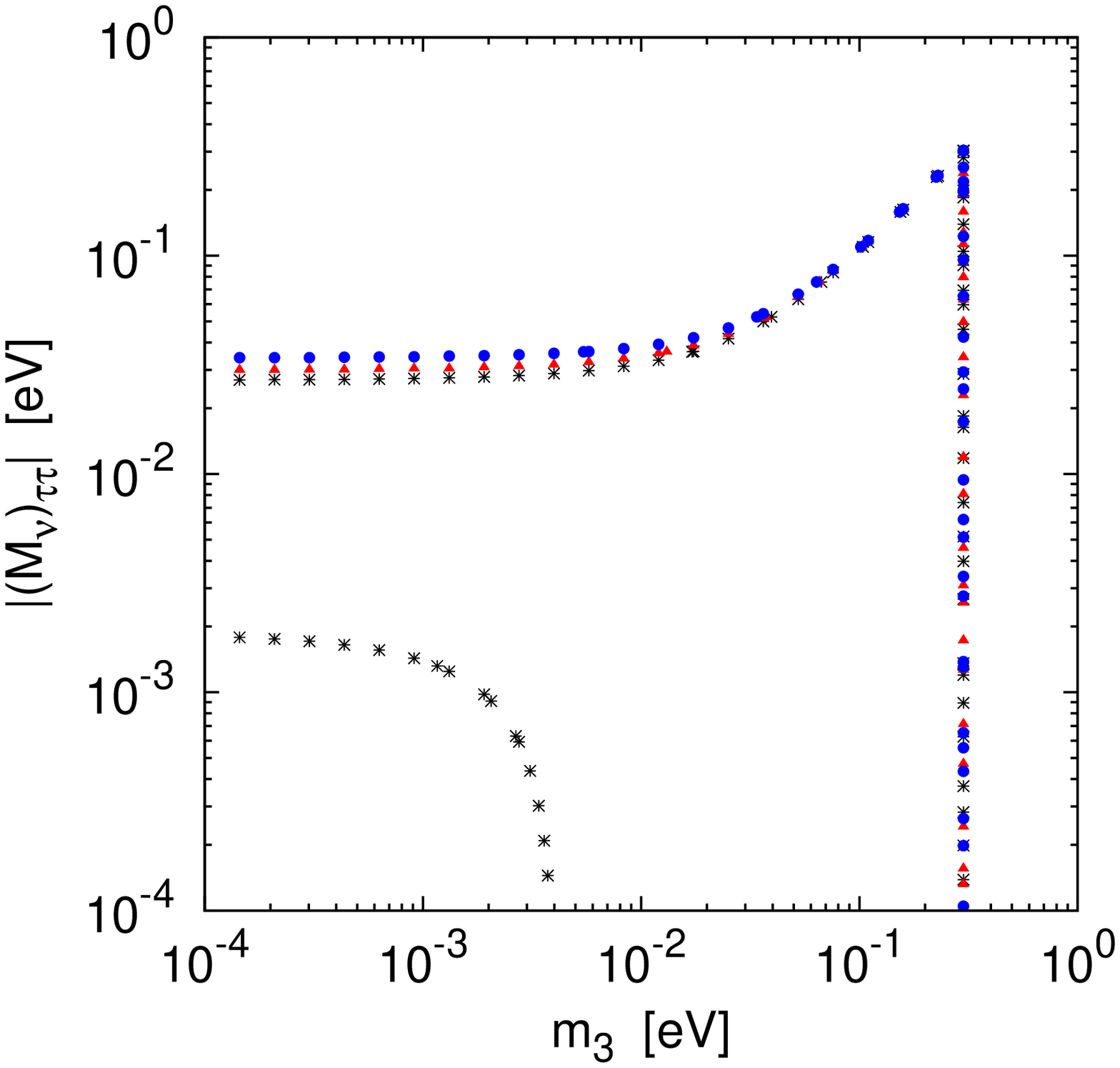}\\
\end{tabular}

\begin{tabular}[t]{ll}
\includegraphics[angle=-90,keepaspectratio=true,scale=\figurescale]
{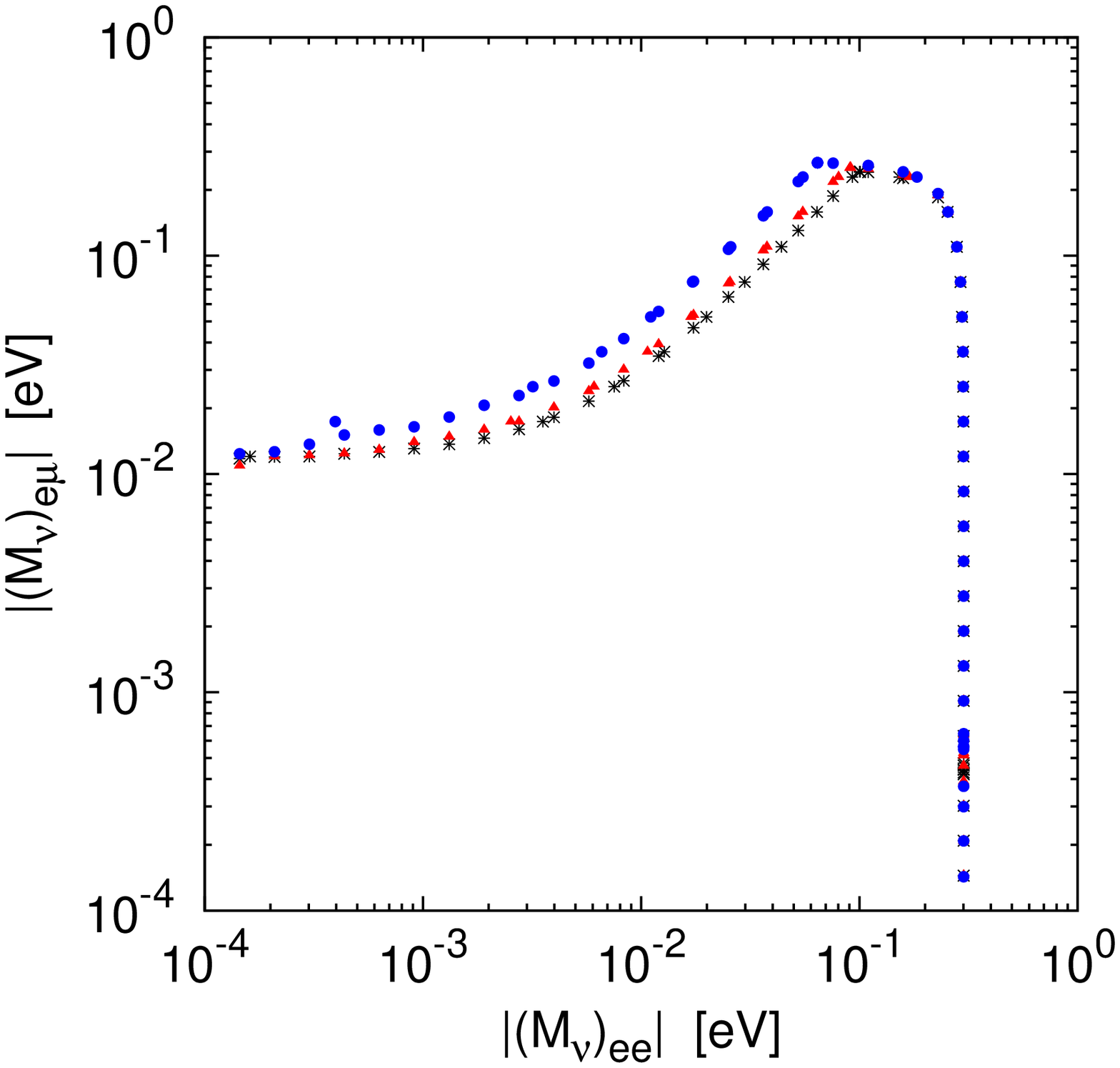} &
\includegraphics[angle=-90,keepaspectratio=true,scale=\figurescale]
{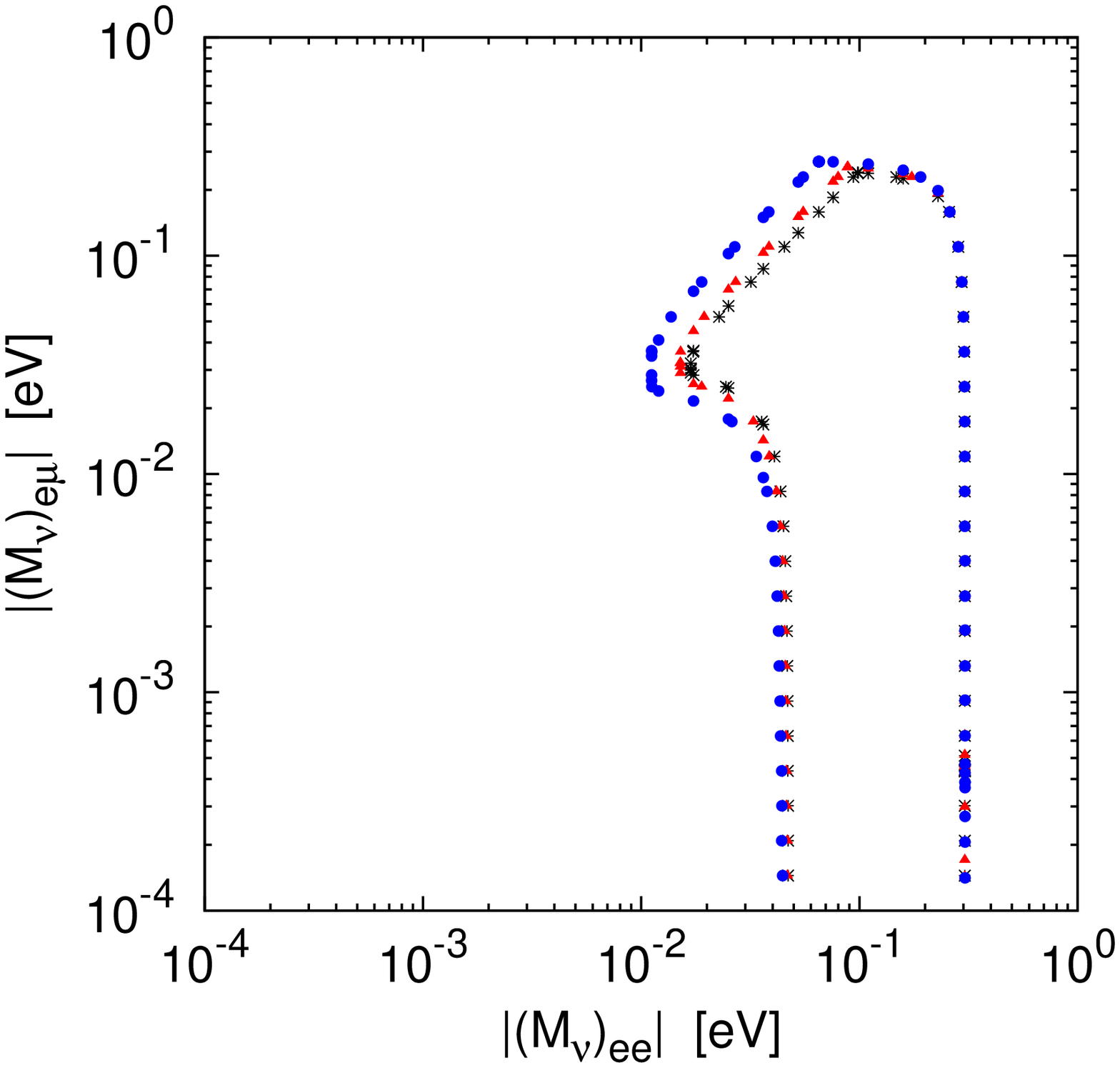}\\
\includegraphics[angle=-90,keepaspectratio=true,scale=\figurescale]
{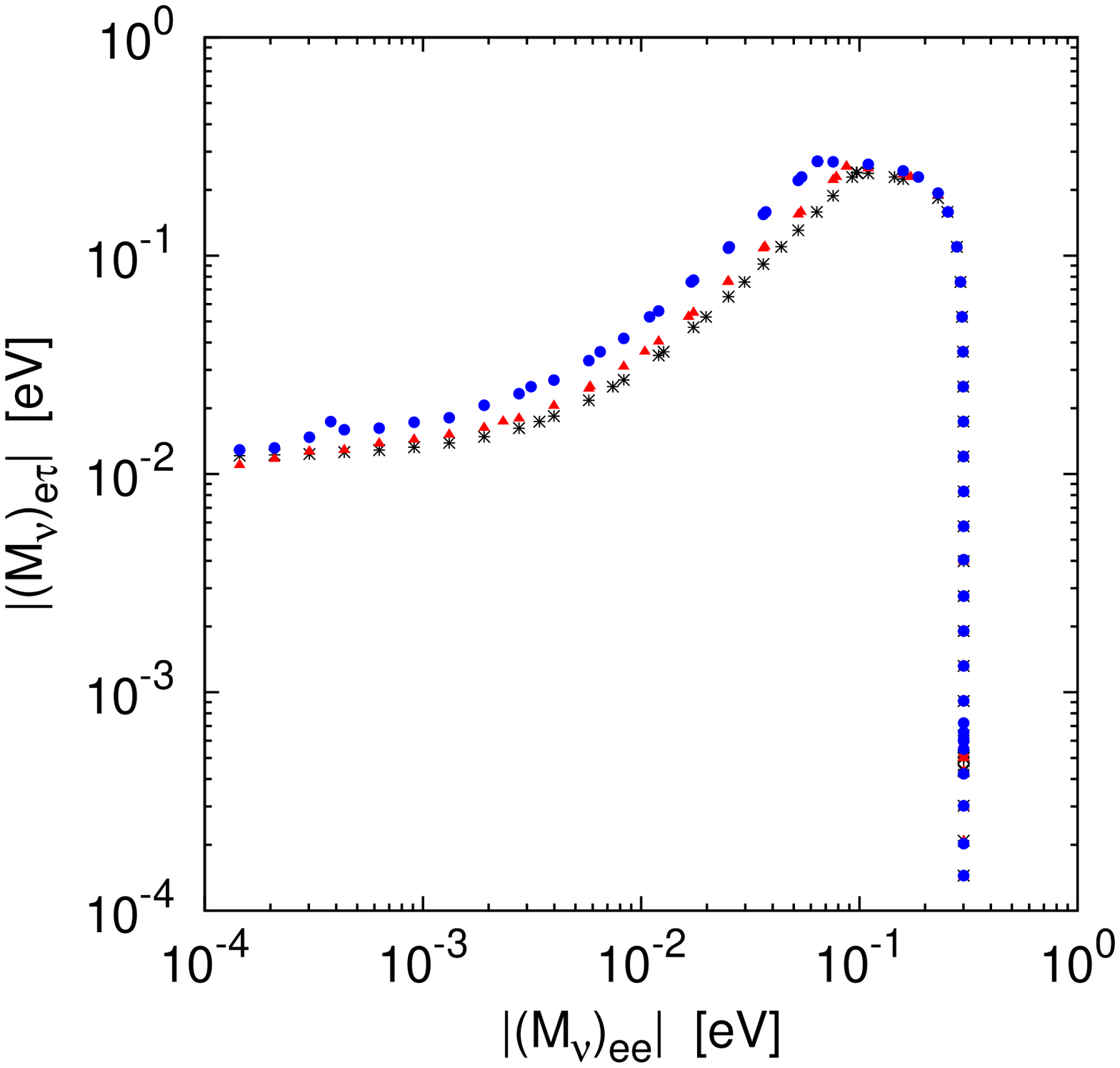} &
\includegraphics[angle=-90,keepaspectratio=true,scale=\figurescale]
{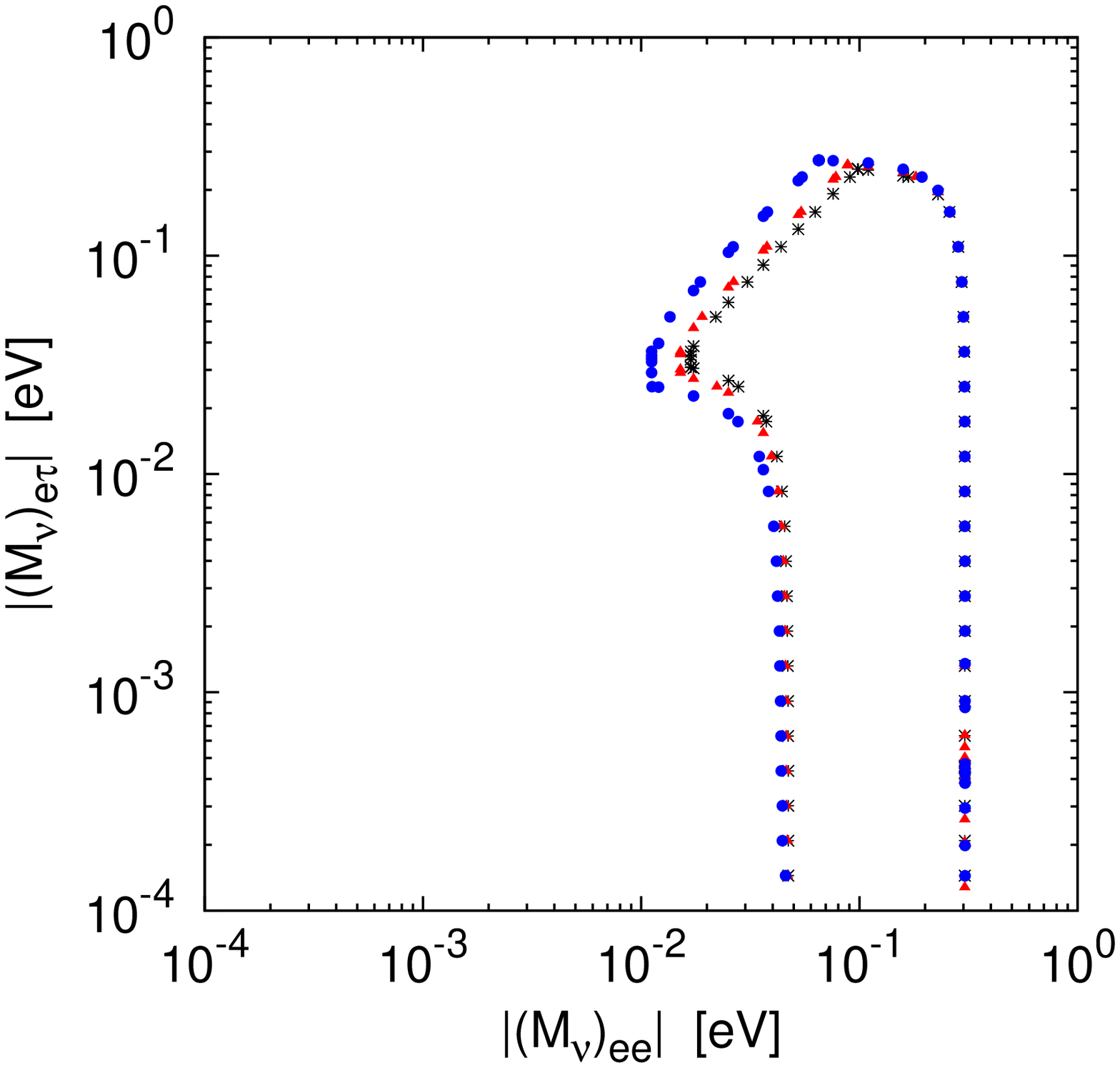}\\
\includegraphics[angle=-90,keepaspectratio=true,scale=\figurescale]
{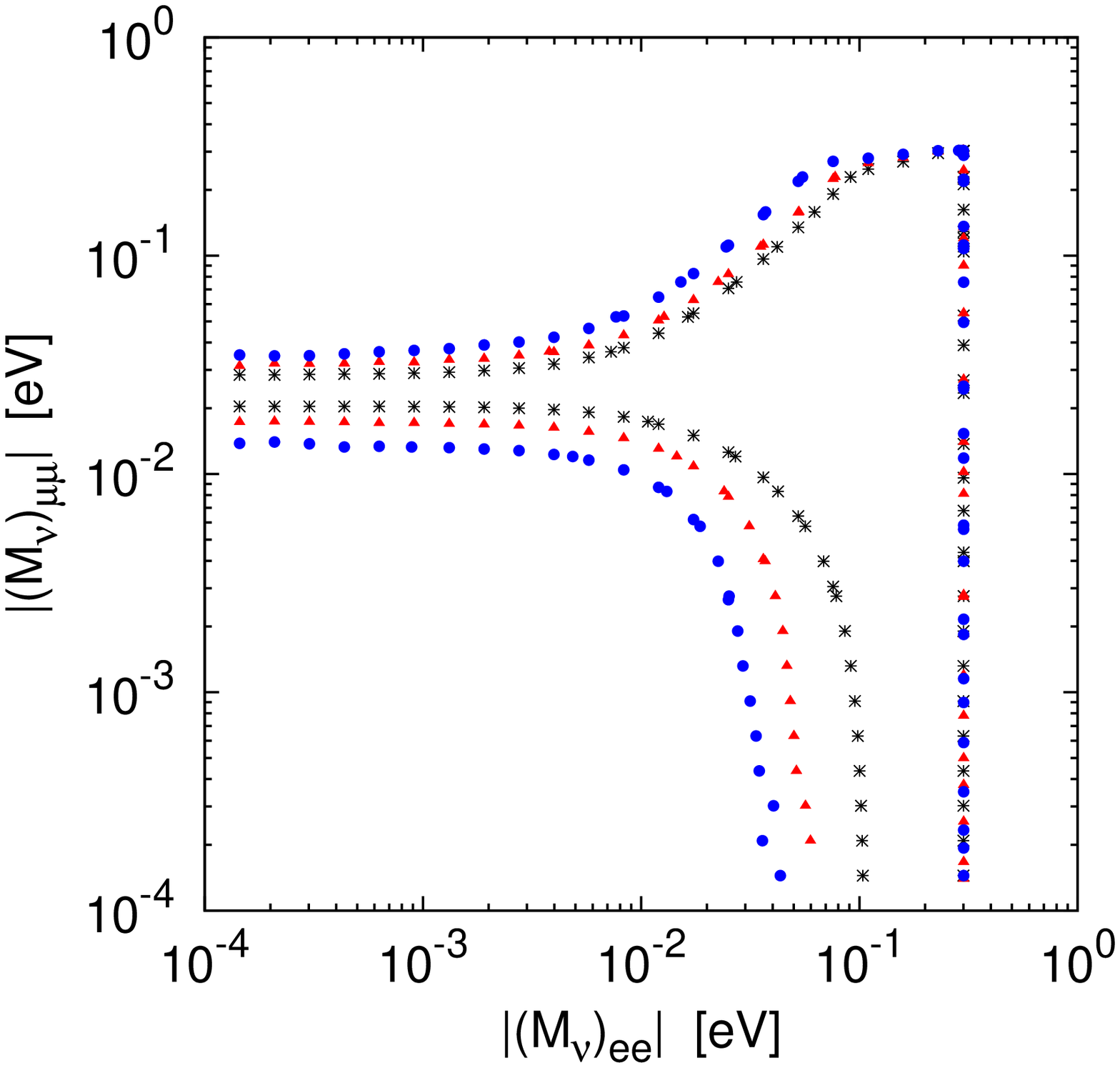} &
\includegraphics[angle=-90,keepaspectratio=true,scale=\figurescale]
{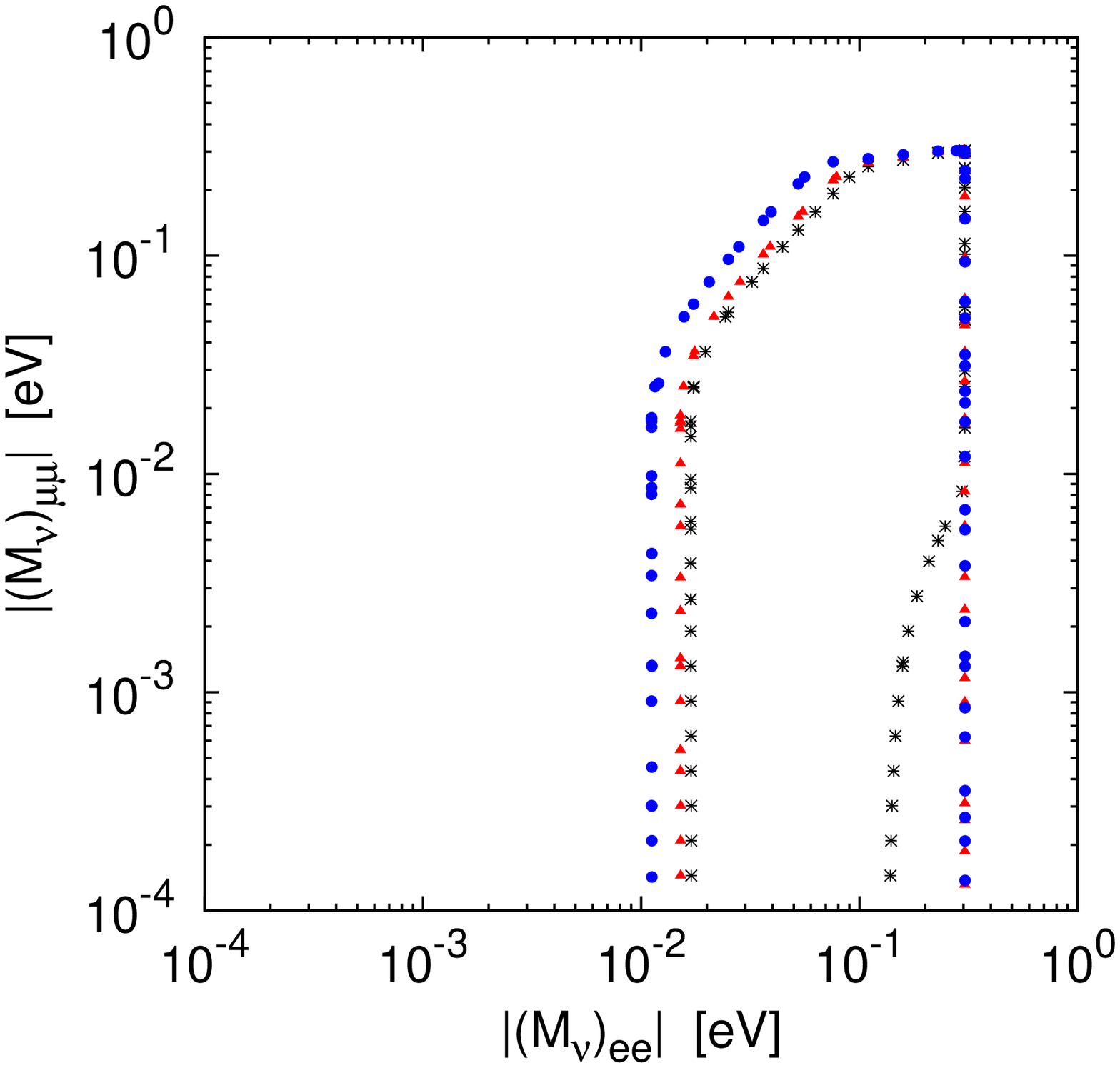}\\
\end{tabular}

\begin{tabular}[t]{ll}
\includegraphics[angle=-90,keepaspectratio=true,scale=\figurescale]
{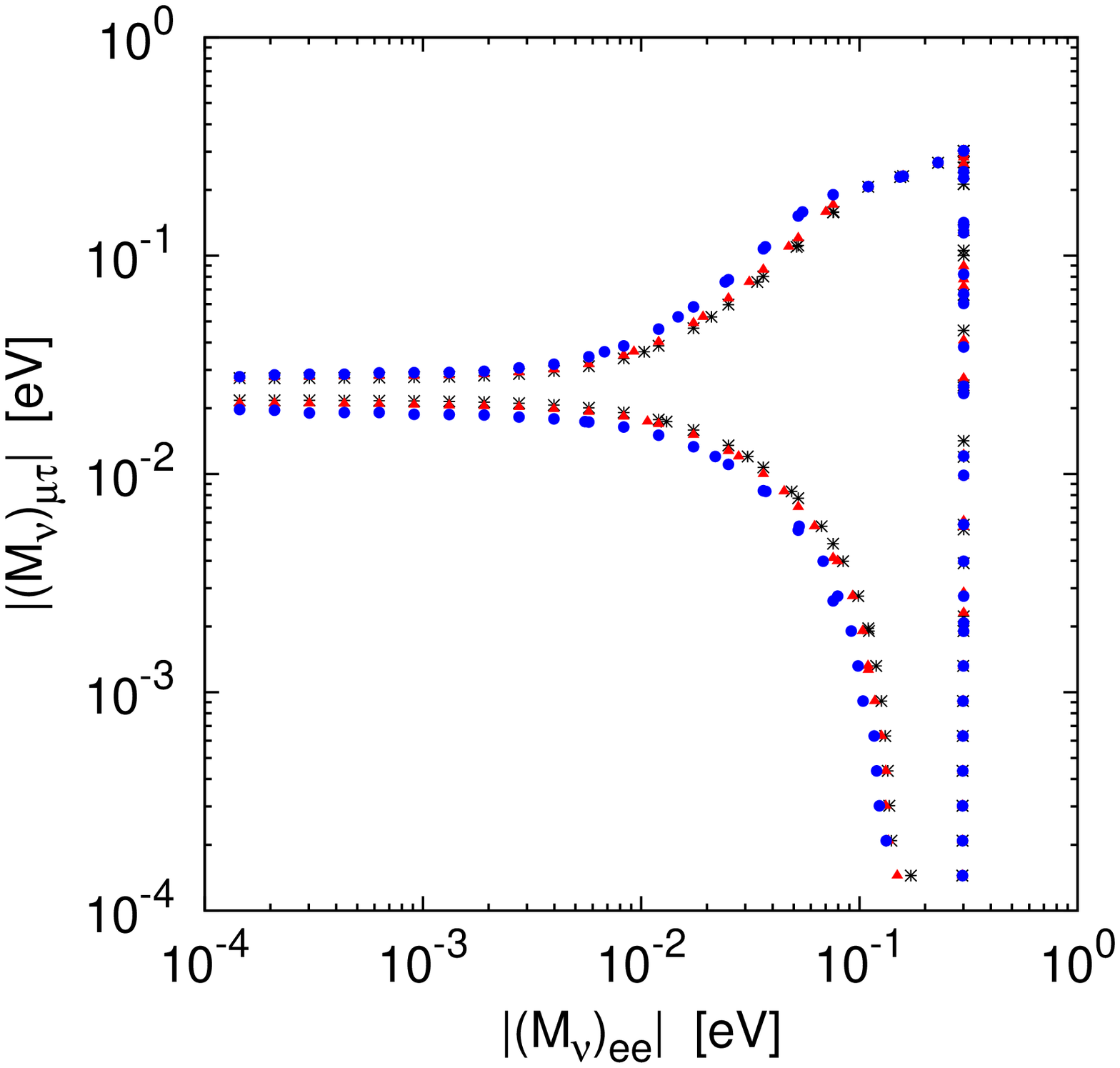} &
\includegraphics[angle=-90,keepaspectratio=true,scale=\figurescale]
{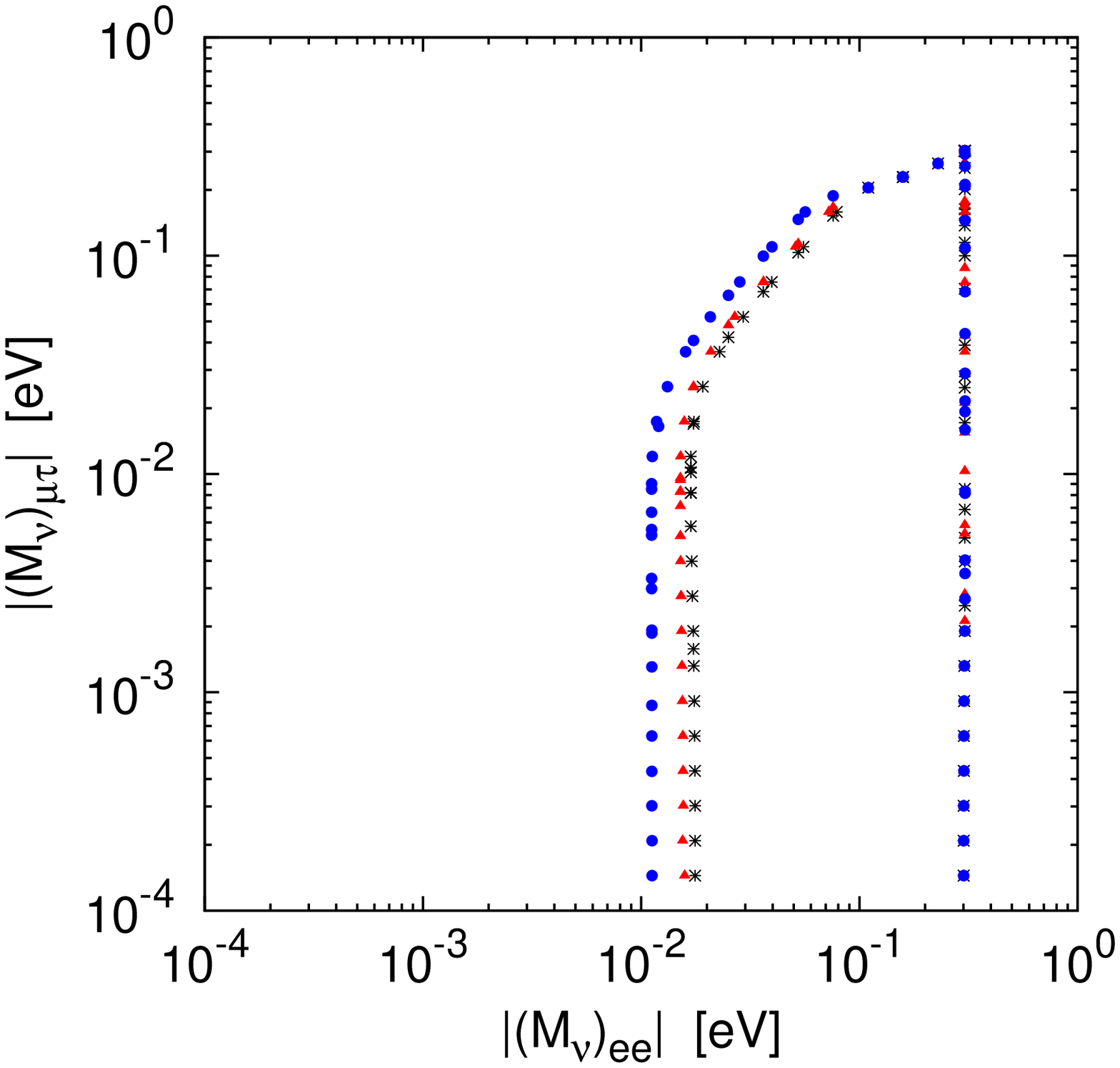}\\
\includegraphics[angle=-90,keepaspectratio=true,scale=\figurescale]
{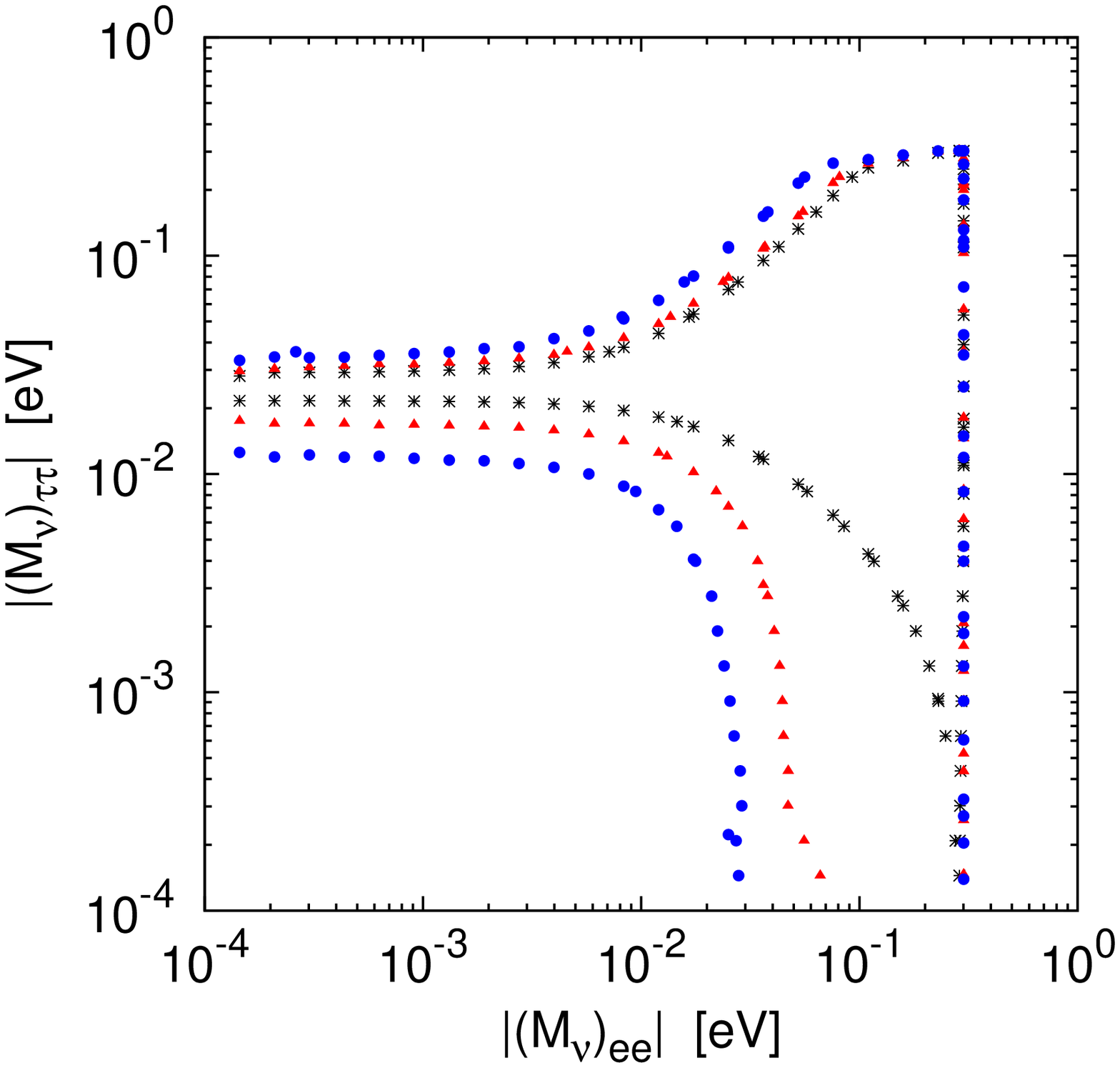} &
\includegraphics[angle=-90,keepaspectratio=true,scale=\figurescale]
{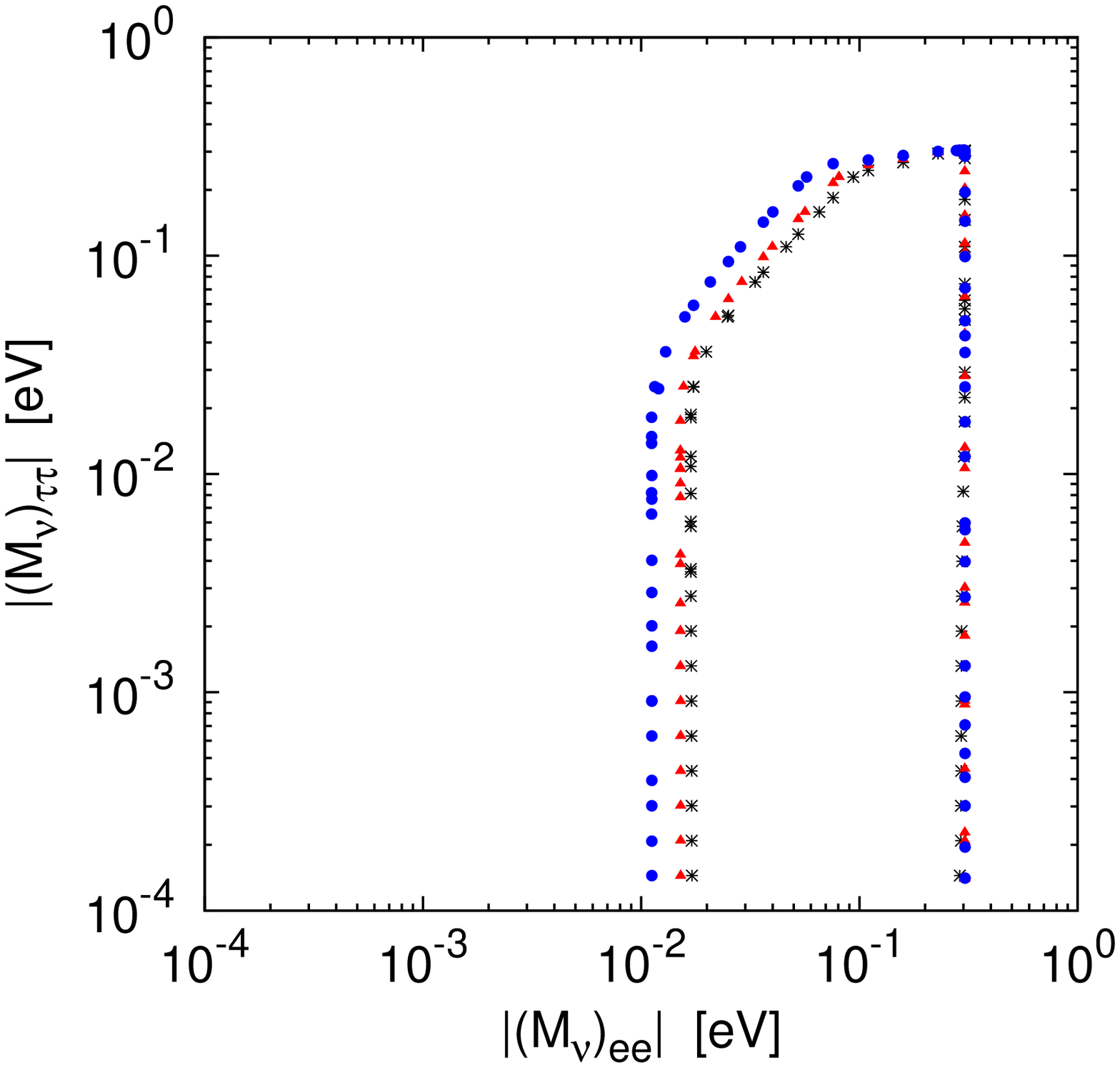}\\
\includegraphics[angle=-90,keepaspectratio=true,scale=\figurescale]
{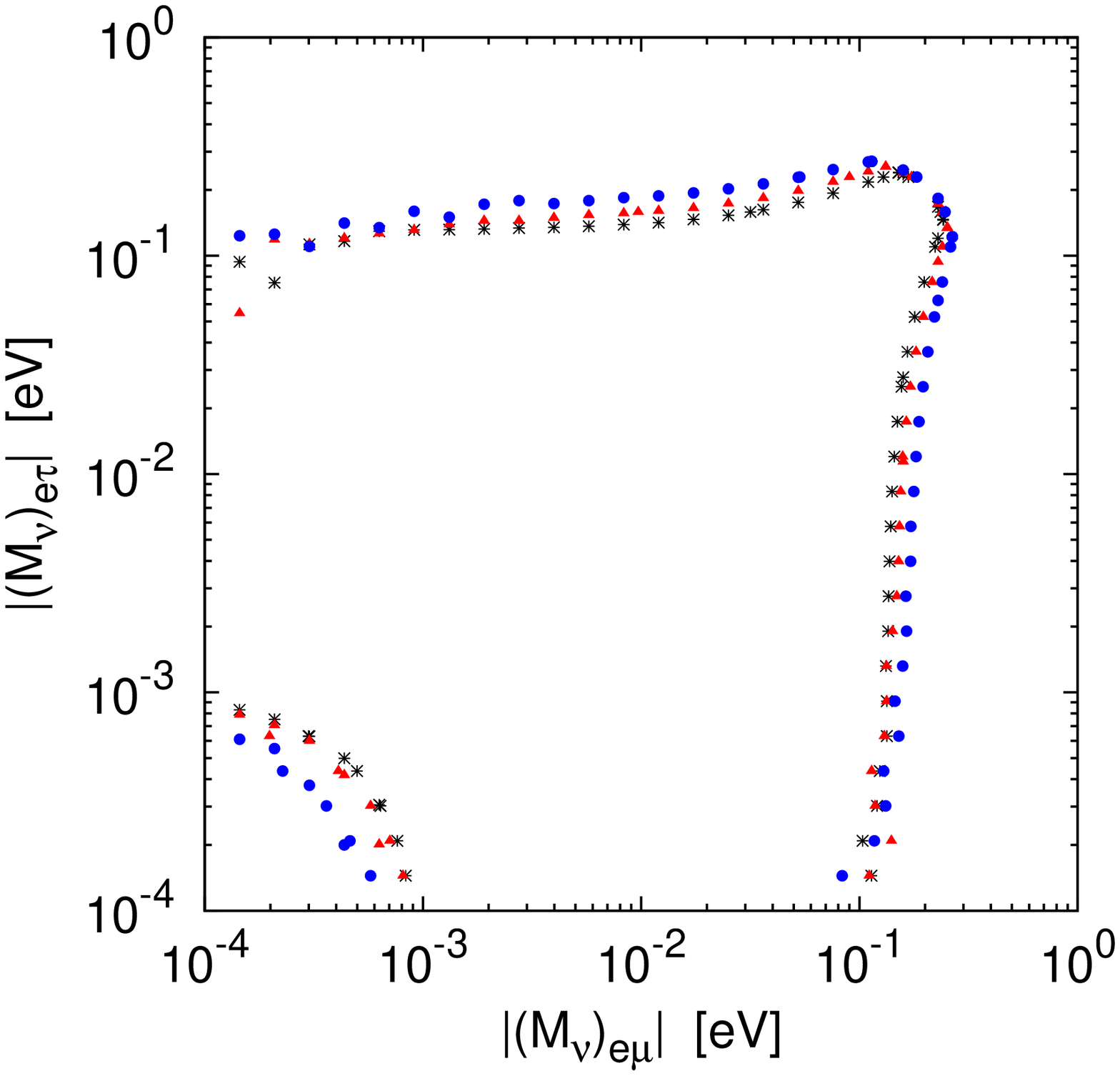} &
\includegraphics[angle=-90,keepaspectratio=true,scale=\figurescale]
{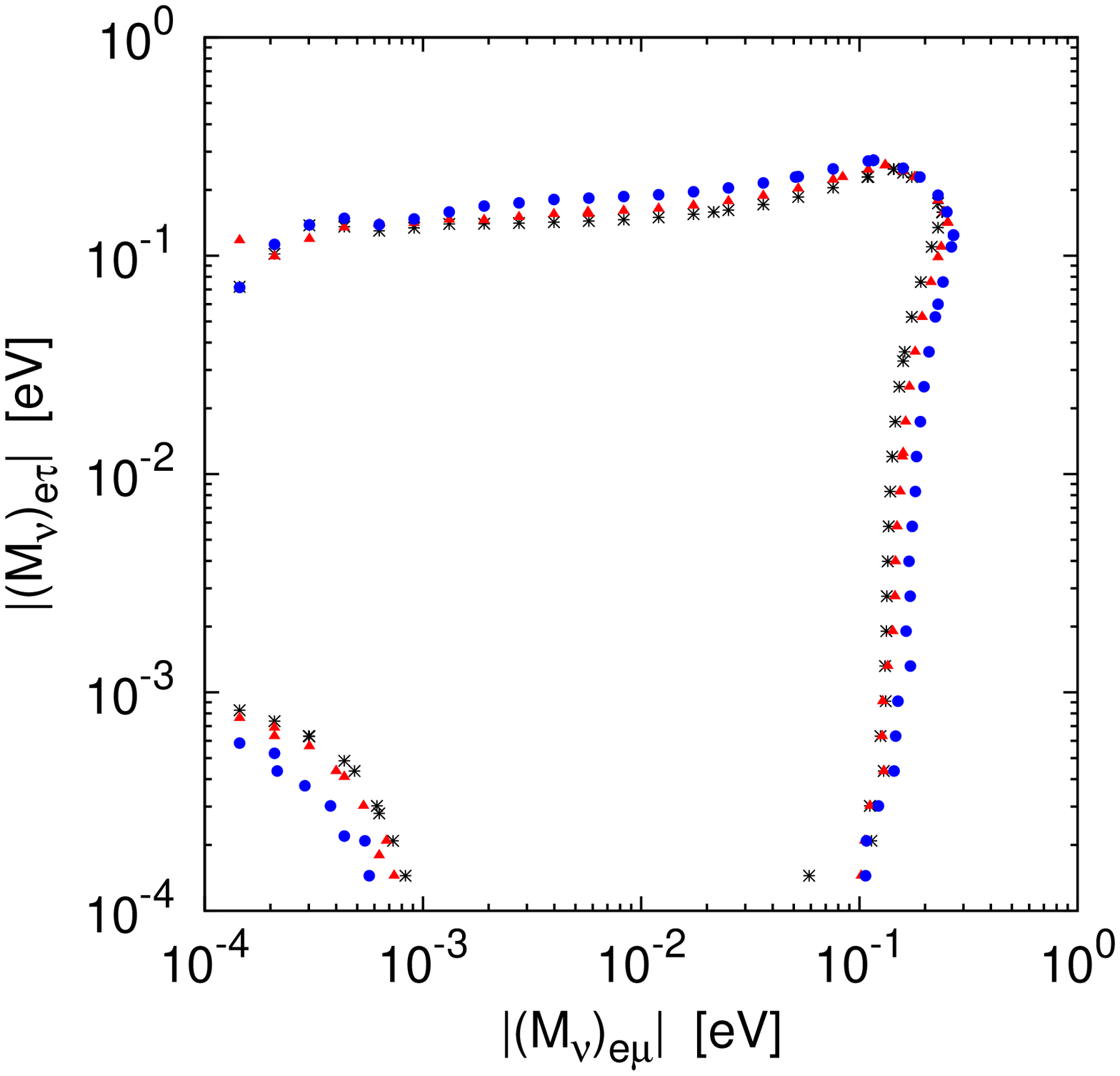}\\
\end{tabular}

\begin{tabular}[t]{ll}
\includegraphics[angle=-90,keepaspectratio=true,scale=\figurescale]
{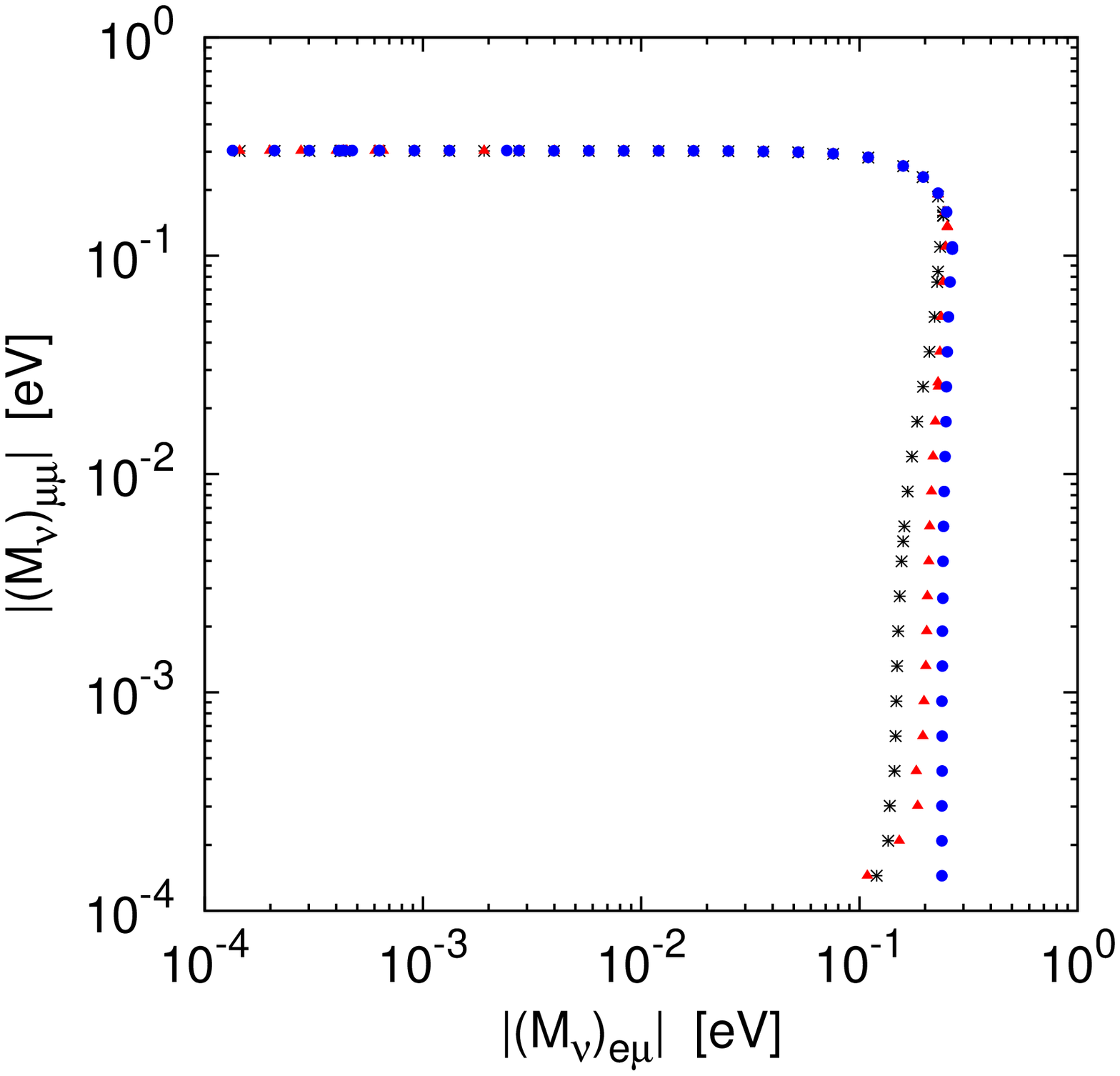} &
\includegraphics[angle=-90,keepaspectratio=true,scale=\figurescale]
{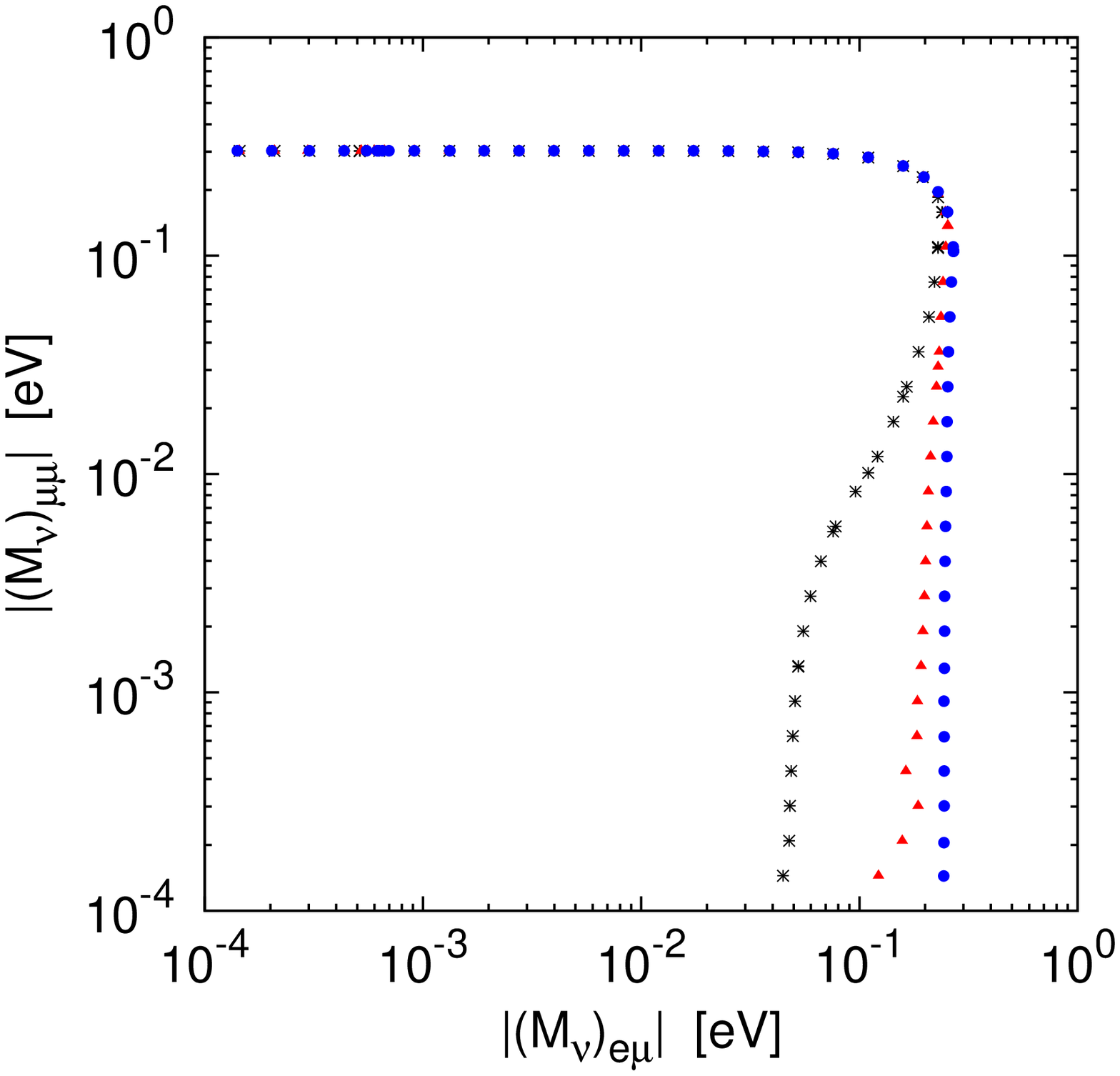}\\
\includegraphics[angle=-90,keepaspectratio=true,scale=\figurescale]
{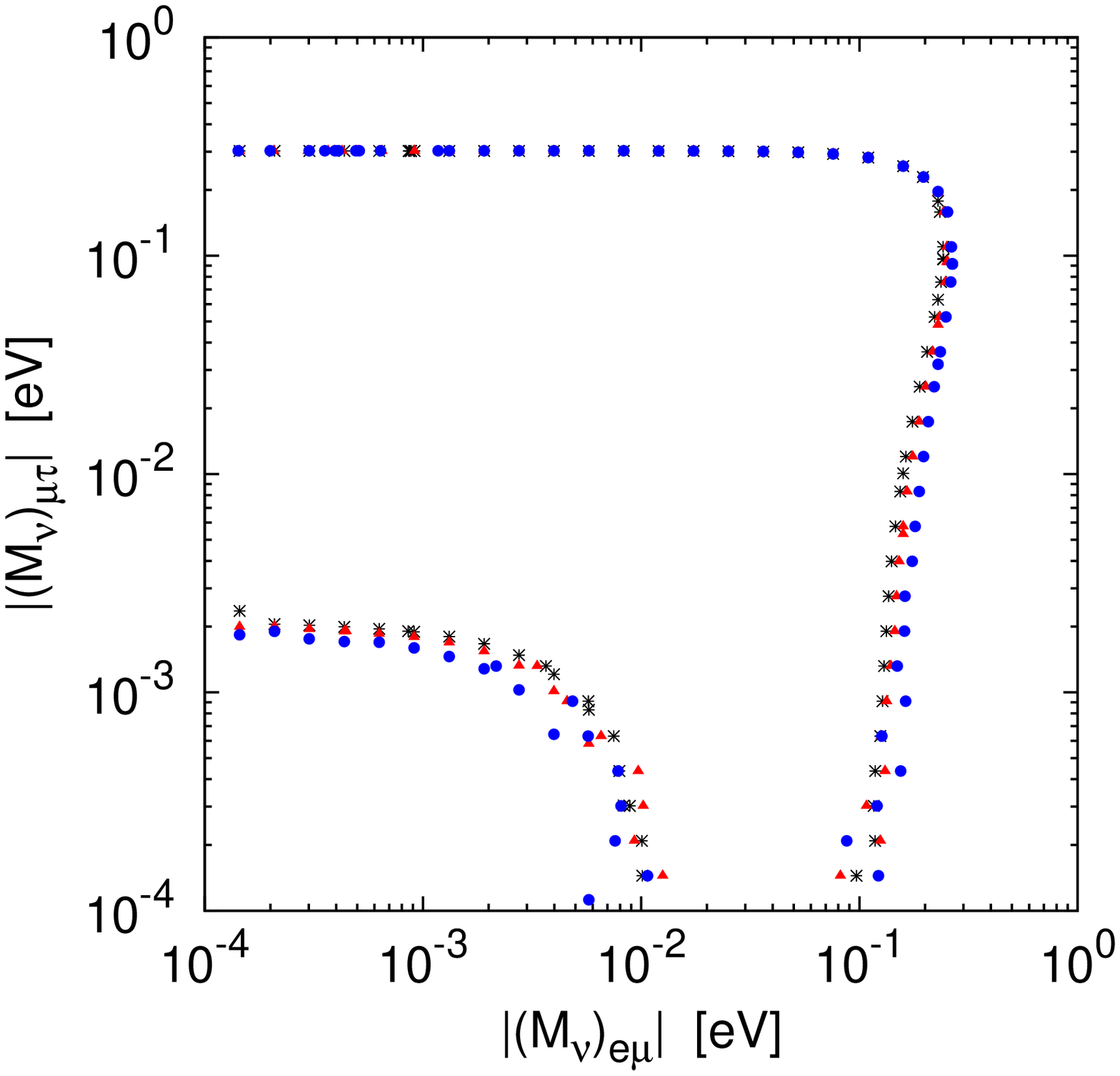} &
\includegraphics[angle=-90,keepaspectratio=true,scale=\figurescale]
{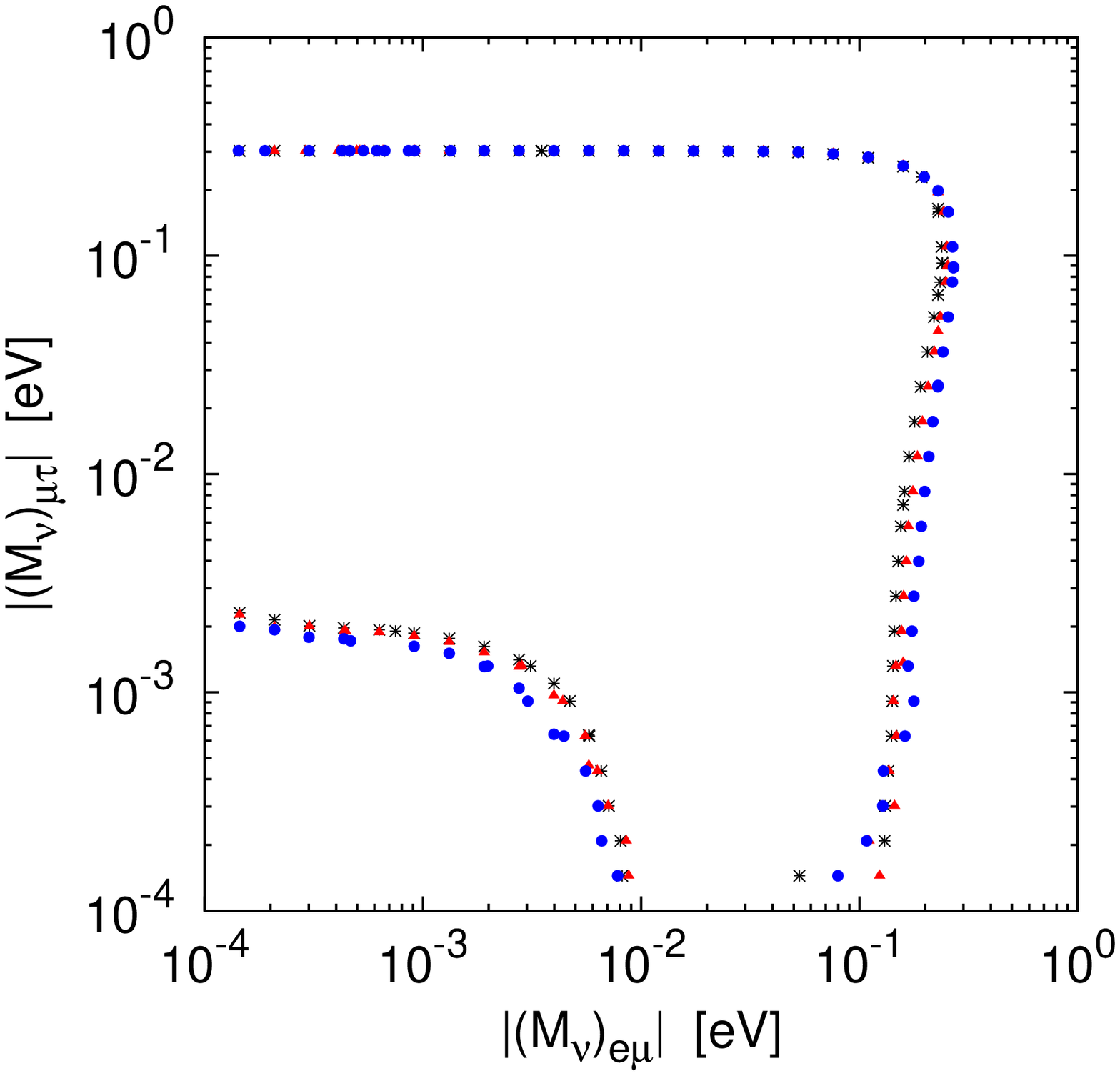}\\
\includegraphics[angle=-90,keepaspectratio=true,scale=\figurescale]
{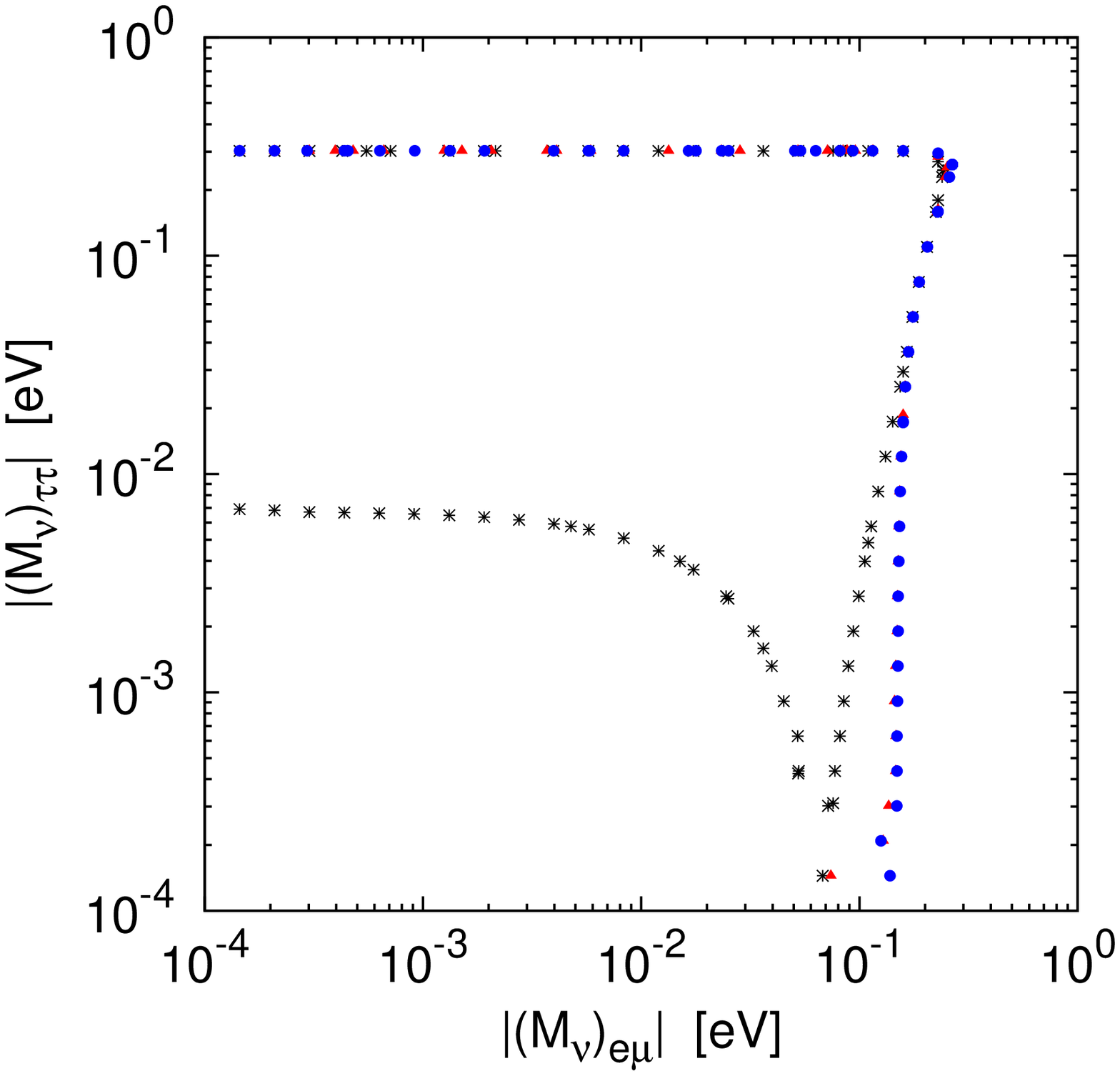} &
\includegraphics[angle=-90,keepaspectratio=true,scale=\figurescale]
{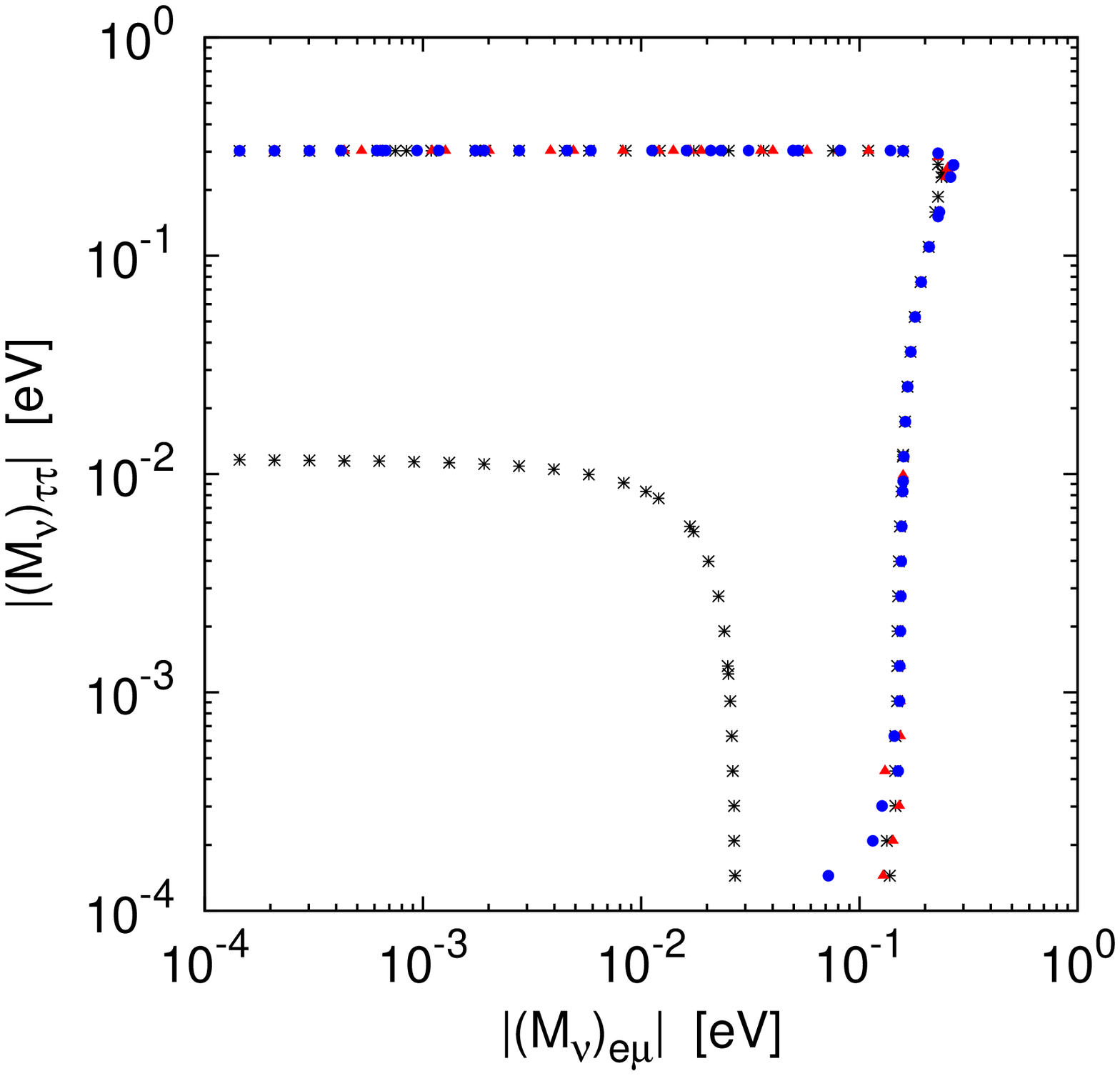}\\
\end{tabular}

\begin{tabular}[t]{ll}
\includegraphics[angle=-90,keepaspectratio=true,scale=\figurescale]
{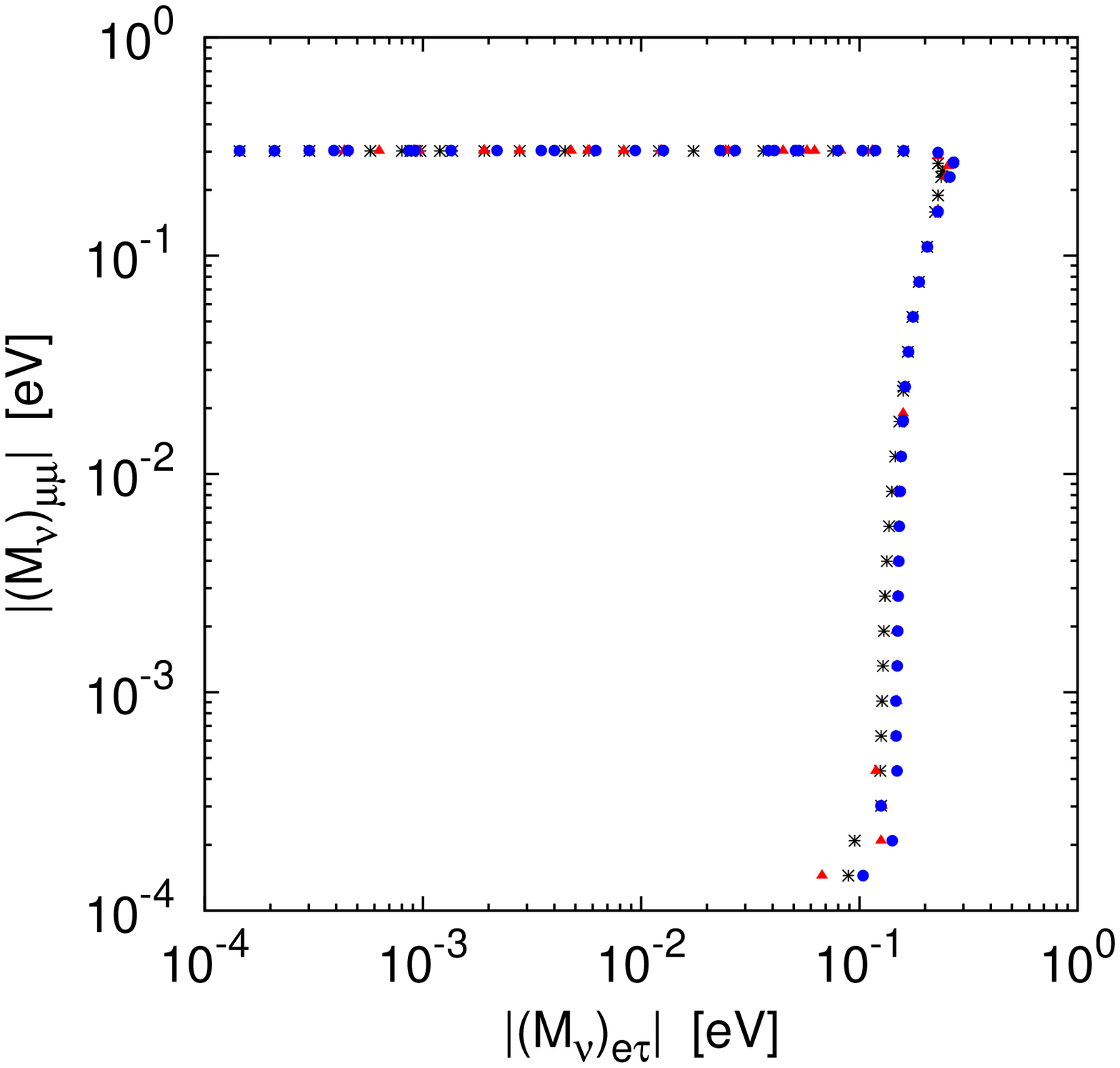} &
\includegraphics[angle=-90,keepaspectratio=true,scale=\figurescale]
{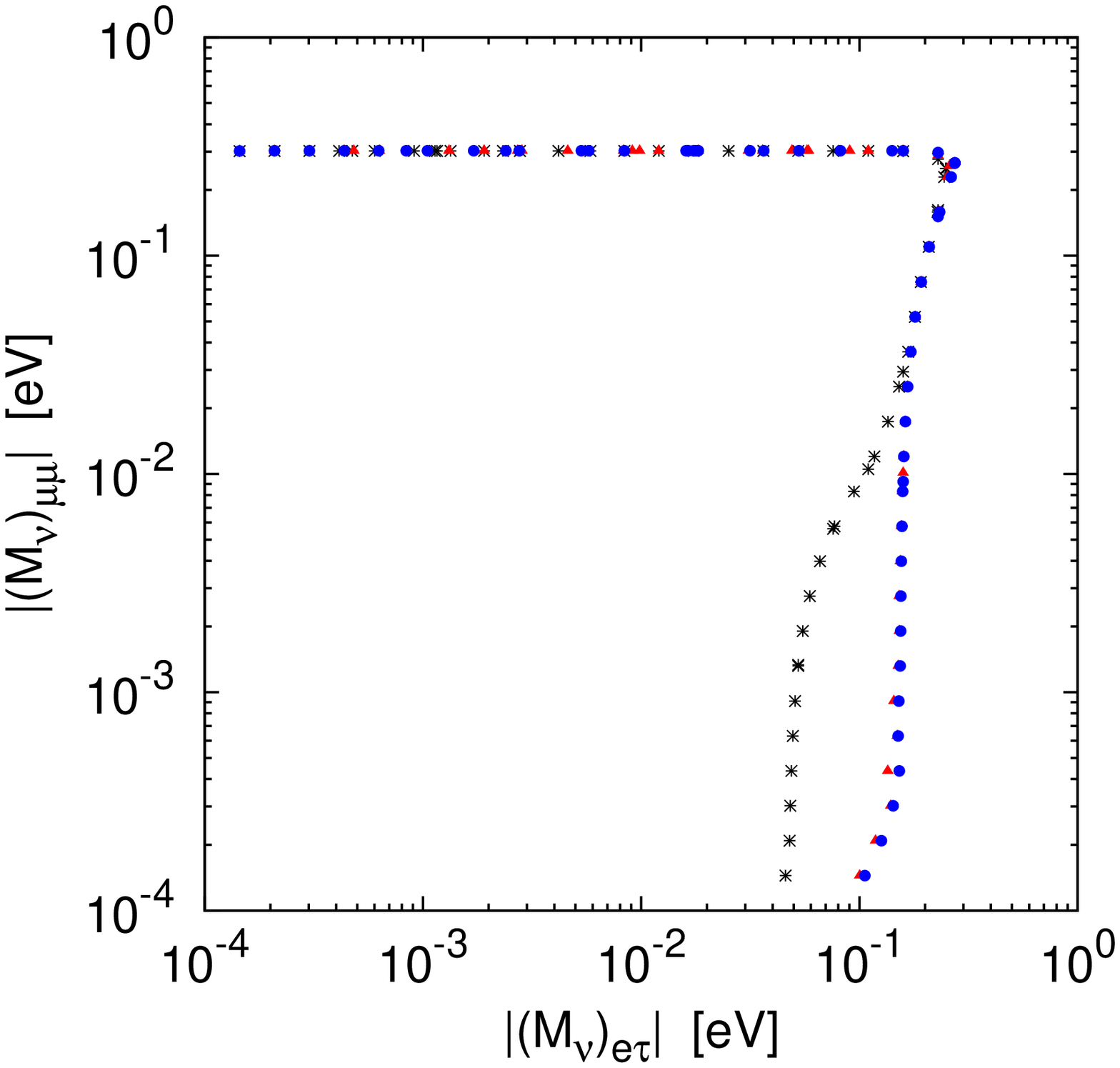}\\
\includegraphics[angle=-90,keepaspectratio=true,scale=\figurescale]
{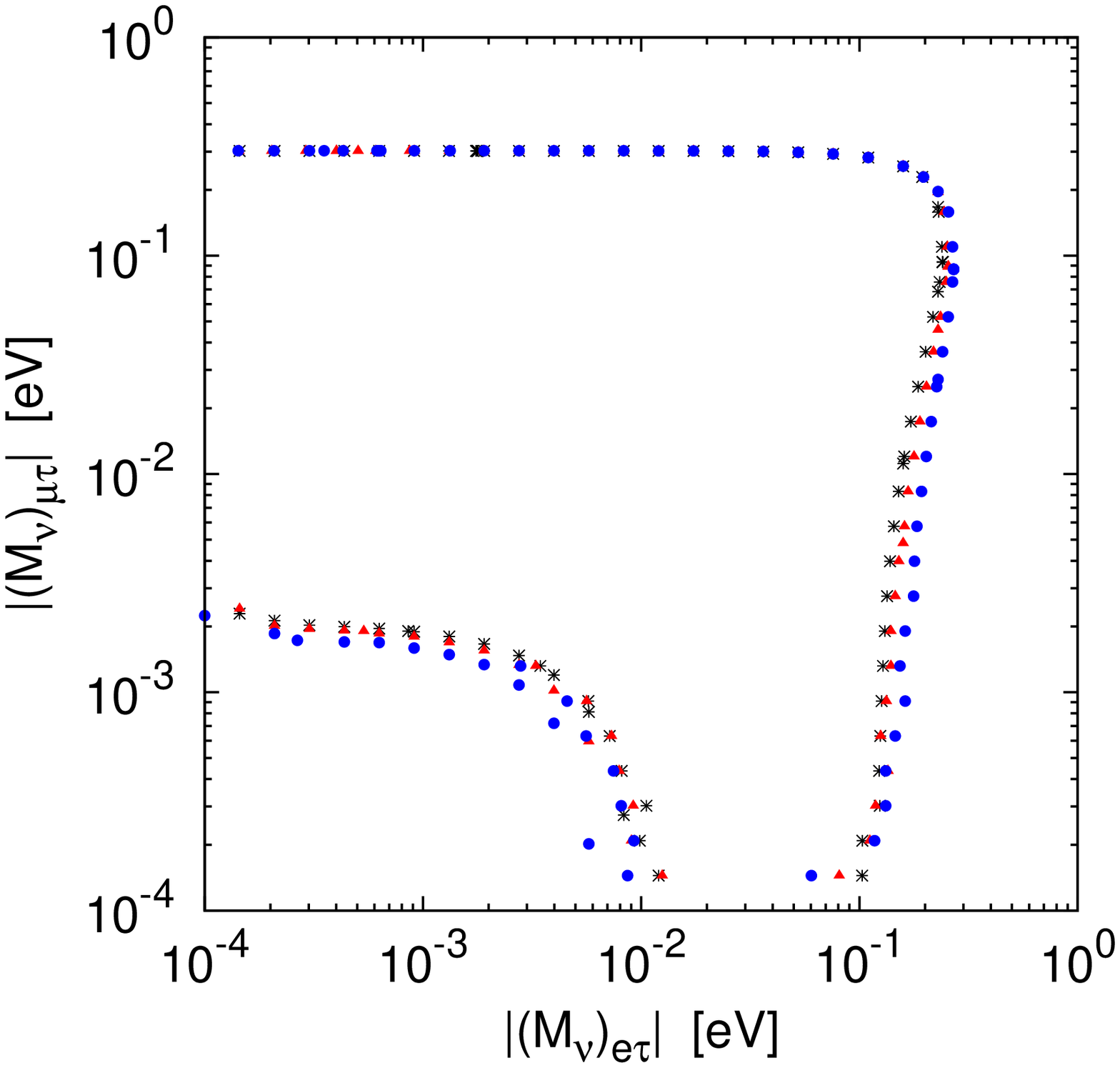} &
\includegraphics[angle=-90,keepaspectratio=true,scale=\figurescale]
{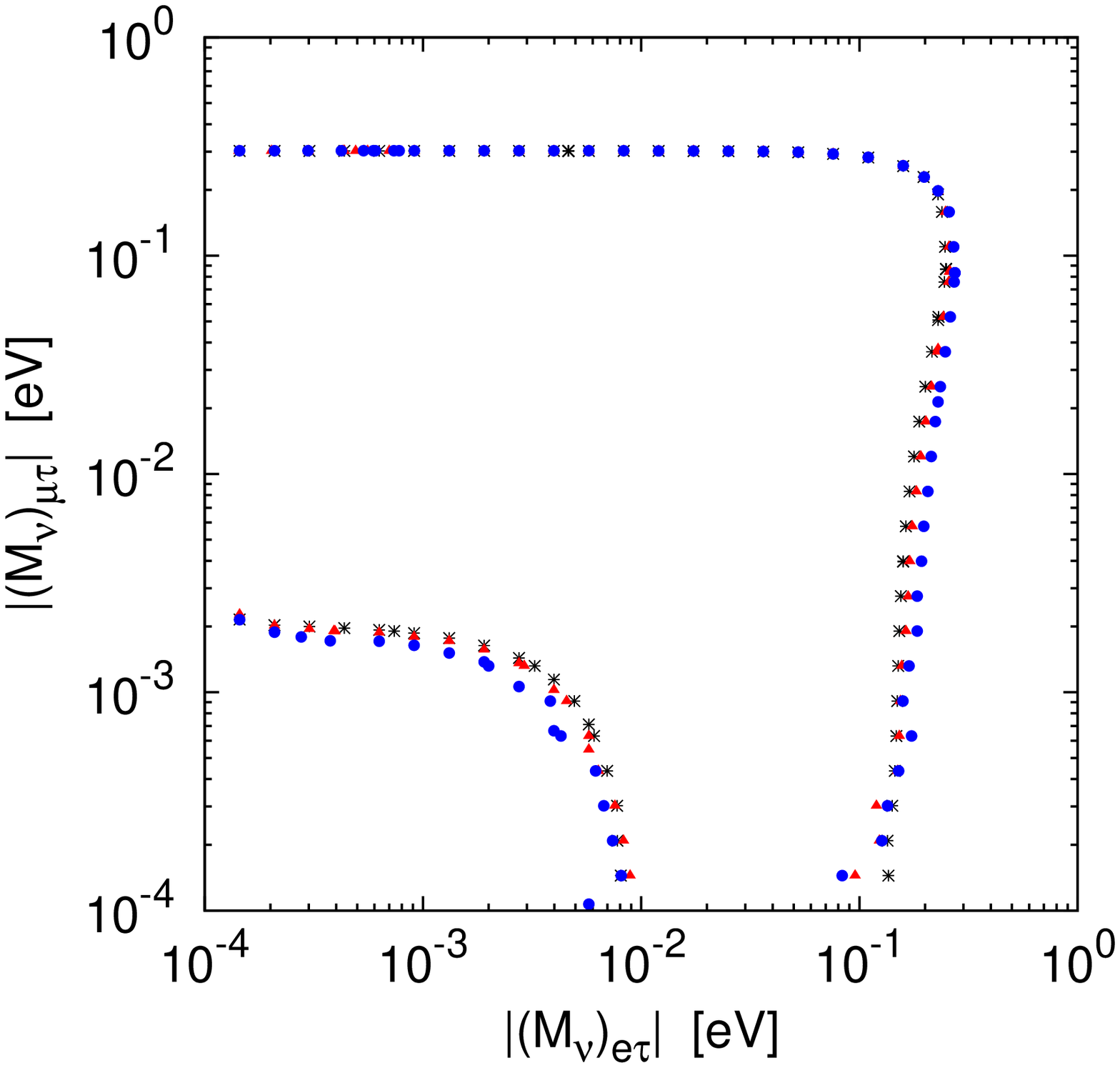}\\
\includegraphics[angle=-90,keepaspectratio=true,scale=\figurescale]
{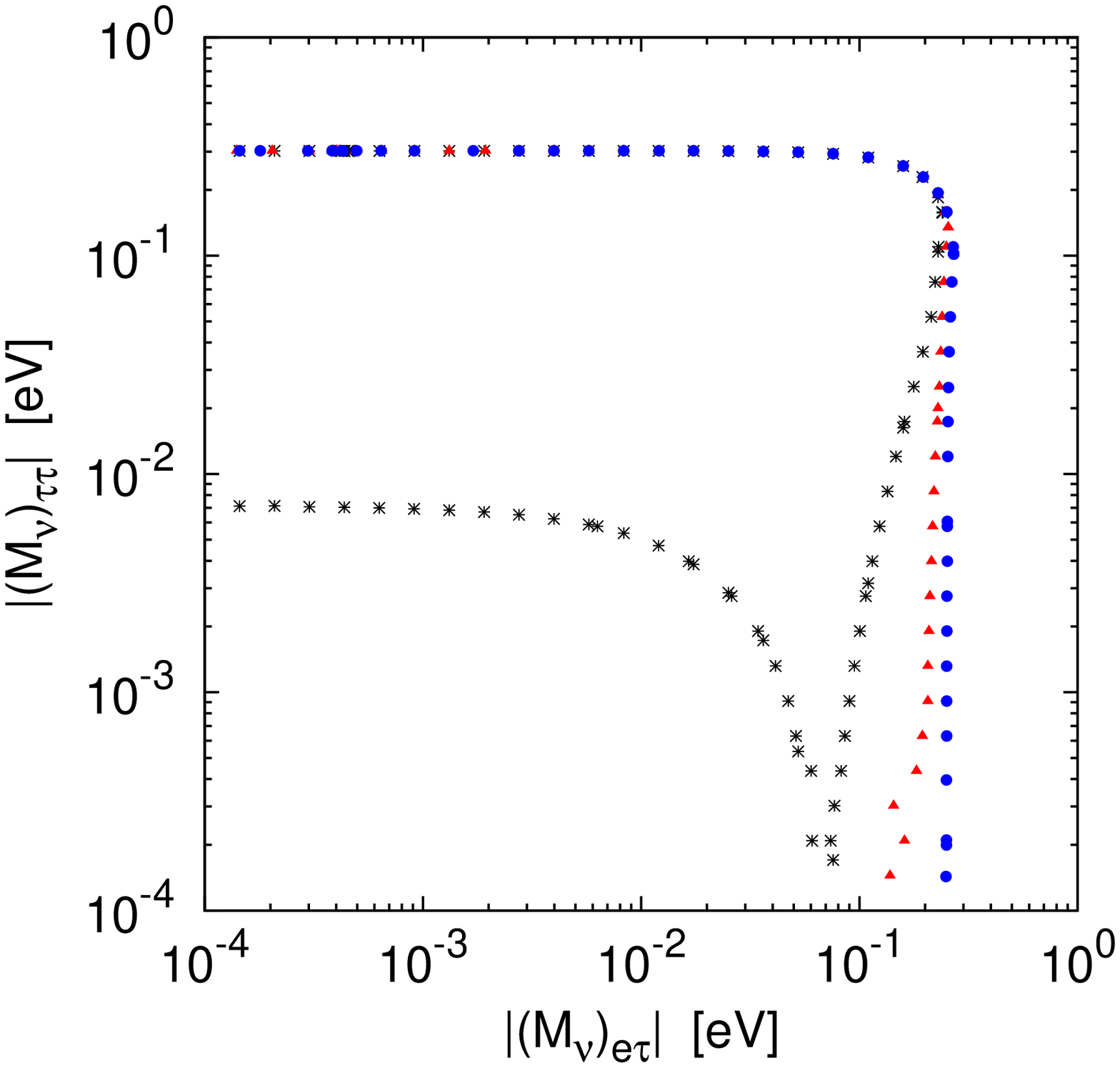} &
\includegraphics[angle=-90,keepaspectratio=true,scale=\figurescale]
{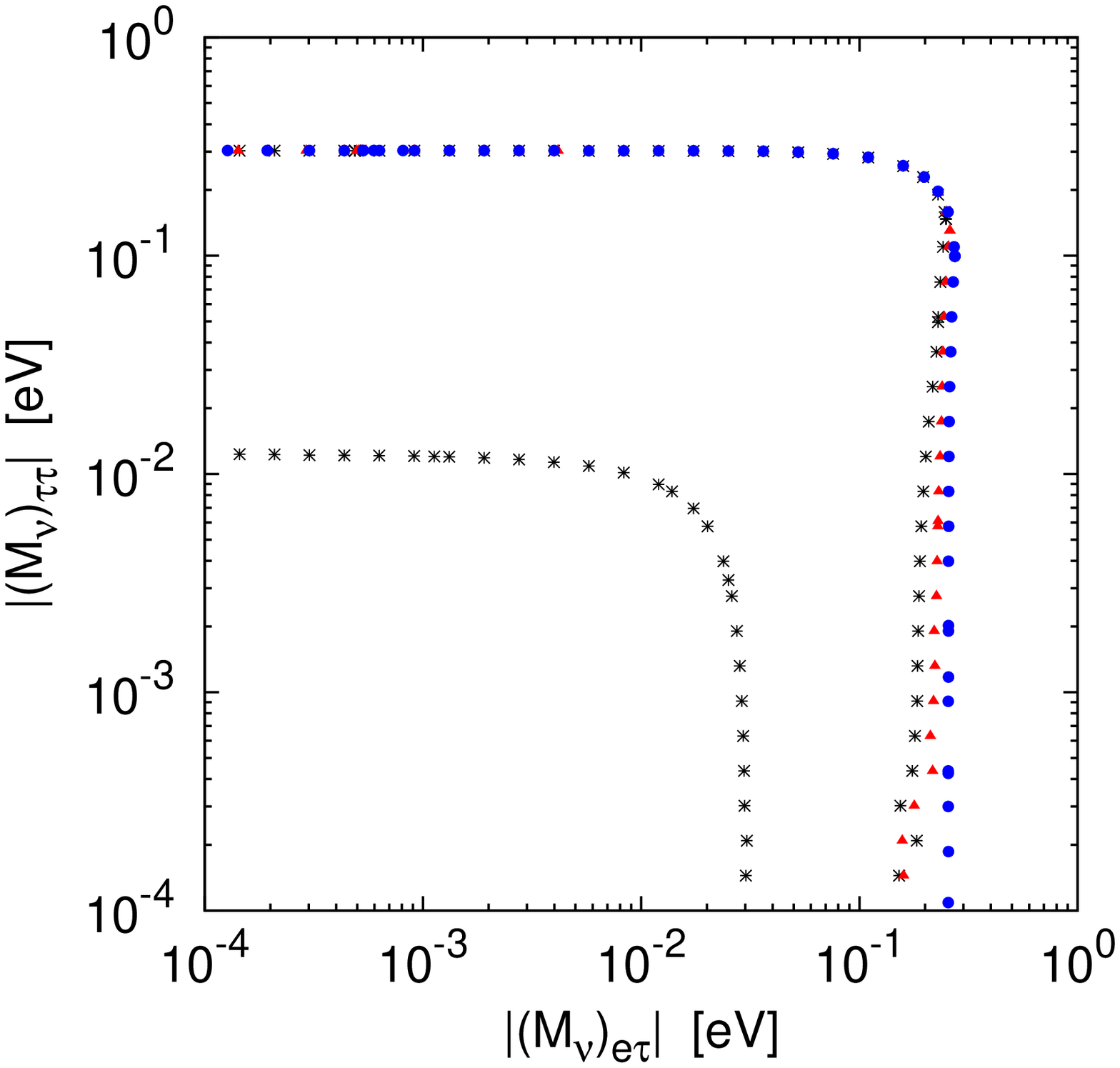}\\
\end{tabular}

\begin{tabular}[t]{ll}
\includegraphics[angle=-90,keepaspectratio=true,scale=\figurescale]
{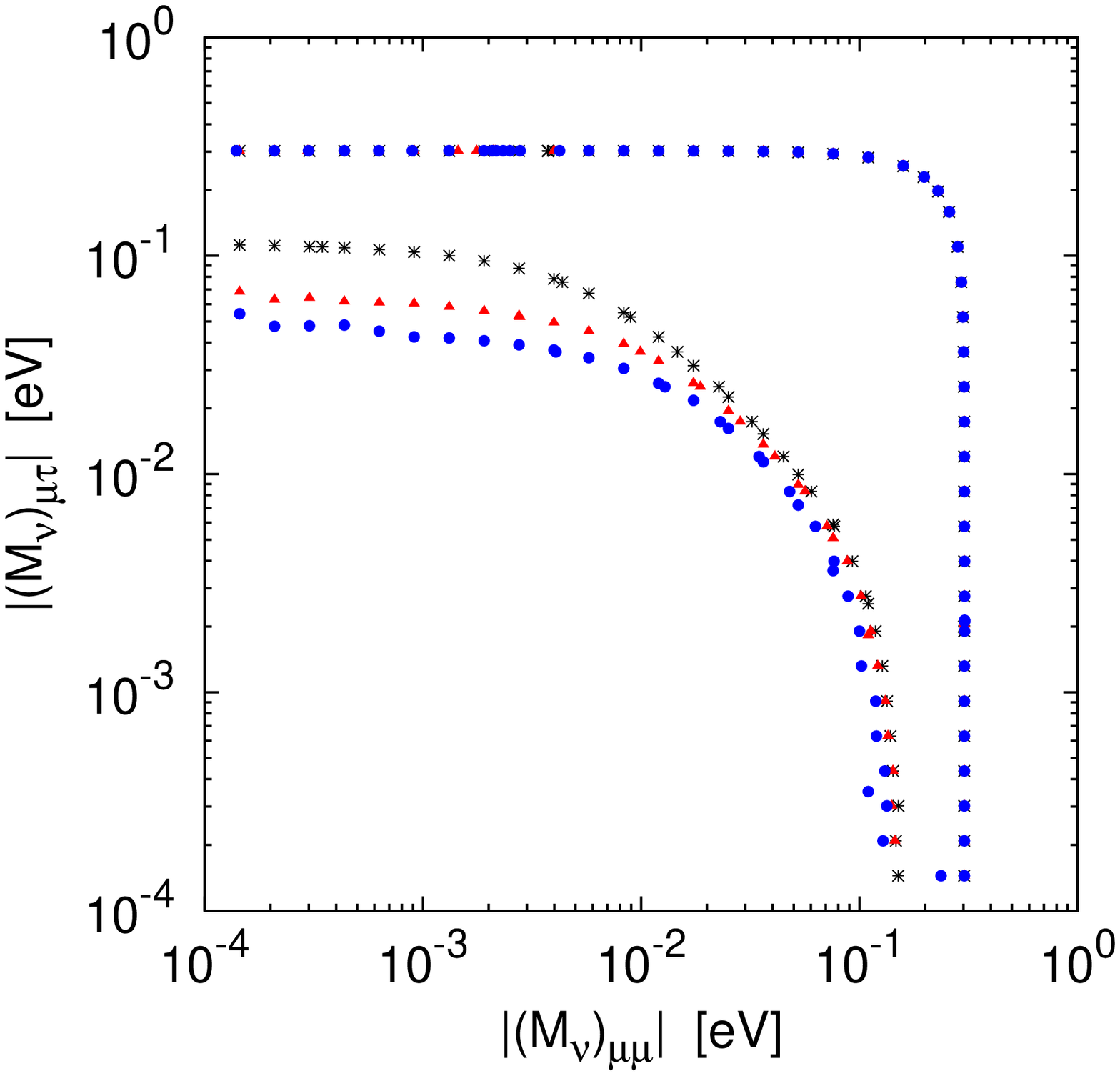} &
\includegraphics[angle=-90,keepaspectratio=true,scale=\figurescale]
{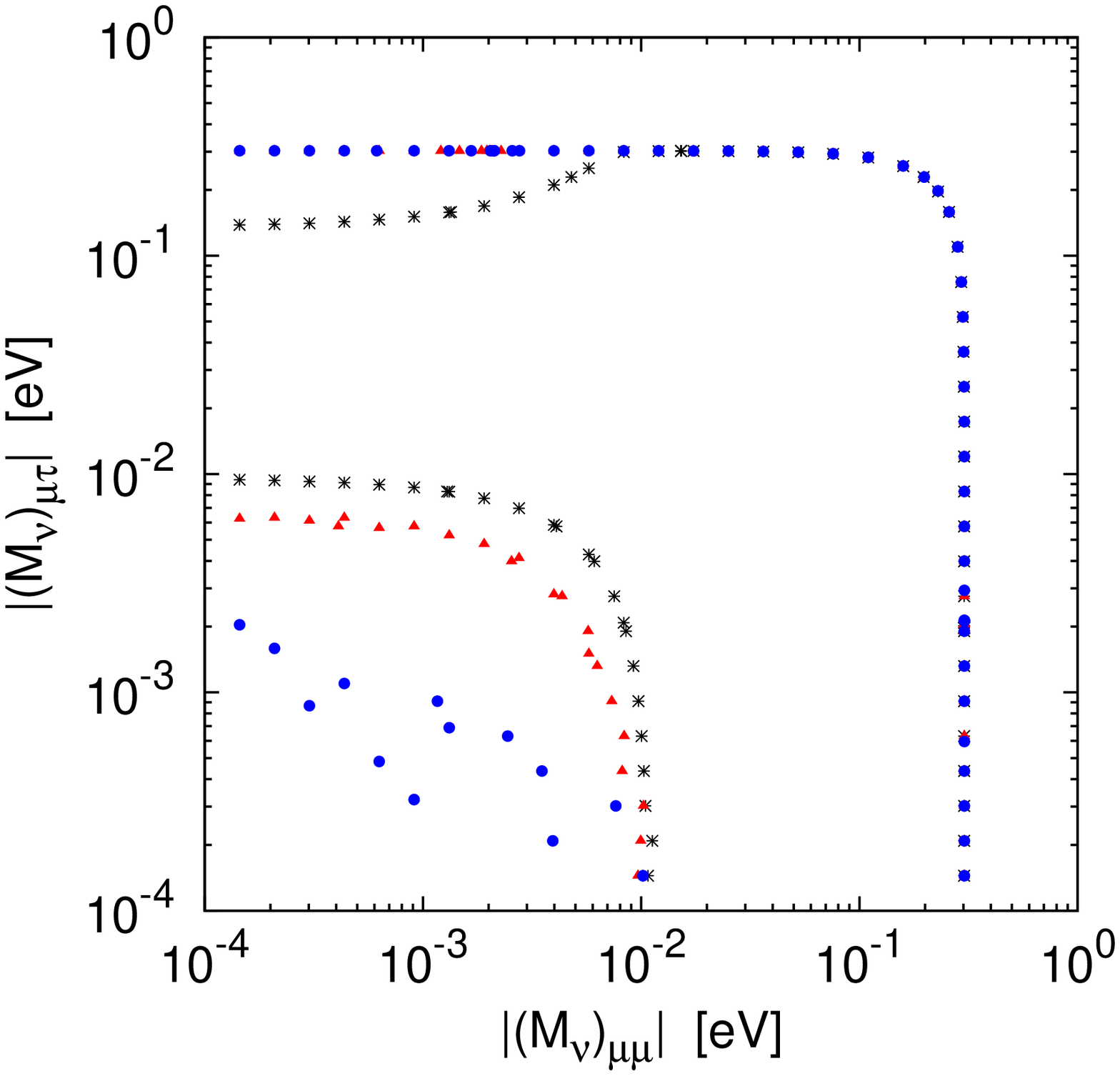}\\
\includegraphics[angle=-90,keepaspectratio=true,scale=\figurescale]
{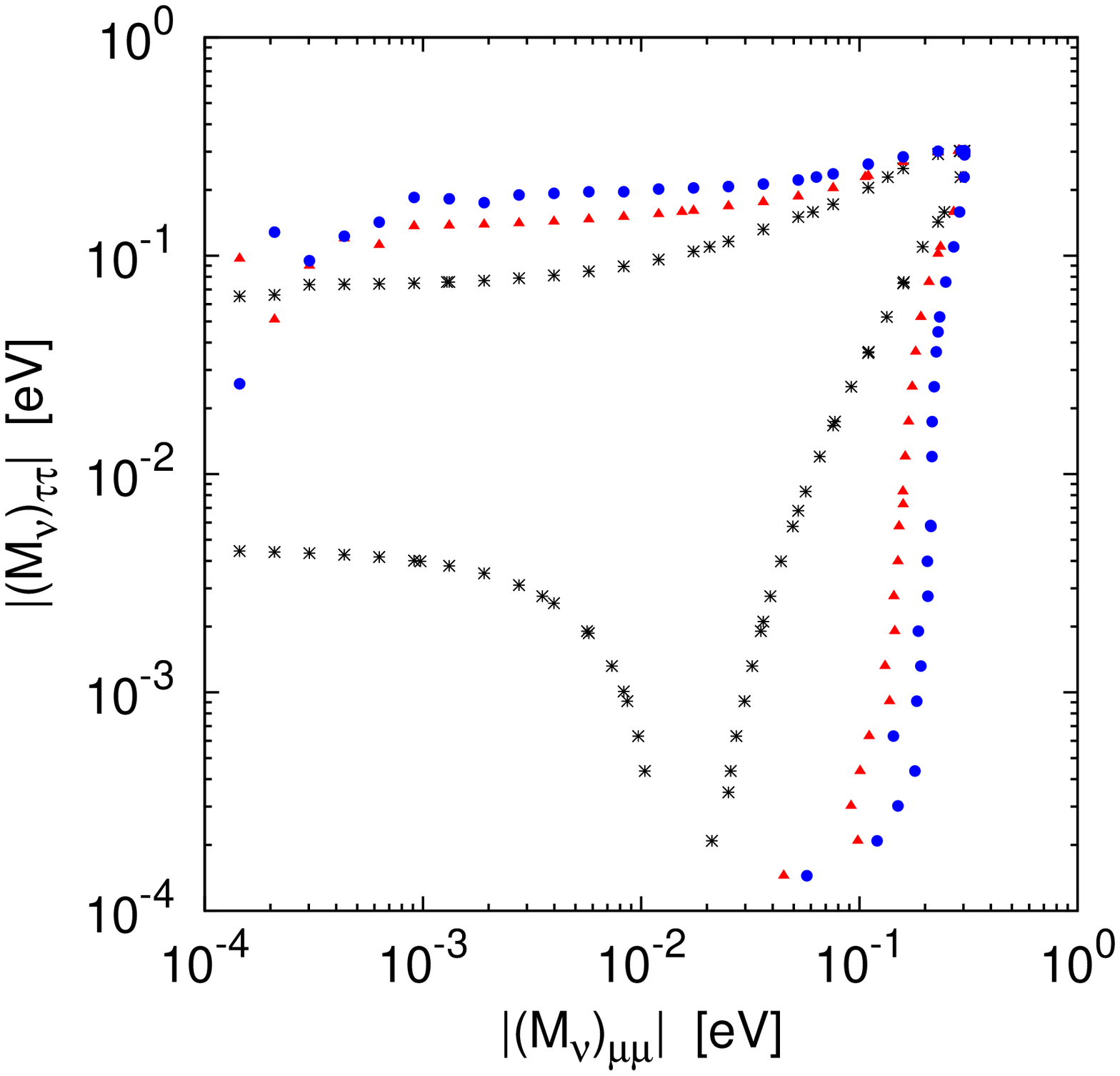} &
\includegraphics[angle=-90,keepaspectratio=true,scale=\figurescale]
{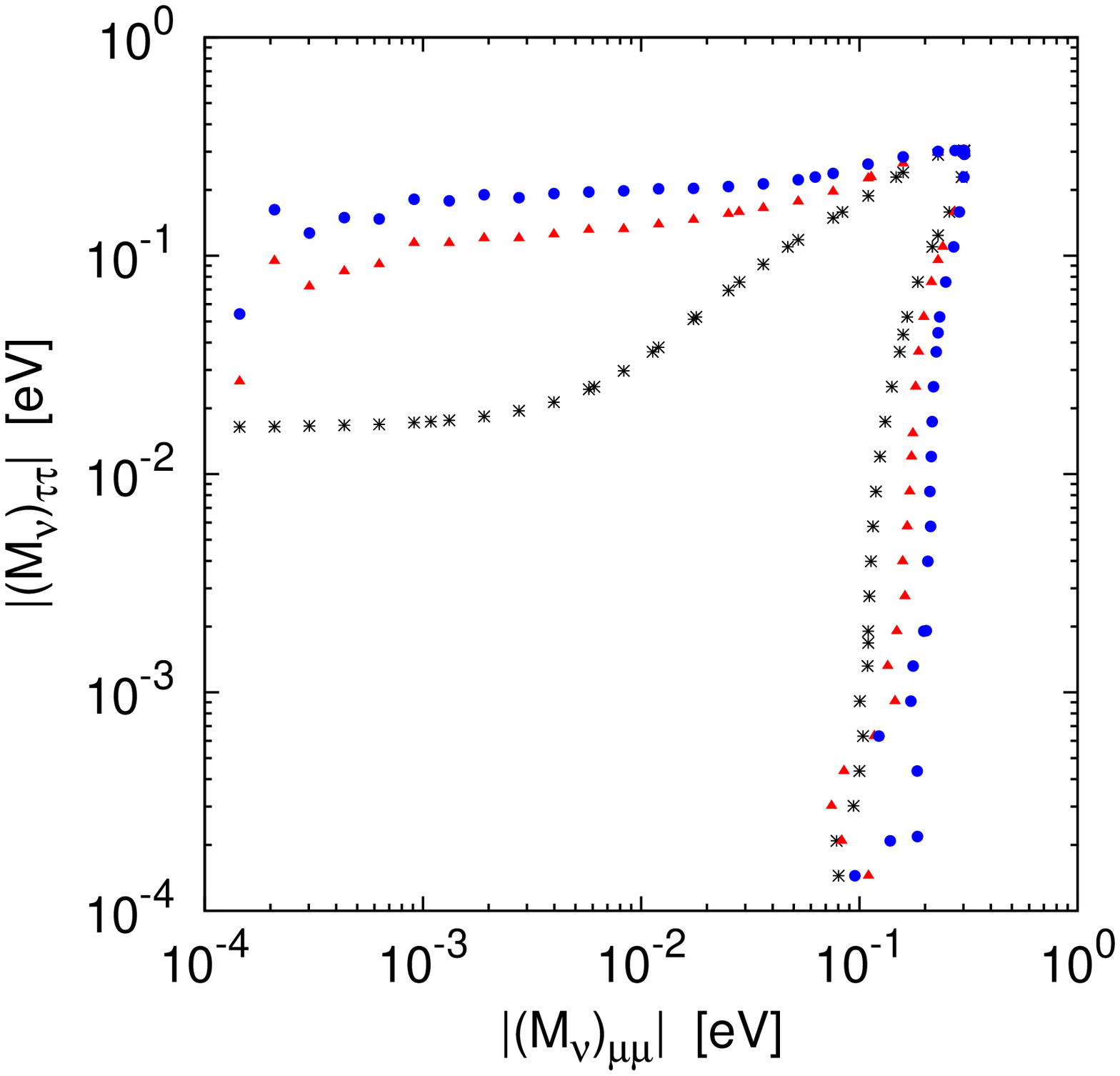}\\
\includegraphics[angle=-90,keepaspectratio=true,scale=\figurescale]
{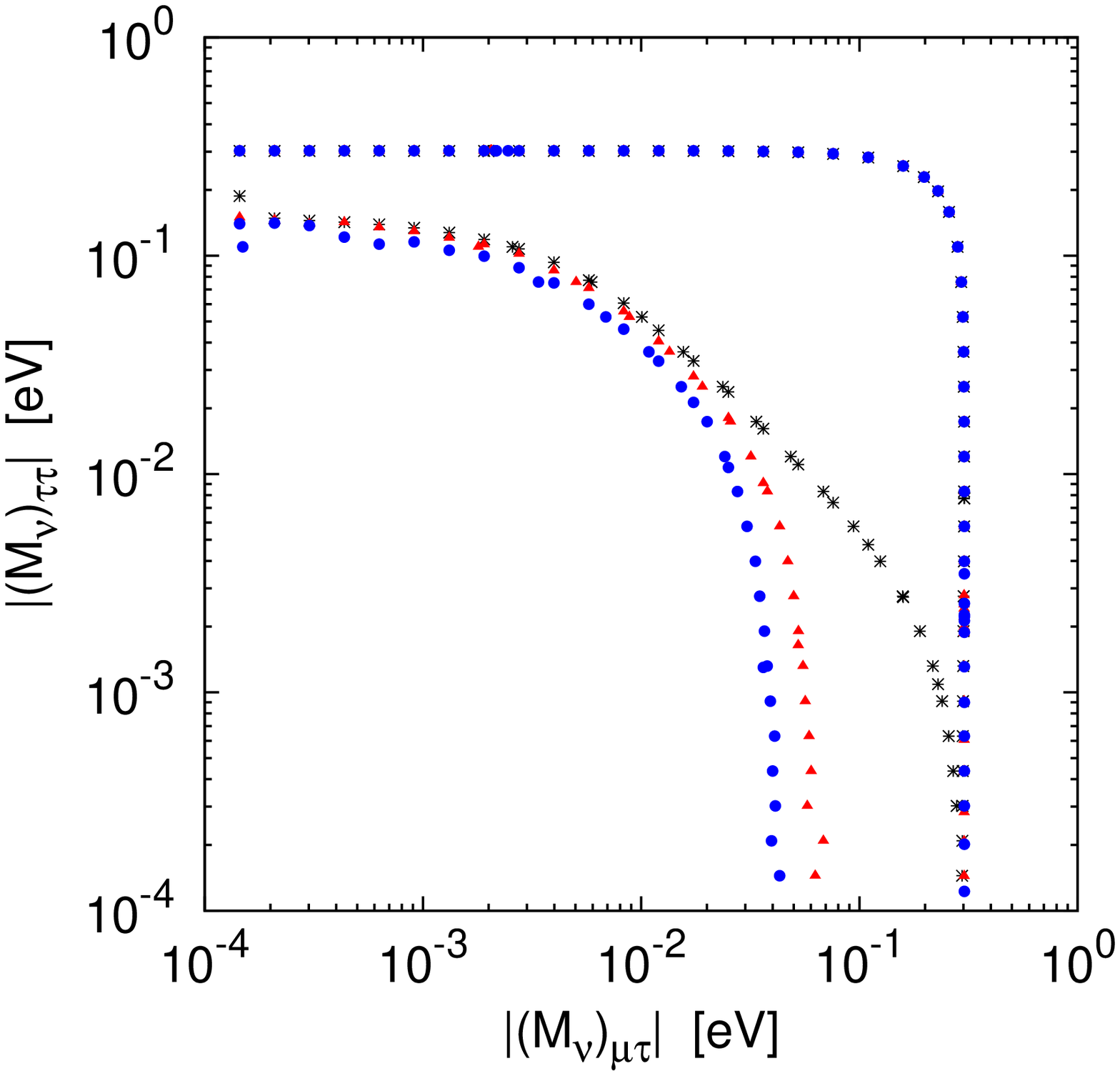} &
\includegraphics[angle=-90,keepaspectratio=true,scale=\figurescale]
{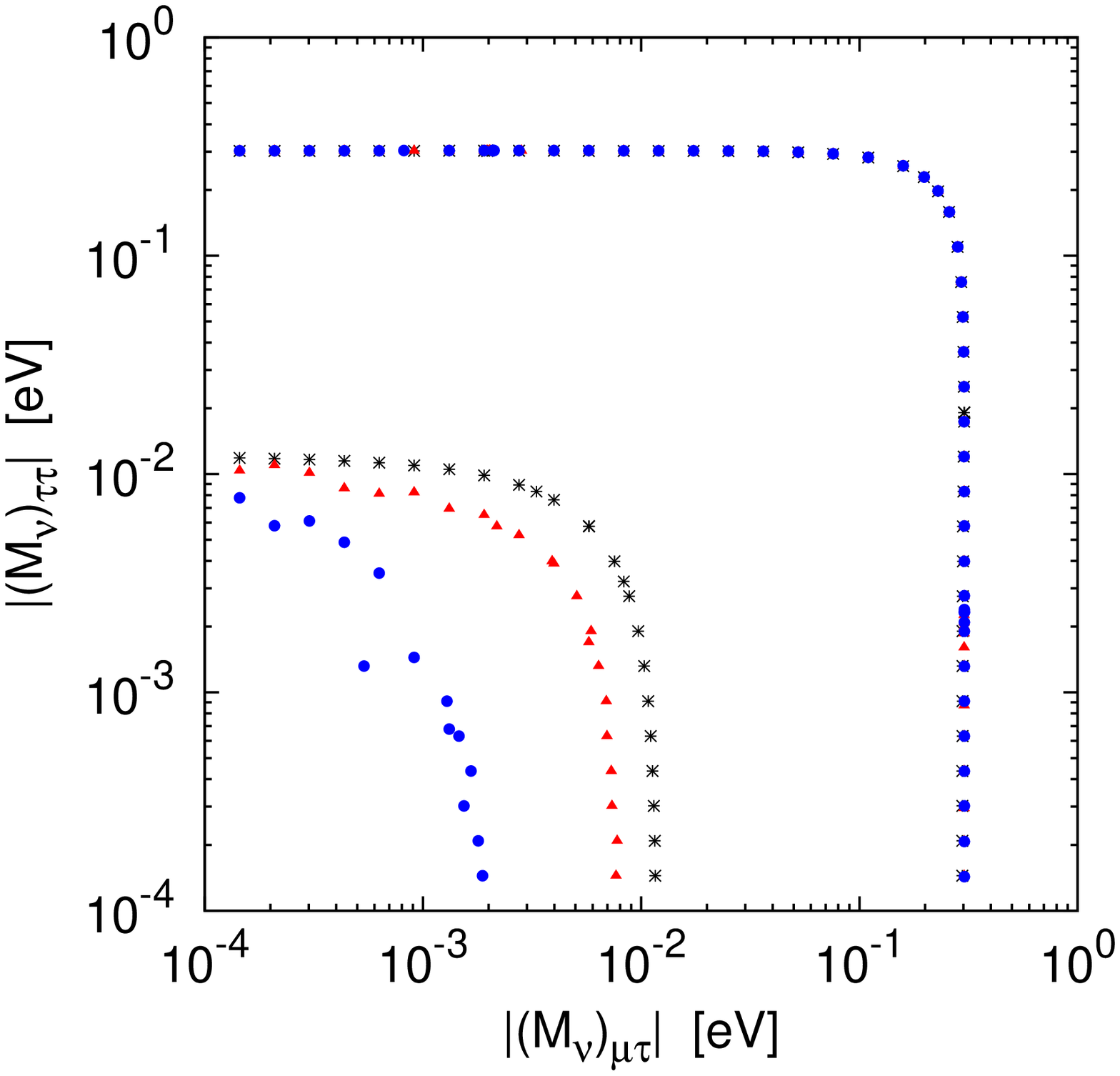}\\
\end{tabular}

\section{Plots based on Forero \textit{et al.} (version 3); \textit{cf.} app.~\ref{fogli}}\label{forero3}

\begin{tabular}[t]{ll}
\includegraphics[angle=-90,keepaspectratio=true,scale=\figurescale]
{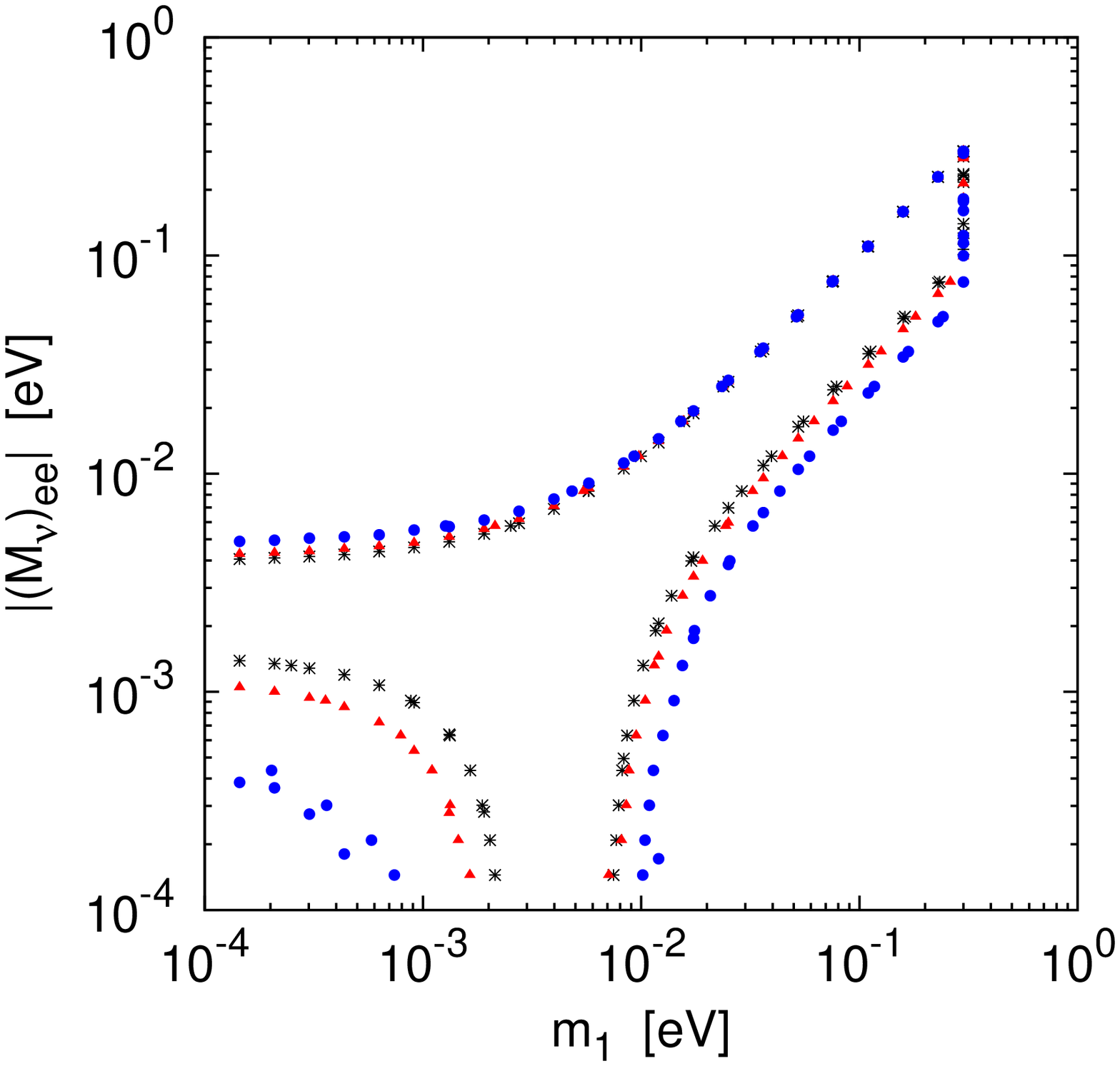} &
\includegraphics[angle=-90,keepaspectratio=true,scale=\figurescale]
{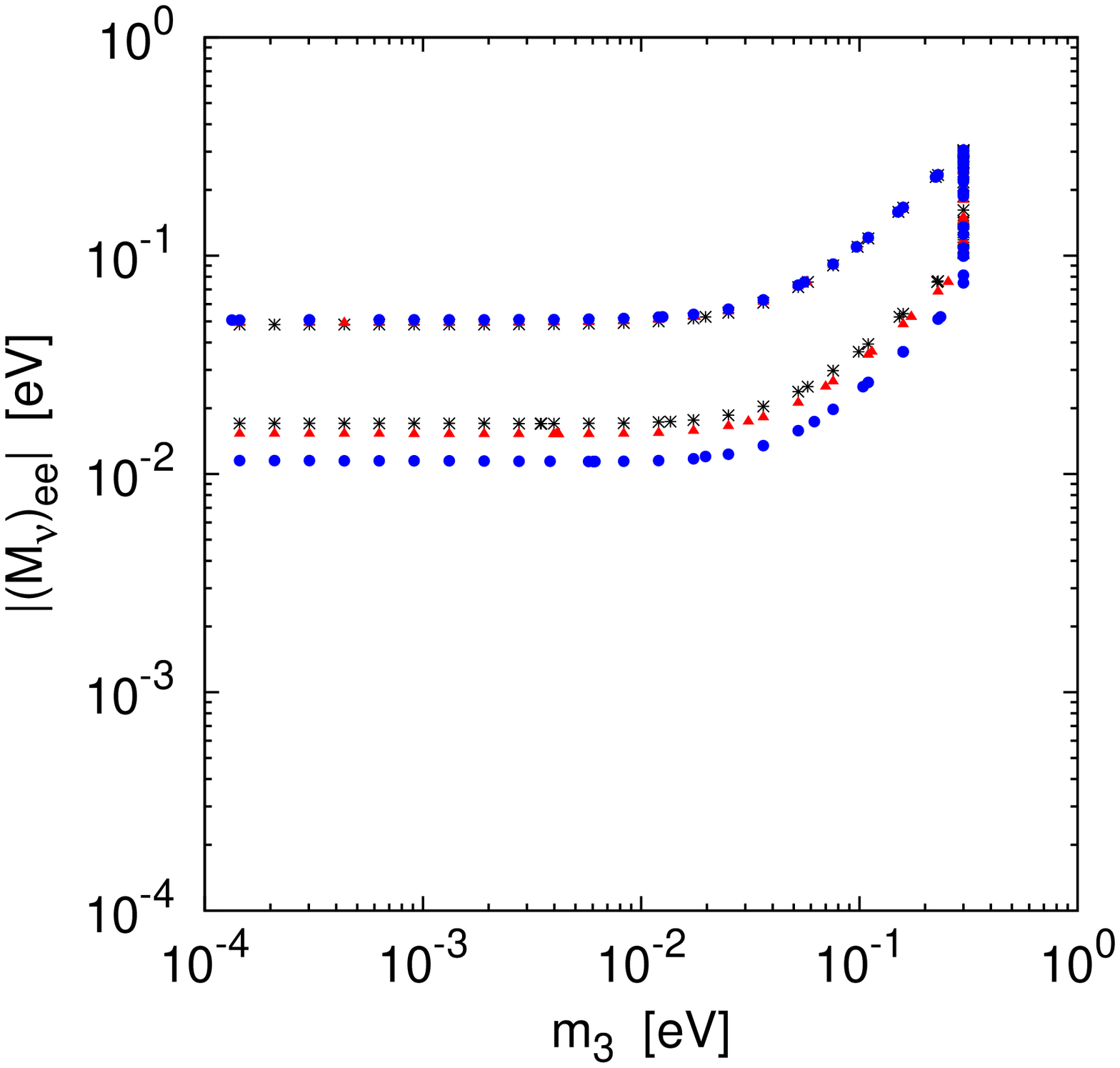}\\
\includegraphics[angle=-90,keepaspectratio=true,scale=\figurescale]
{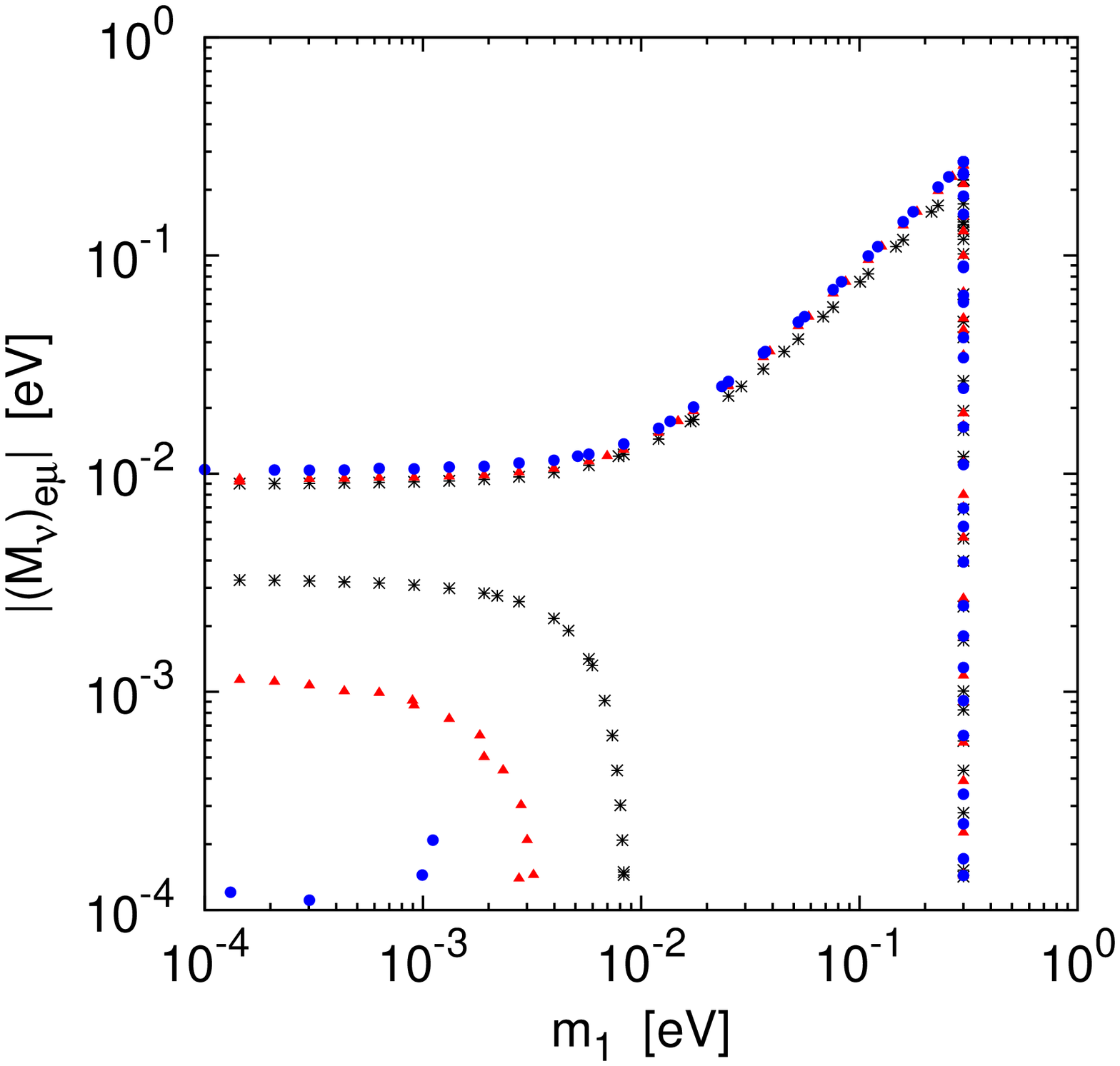} &
\includegraphics[angle=-90,keepaspectratio=true,scale=\figurescale]
{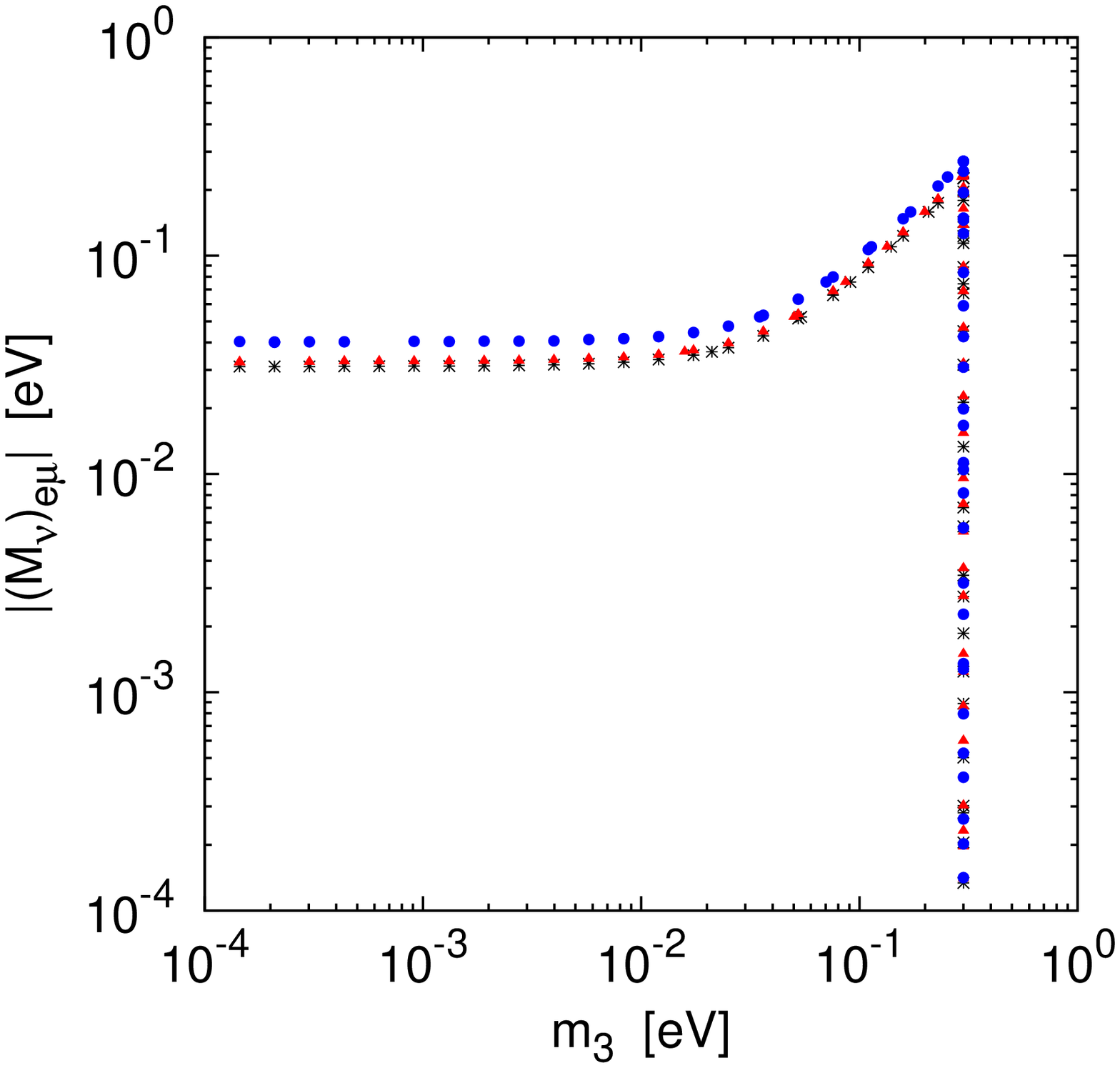}\\
\includegraphics[angle=-90,keepaspectratio=true,scale=\figurescale]
{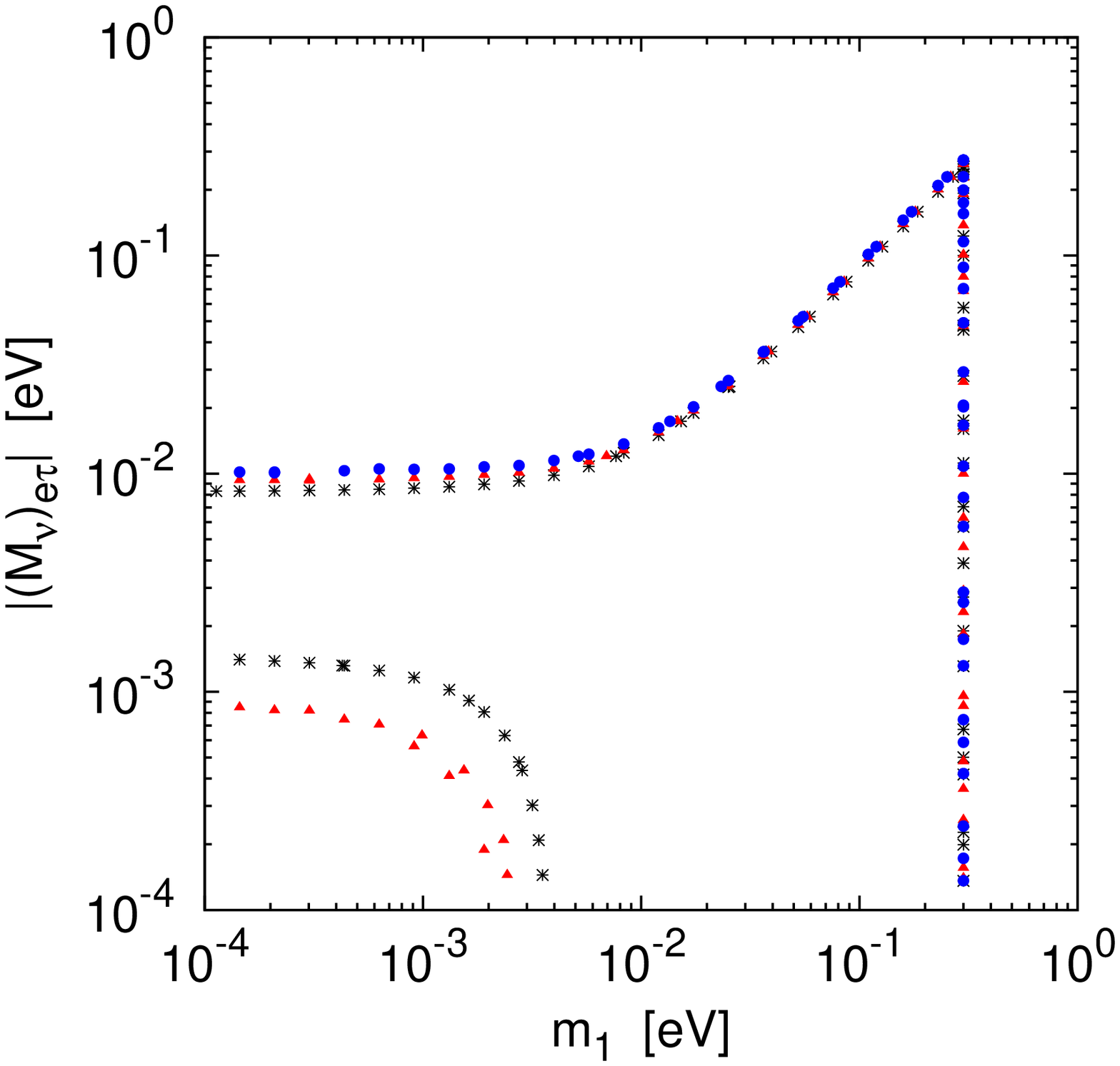} &
\includegraphics[angle=-90,keepaspectratio=true,scale=\figurescale]
{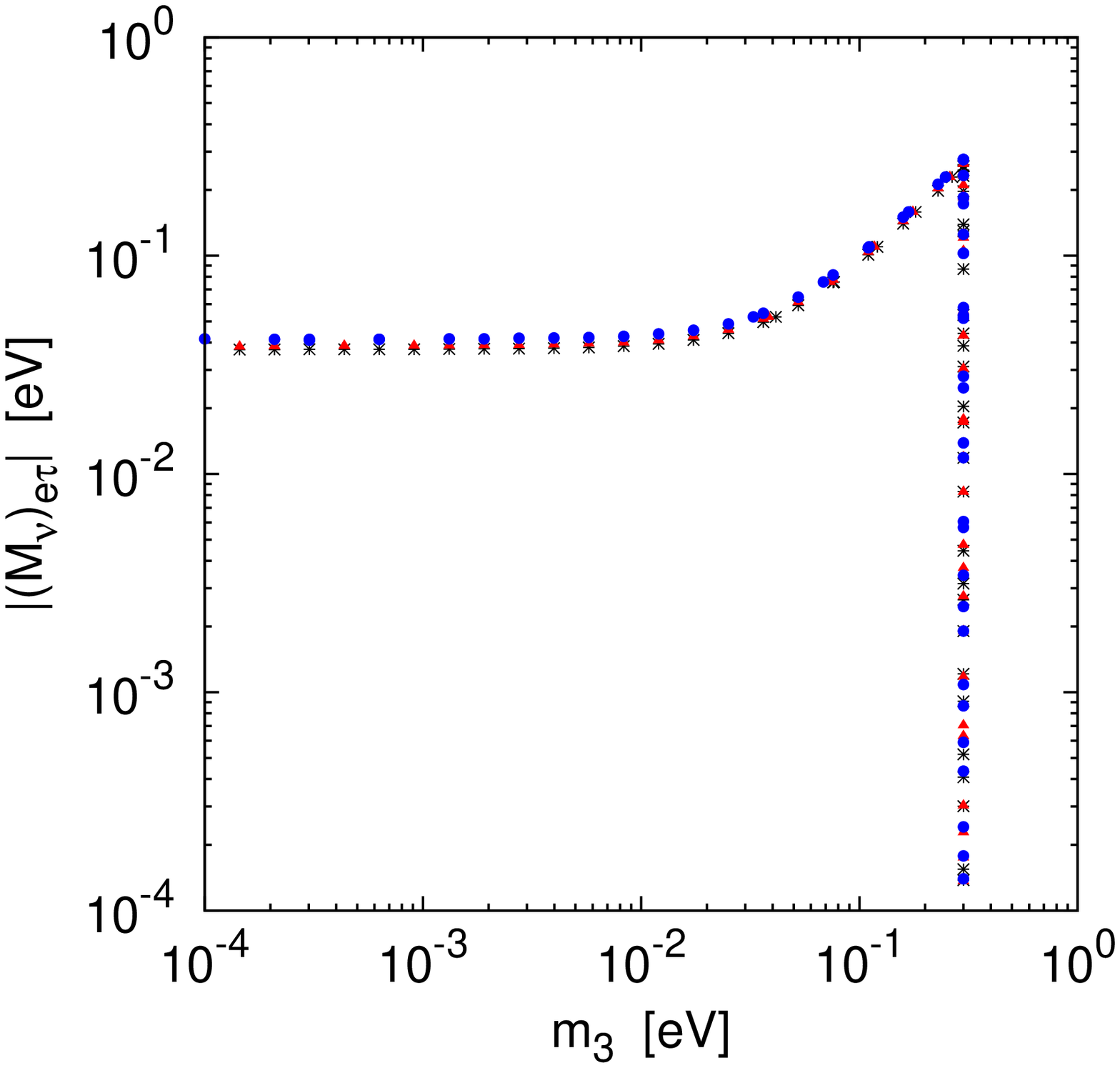}\\
\end{tabular}

\begin{tabular}[t]{ll}
\includegraphics[angle=-90,keepaspectratio=true,scale=\figurescale]
{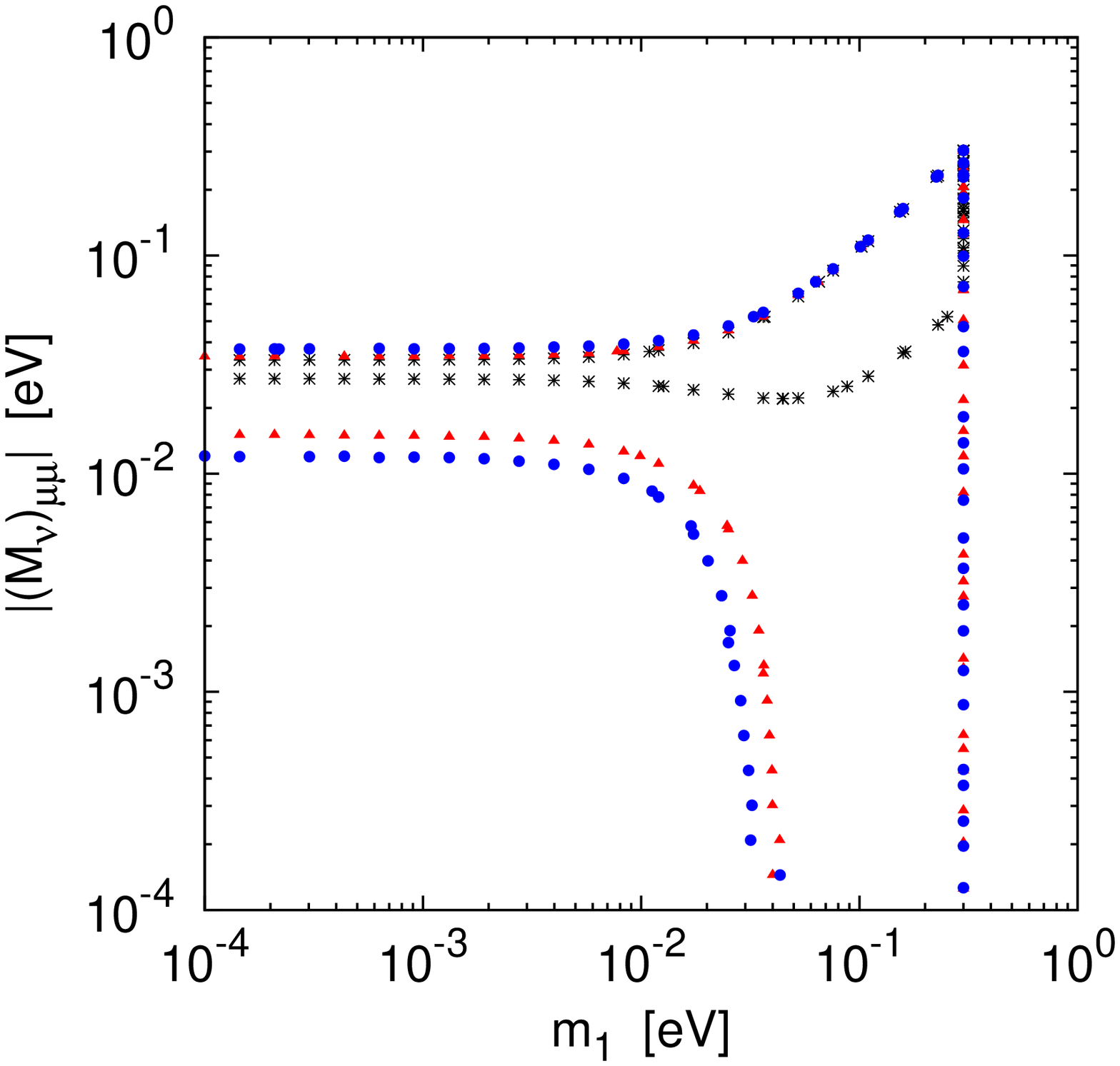} &
\includegraphics[angle=-90,keepaspectratio=true,scale=\figurescale]
{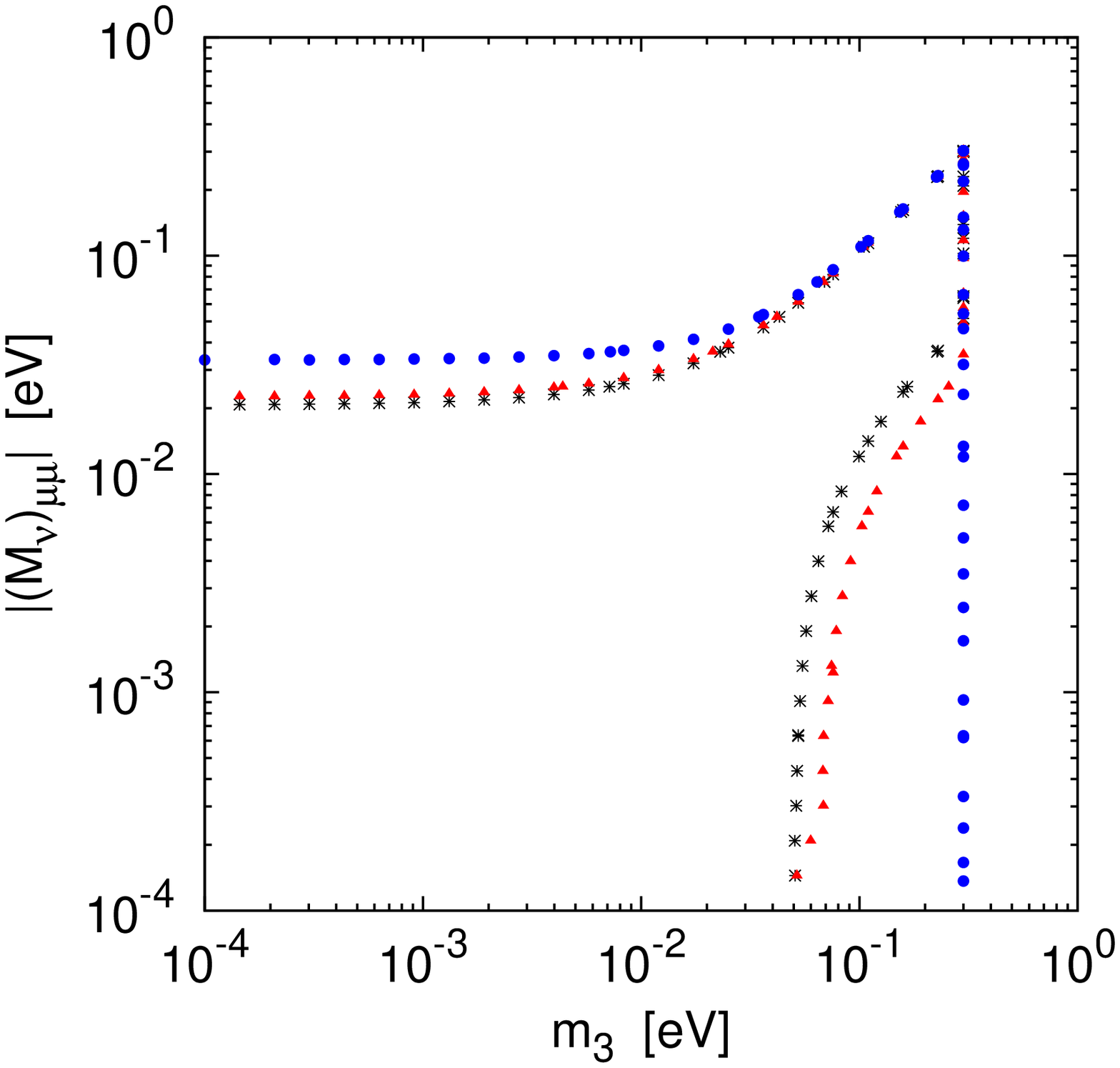}\\
\includegraphics[angle=-90,keepaspectratio=true,scale=\figurescale]
{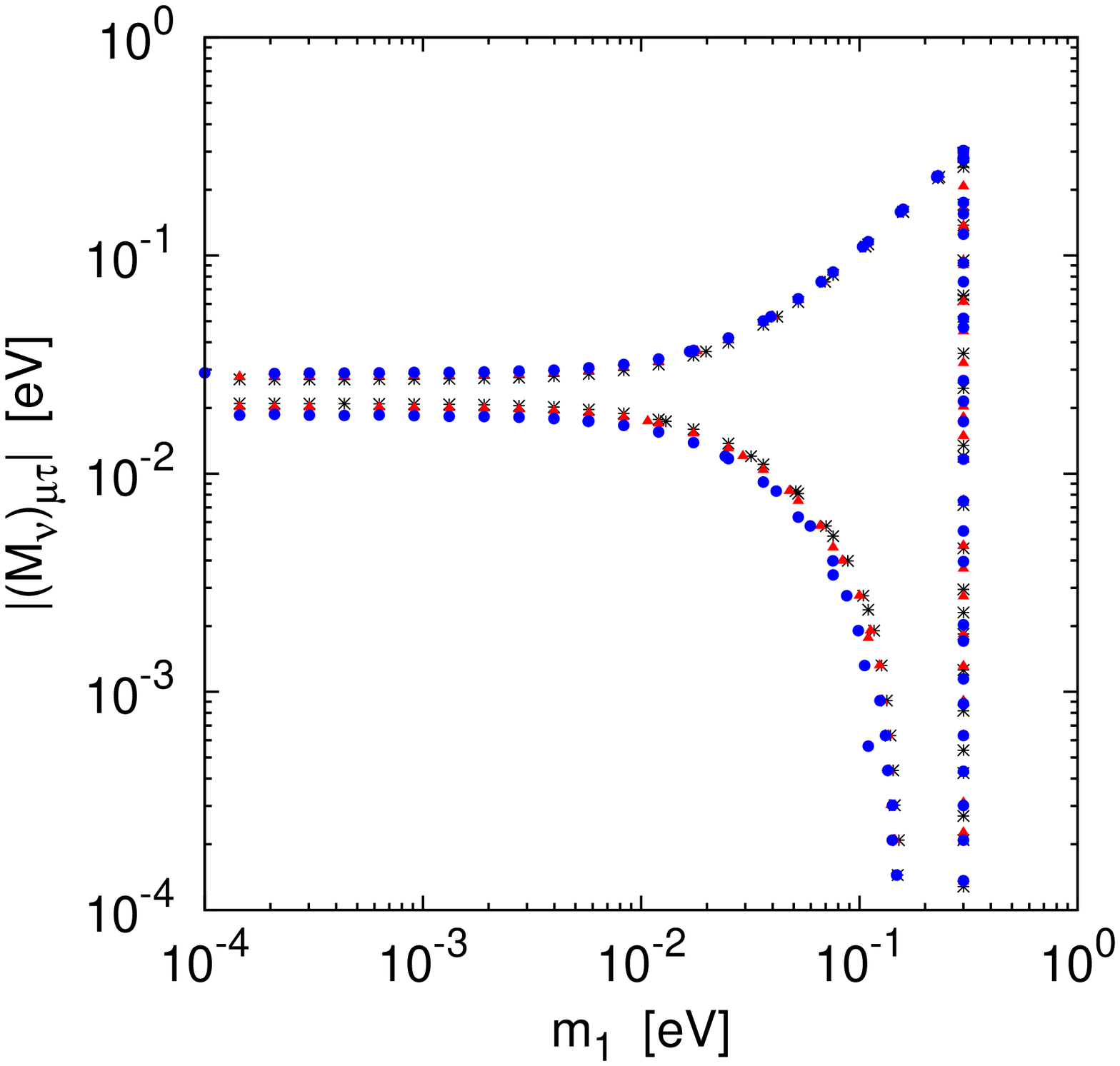} &
\includegraphics[angle=-90,keepaspectratio=true,scale=\figurescale]
{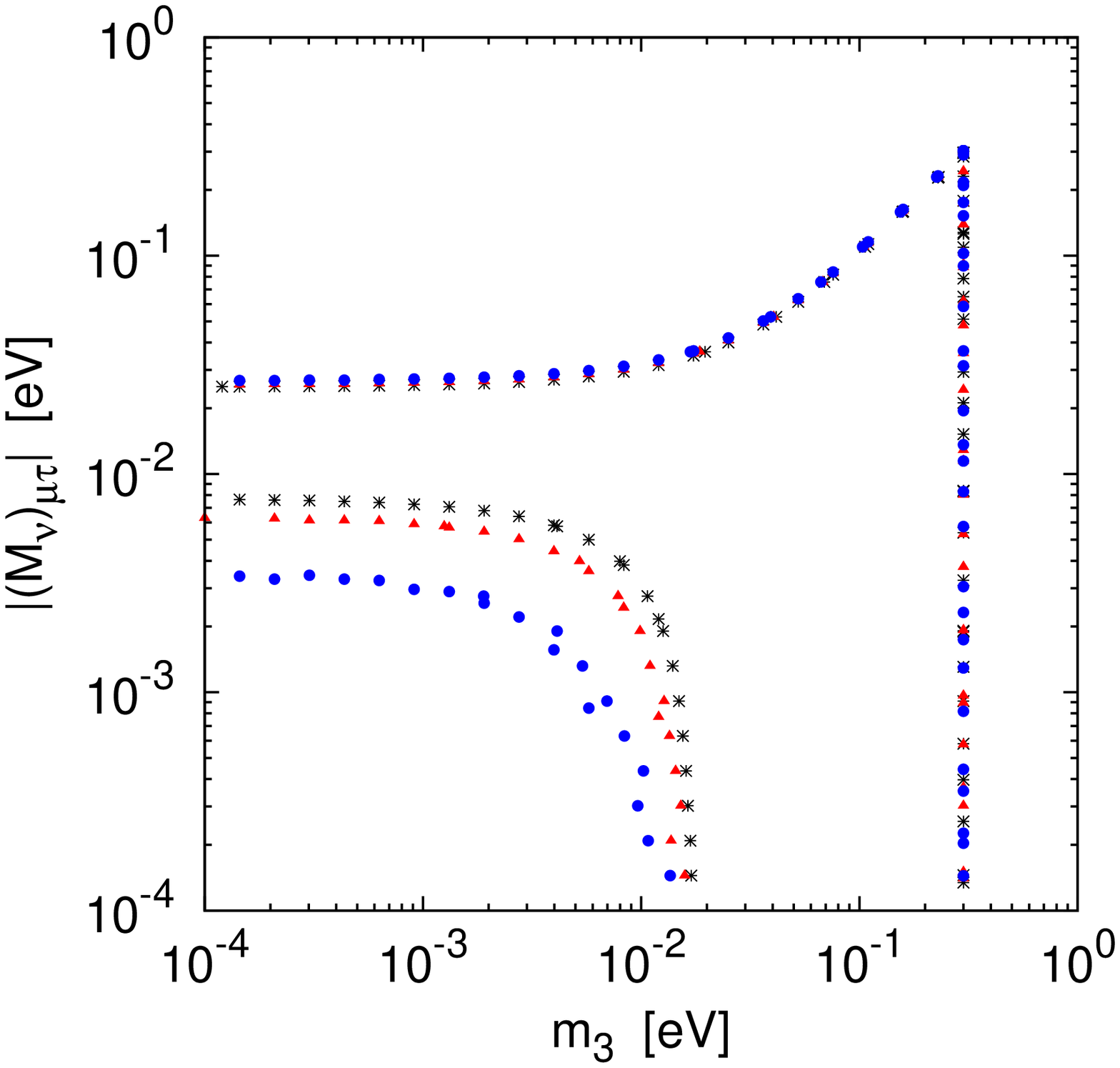}\\
\includegraphics[angle=-90,keepaspectratio=true,scale=\figurescale]
{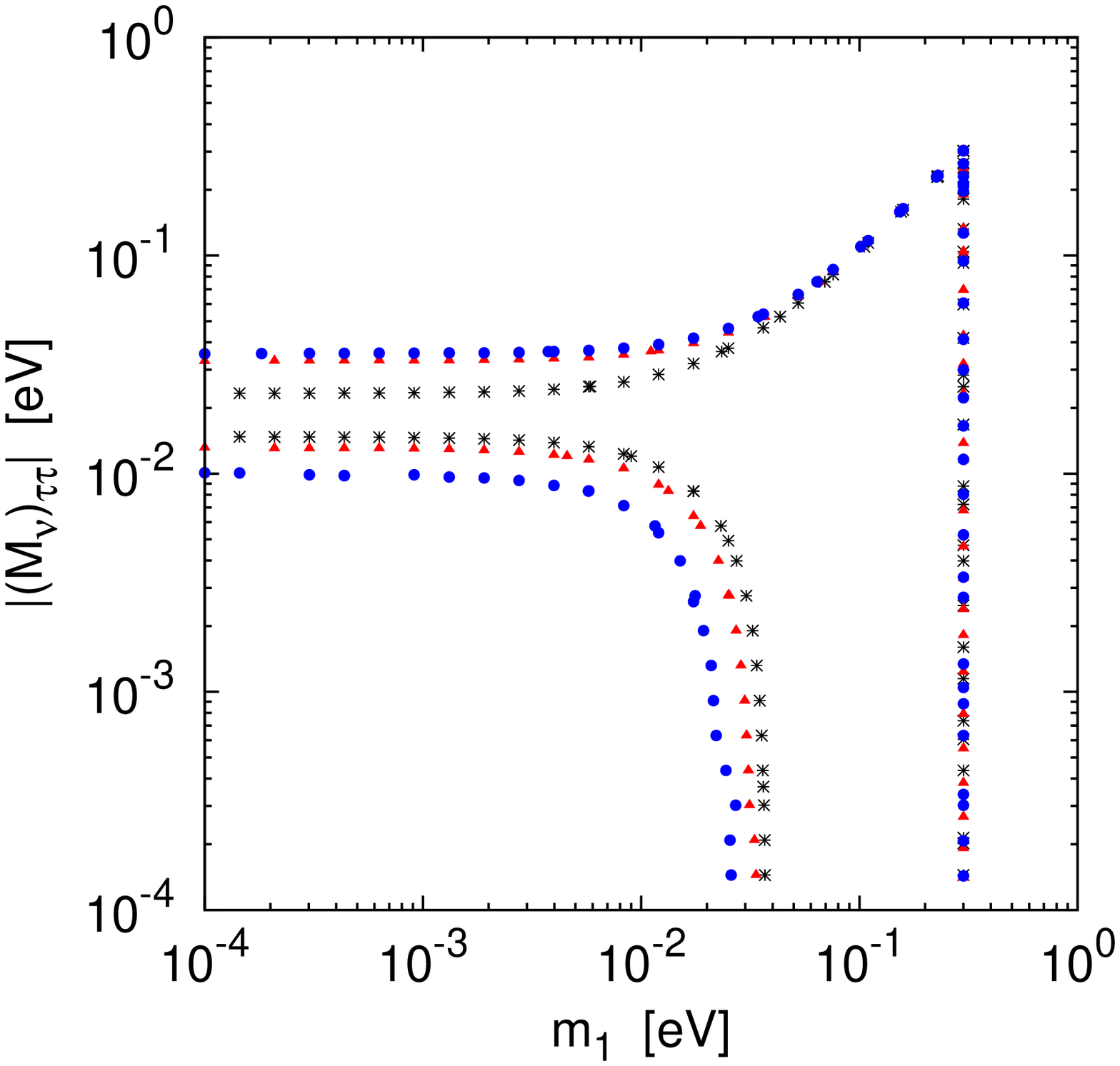} &
\includegraphics[angle=-90,keepaspectratio=true,scale=\figurescale]
{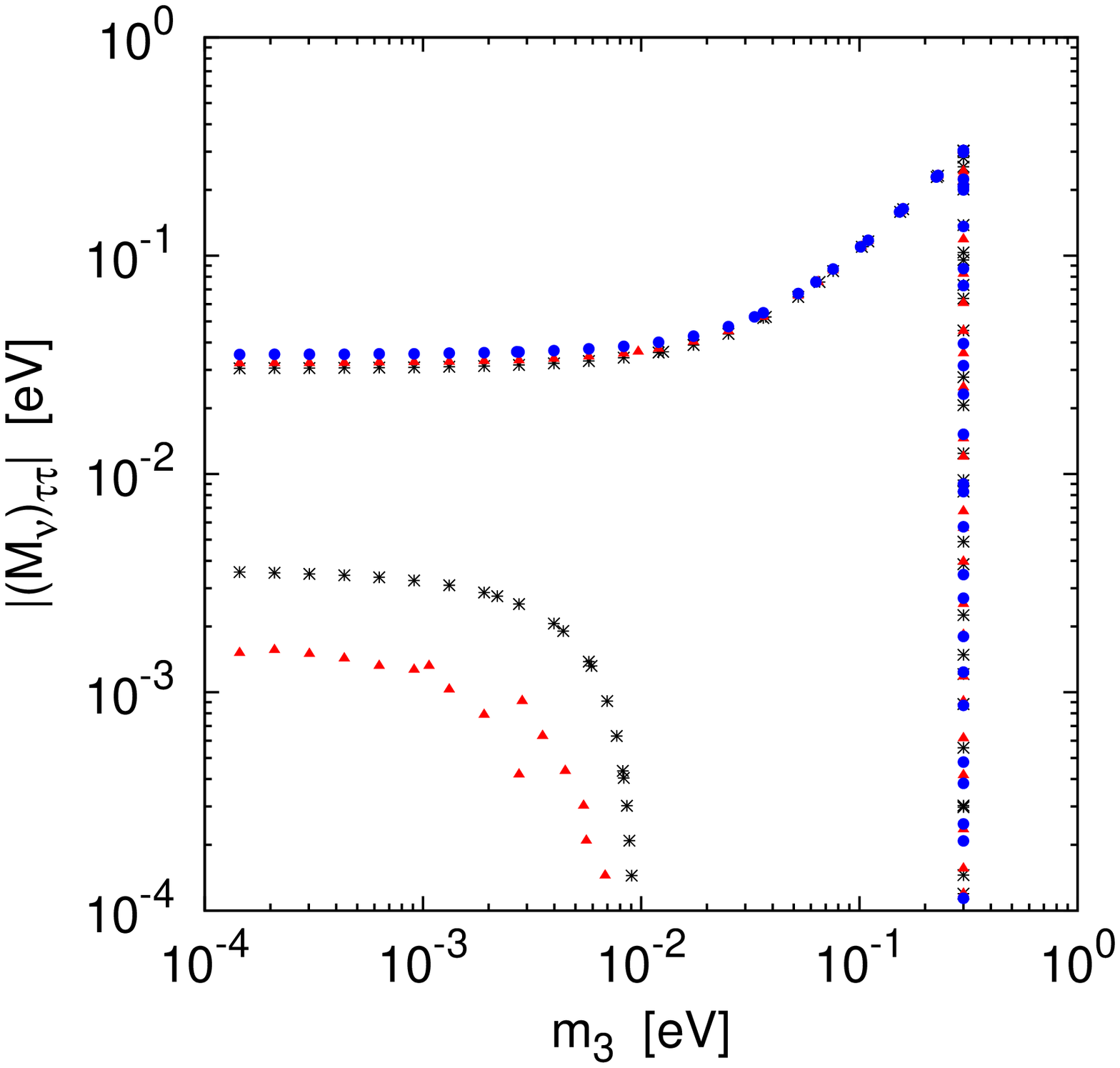}\\
\end{tabular}

\begin{tabular}[t]{ll}
\includegraphics[angle=-90,keepaspectratio=true,scale=\figurescale]
{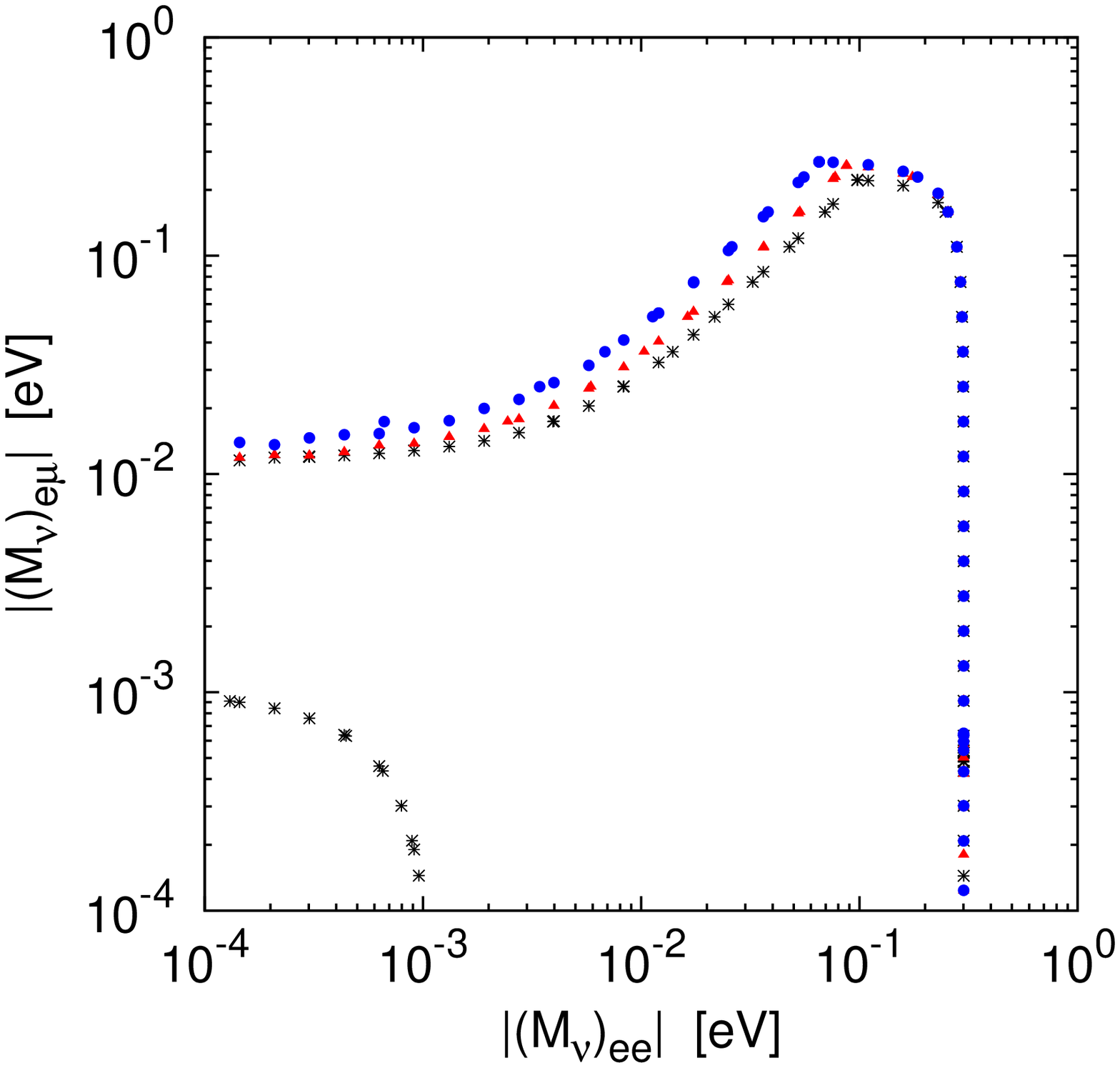} &
\includegraphics[angle=-90,keepaspectratio=true,scale=\figurescale]
{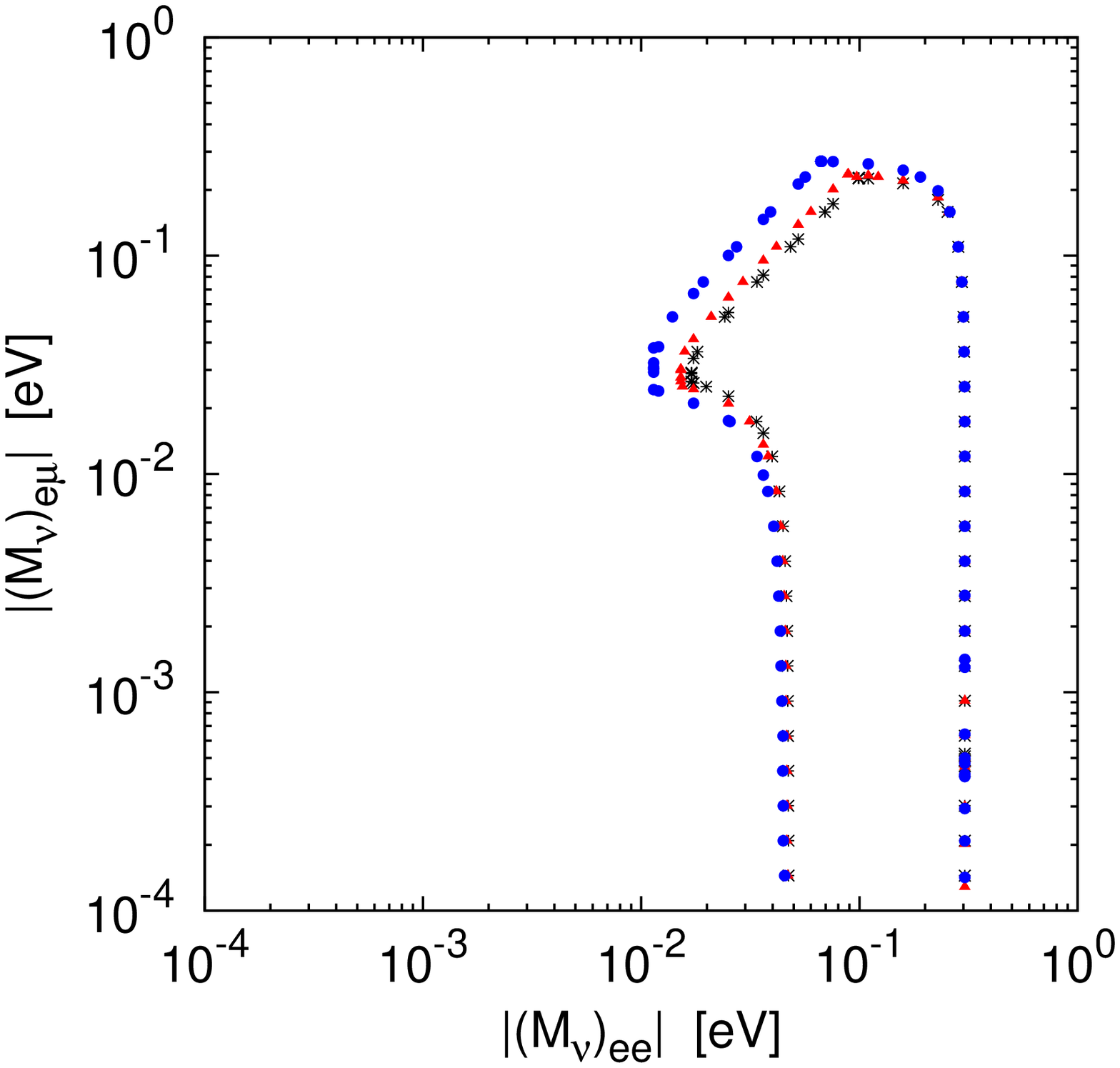}\\
\includegraphics[angle=-90,keepaspectratio=true,scale=\figurescale]
{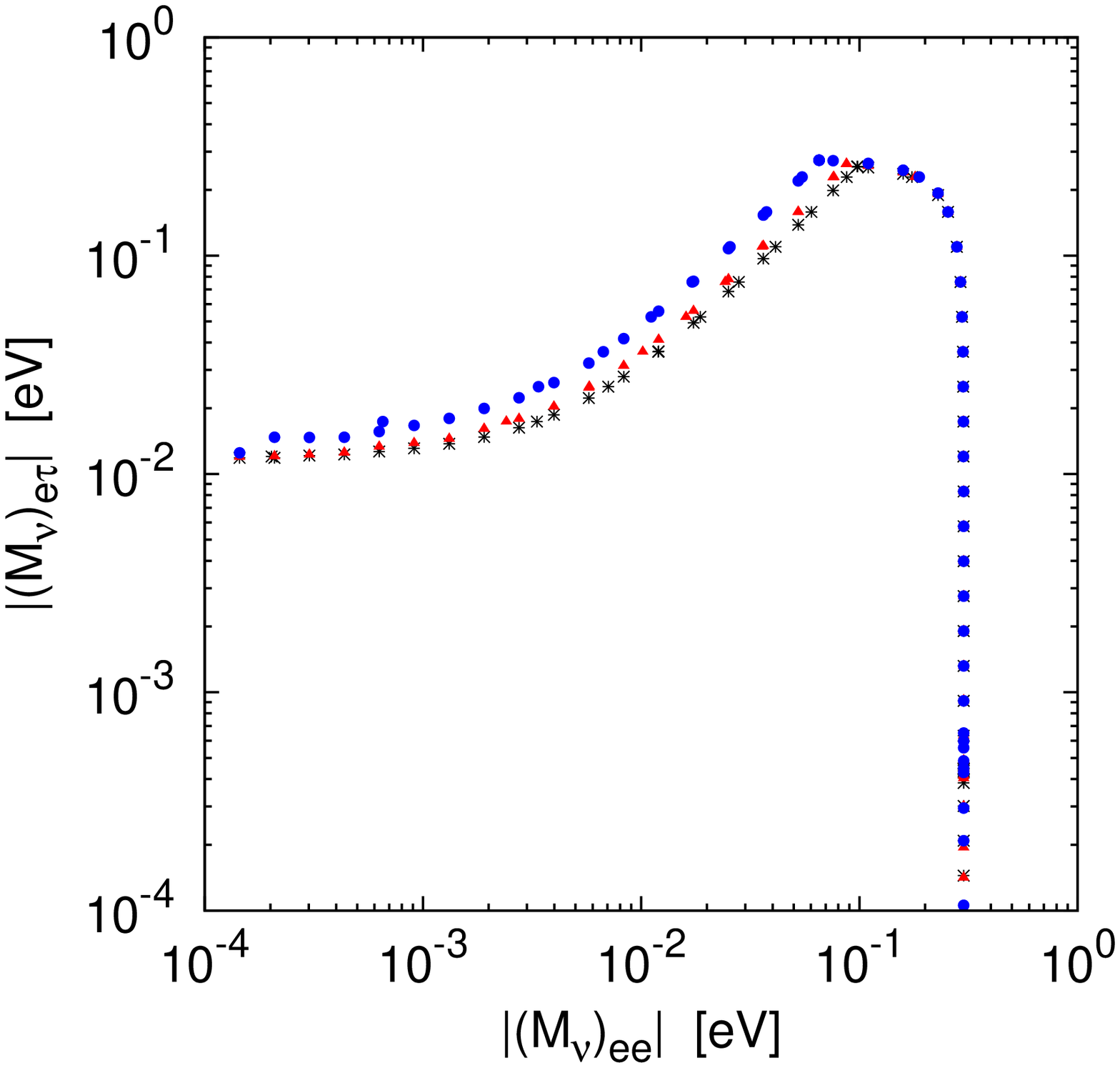} &
\includegraphics[angle=-90,keepaspectratio=true,scale=\figurescale]
{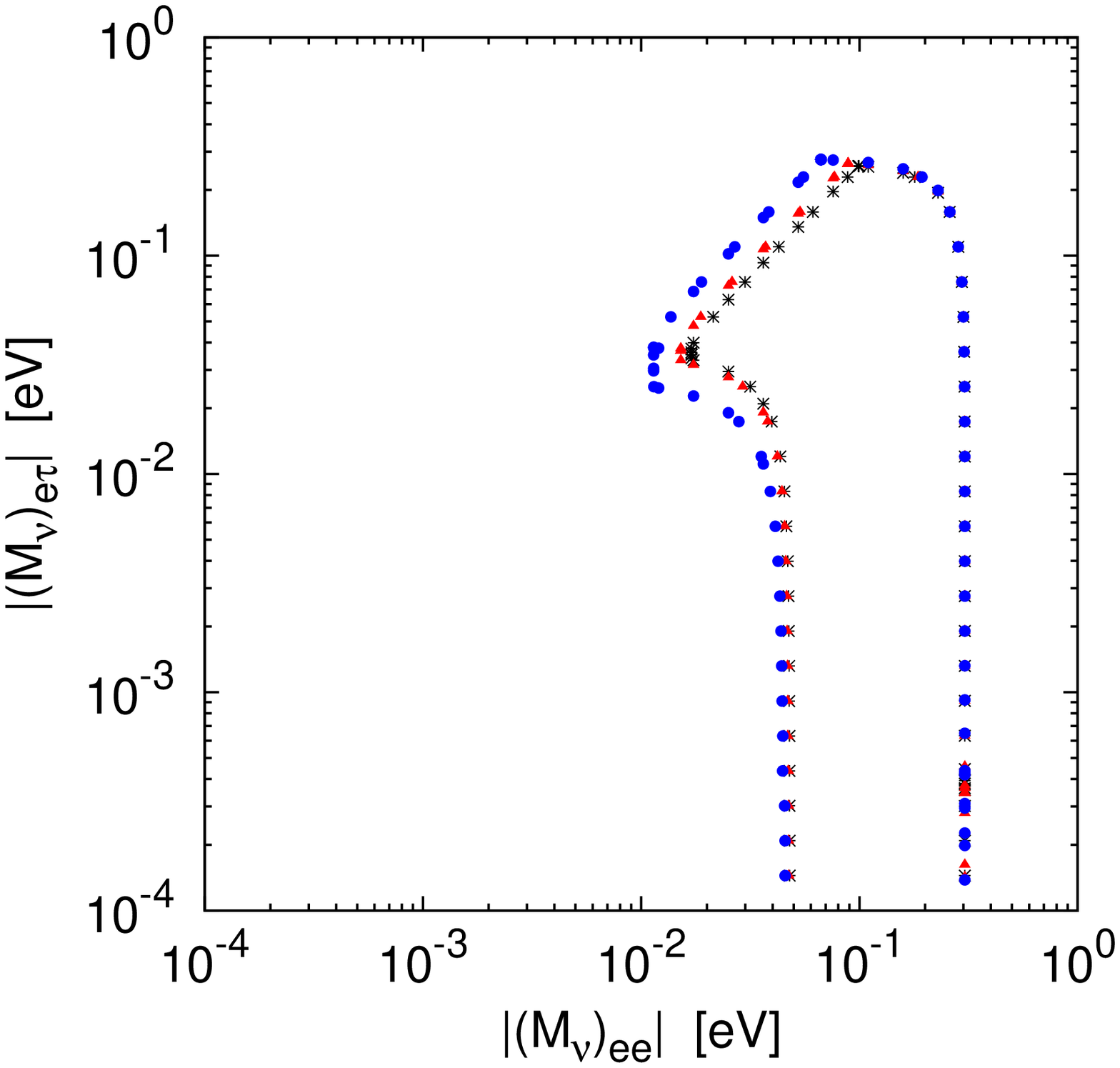}\\
\includegraphics[angle=-90,keepaspectratio=true,scale=\figurescale]
{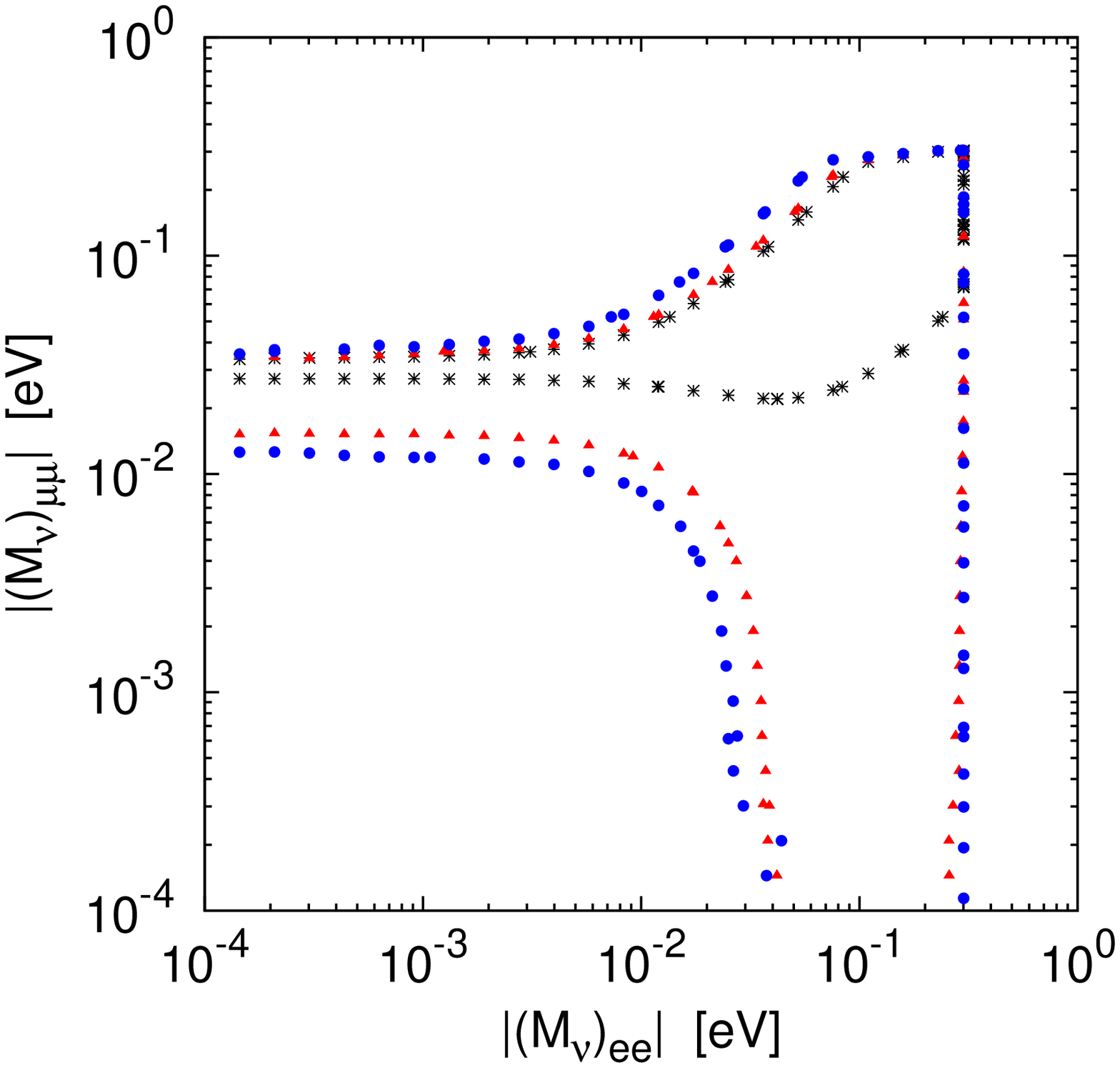} &
\includegraphics[angle=-90,keepaspectratio=true,scale=\figurescale]
{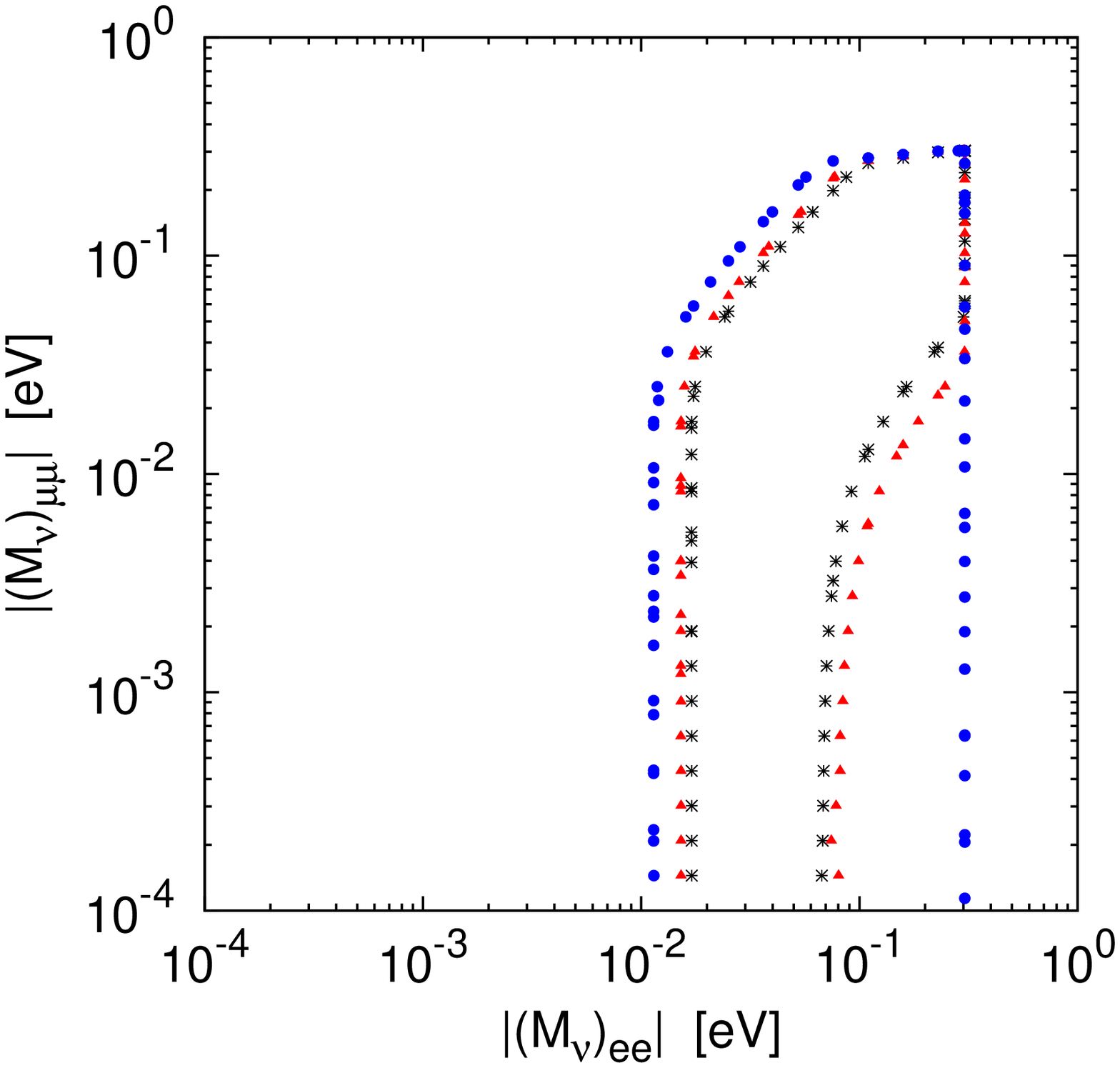}\\
\end{tabular}

\begin{tabular}[t]{ll}
\includegraphics[angle=-90,keepaspectratio=true,scale=\figurescale]
{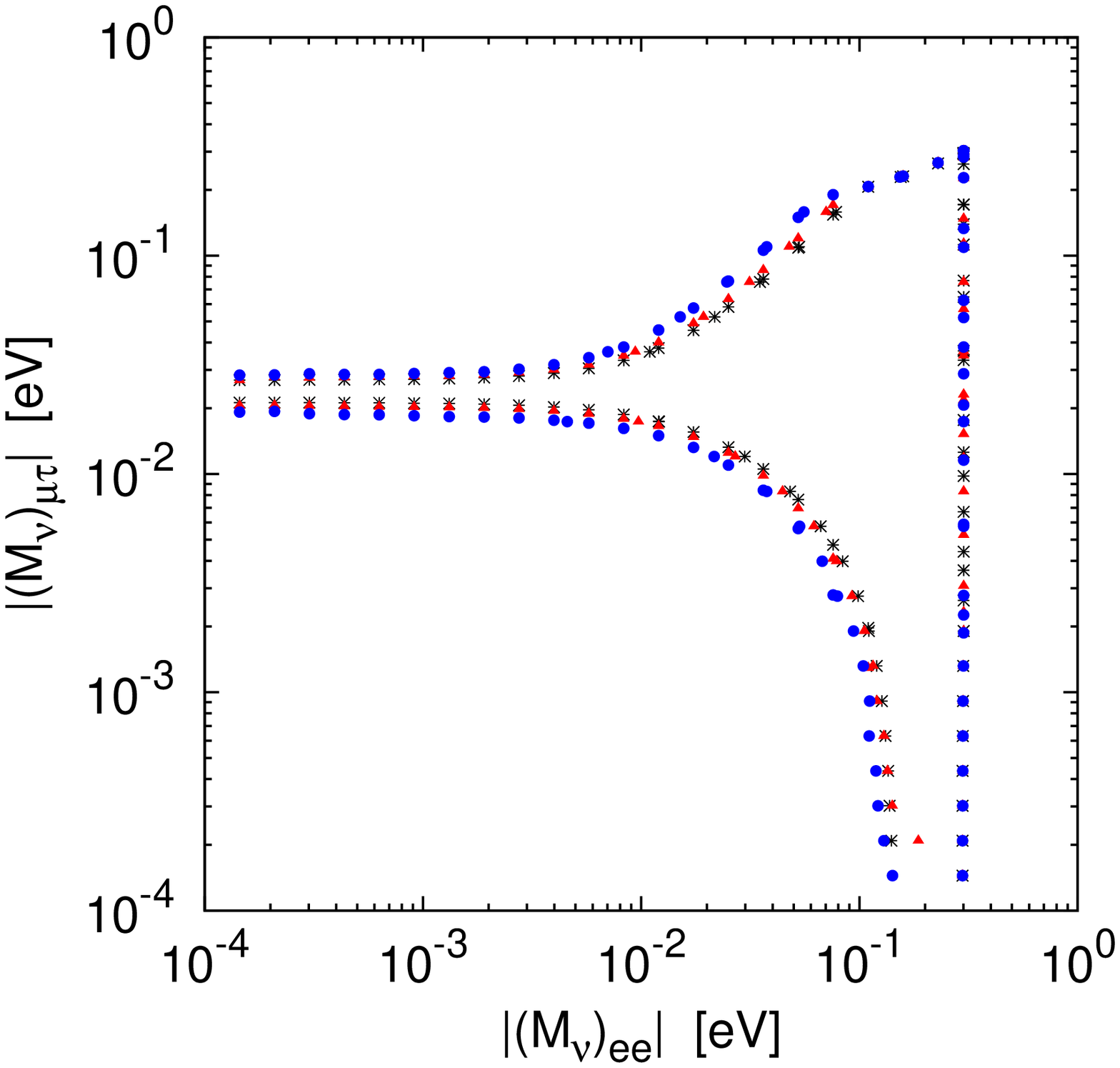} &
\includegraphics[angle=-90,keepaspectratio=true,scale=\figurescale]
{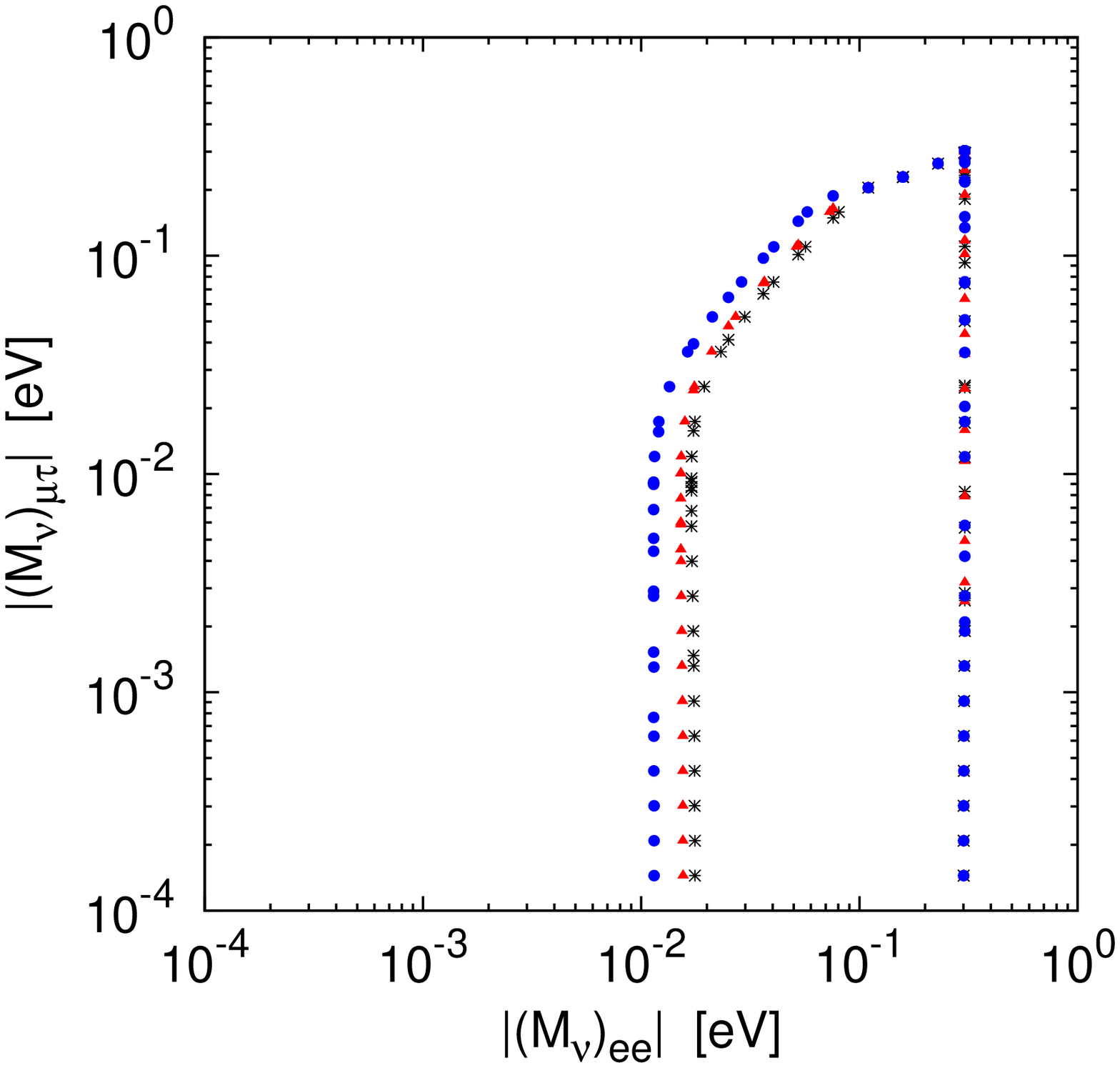}\\
\includegraphics[angle=-90,keepaspectratio=true,scale=\figurescale]
{figures/Forero-new-normal/M11-M33normal.eps} &
\includegraphics[angle=-90,keepaspectratio=true,scale=\figurescale]
{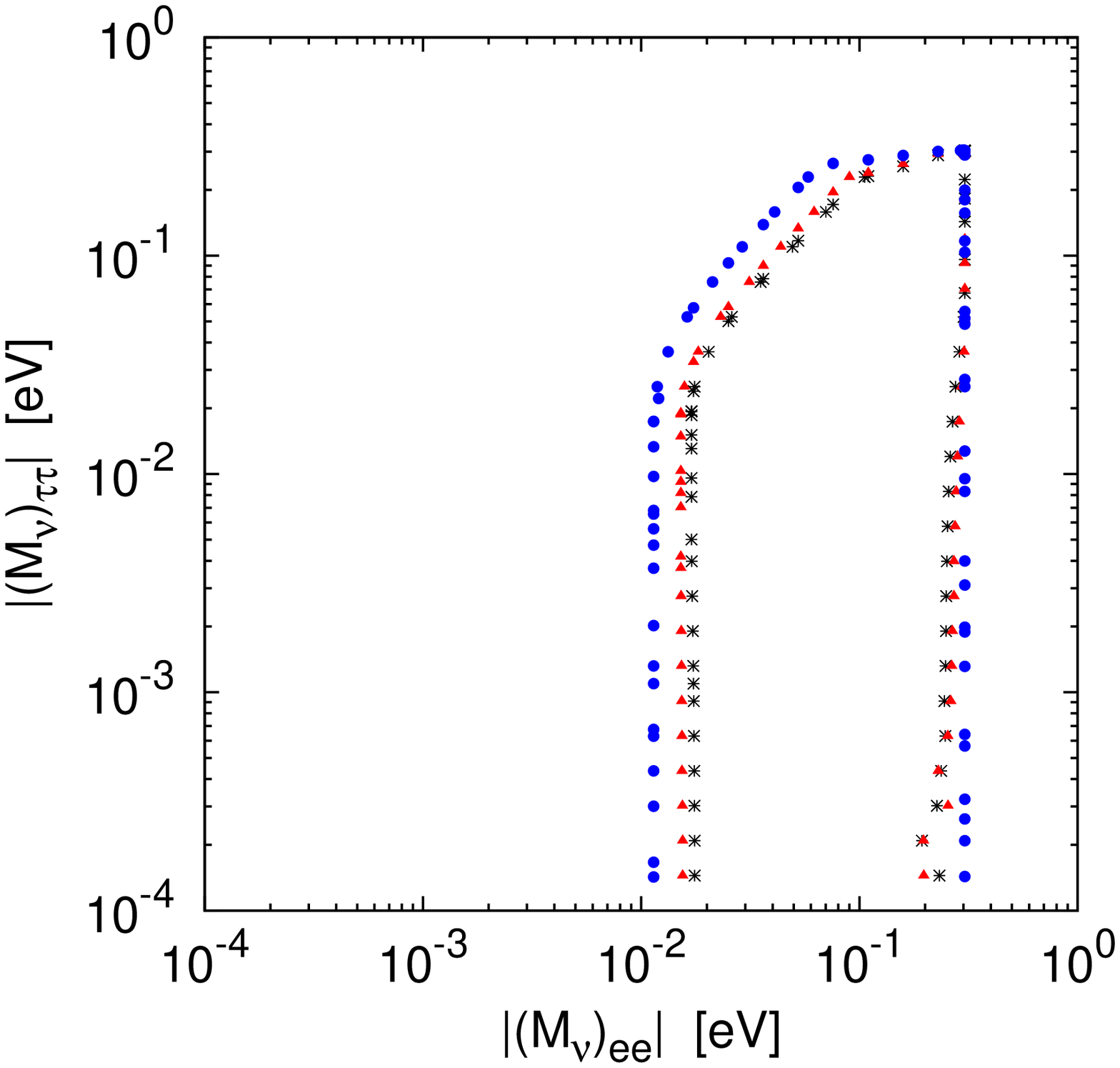}\\
\includegraphics[angle=-90,keepaspectratio=true,scale=\figurescale]
{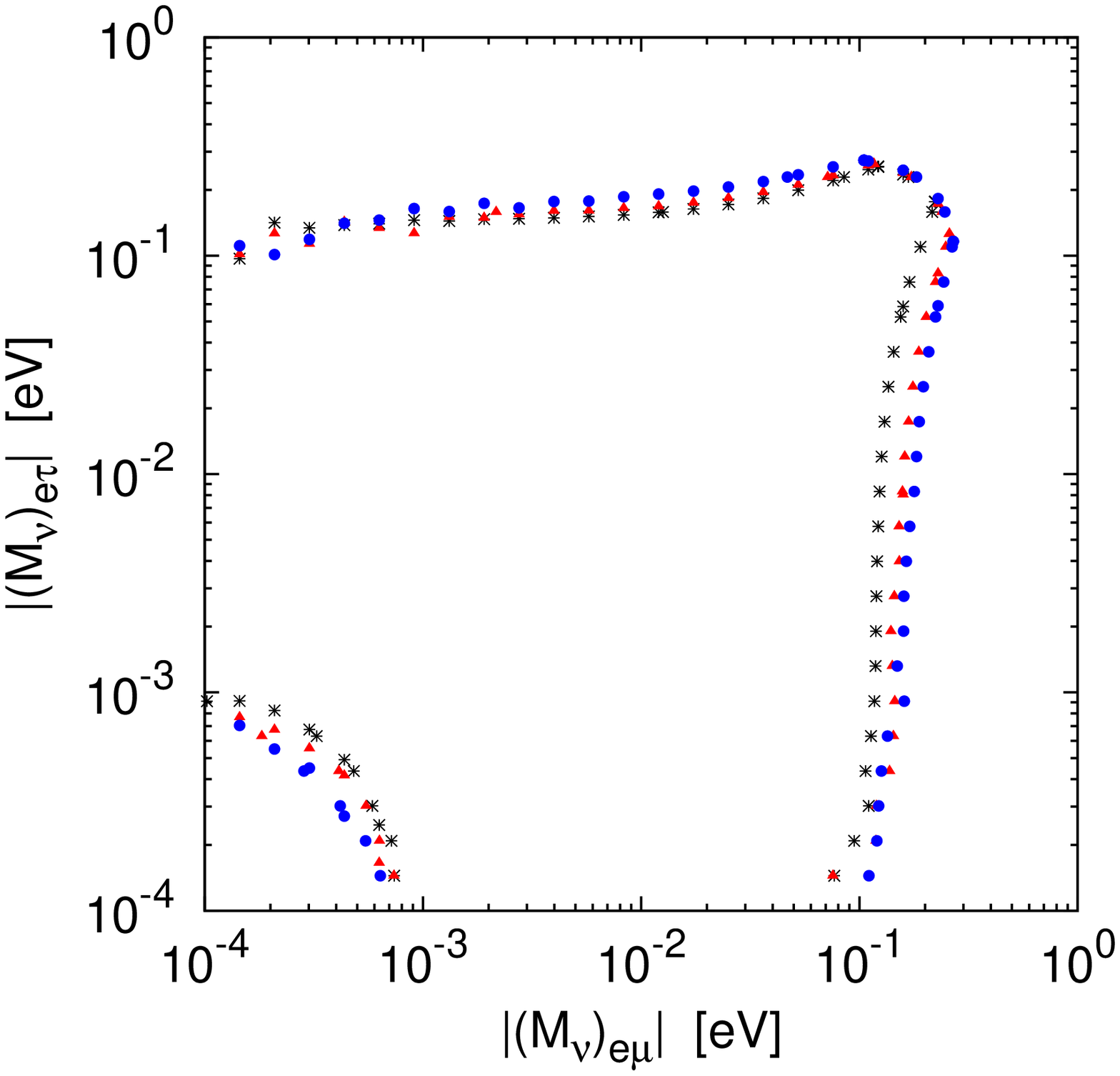} &
\includegraphics[angle=-90,keepaspectratio=true,scale=\figurescale]
{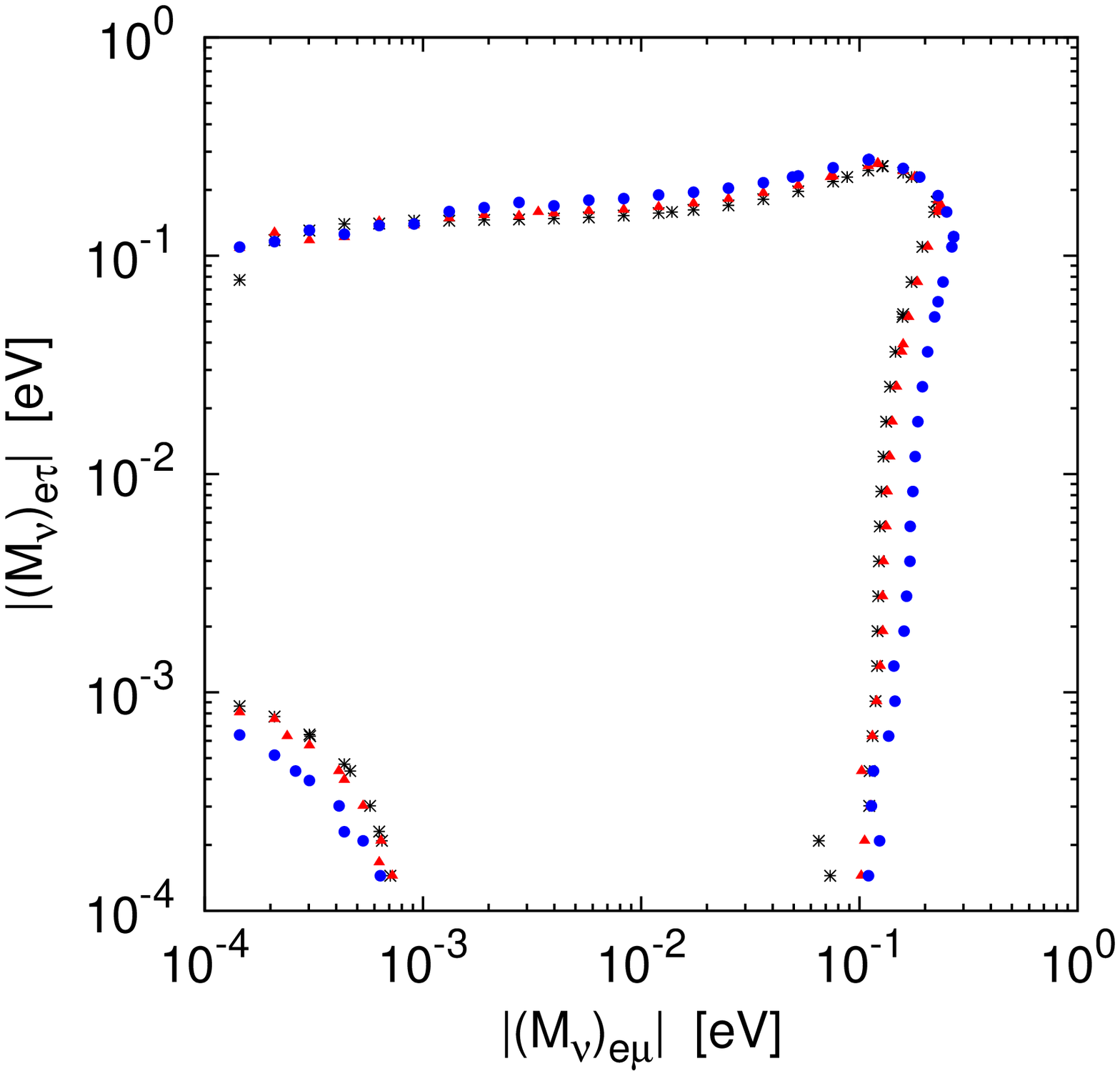}\\
\end{tabular}

\begin{tabular}[t]{ll}
\includegraphics[angle=-90,keepaspectratio=true,scale=\figurescale]
{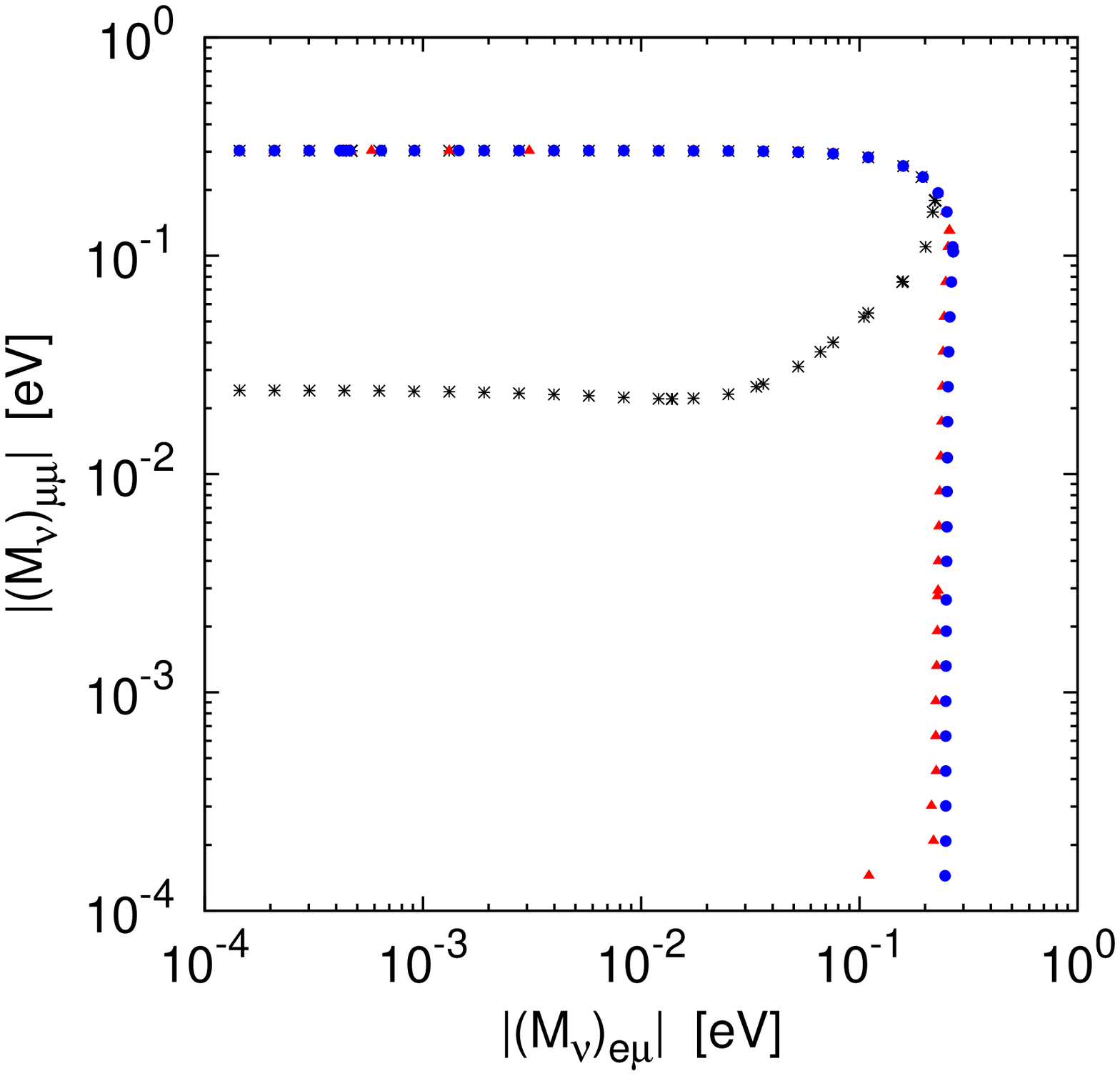} &
\includegraphics[angle=-90,keepaspectratio=true,scale=\figurescale]
{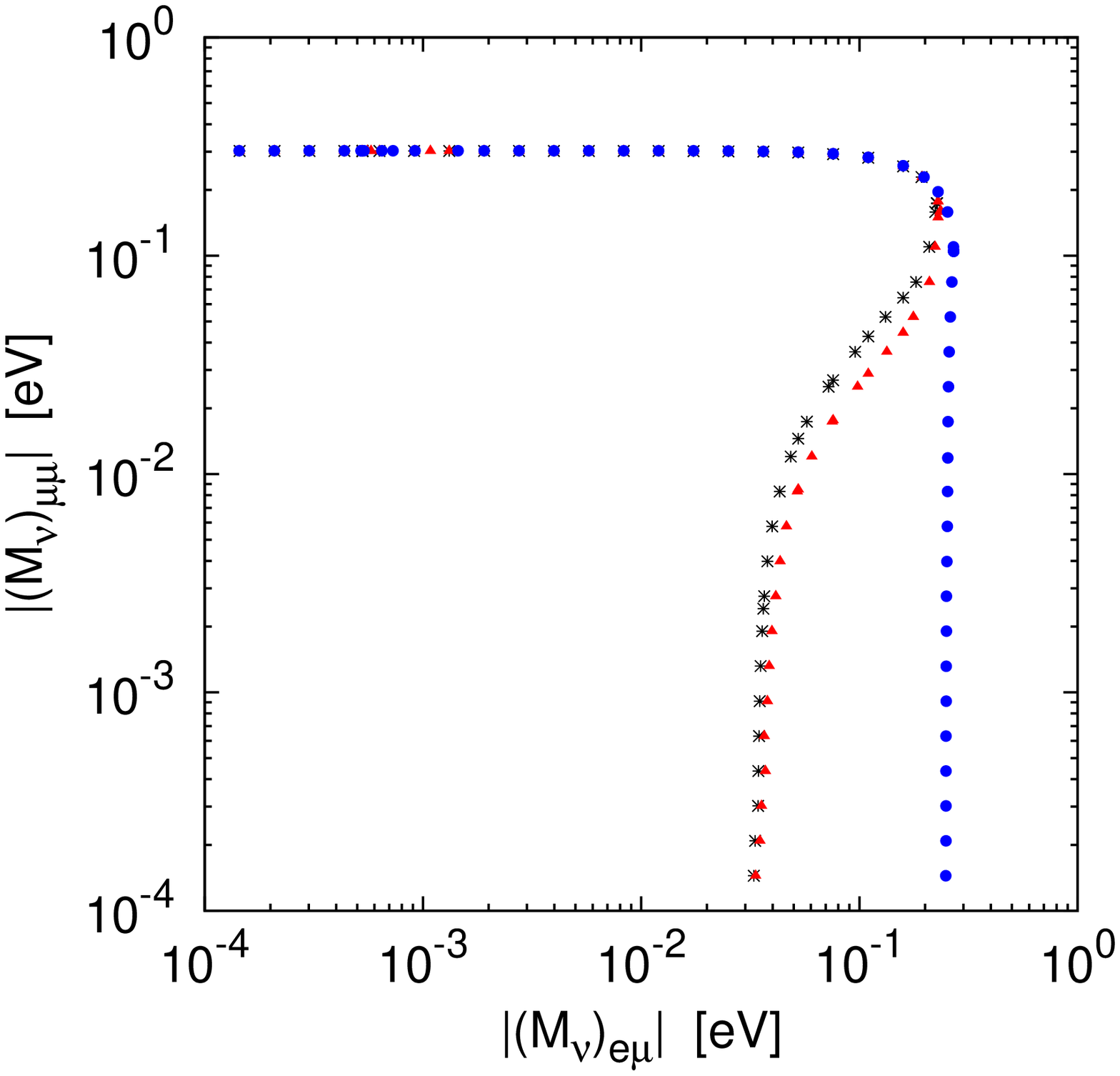}\\
\includegraphics[angle=-90,keepaspectratio=true,scale=\figurescale]
{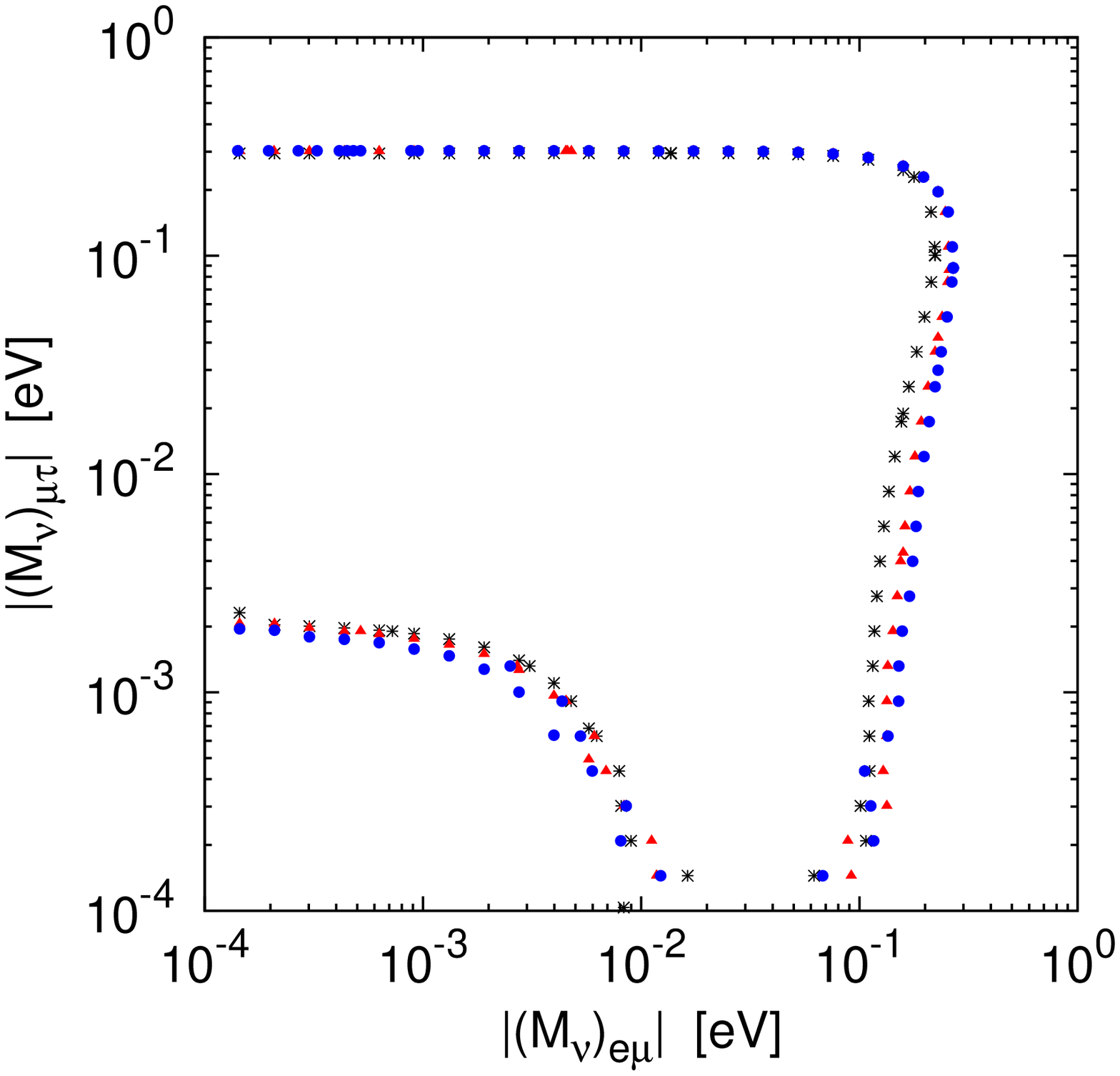} &
\includegraphics[angle=-90,keepaspectratio=true,scale=\figurescale]
{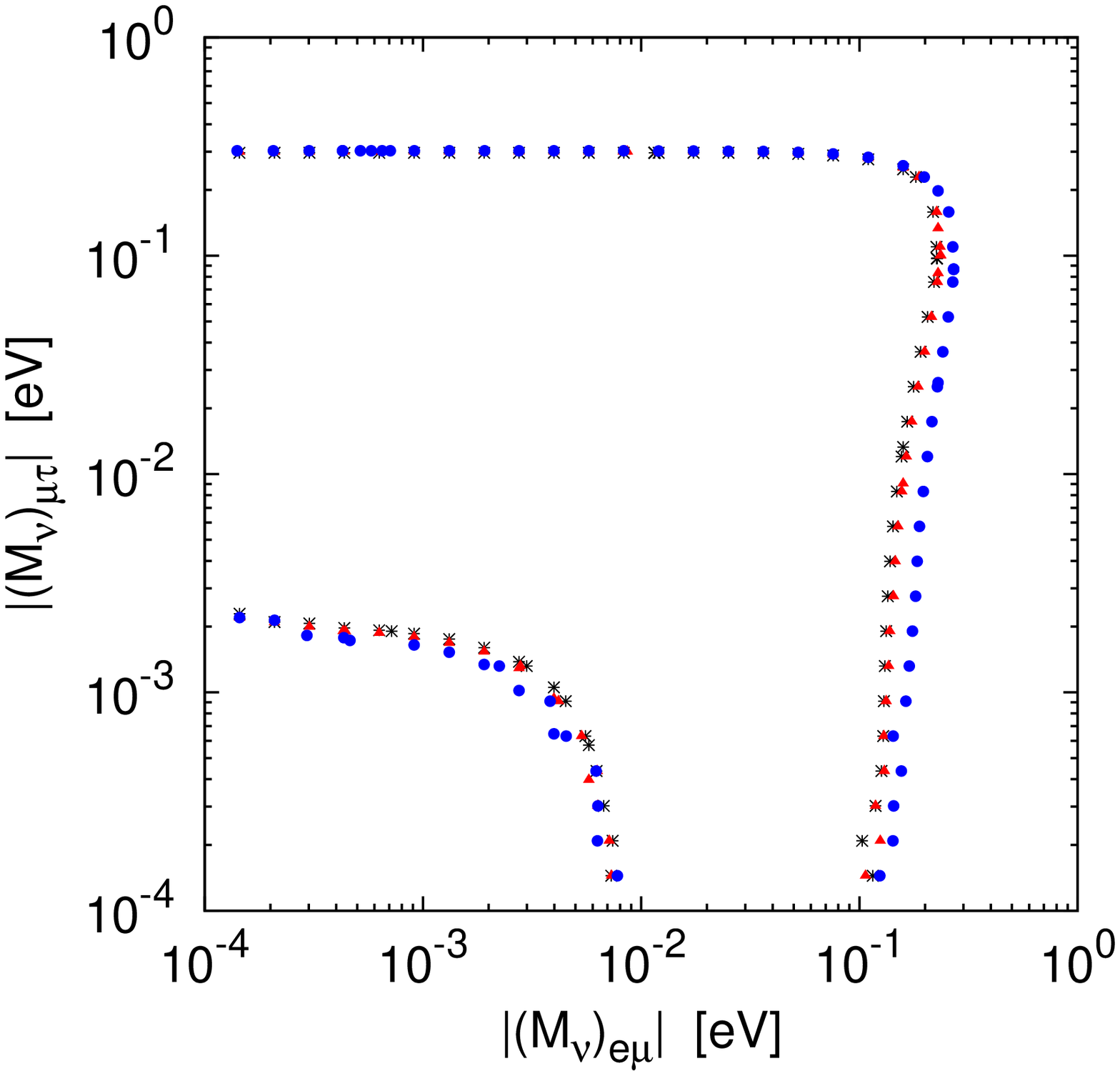}\\
\includegraphics[angle=-90,keepaspectratio=true,scale=\figurescale]
{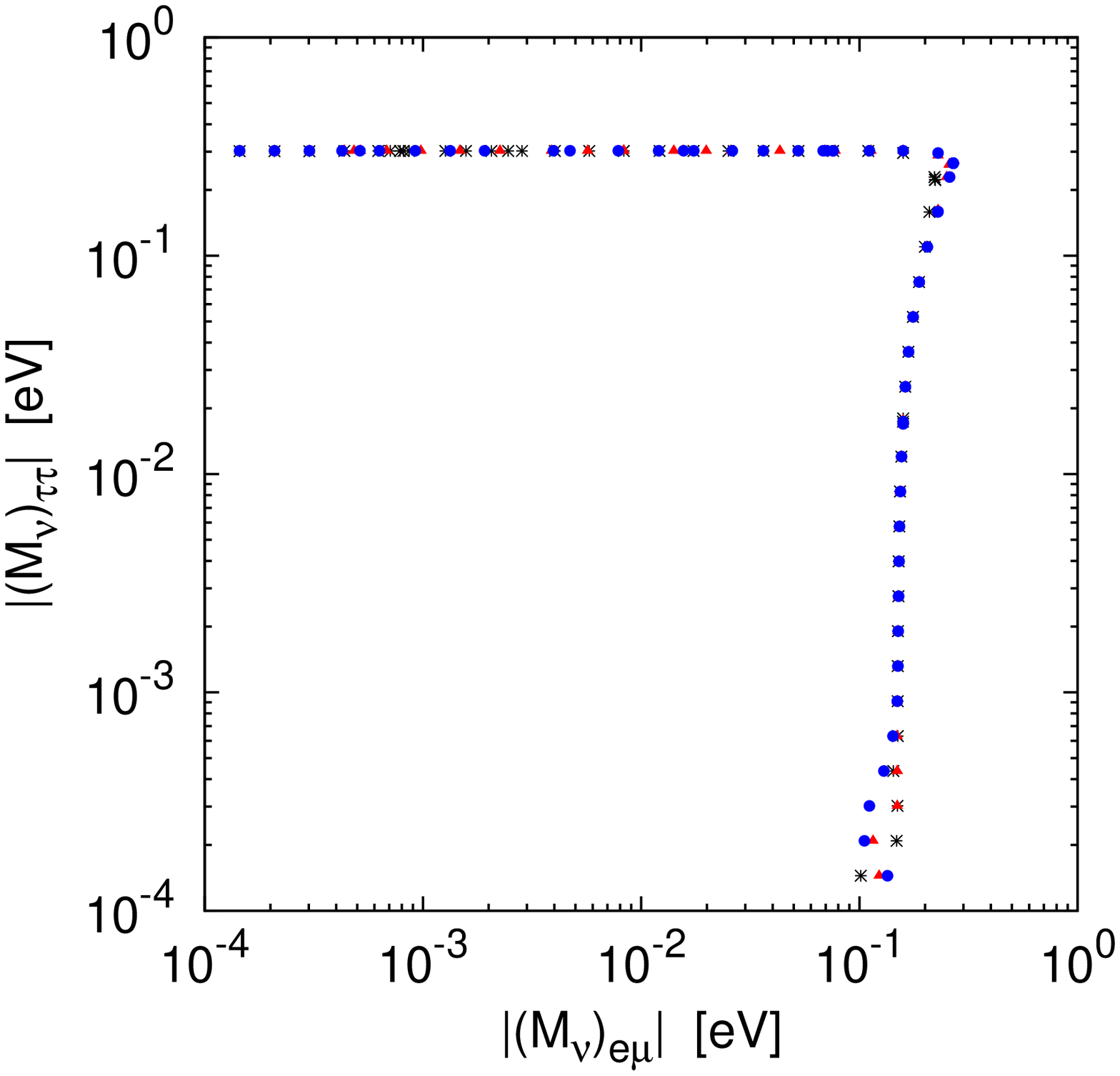} &
\includegraphics[angle=-90,keepaspectratio=true,scale=\figurescale]
{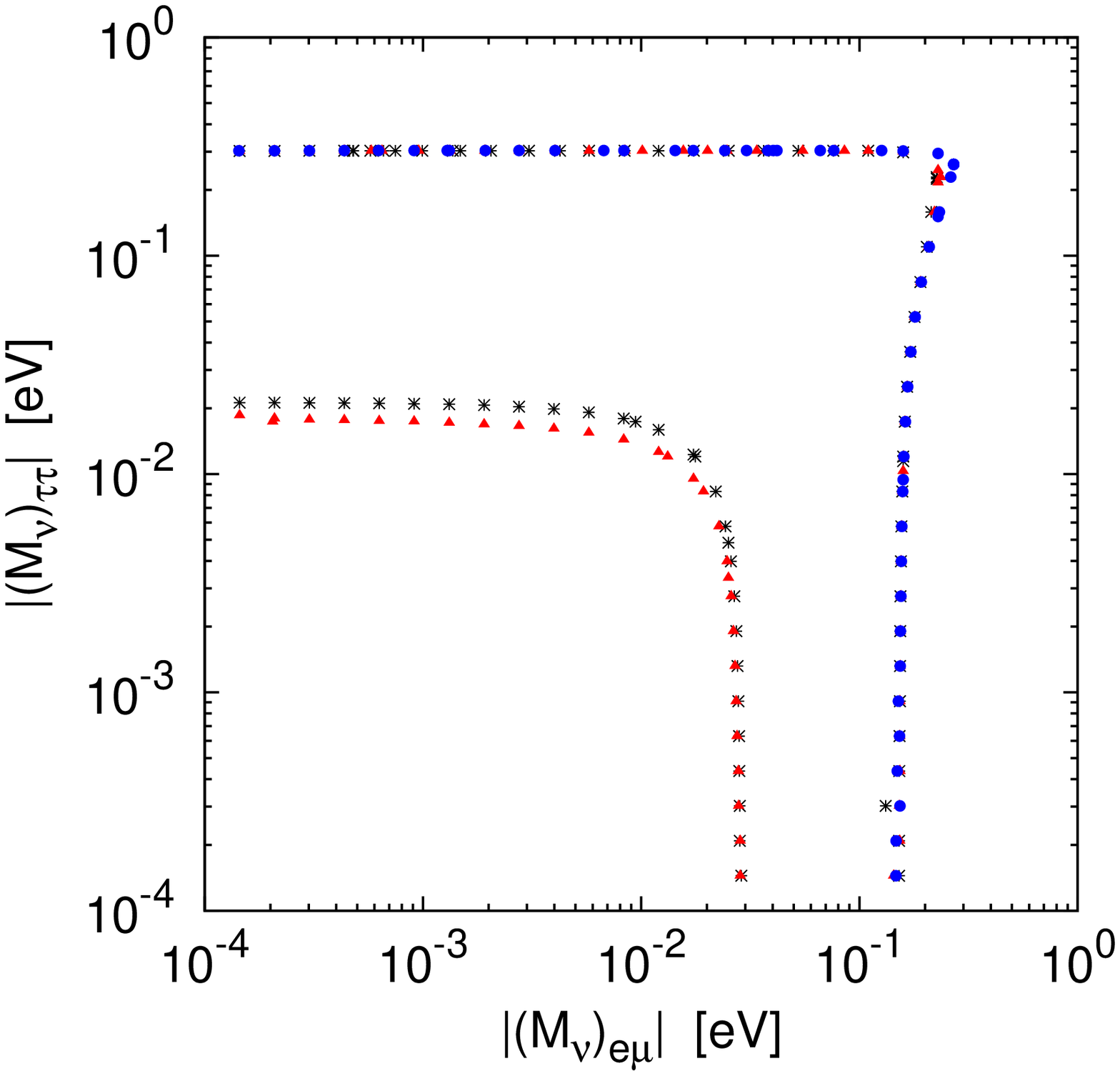}\\
\end{tabular}

\begin{tabular}[t]{ll}
\includegraphics[angle=-90,keepaspectratio=true,scale=\figurescale]
{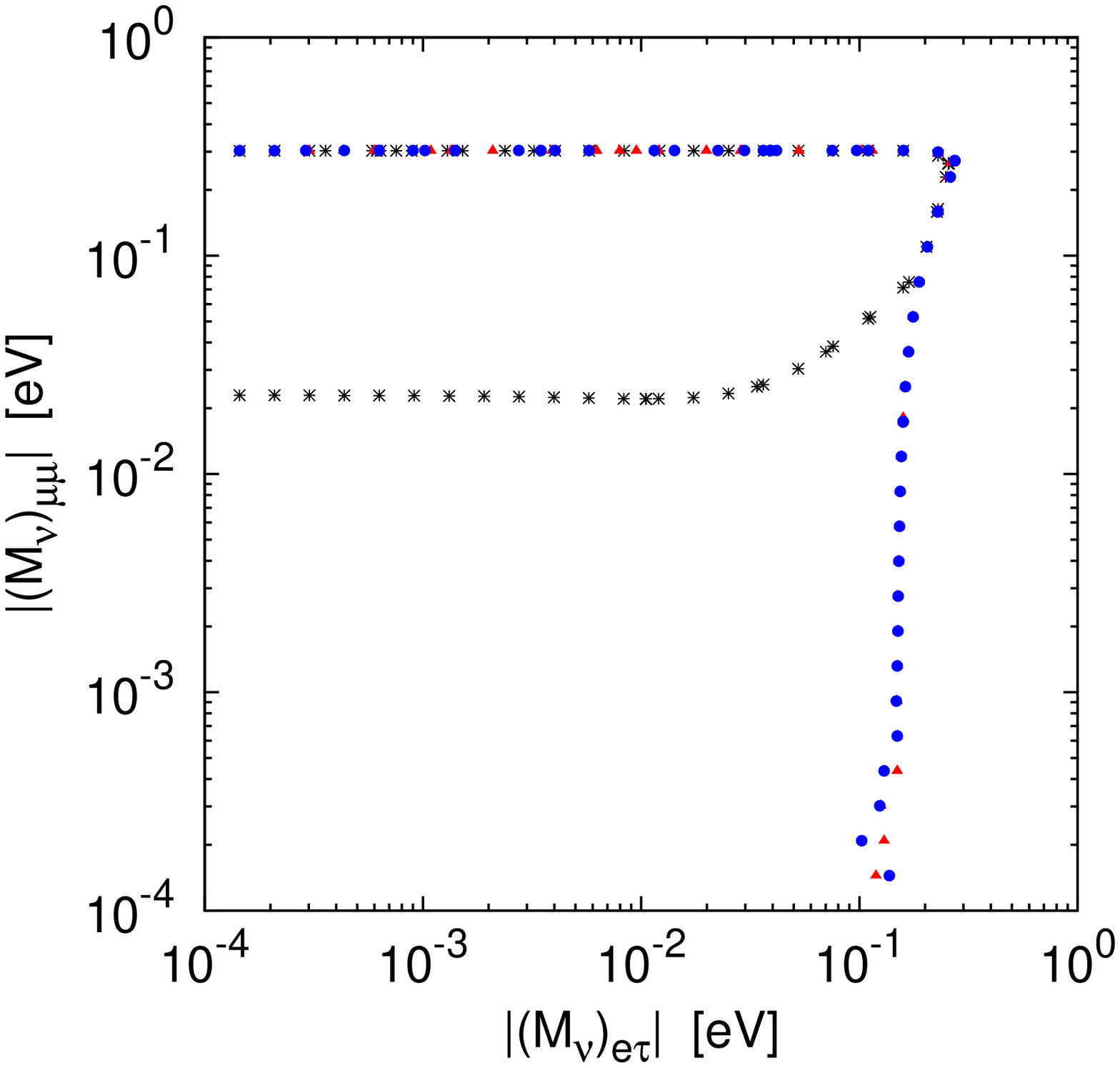} &
\includegraphics[angle=-90,keepaspectratio=true,scale=\figurescale]
{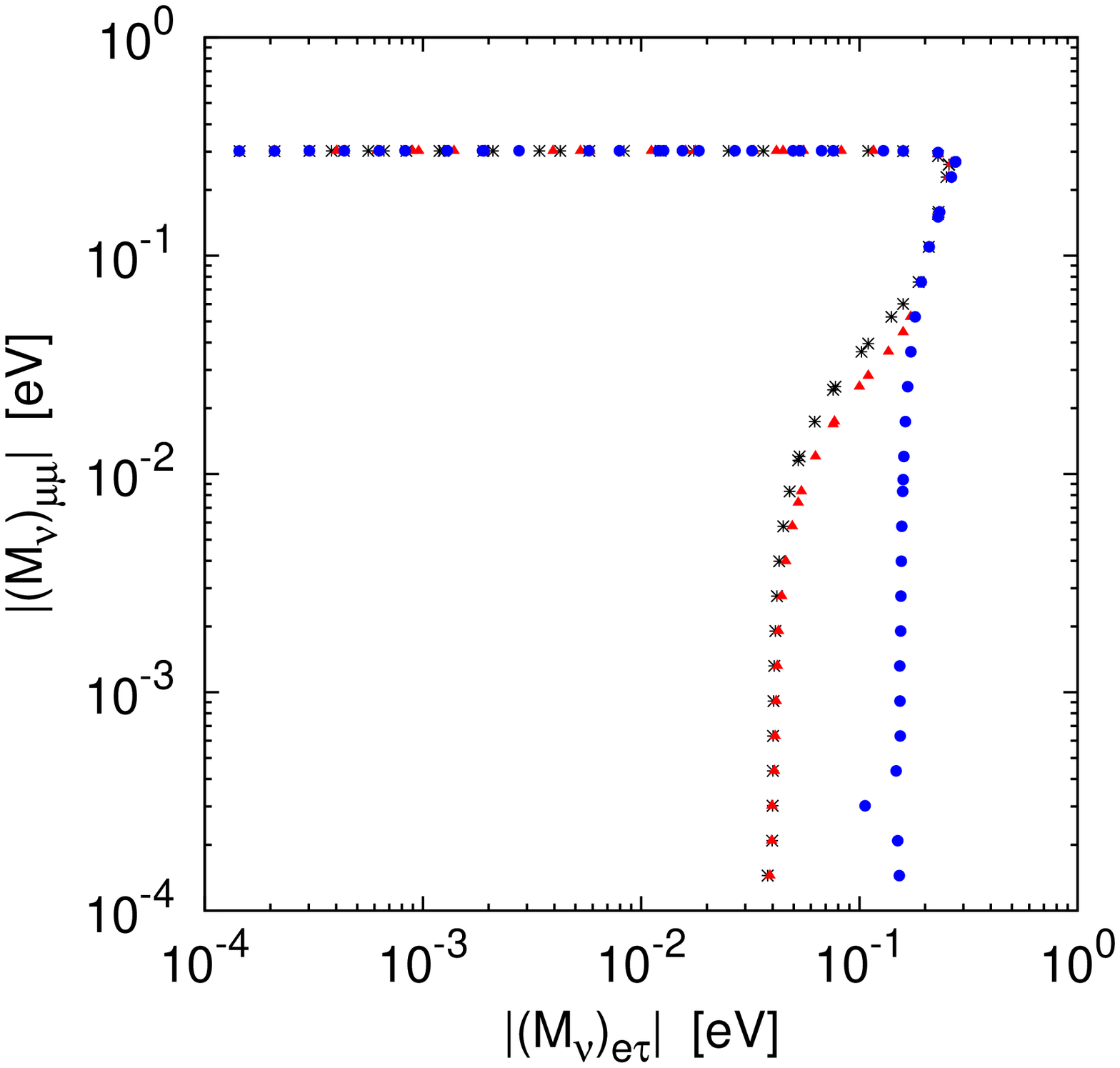}\\
\includegraphics[angle=-90,keepaspectratio=true,scale=\figurescale]
{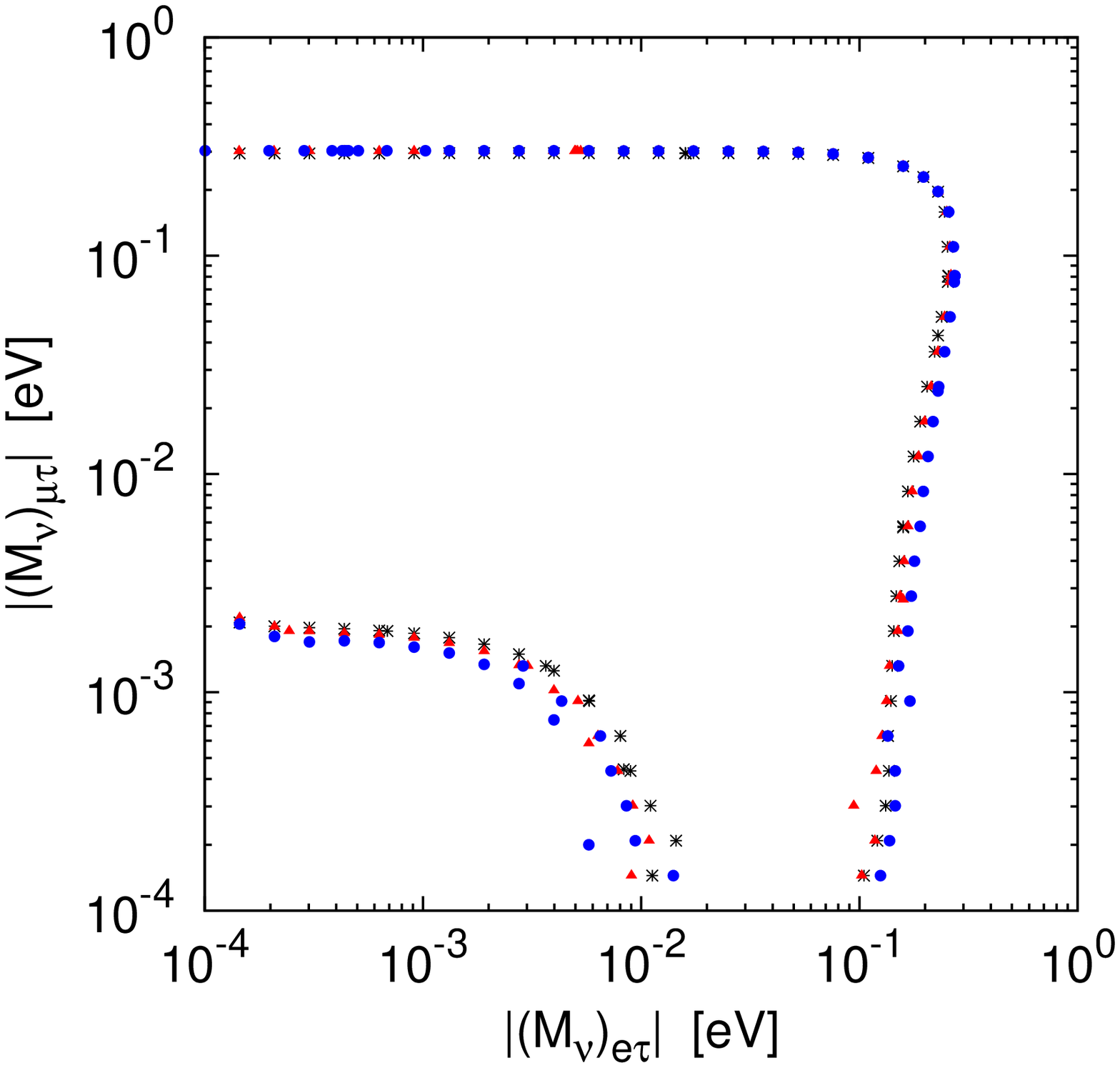} &
\includegraphics[angle=-90,keepaspectratio=true,scale=\figurescale]
{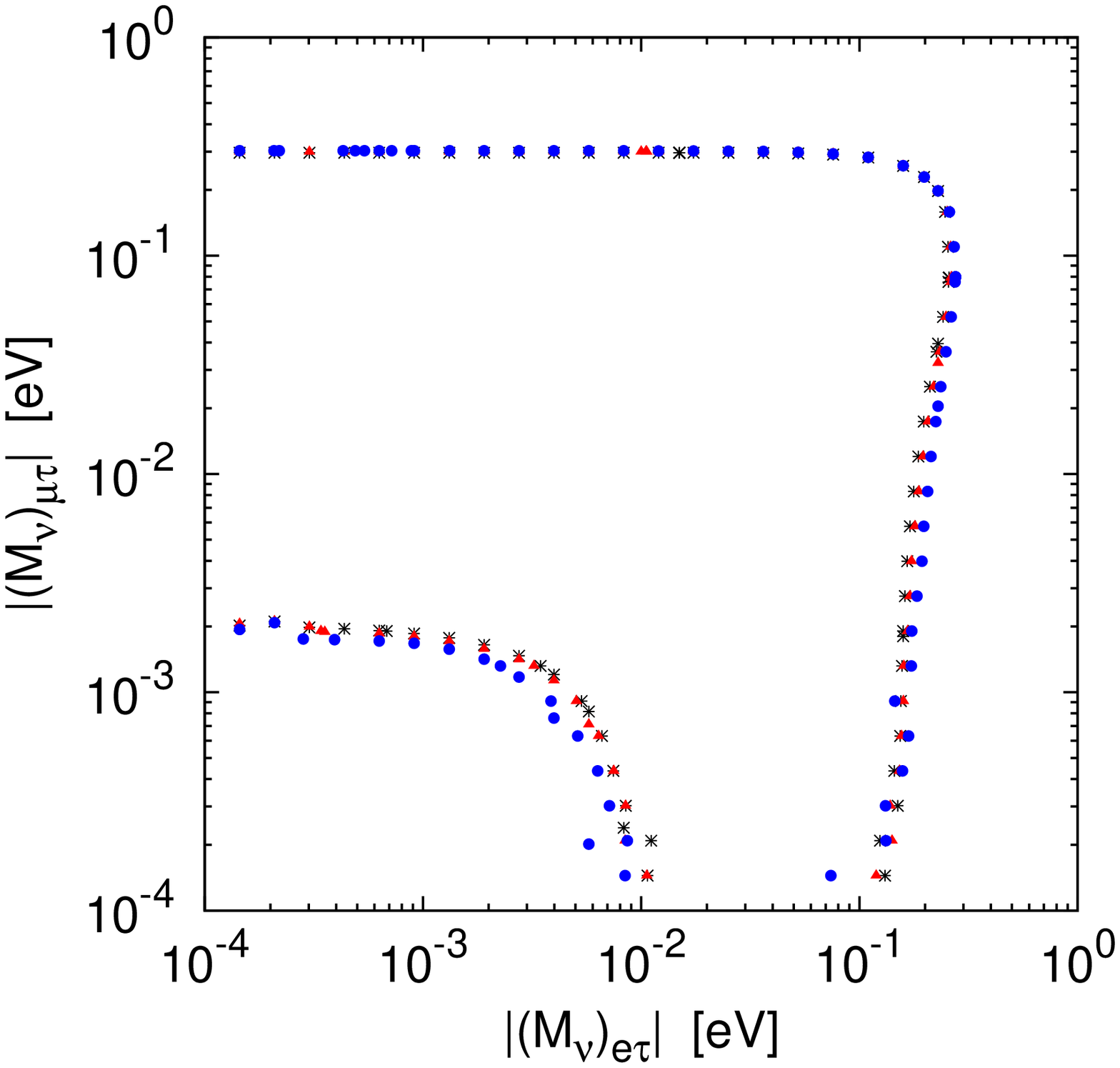}\\
\includegraphics[angle=-90,keepaspectratio=true,scale=\figurescale]
{figures/Forero-new-normal/M13-M33normal.eps} &
\includegraphics[angle=-90,keepaspectratio=true,scale=\figurescale]
{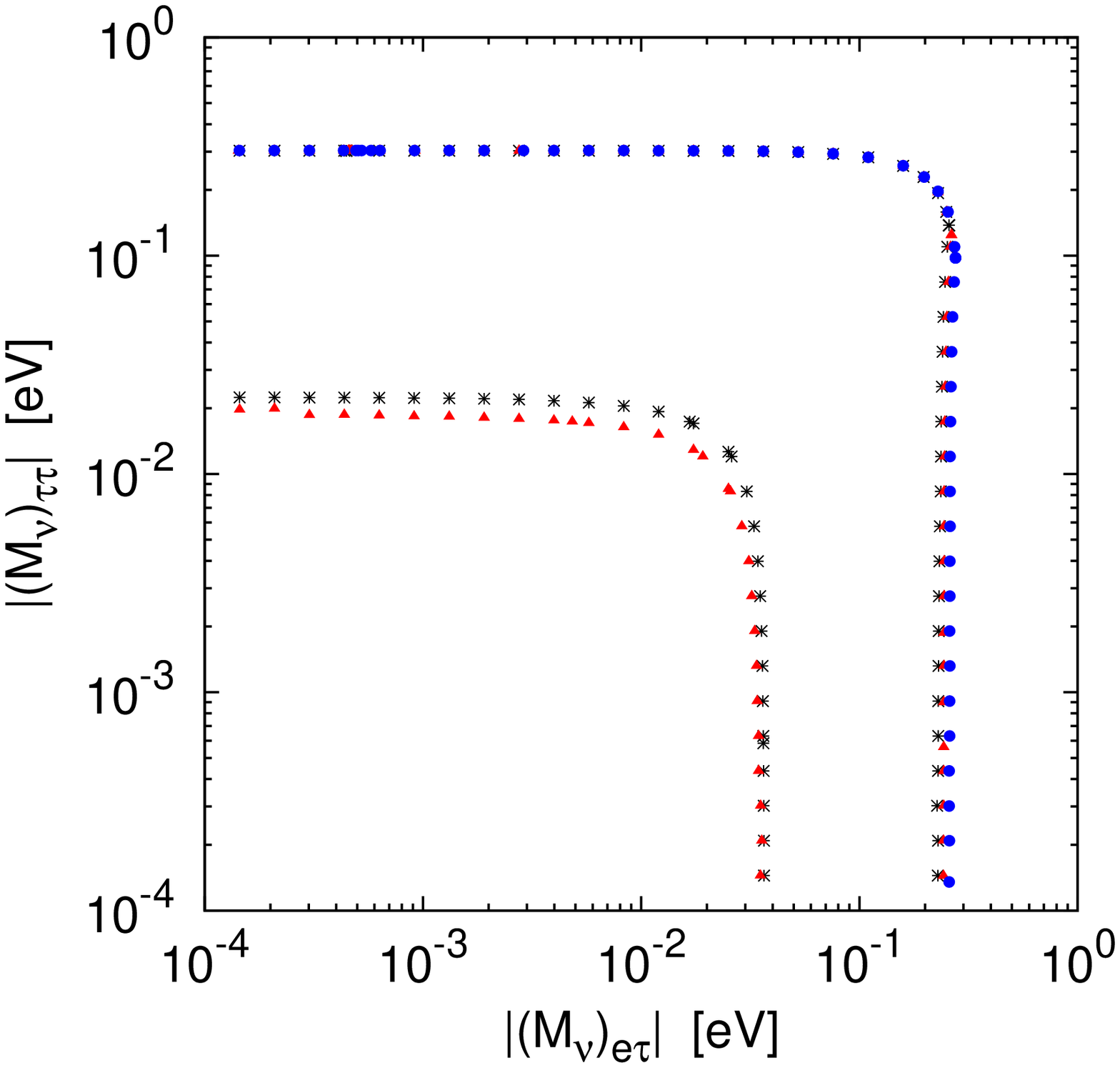}\\
\end{tabular}

\begin{tabular}[t]{ll}
\includegraphics[angle=-90,keepaspectratio=true,scale=\figurescale]
{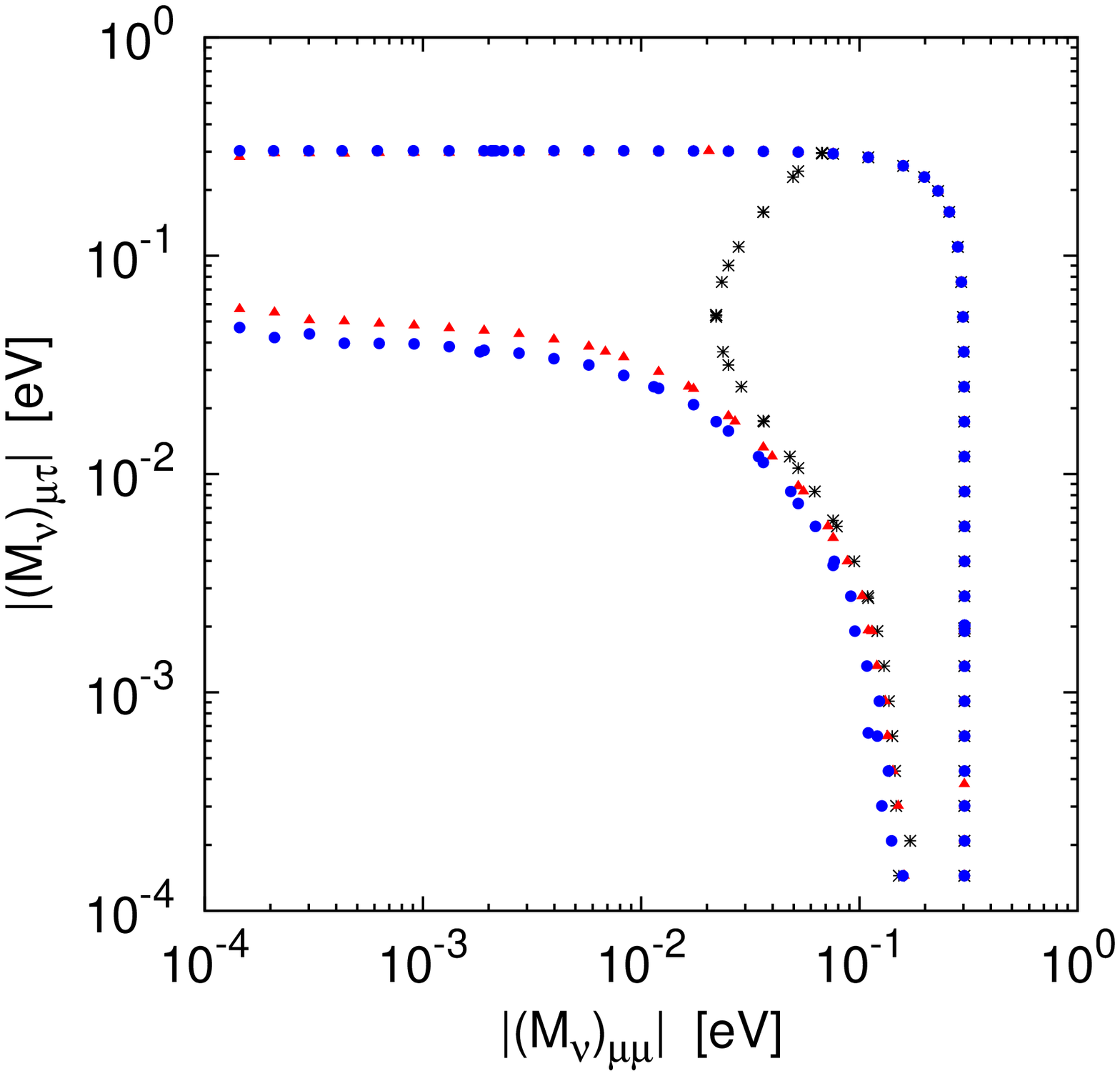} &
\includegraphics[angle=-90,keepaspectratio=true,scale=\figurescale]
{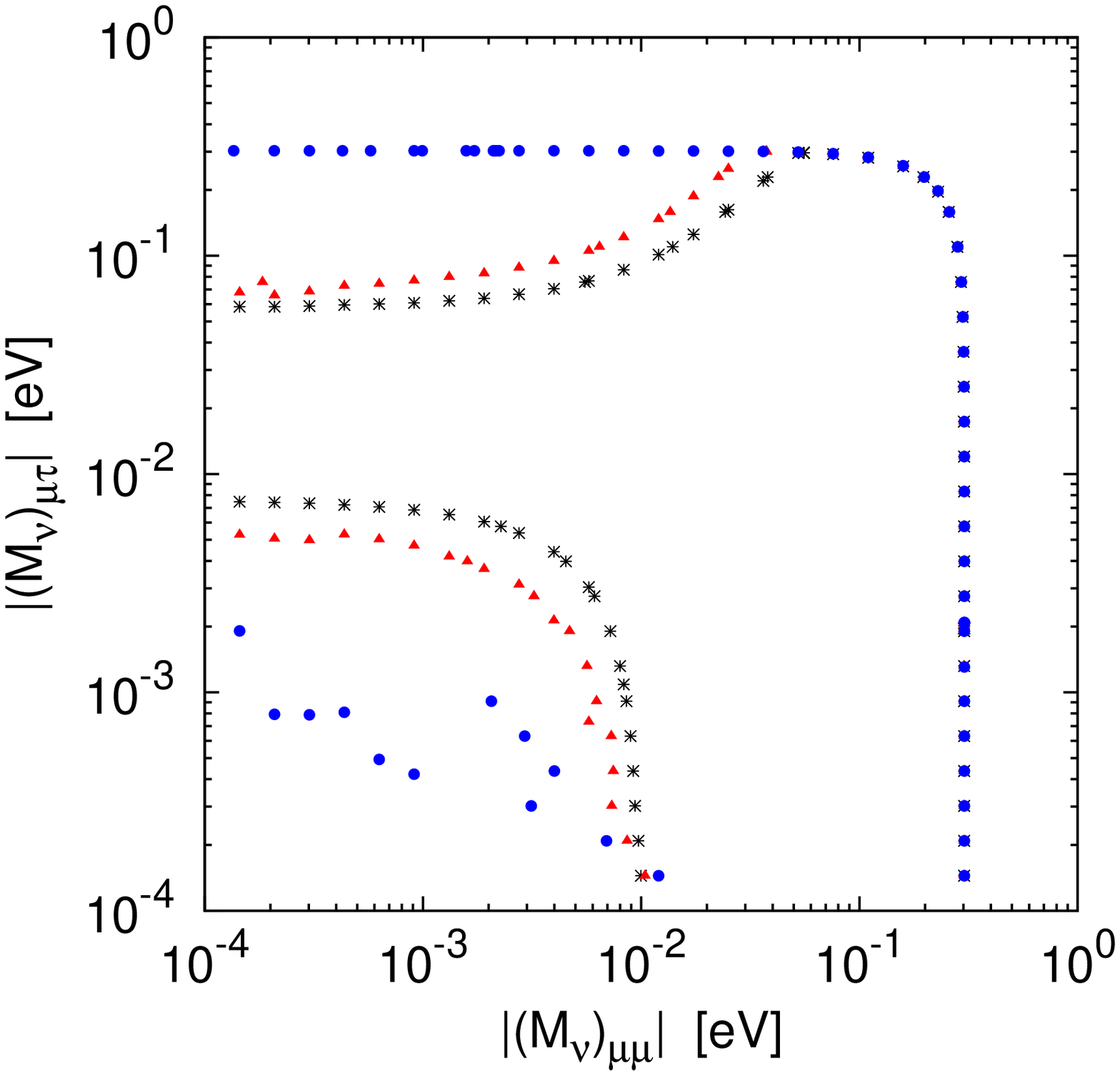}\\
\includegraphics[angle=-90,keepaspectratio=true,scale=\figurescale]
{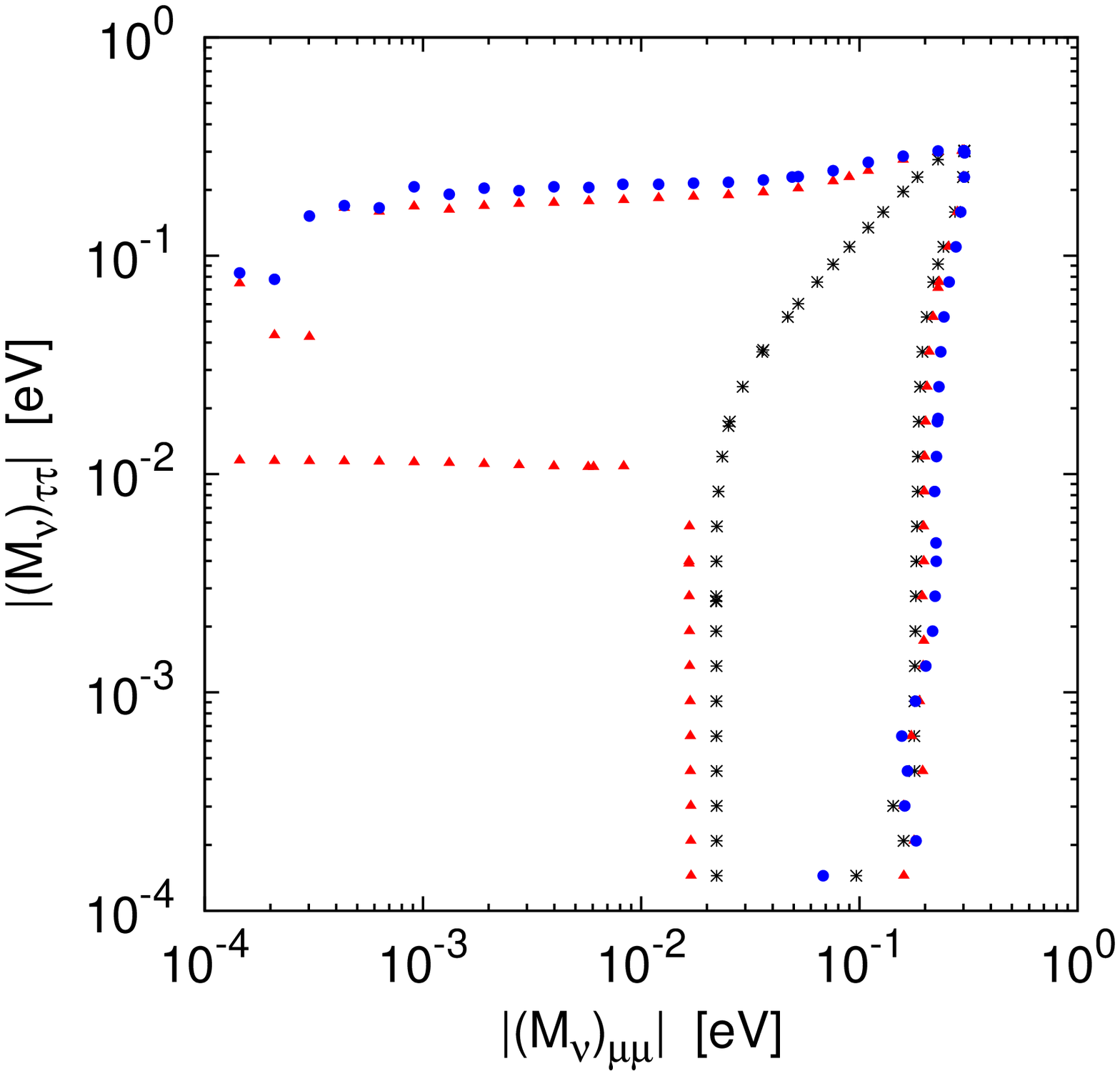} &
\includegraphics[angle=-90,keepaspectratio=true,scale=\figurescale]
{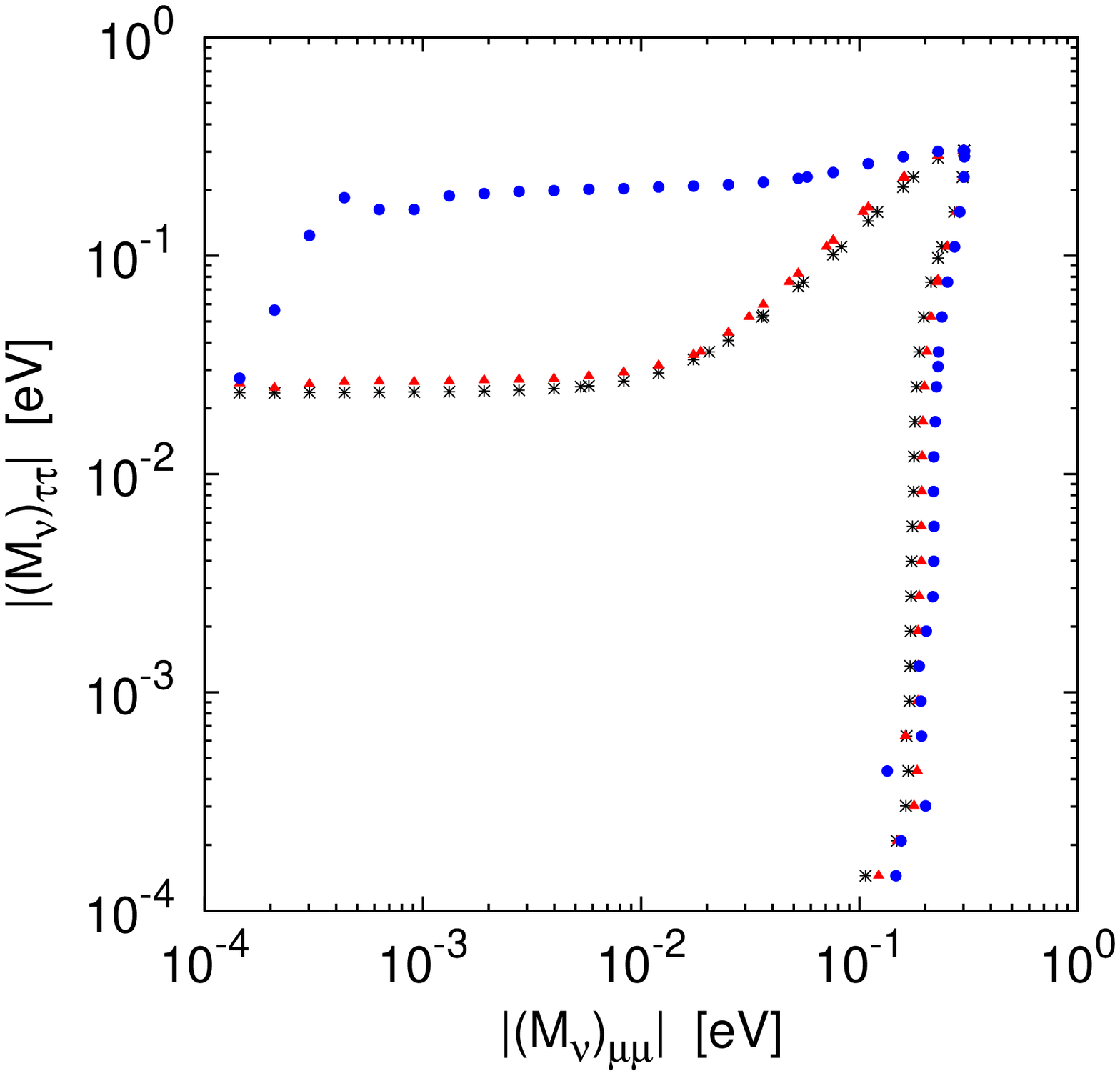}\\
\includegraphics[angle=-90,keepaspectratio=true,scale=\figurescale]
{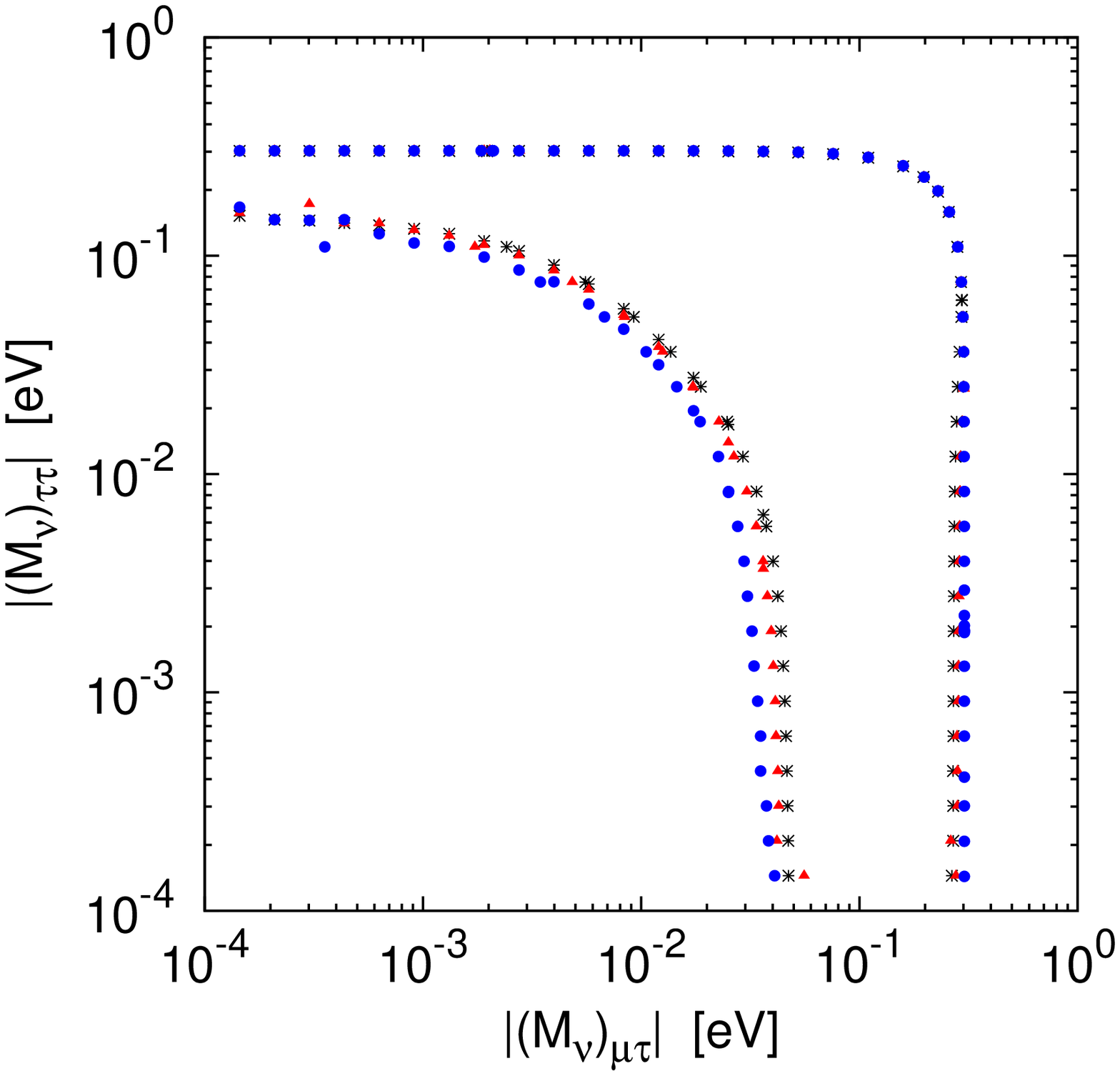} &
\includegraphics[angle=-90,keepaspectratio=true,scale=\figurescale]
{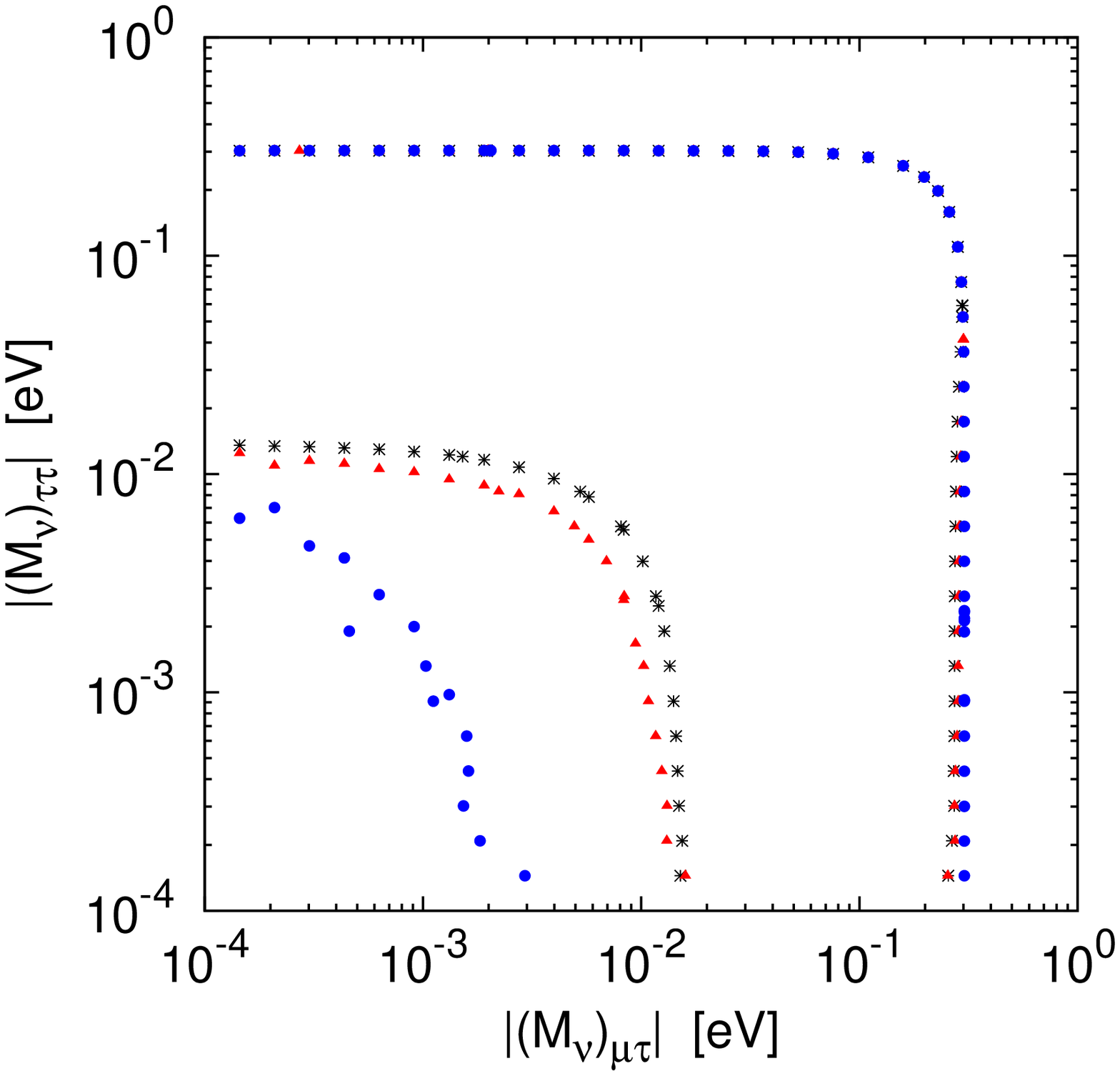}\\
\end{tabular}

\end{appendix}


\begin{thebibliography}{99}

\bibitem{daya-reno}
Y.~Abe \textit{et al.}
(Double Chooz Coll.),
\textit{Indication for the disappearance of reactor electron antineutrinos
in the Double Chooz experiment},
Phys.\ Rev.\ Lett.\ \textbf{108} (2012) 131801
[arXiv:1112.6353]; \\
F.P.~An \textit{et al.} (Daya Bay Coll.),
\textit{Observation of electron-antineutrino disappearance at Daya Bay},
Phys.\ Rev.\ Lett.\ \textbf{108} (2012) 171803
[arXiv:1203.1669]; \\
J.K.~Ahn \textit{et al.} (RENO Coll.),
\textit{Observation of reactor electron antineutrino disappearance in
the RENO experiment},
Phys.\ Rev.\ Lett.\ \textbf{108} (2012) 191802
[arXiv:1204.0626].

\bibitem{T2K-MINOS}
K.~Abe {\it et al.}  (T2K Coll.),
\textit{Indication of electron neutrino appearance from an accelerator-produced
off-axis muon neutrino beam},
Phys.\ Rev.\ Lett.\ \textbf{107} (2011) 041801
[arXiv:1106.2822]; \\
P.~Adamson {\it et al.}  (MINOS Coll.),
\textit{Improved search for muon-neutrino to electron-neutrino oscillations in MINOS},
Phys.\ Rev.\ Lett.\ \textbf{107} (2011) 181802
[arXiv:1108.0015].

\bibitem{forero}
D.V.~Forero, M.~T\'ortola and J.W.F.~Valle,
\textit{Global status of neutrino oscillation parameters after recent
  reactor measurements},
arXiv:1205.4018v2;
\textit{Global status of neutrino oscillation parameters after
  Neutrino-2012},
Phys.\ Rev.\ {\bf D 86} (2012) 073012
[arXiv:1205.4018v3].

\bibitem{fogli}
G.L.~Fogli, E.~Lisi, A.~Marrone, D.~Montanino, A.~Palazzo and
A.M.~Rotunno, 
\textit{Global analysis of neutrino masses, mixings and phases:
  entering the era of leptonic CP violation searches},
Phys.\ Rev.\ {\bf D 86} (2012) 013012
[arXiv:1205.5254v3].

\bibitem{merle}
A.~Merle and W.~Rodejohann,
\textit{The elements of the neutrino mass matrix: Allowed ranges and
  implications of texture zeros},
Phys.\ Rev.\ {\bf D 73} (2006) 073012
[hep-ph/0603111].

\bibitem{rpp}
J.~Beringer \textit{et al.}\
(Particle Data Group),
\textit{Review of particle physics},
Phys.\ Rev.\ \textbf{D~86} (2012) 010001.

\bibitem{CP}
G.C.~Branco, L.~Lavoura and J.P.~Silva,
\textit{CP Violation},
Int.\ Ser.\ Monogr.\ Phys.\  {\bf 103} (1999) 1.

\bibitem{nelder-mead}
J.A.~Nelder and R.~Mead,
\textit{A simplex method for function minimization},
Comput.\ J.\ \textbf{7} (1965) 308.

\bibitem{rodejohann-betabeta}
W.~Rodejohann,
\textit{Neutrinoless double beta decay and neutrino physics},
J.\ Phys.\ {\bf G 39} (2012) 124008
[arXiv:1206.2560].

\bibitem{EXO-200}
M.~Auger {\it et al.}  (EXO Coll.),
\textit{Search for Neutrinoless Double-Beta Decay in $^{136}\mathrm{Xe}$ with EXO-200},
Phys.\ Rev.\ Lett.\ {\bf 109} (2012) 032505
[arXiv:1205.5608].

\bibitem{grimus-ludl}
W.~Grimus and P.O.~Ludl,
\textit{Two-parameter neutrino mass matrices with two texture zeros},
arXiv:1208.4515.

\bibitem{FGM}
P.H.~Frampton, S.L.~Glashow and D.~Marfatia, 
\textit{Zeroes of the neutrino mass matrix},
Phys.\ Lett.\ {\bf B 536} (2002) 79 
[hep-ph/0201008].

\end{thebibliography}
\end{document}